\documentclass[aps,reprint,twocolumn,10pt,superscriptaddress,english,showpacs,longbibliography,nofootinbib,floatfix]{revtex4-1}
\usepackage{amsmath}
\usepackage{amssymb}
\usepackage{amsthm}
\usepackage{amsfonts}
\usepackage{listings}
\usepackage{longtable}
\usepackage{enumerate}
\usepackage{latexsym}
\usepackage{color}
\usepackage{multirow}
\usepackage[figuresright]{rotating}
\usepackage[table]{xcolor}
\usepackage{setspace} 
\usepackage{blindtext}
\usepackage{dcolumn}
\usepackage{braket}

\usepackage{array,etoolbox}
\preto\tabular{\setcounter{magicrownumbers}{0}}
\newcounter{magicrownumbers}
\newcommand\rownumber{\stepcounter{magicrownumbers}\arabic{magicrownumbers}}

\def\ZZ{\mathbb{Z}}
\def\ie{{\it i.e.},\ }
\def\eg{{\it e.g.}\ }
\def\ea{{\it et al.}}

\usepackage{bm}
\usepackage{hyperref}
\hypersetup{
 pdfnewwindow=true, colorlinks=true,
 linkcolor=blue, anchorcolor=blue,
 citecolor=blue, filecolor=blue,
 menucolor=blue, urlcolor=blue}
\definecolor{red}{rgb}{1,0,0}
\definecolor{blue}{rgb}{0,0,1}
\definecolor{black}{rgb}{0,0,0}
\def\red{\color{red}}
\def\blue{\color{blue}}
\def\black{\color{black}}
\newcommand{\beginsupplement}{%
        \setcounter{table}{0}
        \renewcommand{\thetable}{S\arabic{table}}%
        \setcounter{figure}{0}
        \renewcommand{\thefigure}{S\arabic{figure}}%
     }

\newcommand{\rhQ}[1][]{$\rho_{G^Q}^{#1}$}
\usepackage{graphicx}
\usepackage{subfigure}

\usepackage{chemformula}
\let\ce\ch
\newcommand{\webBCSMAG}{\href{http://webbdcrista1.ehu.es/magndata/
}{MAGDATA}}
\newcommand{\webBCS}{\href{https://www.cryst.ehu.es/}{Bilbao Crystallographic Server}}
\newcommand{\webTQCphonon}{\href{https://www.topologicalquantumchemistry.com/topophonons/index.html}{Topological Phonon Database}}

\newcommand{\mpidwebshort}[1]{\href{https://www.topologicalquantumchemistry.com/topophonons/index.html\#/?MPID=mp-#1}{#1}}
\newcommand{\mpidweb}[1]{\href{https://www.topologicalquantumchemistry.com/topophonons/index.html\#/?MPID=mp-#1}{mp-#1}}

\newcommand{\icsdwebshort}[1]{\href{https://www.topologicalquantumchemistry.fr/\#/detail/#1}{#1}}
\newcommand{\icsdweb}[1]{\href{https://www.topologicalquantumchemistry.fr/\#/detail/#1}{ICSD #1}}
\newcommand{\icsdwebdirectlink}[2]{\href{https://www.topologicalquantumchemistry.fr/\#/detail/#1}{#2}}
\newcommand{\webNoICSD}{\href{https://www.topologicalquantumchemistry.fr/}}
\newcommand{\webTQC}{\href{https://www.topologicalquantumchemistry.org/}{Topological Quantum Chemistry website}}
\newcommand{\webMTQC}{\href{https://www.topologicalquantumchemistry.fr/magnetic}{Topological Magnetic Materials website}}
\newcommand{\webflatband}{\href{https://www.topologicalquantumchemistry.fr/flatbands/}{Materials Flatband Database website}}

\newcommand{\bcsidwebshort}[1]{\href{https://www.topologicalquantumchemistry.fr/magnetic/index.html?BCSID=#1}{#1}}
\newcommand{\bcsidweblong}[1]{\href{https://www.topologicalquantumchemistry.fr/magnetic/index.html?BCSID=#1}{BCSID #1}}
\newcommand{\flatwebdirectlink}[2]{\href{https://www.topologicalquantumchemistry.fr/flatbands/index.html?ICSD=#1}{#2}}

\newcommand{\webBCSfull}{\href{https://www.cryst.ehu.es/}{Bilbao Crystallographic Server}}

\newcommand{\webBCSshort}{\href{https://www.cryst.ehu.es/}{BCS}}

\newcommand{\webMP}{\href{https://materialsproject.org/}{Materials Project}}
\newcommand{\webphonondb}{\href{http://phonondb.mtl.kyoto-u.ac.jp/}{Phonon database at Kyoto university}}
\newcommand{\webphonondbshort}{\href{http://phonondb.mtl.kyoto-u.ac.jp/}{PhononDB@kyoto-u}}

\newcommand{\siref}[1]{SI.~\ref{#1}}

\newcommand{\nbrTopMp}{160}
\newcommand{\nbrTopPercentMp}{10.5\%}
\newcommand{\nbrBestTopMp}{23}
\newcommand{\nbrBestSmMp}{53}

\newcommand{\nbrTopKyoto}{2,423}
\newcommand{\nbrTopPercentKyoto}{24.2\%}

\newcommand{\TQCDBNbrICSDs}{73,234}
\newcommand{\TQCDBNbrNoSOCICSDs}{69,730}
\newcommand{\TQCDBPercentNoSOCICSDs}{95.22\%}
\newcommand{\TQCDBNbrFailedNoSOCICSDs}{3,504}
\newcommand{\TQCDBPercentFailedNoSOCICSDs}{4.78\%}
\newcommand{\TQCDBNbrUniqueMaterials}{38,298}
\newcommand{\TQCDBNbrNoSOCUniqueMaterials}{36,163}
\newcommand{\TQCDBPercentageNoSOCUniqueMaterials}{94.43\%}
\newcommand{\TQCDBNbrFailedNoSOCUniqueMaterials}{2,135}
\newcommand{\TQCDBPercentageFailedNoSOCUniqueMaterials}{5.57\%}

\newcommand{\TQCDBNbrMaterialsFElectrons}{10,987}
\newcommand{\TQCDBNbrMaterialsFElectronsPercent}{28.69\%}
\newcommand{\TQCDBNbrNoSOCICSDsNoFElectrons}{52,517}
\newcommand{\TQCDBNbrNoSOCICSDsNoFElectronsPercent}{75.31\%}
\newcommand{\TQCDBNbrMaterialsMagneticMP}{7,124}
\newcommand{\TQCDBNbrMaterialsMagneticMPPercent}{18.60\%}
\newcommand{\TQCDBNbrMaterialsMagneticMPVASP}{13,718}
\newcommand{\TQCDBNbrMaterialsMagneticMPVASPPercent}{35.82\%}
\newcommand{\TQCDBNbrMaterialsMagneticMPVASPFElectrons}{19,987}
\newcommand{\TQCDBNbrMaterialsMagneticMPVASPFElectronsPercent}{52.19\%}

\newcommand{\TQCDBNbrMaterialsTI}{6,128}
\newcommand{\TQCDBNbrMaterialsTIPercent}{16.00\%}
\newcommand{\TQCDBNbrMaterialsSM}{14,037}
\newcommand{\TQCDBNbrMaterialsSMPercent}{36.65\%}
\newcommand{\TQCDBNbrMaterialstrivial}{18,133}
\newcommand{\TQCDBNbrMaterialstrivialPercent}{47.35\%}

\newcommand{\TQCDBNbrMaterialsNLC}{3,000}
\newcommand{\TQCDBNbrMaterialsNLCPercent}{7.83\%}
\newcommand{\TQCDBNbrMaterialsSEBR}{3,128}
\newcommand{\TQCDBNbrMaterialsSEBRPercent}{8.17\%}
\newcommand{\TQCDBNbrMaterialsES}{4,102}
\newcommand{\TQCDBNbrMaterialsESPercent}{10.71\%}
\newcommand{\TQCDBNbrMaterialsESFD}{9,935}
\newcommand{\TQCDBNbrMaterialsESFDPercent}{25.94\%}
\newcommand{\TQCDBNbrMaterialsLCEBR}{18,133}
\newcommand{\TQCDBNbrMaterialsLCEBRPercent}{47.35\%}
\newcommand{\TQCDBNbrTopologicalMaterials}{20,165}
\newcommand{\TQCDBNbrTopologicalMaterialsPercent}{52.65\%}

\newcommand{\TQCDBNbrNoSOCMaterialsSM}{20,298}
\newcommand{\TQCDBNbrNoSOCMaterialsSMPercent}{56.13\%}
\newcommand{\TQCDBNbrNoSOCMaterialstrivial}{15,865}
\newcommand{\TQCDBNbrNoSOCMaterialstrivialPercent}{43.87\%}
\newcommand{\TQCDBNbrNoSOCMaterialsES}{6,006}
\newcommand{\TQCDBNbrNoSOCMaterialsESPercent}{16.61\%}
\newcommand{\TQCDBNbrNoSOCMaterialsESFD}{13,997}
\newcommand{\TQCDBNbrNoSOCMaterialsESFDPercent}{38.71\%}
\newcommand{\TQCDBNbrNoSOCMaterialsLCEBR}{15,865}
\newcommand{\TQCDBNbrNoSOCMaterialsLCEBRPercent}{43.87\%}
\newcommand{\TQCDBNbrNoSOCMaterialsNLCSM}{251}
\newcommand{\TQCDBNbrNoSOCMaterialsNLCSMPercent}{0.69\%}
\newcommand{\TQCDBNbrNoSOCMaterialsSEBRSM}{44}
\newcommand{\TQCDBNbrNoSOCMaterialsSEBRSMPercent}{0.12\%}
\newcommand{\TQCDBNbrNoSOCMaterialsNLCSMES}{6,257}
\newcommand{\TQCDBNbrNoSOCMaterialsSEBRSMESFD}{14,041}
\newcommand{\TQCDBNbrNoSOCMaterialsNLCSMSEBRSM}{295}

\newcommand{\TQCDBNbrMaterialsWithSOCNoSOCTI}{5,382}
\newcommand{\TQCDBNbrMaterialsWithSOCNoSOCTIPercent}{14.88\%}
\newcommand{\TQCDBNbrMaterialsWithSOCNoSOCSM}{13,270}
\newcommand{\TQCDBNbrMaterialsWithSOCNoSOCSMPercent}{36.69\%}
\newcommand{\TQCDBNbrMaterialsWithSOCNoSOCtrivial}{17,511}
\newcommand{\TQCDBNbrMaterialsWithSOCNoSOCtrivialPercent}{48.42\%}

\newcommand{\TQCDBNbrMaterialsWithSOCNoSOCNLC}{2,568}
\newcommand{\TQCDBNbrMaterialsWithSOCNoSOCNLCPercent}{7.10\%}
\newcommand{\TQCDBNbrMaterialsWithSOCNoSOCSEBR}{2,814}
\newcommand{\TQCDBNbrMaterialsWithSOCNoSOCSEBRPercent}{7.78\%}
\newcommand{\TQCDBNbrMaterialsWithSOCNoSOCES}{3,785}
\newcommand{\TQCDBNbrMaterialsWithSOCNoSOCESPercent}{10.47\%}
\newcommand{\TQCDBNbrMaterialsWithSOCNoSOCESFD}{9,485}
\newcommand{\TQCDBNbrMaterialsWithSOCNoSOCESFDPercent}{26.23\%}
\newcommand{\TQCDBNbrMaterialsWithSOCNoSOCLCEBR}{17,511}
\newcommand{\TQCDBNbrMaterialsWithSOCNoSOCLCEBRPercent}{48.42\%}
\newcommand{\TQCDBNbrMaterialsWithSOCNoSOCNLCSEBR}{5,382}
\newcommand{\TQCDBNbrMaterialsWithSOCNoSOCSEBRESFD}{12,299}
\newcommand{\TQCDBNbrMaterialsWithSOCNoSOCNLCES}{6,353}

\newcommand{\TQCDBNbrSTopo}{769}
\newcommand{\TQCDBNbrSTopoPercentage}{2.01\%}
\newcommand{\TQCDBNbrSTopoNLC}{37}
\newcommand{\TQCDBNbrSTopoNLCPercent}{0.10\%}
\newcommand{\TQCDBNbrSTopoSEBR}{109}
\newcommand{\TQCDBNbrSTopoSEBRPercent}{0.28\%}
\newcommand{\TQCDBNbrSTopoES}{178}
\newcommand{\TQCDBNbrSTopoESPercent}{0.46\%}
\newcommand{\TQCDBNbrSTopoESFD}{339}
\newcommand{\TQCDBNbrSTopoESFDPercent}{0.89\%}
\newcommand{\TQCDBNbrSTopoLCEBR}{106}
\newcommand{\TQCDBNbrSTopoLCEBRPercent}{0.28\%}

\newcommand{\TQCDBNbrTopoBand}{33,698}
\newcommand{\TQCDBNbrTopoBandPercent}{87.99\%}

\newcommand{\TQCDBNbrSMetal}{17}
\newcommand{\TQCDBNbrSMetalPercent}{0.04\%}
\newcommand{\TQCDBNbrSMetalICSDs}{65}
\newcommand{\TQCDBNbrSMetalPercentICSDs}{0.09\%}

\newcommand{\TQCDBBandSetsUniqueMaterials}{1,996,728}
\newcommand{\TQCDBNbrLCEBRBandSetsUniqueMaterials}{750,504}
\newcommand{\TQCDBPercentLCEBRBandSetsUniqueMaterials}{37.59\%}
\newcommand{\TQCDBNbrNLCBandSetsUniqueMaterials}{859,606}
\newcommand{\TQCDBPercentNLCBandSetsUniqueMaterials}{43.05\%}
\newcommand{\TQCDBNbrSEBRBandSetsUniqueMaterials}{379,321}
\newcommand{\TQCDBPercentSEBRBandSetsUniqueMaterials}{19.00\%}
\newcommand{\TQCDBNbrStrongBandSetsUniqueMaterials}{1,238,927}
\newcommand{\TQCDBPercentStrongBandSetsUniqueMaterials}{62.05\%}
\newcommand{\TQCDBNbrFragileBandSetsUniqueMaterials}{7,297}
\newcommand{\TQCDBPercentFragileBandSetsUniqueMaterials}{0.37\%}

\newcommand{\TQCDBNbrNoSOCSTopo}{28}
\newcommand{\TQCDBNbrNoSOCSTopoPercentage}{0.08\%}
\newcommand{\TQCDBNbrNoSOCSTopoNLC}{0}
\newcommand{\TQCDBNbrNoSOCSTopoNLCPercent}{0.00\%}
\newcommand{\TQCDBNbrNoSOCSTopoSEBR}{0}
\newcommand{\TQCDBNbrNoSOCSTopoSEBRPercent}{0.00\%}
\newcommand{\TQCDBNbrNoSOCSTopoES}{5}
\newcommand{\TQCDBNbrNoSOCSTopoESPercent}{0.01\%}
\newcommand{\TQCDBNbrNoSOCSTopoESFD}{14}
\newcommand{\TQCDBNbrNoSOCSTopoESFDPercent}{0.04\%}
\newcommand{\TQCDBNbrNoSOCSTopoLCEBR}{8}
\newcommand{\TQCDBNbrNoSOCSTopoLCEBRPercent}{0.02\%}

\newcommand{\TQCDBNbrNoSOCTopoBand}{10,001}
\newcommand{\TQCDBNbrNoSOCTopoBandPercent}{27.66\%}

\newcommand{\TQCDBNbrNoSOCSMetal}{1,138}
\newcommand{\TQCDBNbrNoSOCSMetalPercent}{3.15\%}
\newcommand{\TQCDBNbrNoSOCSMetalICSDs}{3,495}
\newcommand{\TQCDBNbrNoSOCSMetalPercentICSDs}{5.01\%}

\newcommand{\TQCDBVASPICPUhSOC}{9,500,801.70}
\newcommand{\TQCDBVASPIICPUhSOC}{170,633.80}
\newcommand{\TQCDBVASPIIICPUhSOC}{2,560,882.60}
\newcommand{\TQCDBVASPIVCPUhSOC}{5,804,948.80}
\newcommand{\TQCDBVASPTotalCPUhSOC}{18,037,266.90}

\newcommand{\TQCDBVASPICPUhNoSOC}{2,298,583.30}
\newcommand{\TQCDBVASPIICPUhNoSOC}{56,520.70}
\newcommand{\TQCDBVASPIIICPUhNoSOC}{773,019.50}
\newcommand{\TQCDBVASPIVCPUhNoSOC}{1,397,550.30}
\newcommand{\TQCDBVASPTotalCPUhNoSOC}{4,525,673.80}

\newcommand{\TQCDBVASPTotalCPUTimeAllHours}{22,562,940.70}
\newcommand{\TQCDBVASPTotalCPUTimeAllMillionHours}{22.60}

\newcommand{\TQCDBTotalStorage}{2,037.80Gb}

\newcommand{\TQCDBNoSOCTotalStorage}{343.30Gb}
\newcommand{\TQCDBPercentNoSOCTotalStorage}{16.85\%}
\newcommand{\TQCDBNoSOCPROCARStorage}{173.60Gb}
\newcommand{\TQCDBNoSOCCHGCARStorage}{119.00Gb}

\newcommand{\TQCDBTotalSOCStorage}{1,694.50Gb}
\newcommand{\TQCDBPercentSOCTotalStorage}{83.15\%}
\newcommand{\TQCDBPROCARSOCStorage}{1,013.80Gb}
\newcommand{\TQCDBPercentPROCARSOCStorage}{49.75\%}
\newcommand{\TQCDBCHGCARSOCStorage}{559.30Gb}
\newcommand{\TQCDBPercentCHGCARSOCStorage}{27.45\%}

\newcommand{\TQCDBNbrSkippedICSDs}{22,996}
\newcommand{\TQCDBNbrFailedComputedICSDs}{11,315}
\newcommand{\TQCDBNbrFailedVASPToTraceICSDs}{5,014}
\newcommand{\TQCDBNbrFailedVASPToTraceBadTRICSDs}{1,703}
\newcommand{\TQCDBNbrFailedVASPToTraceBadTracesICSDs}{655}
\newcommand{\TQCDBNbrFailedVASPToTraceAccidentalFermiICSDs}{2,647}
\newcommand{\TPDBNbrEntries}{10,249}
\newcommand{\TPDBNbrKyotoMaterials}{9,991}
\newcommand{\TPDBNbrMPMaterials}{1,516}
\newcommand{\TPDBNbrKyotoMPMaterials}{11,507}

\newcommand{\TPDBNbrKyotoMaterialsWoNACNoNegative}{7,809}
\newcommand{\TPDBPercentKyotoMaterialsWoNACNoNegative}{78.16}
\newcommand{\TPDBNbrKyotoMaterialsWithNACNoNegative}{7,983}
\newcommand{\TPDBPercentKyotoMaterialsWithNACNoNegative}{79.90}
\newcommand{\TPDBNbrMPMaterialsWoNACNoNegative}{1,389}
\newcommand{\TPDBPercentMPMaterialsWoNACNoNegative}{91.62}
\newcommand{\TPDBNbrMPMaterialsWithNACNoNegative}{1,394}
\newcommand{\TPDBPercentMPMaterialsWithNACNoNegative}{91.95}

\newcommand{\TPDBNbrKyotoMaterialsNegativeBands}{2,008}
\newcommand{\TPDBPercentKyotoMaterialsNegativeBands}{20.10}
\newcommand{\TPDBNbrKyotoMaterialsDiscarded}{1,707}
\newcommand{\TPDBPercentKyotoMaterialsDiscarded}{17.09}
\newcommand{\TPDBNbrKyotoMaterialsInterpolationErrors}{76}
\newcommand{\TPDBPercentKyotoMaterialsInterpolationErrors}{0.76}
\newcommand{\TPDBNbrKyotoMaterialsInstabilities}{225}
\newcommand{\TPDBPercentKyotoMaterialsInstabilities}{2.25}

\newcommand{\TPDBNbrMPMaterialsNegativeBands}{122}
\newcommand{\TPDBPercentMPMaterialsNegativeBands}{8.05}
\newcommand{\TPDBNbrMPMaterialsDiscarded}{94}
\newcommand{\TPDBPercentMPMaterialsDiscarded}{6.20}
\newcommand{\TPDBNbrMPMaterialsInterpolationErrors}{4}
\newcommand{\TPDBPercentMPMaterialsInterpolationErrors}{0.26}
\newcommand{\TPDBNbrMPMaterialsInstabilities}{24}
\newcommand{\TPDBPercentMPMaterialsInstabilities}{0.24}

\newcommand{\TPDBNbrKyotoMaterialsWoWNACBandChanges}{890}
\newcommand{\TPDBPercentKyotoMaterialsWoWNACBandChanges}{8.91}
\newcommand{\TPDBNbrMPMaterialsWoWNACBandChanges}{84}
\newcommand{\TPDBPercentMPMaterialsWoWNACBandChanges}{5.54}

\newcommand{\TPDBNbrMaterialsWoNACNonTrivialNonAtomic}{6,194}
\newcommand{\TPDBPercentMaterialsWoNACNonTrivialNonAtomic}{60.44}
\newcommand{\TPDBNbrMaterialsWoNACOnlyTrivial}{4,055}
\newcommand{\TPDBPercentMaterialsWoNACOnlyTrivial}{39.56}
\newcommand{\TPDBNbrMaterialsWithNACNonTrivialNonAtomic}{6,189}
\newcommand{\TPDBPercentMaterialsWithNACNonTrivialNonAtomic}{60.39}
\newcommand{\TPDBNbrMaterialsWithNACOnlyTrivial}{4,060}
\newcommand{\TPDBPercentMaterialsWithNACOnlyTrivial}{39.61}
\newcommand{\TPDBNbrMaterialsWoNACNonTrivialNonAtomicNoNegative}{4,873}
\newcommand{\TPDBPercentMaterialsWoNACNonTrivialNonAtomicNoNegative}{52.98}
\newcommand{\TPDBNbrMaterialsWithNACNonTrivialNonAtomicNoNegative}{4,945}
\newcommand{\TPDBPercentMaterialsWithNACNonTrivialNonAtomicNoNegative}{52.74}

\newcommand{\TPDBNbrKyotoMaterialsWoNACNonTrivialNonAtomic}{5,456}
\newcommand{\TPDBPercentKyotoMaterialsWoNACNonTrivialNonAtomic}{54.61}
\newcommand{\TPDBNbrKyotoMaterialsWoNACOnlyTrivial}{4,535}
\newcommand{\TPDBPercentKyotoMaterialsWoNACOnlyTrivial}{45.39}
\newcommand{\TPDBNbrKyotoMaterialsWithNACNonTrivialNonAtomic}{5,453}
\newcommand{\TPDBPercentKyotoMaterialsWithNACNonTrivialNonAtomic}{54.58}
\newcommand{\TPDBNbrKyotoMaterialsWithNACOnlyTrivial}{4,538}
\newcommand{\TPDBPercentKyotoMaterialsWithNACOnlyTrivial}{45.42}
\newcommand{\TPDBNbrMPMaterialsWoNACNonTrivialNonAtomic}{738}
\newcommand{\TPDBPercentMPMaterialsWoNACNonTrivialNonAtomic}{48.68}
\newcommand{\TPDBNbrMPMaterialsWoNACOnlyTrivial}{778}
\newcommand{\TPDBPercentMPMaterialsWoNACOnlyTrivial}{51.32}
\newcommand{\TPDBNbrMPMaterialsWithNACNonTrivialNonAtomic}{736}
\newcommand{\TPDBPercentMPMaterialsWithNACNonTrivialNonAtomic}{48.55}
\newcommand{\TPDBNbrMPMaterialsWithNACOnlyTrivial}{780}
\newcommand{\TPDBPercentMPMaterialsWithNACOnlyTrivial}{51.45}

\newcommand{\TPDBNbrKyotoMaterialsWoNACTopologicalNoNegative}{1,780}
\newcommand{\TPDBPercentKyotoMaterialsWoNACTopologicalNoNegative}{22.79}
\newcommand{\TPDBNbrKyotoMaterialsWithNACTopologicalNoNegative}{1,831}
\newcommand{\TPDBPercentKyotoMaterialsWithNACTopologicalNoNegative}{22.94}
\newcommand{\TPDBNbrKyotoMaterialsWoNACFragileNoNegative}{16}
\newcommand{\TPDBPercentKyotoMaterialsWoNACFragileNoNegative}{0.20}
\newcommand{\TPDBNbrKyotoMaterialsWithNACFragileNoNegative}{15}
\newcommand{\TPDBPercentKyotoMaterialsWithNACFragileNoNegative}{0.19}
\newcommand{\TPDBNbrKyotoMaterialsWoNACOABRNoNegative}{2,907}
\newcommand{\TPDBPercentKyotoMaterialsWoNACOABRNoNegative}{37.23}
\newcommand{\TPDBNbrKyotoMaterialsWithNACOABRNoNegative}{2,950}
\newcommand{\TPDBPercentKyotoMaterialsWithNACOABRNoNegative}{36.95}
\newcommand{\TPDBNbrKyotoMaterialsWoNACOOABRNoNegative}{1,129}
\newcommand{\TPDBPercentKyotoMaterialsWoNACOOABRNoNegative}{14.46}
\newcommand{\TPDBNbrKyotoMaterialsWithNACOOABRNoNegative}{1,149}
\newcommand{\TPDBPercentKyotoMaterialsWithNACOOABRNoNegative}{14.39}
\newcommand{\TPDBNbrKyotoMaterialsWoNACNonTrivialNoNegative}{1,790}
\newcommand{\TPDBPercentKyotoMaterialsWoNACNonTrivialNoNegative}{22.92}
\newcommand{\TPDBNbrKyotoMaterialsWithNACNonTrivialNoNegative}{1,840}
\newcommand{\TPDBPercentKyotoMaterialsWithNACNonTrivialNoNegative}{23.05}
\newcommand{\TPDBNbrKyotoMaterialsWoNACNonTrivialNonAtomicNoNegative}{4,211}
\newcommand{\TPDBPercentKyotoMaterialsWoNACNonTrivialNonAtomicNoNegative}{53.92}
\newcommand{\TPDBNbrKyotoMaterialsWithNACNonTrivialNonAtomicNoNegative}{4,282}
\newcommand{\TPDBPercentKyotoMaterialsWithNACNonTrivialNonAtomicNoNegative}{53.64}
\newcommand{\TPDBNbrKyotoMaterialsWoNACOnlyTrivialNoNegative}{3,598}
\newcommand{\TPDBPercentKyotoMaterialsWoNACOnlyTrivialNoNegative}{46.08}
\newcommand{\TPDBNbrKyotoMaterialsWithNACOnlyTrivialNoNegative}{3,701}
\newcommand{\TPDBPercentKyotoMaterialsWithNACOnlyTrivialNoNegative}{46.36}
\newcommand{\TPDBNbrMPMaterialsWoNACTopologicalNoNegative}{151}
\newcommand{\TPDBPercentMPMaterialsWoNACTopologicalNoNegative}{10.87}
\newcommand{\TPDBNbrMPMaterialsWithNACTopologicalNoNegative}{153}
\newcommand{\TPDBPercentMPMaterialsWithNACTopologicalNoNegative}{10.98}
\newcommand{\TPDBNbrMPMaterialsWoNACFragileNoNegative}{1}
\newcommand{\TPDBPercentMPMaterialsWoNACFragileNoNegative}{0.07}
\newcommand{\TPDBNbrMPMaterialsWithNACFragileNoNegative}{1}
\newcommand{\TPDBPercentMPMaterialsWithNACFragileNoNegative}{0.07}
\newcommand{\TPDBNbrMPMaterialsWoNACOABRNoNegative}{331}
\newcommand{\TPDBPercentMPMaterialsWoNACOABRNoNegative}{23.83}
\newcommand{\TPDBNbrMPMaterialsWithNACOABRNoNegative}{332}
\newcommand{\TPDBPercentMPMaterialsWithNACOABRNoNegative}{23.82}
\newcommand{\TPDBNbrMPMaterialsWoNACOOABRNoNegative}{286}
\newcommand{\TPDBPercentMPMaterialsWoNACOOABRNoNegative}{20.59}
\newcommand{\TPDBNbrMPMaterialsWithNACOOABRNoNegative}{288}
\newcommand{\TPDBPercentMPMaterialsWithNACOOABRNoNegative}{20.66}
\newcommand{\TPDBNbrMPMaterialsWoNACNonTrivialNoNegative}{152}
\newcommand{\TPDBPercentMPMaterialsWoNACNonTrivialNoNegative}{10.94}
\newcommand{\TPDBNbrMPMaterialsWithNACNonTrivialNoNegative}{154}
\newcommand{\TPDBPercentMPMaterialsWithNACNonTrivialNoNegative}{11.05}
\newcommand{\TPDBNbrMPMaterialsWoNACNonTrivialNonAtomicNoNegative}{662}
\newcommand{\TPDBPercentMPMaterialsWoNACNonTrivialNonAtomicNoNegative}{47.66}
\newcommand{\TPDBNbrMPMaterialsWithNACNonTrivialNonAtomicNoNegative}{663}
\newcommand{\TPDBPercentMPMaterialsWithNACNonTrivialNonAtomicNoNegative}{47.56}
\newcommand{\TPDBNbrMPMaterialsWoNACOnlyTrivialNoNegative}{727}
\newcommand{\TPDBPercentMPMaterialsWoNACOnlyTrivialNoNegative}{52.34}
\newcommand{\TPDBNbrMPMaterialsWithNACOnlyTrivialNoNegative}{731}
\newcommand{\TPDBPercentMPMaterialsWithNACOnlyTrivialNoNegative}{52.44}

\newcommand{\TPDBNbrKyotoMPMaterialsWithNACStrictNegative}{9,105}
\newcommand{\TPDBPercentKyotoMPMaterialsWithNACStrictNegative}{79.13}

\newcommand{\TPDBNbrKyotoMaterialsWoNACTopological}{2,361}
\newcommand{\TPDBPercentKyotoMaterialsWoNACTopological}{23.63}
\newcommand{\TPDBNbrKyotoMaterialsWithNACTopological}{2,374}
\newcommand{\TPDBPercentKyotoMaterialsWithNACTopological}{23.76}
\newcommand{\TPDBNbrKyotoMaterialsWoNACFragile}{21}
\newcommand{\TPDBPercentKyotoMaterialsWoNACFragile}{0.21}
\newcommand{\TPDBNbrKyotoMaterialsWithNACFragile}{20}
\newcommand{\TPDBPercentKyotoMaterialsWithNACFragile}{0.20}
\newcommand{\TPDBNbrKyotoMaterialsWoNACOABR}{3,736}
\newcommand{\TPDBPercentKyotoMaterialsWoNACOABR}{37.39}
\newcommand{\TPDBNbrKyotoMaterialsWithNACOABR}{3,736}
\newcommand{\TPDBPercentKyotoMaterialsWithNACOABR}{37.39}
\newcommand{\TPDBNbrKyotoMaterialsWoNACOOABR}{1,562}
\newcommand{\TPDBPercentKyotoMaterialsWoNACOOABR}{15.63}
\newcommand{\TPDBNbrKyotoMaterialsWithNACOOABR}{1,562}
\newcommand{\TPDBPercentKyotoMaterialsWithNACOOABR}{15.63}
\newcommand{\TPDBNbrMPMaterialsWoNACTopological}{160}
\newcommand{\TPDBPercentMPMaterialsWoNACTopological}{10.55}
\newcommand{\TPDBNbrMPMaterialsWithNACTopological}{162}
\newcommand{\TPDBPercentMPMaterialsWithNACTopological}{10.69}
\newcommand{\TPDBNbrMPMaterialsWoNACFragile}{1}
\newcommand{\TPDBPercentMPMaterialsWoNACFragile}{0.07}
\newcommand{\TPDBNbrMPMaterialsWithNACFragile}{1}
\newcommand{\TPDBPercentMPMaterialsWithNACFragile}{0.07}
\newcommand{\TPDBNbrMPMaterialsWoNACOABR}{351}
\newcommand{\TPDBPercentMPMaterialsWoNACOABR}{23.15}
\newcommand{\TPDBNbrMPMaterialsWithNACOABR}{351}
\newcommand{\TPDBPercentMPMaterialsWithNACOABR}{23.15}
\newcommand{\TPDBNbrMPMaterialsWoNACOOABR}{342}
\newcommand{\TPDBPercentMPMaterialsWoNACOOABR}{22.56}
\newcommand{\TPDBNbrMPMaterialsWithNACOOABR}{342}
\newcommand{\TPDBPercentMPMaterialsWithNACOOABR}{22.56}

\newcommand{\TPDBNbrIdealTopoKyotoMaterials}{158}
\newcommand{\TPDBPercentIdealTopoKyotoMaterials}{1.58}
\newcommand{\TPDBNbrIdealTopoMPMaterials}{29}
\newcommand{\TPDBPercentIdealTopoMPMaterials}{1.91}
\newcommand{\TPDBNbrIdealTopoMaterials}{187}
\newcommand{\TPDBPercentIdealTopoMaterials}{1.63}
\newcommand{\TPDBNbrIdealFragileKyotoMaterials}{0}
\newcommand{\TPDBPercentIdealFragileKyotoMaterials}{0.00}
\newcommand{\TPDBNbrIdealFragileMPMaterials}{0}
\newcommand{\TPDBPercentIdealFragileMPMaterials}{0.00}
\newcommand{\TPDBNbrIdealFragileMaterials}{0}
\newcommand{\TPDBPercentIdealFragileMaterials}{0.00}
\newcommand{\TPDBNbrIdealSMKyotoMaterials}{790}
\newcommand{\TPDBPercentIdealSMKyotoMaterials}{7.91}
\newcommand{\TPDBNbrIdealSMMPMaterials}{56}
\newcommand{\TPDBPercentIdealSMMPMaterials}{3.69}
\newcommand{\TPDBNbrIdealSMMaterials}{846}
\newcommand{\TPDBPercentIdealSMMaterials}{7.35}
\newcommand{\TPDBNbrIdealNonTrivialKyotoMaterials}{935}
\newcommand{\TPDBPercentIdealNonTrivialKyotoMaterials}{9.36}
\newcommand{\TPDBNbrIdealNonTrivialMPMaterials}{85}
\newcommand{\TPDBPercentIdealNonTrivialMPMaterials}{5.61}
\newcommand{\TPDBNbrIdealNonTrivialMaterials}{1,020}
\newcommand{\TPDBPercentIdealNonTrivialMaterials}{8.86}
\newcommand{\TPDBNbrIdealOABRKyotoMaterials}{1,323}
\newcommand{\TPDBPercentIdealOABRKyotoMaterials}{13.24}
\newcommand{\TPDBNbrIdealOABRMPMaterials}{107}
\newcommand{\TPDBPercentIdealOABRMPMaterials}{7.06}
\newcommand{\TPDBNbrIdealOABRMaterials}{1,430}
\newcommand{\TPDBPercentIdealOABRMaterials}{12.43}
\newcommand{\TPDBNbrIdealOOABRKyotoMaterials}{497}
\newcommand{\TPDBPercentIdealOOABRKyotoMaterials}{4.97}
\newcommand{\TPDBNbrIdealOOABRMPMaterials}{126}
\newcommand{\TPDBPercentIdealOOABRMPMaterials}{8.31}
\newcommand{\TPDBNbrIdealOOABRMaterials}{623}
\newcommand{\TPDBPercentIdealOOABRMaterials}{5.41}
\usepackage{appendix}

\begin{document}

\newcommand{\sgsymb}[1]{\ifnum#1=1
$P1$\else
\ifnum#1=2
$P\bar{1}$\else
\ifnum#1=3
$P2$\else
\ifnum#1=4
$P2_1$\else
\ifnum#1=5
$C2$\else
\ifnum#1=6
$Pm$\else
\ifnum#1=7
$Pc$\else
\ifnum#1=8
$Cm$\else
\ifnum#1=9
$Cc$\else
\ifnum#1=10
$P2/m$\else
\ifnum#1=11
$P2_1/m$\else
\ifnum#1=12
$C2/m$\else
\ifnum#1=13
$P2/c$\else
\ifnum#1=14
$P2_1/c$\else
\ifnum#1=15
$C2/c$\else
\ifnum#1=16
$P222$\else
\ifnum#1=17
$P222_1$\else
\ifnum#1=18
$P2_12_12$\else
\ifnum#1=19
$P2_12_12_1$\else
\ifnum#1=20
$C222_1$\else
\ifnum#1=21
$C222$\else
\ifnum#1=22
$F222$\else
\ifnum#1=23
$I222$\else
\ifnum#1=24
$I2_12_12_1$\else
\ifnum#1=25
$Pmm2$\else
\ifnum#1=26
$Pmc2_1$\else
\ifnum#1=27
$Pcc2$\else
\ifnum#1=28
$Pma2$\else
\ifnum#1=29
$Pca2_1$\else
\ifnum#1=30
$Pnc2$\else
\ifnum#1=31
$Pmn2_1$\else
\ifnum#1=32
$Pba2$\else
\ifnum#1=33
$Pna2_1$\else
\ifnum#1=34
$Pnn2$\else
\ifnum#1=35
$Cmm2$\else
\ifnum#1=36
$Cmc2_1$\else
\ifnum#1=37
$Ccc2$\else
\ifnum#1=38
$Amm2$\else
\ifnum#1=39
$Aem2$\else
\ifnum#1=40
$Ama2$\else
\ifnum#1=41
$Aea2$\else
\ifnum#1=42
$Fmm2$\else
\ifnum#1=43
$Fdd2$\else
\ifnum#1=44
$Imm2$\else
\ifnum#1=45
$Iba2$\else
\ifnum#1=46
$Ima2$\else
\ifnum#1=47
$Pmmm$\else
\ifnum#1=48
$Pnnn$\else
\ifnum#1=49
$Pccm$\else
\ifnum#1=50
$Pban$\else
\ifnum#1=51
$Pmma$\else
\ifnum#1=52
$Pnna$\else
\ifnum#1=53
$Pmna$\else
\ifnum#1=54
$Pcca$\else
\ifnum#1=55
$Pbam$\else
\ifnum#1=56
$Pccn$\else
\ifnum#1=57
$Pbcm$\else
\ifnum#1=58
$Pnnm$\else
\ifnum#1=59
$Pmmn$\else
\ifnum#1=60
$Pbcn$\else
\ifnum#1=61
$Pbca$\else
\ifnum#1=62
$Pnma$\else
\ifnum#1=63
$Cmcm$\else
\ifnum#1=64
$Cmce$\else
\ifnum#1=65
$Cmmm$\else
\ifnum#1=66
$Cccm$\else
\ifnum#1=67
$Cmme$\else
\ifnum#1=68
$Ccce$\else
\ifnum#1=69
$Fmmm$\else
\ifnum#1=70
$Fddd$\else
\ifnum#1=71
$Immm$\else
\ifnum#1=72
$Ibam$\else
\ifnum#1=73
$Ibca$\else
\ifnum#1=74
$Imma$\else
\ifnum#1=75
$P4$\else
\ifnum#1=76
$P4_1$\else
\ifnum#1=77
$P4_2$\else
\ifnum#1=78
$P4_3$\else
\ifnum#1=79
$I4$\else
\ifnum#1=80
$I4_1$\else
\ifnum#1=81
$P\bar{4}$\else
\ifnum#1=82
$I\bar{4}$\else
\ifnum#1=83
$P4/m$\else
\ifnum#1=84
$P4_2/m$\else
\ifnum#1=85
$P4/n$\else
\ifnum#1=86
$P4_2/n$\else
\ifnum#1=87
$I4/m$\else
\ifnum#1=88
$I4_1/a$\else
\ifnum#1=89
$P422$\else
\ifnum#1=90
$P42_12$\else
\ifnum#1=91
$P4_122$\else
\ifnum#1=92
$P4_12_12$\else
\ifnum#1=93
$P4_222$\else
\ifnum#1=94
$P4_22_12$\else
\ifnum#1=95
$P4_322$\else
\ifnum#1=96
$P4_32_12$\else
\ifnum#1=97
$I422$\else
\ifnum#1=98
$I4_122$\else
\ifnum#1=99
$P4mm$\else
\ifnum#1=100
$P4bm$\else
\ifnum#1=101
$P4_2cm$\else
\ifnum#1=102
$P4_2nm$\else
\ifnum#1=103
$P4cc$\else
\ifnum#1=104
$P4nc$\else
\ifnum#1=105
$P4_2mc$\else
\ifnum#1=106
$P4_2bc$\else
\ifnum#1=107
$I4mm$\else
\ifnum#1=108
$I4cm$\else
\ifnum#1=109
$I4_1md$\else
\ifnum#1=110
$I4_1cd$\else
\ifnum#1=111
$P\bar{4}2m$\else
\ifnum#1=112
$P\bar{4}2c$\else
\ifnum#1=113
$P\bar{4}2_1m$\else
\ifnum#1=114
$P\bar{4}2_1c$\else
\ifnum#1=115
$P\bar{4}m2$\else
\ifnum#1=116
$P\bar{4}c2$\else
\ifnum#1=117
$P\bar{4}b2$\else
\ifnum#1=118
$P\bar{4}n2$\else
\ifnum#1=119
$I\bar{4}m2$\else
\ifnum#1=120
$I\bar{4}c2$\else
\ifnum#1=121
$I\bar{4}2m$\else
\ifnum#1=122
$I\bar{4}2d$\else
\ifnum#1=123
$P4/mmm$\else
\ifnum#1=124
$P4/mcc$\else
\ifnum#1=125
$P4/nbm$\else
\ifnum#1=126
$P4/nnc$\else
\ifnum#1=127
$P4/mbm$\else
\ifnum#1=128
$P4/mnc$\else
\ifnum#1=129
$P4/nmm$\else
\ifnum#1=130
$P4/ncc$\else
\ifnum#1=131
$P4_2/mmc$\else
\ifnum#1=132
$P4_2/mcm$\else
\ifnum#1=133
$P4_2/nbc$\else
\ifnum#1=134
$P4_2/nnm$\else
\ifnum#1=135
$P4_2/mbc$\else
\ifnum#1=136
$P4_2/mnm$\else
\ifnum#1=137
$P4_2/nmc$\else
\ifnum#1=138
$P4_2/ncm$\else
\ifnum#1=139
$I4/mmm$\else
\ifnum#1=140
$I4/mcm$\else
\ifnum#1=141
$I4_1/amd$\else
\ifnum#1=142
$I4_1/acd$\else
\ifnum#1=143
$P3$\else
\ifnum#1=144
$P3_1$\else
\ifnum#1=145
$P3_2$\else
\ifnum#1=146
$R3$\else
\ifnum#1=147
$P\bar{3}$\else
\ifnum#1=148
$R\bar{3}$\else
\ifnum#1=149
$P312$\else
\ifnum#1=150
$P321$\else
\ifnum#1=151
$P3_112$\else
\ifnum#1=152
$P3_121$\else
\ifnum#1=153
$P3_212$\else
\ifnum#1=154
$P3_221$\else
\ifnum#1=155
$R32$\else
\ifnum#1=156
$P3m1$\else
\ifnum#1=157
$P31m$\else
\ifnum#1=158
$P3c1$\else
\ifnum#1=159
$P31c$\else
\ifnum#1=160
$R3m$\else
\ifnum#1=161
$R3c$\else
\ifnum#1=162
$P\bar{3}1m$\else
\ifnum#1=163
$P\bar{3}1c$\else
\ifnum#1=164
$P\bar{3}m1$\else
\ifnum#1=165
$P\bar{3}c1$\else
\ifnum#1=166
$R\bar{3}m$\else
\ifnum#1=167
$R\bar{3}c$\else
\ifnum#1=168
$P6$\else
\ifnum#1=169
$P6_1$\else
\ifnum#1=170
$P6_5$\else
\ifnum#1=171
$P6_2$\else
\ifnum#1=172
$P6_4$\else
\ifnum#1=173
$P6_3$\else
\ifnum#1=174
$P\bar{6}$\else
\ifnum#1=175
$P6/m$\else
\ifnum#1=176
$P6_3/m$\else
\ifnum#1=177
$P622$\else
\ifnum#1=178
$P6_122$\else
\ifnum#1=179
$P6_522$\else
\ifnum#1=180
$P6_222$\else
\ifnum#1=181
$P6_422$\else
\ifnum#1=182
$P6_322$\else
\ifnum#1=183
$P6mm$\else
\ifnum#1=184
$P6cc$\else
\ifnum#1=185
$P6_3cm$\else
\ifnum#1=186
$P6_3mc$\else
\ifnum#1=187
$P\bar{6}m2$\else
\ifnum#1=188
$P\bar{6}c2$\else
\ifnum#1=189
$P\bar{6}2m$\else
\ifnum#1=190
$P\bar{6}2c$\else
\ifnum#1=191
$P6/mmm$\else
\ifnum#1=192
$P6/mcc$\else
\ifnum#1=193
$P6_3/mcm$\else
\ifnum#1=194
$P6_3/mmc$\else
\ifnum#1=195
$P23$\else
\ifnum#1=196
$F23$\else
\ifnum#1=197
$I23$\else
\ifnum#1=198
$P2_13$\else
\ifnum#1=199
$I2_13$\else
\ifnum#1=200
$Pm\bar{3}$\else
\ifnum#1=201
$Pn\bar{3}$\else
\ifnum#1=202
$Fm\bar{3}$\else
\ifnum#1=203
$Fd\bar{3}$\else
\ifnum#1=204
$Im\bar{3}$\else
\ifnum#1=205
$Pa\bar{3}$\else
\ifnum#1=206
$Ia\bar{3}$\else
\ifnum#1=207
$P432$\else
\ifnum#1=208
$P4_232$\else
\ifnum#1=209
$F432$\else
\ifnum#1=210
$F4_132$\else
\ifnum#1=211
$I432$\else
\ifnum#1=212
$P4_332$\else
\ifnum#1=213
$P4_132$\else
\ifnum#1=214
$I4_132$\else
\ifnum#1=215
$P\bar{4}3m$\else
\ifnum#1=216
$F\bar{4}3m$\else
\ifnum#1=217
$I\bar{4}3m$\else
\ifnum#1=218
$P\bar{4}3n$\else
\ifnum#1=219
$F\bar{4}3c$\else
\ifnum#1=220
$I\bar{4}3d$\else
\ifnum#1=221
$Pm\bar{3}m$\else
\ifnum#1=222
$Pn\bar{3}n$\else
\ifnum#1=223
$Pm\bar{3}n$\else
\ifnum#1=224
$Pn\bar{3}m$\else
\ifnum#1=225
$Fm\bar{3}m$\else
\ifnum#1=226
$Fm\bar{3}c$\else
\ifnum#1=227
$Fd\bar{3}m$\else
\ifnum#1=228
$Fd\bar{3}c$\else
\ifnum#1=229
$Im\bar{3}m$\else
\ifnum#1=230
$Ia\bar{3}d$\else
{\color{red}{Invalid SG number}}
\fi
\fi
\fi
\fi
\fi
\fi
\fi
\fi
\fi
\fi
\fi
\fi
\fi
\fi
\fi
\fi
\fi
\fi
\fi
\fi
\fi
\fi
\fi
\fi
\fi
\fi
\fi
\fi
\fi
\fi
\fi
\fi
\fi
\fi
\fi
\fi
\fi
\fi
\fi
\fi
\fi
\fi
\fi
\fi
\fi
\fi
\fi
\fi
\fi
\fi
\fi
\fi
\fi
\fi
\fi
\fi
\fi
\fi
\fi
\fi
\fi
\fi
\fi
\fi
\fi
\fi
\fi
\fi
\fi
\fi
\fi
\fi
\fi
\fi
\fi
\fi
\fi
\fi
\fi
\fi
\fi
\fi
\fi
\fi
\fi
\fi
\fi
\fi
\fi
\fi
\fi
\fi
\fi
\fi
\fi
\fi
\fi
\fi
\fi
\fi
\fi
\fi
\fi
\fi
\fi
\fi
\fi
\fi
\fi
\fi
\fi
\fi
\fi
\fi
\fi
\fi
\fi
\fi
\fi
\fi
\fi
\fi
\fi
\fi
\fi
\fi
\fi
\fi
\fi
\fi
\fi
\fi
\fi
\fi
\fi
\fi
\fi
\fi
\fi
\fi
\fi
\fi
\fi
\fi
\fi
\fi
\fi
\fi
\fi
\fi
\fi
\fi
\fi
\fi
\fi
\fi
\fi
\fi
\fi
\fi
\fi
\fi
\fi
\fi
\fi
\fi
\fi
\fi
\fi
\fi
\fi
\fi
\fi
\fi
\fi
\fi
\fi
\fi
\fi
\fi
\fi
\fi
\fi
\fi
\fi
\fi
\fi
\fi
\fi
\fi
\fi
\fi
\fi
\fi
\fi
\fi
\fi
\fi
\fi
\fi
\fi
\fi
\fi
\fi
\fi
\fi
\fi
\fi
\fi
\fi
\fi
\fi
\fi
\fi
\fi
\fi
\fi
\fi
\fi
\fi
\fi
\fi
\fi
\fi
\fi
\fi
\fi
\fi
\fi
\fi}

\newcommand{\sgsymbnum}[1]{SG #1 (\sgsymb{#1})}

\title{Catalogue of topological phonon materials}

\author{Yuanfeng Xu}
\affiliation{Department of Physics, Princeton University, Princeton, New Jersey 08544, USA}
\affiliation{Center for Correlated Matter and School of Physics, Zhejiang University, Hangzhou 310058, China}

\author{M. G. Vergniory}
\affiliation{Donostia International Physics Center, P. Manuel de Lardizabal 4, 20018 Donostia-San Sebastian, Spain}
\affiliation{Max Planck Institute for Chemical Physics of Solids, 01309 Dresden, Germany}

\author{Da-Shuai Ma}
\affiliation{Institute for Structure and Function, Department of Physics, Chongqing Key Laboratory for Strongly Coupled Physics, Chongqing University, Chongqing 400044, China}

\author{Juan L. Ma\~nes}
\affiliation{Department of Physics, University of the Basque Country UPV/EHU, Apartado 644, 48080 Bilbao, Spain}

\author{Zhi-Da Song}
\affiliation{Department of Physics, Princeton University, Princeton, New Jersey 08544, USA}
\affiliation{International Center for Quantum Materials, School of Physics, Peking University, Beijing 100871, China}

\author{B. Andrei Bernevig }
\email{bernevig@princeton.edu}
\affiliation{Department of Physics, Princeton University, Princeton, New Jersey 08544, USA}
\affiliation{Donostia International Physics Center, P. Manuel de Lardizabal 4, 20018 Donostia-San Sebastian, Spain}
\affiliation{IKERBASQUE, Basque Foundation for Science, Bilbao, Spain}

\author{Nicolas Regnault}
\affiliation{Laboratoire de Physique de l'Ecole normale sup\'{e}rieure, ENS, Universit\'{e} PSL, CNRS, Sorbonne Universit\'{e}, Universit\'{e} Paris-Diderot, Sorbonne Paris Cit\'{e}, 75005 Paris, France}
\affiliation{Department of Physics, Princeton University, Princeton, New Jersey 08544, USA}

\author{Luis Elcoro}
\affiliation{Department of Physics, University of the Basque Country UPV/EHU, Apartado 644, 48080 Bilbao, Spain}

\begin{abstract}
Phonons play a crucial role in many properties of solid state systems, such as thermal and electrical conductivity, neutron scattering and associated effects or superconductivity. Hence, it is expected that topological phonons will also lead to rich and unconventional physics and the search of materials hosting topological phonons becomes a priority in the field.
In electronic crystalline materials, a large part of the topological properties of Bloch states can be indicated by their symmetry eigenvalues in reciprocal space. This has been adapted to the high-throughput calculations of topological materials, and more than half of the stoichiometric materials on the databases are found to be topological insulators or semi-metals. 
Based on the existing phonon materials databases, here we have performed the first catalogue of topological phonon bands for more than ten thousand  three-dimensional crystalline materials. Using topological quantum chemistry, we calculate the band representations, compatibility relations, and band topologies of each isolated set of phonon bands for the materials in the phonon databases. We have also calculated the real space invariants for all the topologically trivial bands and classified them as atomic and obstructed atomic bands. In particular, surface phonon modes (dispersion) are calculated on different cleavage planes for all the materials. 
Remarkably, we select more than one thousand ``ideal'' non-trivial phonon materials to fascinate the future experimental studies. All the data-sets obtained in the the high-throughput calculations are used to build a \webTQCphonon.
\end{abstract}

\maketitle

\section{Introduction}

A crystalline electronic material is topologically non-trivial if its reciprocal-space Bloch states cannot be expressed in terms of localized Wannier functions (they are not wannierizable) in real space during an  adiabatic process that preserves the symmetry. In the presence of crystalline or time reversal symmetries, most of the band topologies can be diagnosed by the Fu-Kane-like topological invariants \cite{fu2007} in terms of the band representations at the high-symmetry momenta. Recently, symmetry protected topological phases in both the 230 paramagnetic and the 1421 magnetic space groups were exhausted by two methods: topological quantum chemistry (TQC) \cite{bradlyn_topological_2017,MTQC} and symmetry-based indicators \cite{po_symmetry-based_2017,watanabe2018structure,SlagerSymmetry,song_d-2-dimensional_2017}, which facilitated the construction of catalogues of topological bands in both paramagnetic and magnetic electronic materials \cite{vergniory_complete_2019,zhang2019catalogue,tang2019comprehensive,xu2020high,Vergniory2021}. High-throughput calculations found that more than 50\% of the stoichiometric materials in the Inorganic Crystal Structure Database \cite{ICSD} are diagnosed as topological insulators or semi-metals at the Fermi level and about 88\% had at least one topological band somewhere in the energy spectrum \cite{Vergniory2021}. 
As well as the electronic properties analyzed so far, the phonon structure is another platform for the realization of nontrivial band topologies in solid-state materials. The gapless nodes of both higher Chern number \cite{PhysRevLett.120.016401,PhysRevLett.121.035302,PhysRevLett.123.245302} and non-Abelian charges were predicted in several phonon materials \cite{peng2022phonons,peng2022}.
The existence of symmetry-protected topological phases has not been systematically proved in crystalline phonon band structures except for a few isolated cases \cite{PhysRevLett.120.016401,PhysRevLett.121.035302,PhysRevLett.123.245302,liu2021straight,zhu2022symmetry} and a brute-force screening of band nodes by calculating the band gaps along high-symmetry paths in phonon band structures \cite{li2021computation}. 
The methods of TQC were used to explore the topology of phonons on a particular two-dimensional lattice~\cite{manes_fragile_2019}, but no material realizations were found. In the present work, we fully extend the application of TQC to phonon systems in all the 3D space groups and systematically complete a catalogue of topological phonon bands.
Based on the dynamical matrices of more than ten thousand  materials that were obtained from {\it ab initio} calculations and stored on the databases \webphonondb\ (\webphonondbshort) and \webMP\ \cite{petretto2018high}, we have performed a high-throughput calculation of phonon irreducible representations (irreps), compatibility relations, topological indices and TQC real space invariants (RSIs) beyond symmetry indicators \cite{song2020,xu2021three}. We also compute the surface phonon dispersion  on different surfaces in all the materials.
Every isolated set of fully connected bands along all the high symmetry lines and planes in momentum space is finally diagnosed as topologically nontrivial (with strong or fragile topology), as obstructed atomic band representation (OABR, a wannierizable set of bands with at least one Wannier function out of the occupied atomic positions), as orbital-selected OABR (OOABR) or as an atomic (trivial) set of bands (wannierizable bands all located at the occupied atomic positions). 
Combined with a numerical analysis of the phonon band structure, we have chosen \TPDBNbrIdealNonTrivialKyotoMaterials\ and \TPDBNbrIdealNonTrivialMPMaterials\ ``ideal'' phonon materials from the \webphonondbshort\ and \webMP, respectively. These selected materials have at least one non-trivial topology: strong or fragile topological bands, an OABR or OOABR topology, or symmetry-enforced band nodes.
Finally, the full list of phonon materials and a wide array of data, including band structures, density of states (DOS), sets of irreps, topology classification, topological indices and surface states, are collected to build a \webTQCphonon.

\section{Workflow}\label{sec:workflow}
\begin{figure*}
    \centering
    \includegraphics[width=6.5in]{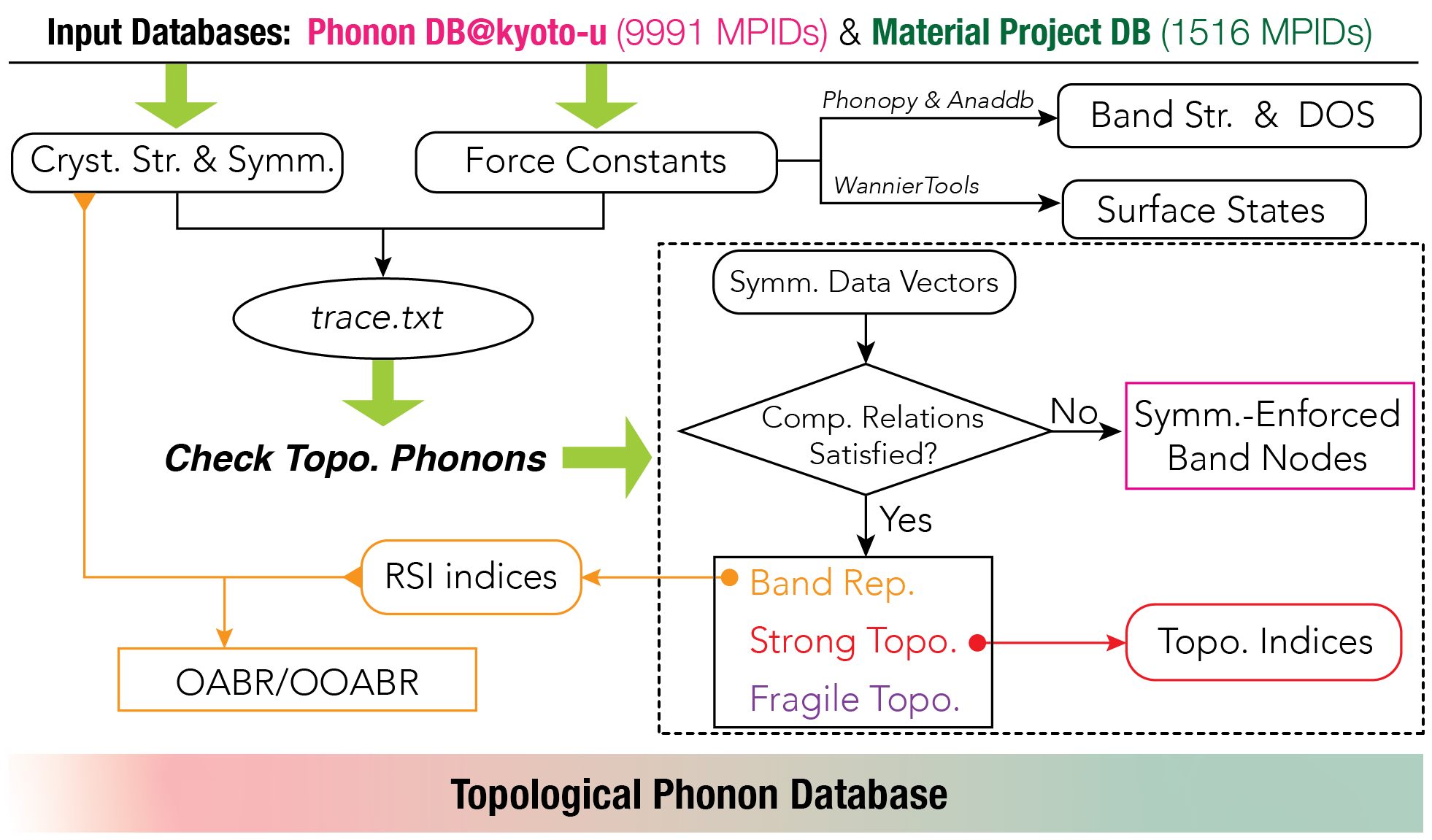}
    \caption{Workflow scheme of the construction of the \webTQCphonon. The high-throughput calculations rely on the dynamical matrices of  \TPDBNbrKyotoMaterials\ and \TPDBNbrMPMaterials\ MPID entries on the databases \webphonondbshort\ and \webMP, respectively. We first analyze the crystal structure and symmetry properties for each entry and filter out the entries that have symmetry inconsistencies in its dynamical matrix. Through the Fourier transform of the dynamical matrix we obtain the force constants which are used to calculate the phonon band structure, the DOS, the symmetry eigenvalues at the high-symmetry momenta ({\it trace.txt} file), and the surface states. The bulk calculations are performed by Phonopy \cite{phonopy} and Anaddb \cite{gonze2002first,gonze2009abinit,gonze2016recent} software packages, and the surface states are calculated with WannierTools package\cite{WU2017}. Lastly, the {\it trace.txt} files are fed into the new program \href{https://www.cryst.ehu.es/cryst/checktopologicalphonons.html}{Check Topological Phonons} to obtain the symmetry data vectors, the isolated sets of bands that satisfy the compatibility relations, the topological properties, and the topological indices. All the materials' data are collected to build a \webTQCphonon.}
    \label{fig:workflow}
\end{figure*}

The high-throughput calculations were performed using as input data the two phonon material databases, \webphonondbshort\ and \webMP, which collect the dynamical matrices for \TPDBNbrKyotoMaterials\ and \TPDBNbrMPMaterials\ compounds of different Material Project IDentification number(MPID), respectively. A brief overview of the two databases is provided in the Supplementary Information (SI) \ref{app:phonondatabases}. By filtering out the MPIDs that have symmetry inconsistencies, \ie the dynamical matrix is incompatible with the associated space group, we have designed several modules (as schematically illustrated in Fig. \ref{fig:workflow}) to calculate the phonon irreps and to diagnose the phonon topology of the remaining \TPDBNbrKyotoMaterials\ and \TPDBNbrMPMaterials\ MPIDs. 
In this section we give only an outline of the operation of each module. Further details can be found in the \siref{app:highthroughput}.

In step A, the force constants data of each MPID entry were obtained through the inverse Fourier transform of the dynamical matrix, \ie the second order derivatives of the total energy with respect to atomic displacements on a regular grid of the first Brillouin zone used in the {\it ab initio} calculations. Based on the force constants data, the band structures along all high-symmetry paths in the Brillouin zone and DOS were calculated with an interpolation method. 

In step B, we calculated the symmetry eigenvalues of phonon vibrational modes at all the high-symmetry momenta (the maximal {\bf q}-vectors \cite{bradlyn_topological_2017}) and generate the {\it trace.txt} file collecting all the symmetry eigenvalues. More details about the calculation of symmetry eigenvalues are provided in \siref{app:trace}.
Then, we fed all the {\it trace.txt} files into the recently implemented utility \href{https://www.cryst.ehu.es/cryst/checktopologicalphonons.html}{Check Topological Phonons} (see \siref{app:checktop}) closely related with the program \href{https://www.cryst.ehu.es/cryst/checktopologicalmat.html}{Check Topological Mat.} for electronic bands \cite{vergniory_complete_2019,Vergniory2021}, which is available online at the \webBCS, and for each file we performed the steps C, D, and E.

In step C, we identified the single-valued irreps (\ie the irreps without spin-orbit coupling) at the maximal {\bf q}-vectors for all the {\it trace.txt} files. The symmetry properties of a set of phonon bands are characterized by the multiplicities of irreps at these maximal {\bf q}-vectors, which form a \emph{symmetry data vector}.

In step D we checked the band connectivity along all the high-symmetry paths in the Brillouin zone.
A set of states connected along the Brillouin zone built an isolated set of bands if its symmetry data vector satisfies all the compatibility relations, \ie for every pair ${\bf q}_1$ and ${\bf q}_2$ of maximal {\bf q}-vectors and the intermediate path ${\bf q}_p$ that connects both points, the irreps at ${\bf q}_1$ and ${\bf q}_2$ subduce into the same set of irreps at ${\bf q}_p$. Otherwise, the set of eigenstates necessarily has symmetry-enforced band crossings (nodes) with other bands above or below the given set of states and, therefore, do not form an isolated set of bands. In our calculations we identify all the isolated sets of bands that cannot be further split into subsets that also form isolated bands. Together with the identification of each isolated band by its symmetry data vector, we also store the symmetry data vector of the \emph{cumulative set} of bands, \ie, the whole set of bands starting from the band with lowest energy up to a given band.

In step E, using the TQC method \cite{bradlyn_topological_2017, MTQC}, we identified the topology of each isolated set of bands and also the topology of the cumulative band set. In both cases, we diagnosed them as strong topological with topological indices (or symmetry indicators), as fragile topological or as a band representation (trivial). 

In step F, for all the topologically trivial sets labeled as band representation, we calculated their RSIs which indicates the location of the corresponding Wannier functions in real space. A set of bands is referred to as OABR when its RSI is non-zero at an Wyckoff position that is not occupied by atoms \cite{bradlyn_topological_2017,song2020,xu2021filling,xu2021three,gao2021unconventional} . If the non-zero RSIs are located at occupied Wyckoff positions, but the orbitals that define the RSI do not correspond to the  vector representation of the site-symmetry group (the representation of the $p$ orbitals), the related set of bands are labelled as an OOABR \cite{xu2021three}.

For semiconductor materials of polar crystal structures, the long-range macroscopic electric field induced by long wavelength longitudinal optical (LO) phonons is usually non-negligible \cite{RevModPhys.73.515}. The coupling between LO mode and electric fields, is a non-analytical correction (NAC) to the dynamical matrix \cite{gonze1997dynamical,wang2010mixed}, resulting in an energy difference between the LO mode and the transverse optical (TO) mode (see \siref{app:NAC} for a formal definition of NAC). 
In general, this LO-TO splitting destroys the continuity of the phonon spectrum when approaching to the center of the Brillouin zone ($\bf q=0$) along different directions.
For a special case, the energy order of two bands (of different band representations) could be inverted by the LO-TO splitting at a finite momentum close to the $\Gamma$ point with respect to the band order at $\Gamma$  (this inversion can happen due to the mentioned discontinuity). In the TQC high-throughput method, only the information obtained at the maximal {\bf q}-vectors is used to identify the topology of a given set of bands. As a consequence, the potential band inversion due to the LO-TO splitting can be missed and the 
topological classification might be wrong. Even so, we have applied the TQC method also to phonons with NAC and provided the results for reference. 

Although the topological surface states in electronic materials have been detected by several spectroscopic and transport experiments, the related experiments on topological phonon surface states are rare. First, the experimental technique of surface phonon detection is limited to the electron energy loss spectroscopy, whose energy resolution is down to $5\sim10$meV so that the surface phonon modes are difficult to be resolved. Second, a topological phonon material database with surface phonon spectrum to serve as a guideline to experiments was so far unavailable. In the present work, we have also developed a high-throughput algorithm for the calculation of phonon surface states. The method is detailed in \siref{app:surfacestates} and the algorithm has been embedded with the WannierTools soft package\cite{WU2017}. For each phonon material, we select three non-equivalent cleavage planes and calculate their surface states with a finite-size slab structure. For further detailed calculations and analysis of the phonon surface states, we have provided the WannierTools input files for each MPID entry on the \webTQCphonon.

\section{Results}

\subsection{Material Database and statistics}

By applying the above high-throughput screening method, we have successfully identified the band representations and band topologies for \TPDBNbrKyotoMaterials\ MPID entries in the \webphonondbshort\ and \TPDBNbrMPMaterials\ MPID entries in the \webMP. As the {\it ab initio} calculations were performed on a finite-size supercell structure (for \webphonondbshort) or a less dense mesh grid in the Brillouin zone, the resulting phonon band structures usually host negative frequencies in a partial region of the Brillouin zone. In \siref{app:negativenergies}, we tag the materials with a minimum frequency lower than $-5$meV as MPID entries with ``negativity'' issues. Depending on the features of these negative frequencies in the phonon spectrum with NAC, we manually divide the MPIDs with ``negativities'' into three categories: \emph{interpolation error} when a weak negativity over part of the Brillouin zone results from interpolation errors, \emph{instability} indicates that a potential structural phase transition might occur at lower temperature and \emph{discarded} when a strong negativity is present over the whole (or most of the) Brillouin zone.

\begin{figure*}
    \centering
    \includegraphics[width=7.0in]{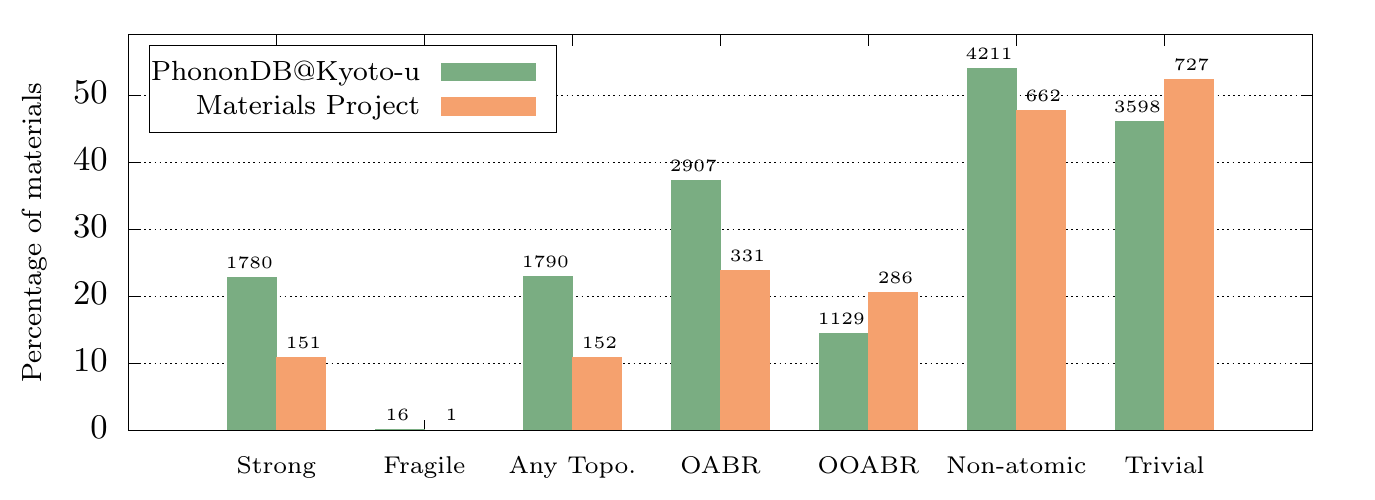}
    \caption[Statistics of non-trivial phonon band sets for the phonon materials from \webphonondbshort\ and \webMP(MP)]{Statistics of non-trivial phonon band sets for the phonon materials from \webphonondbshort\ (green) and \webMP (orange). All numbers and percentages are from the calculations without NAC and without the materials tagged as having negative phonon frequencies. Numbers above each bar is the number of materials for each category and data source. From left to right: {\it Strong} and {\it Fragile} refer to materials hosting at least one cumulative band set of strong and fragile topology, respectively.  {\it Any Topo.} corresponds to materials hosting at least one cumulative band set that is either strong or fragile topology. {\it OABR} and {\it OOABR} gives similar figures for OABR and OOABR band sets, while {\it Non-atomic} stands for the number of materials hosting at least one non-atomic cumulative band sets, \ie strong-topology, fragile-topology, OABR or OOABR. Note that a material can host several types of non-trivial or non-atomic band sets. Thus the percentages do not add up. Finally, {\it Trivial} gives the number of materials having solely atomic cumulative band sets.}
    \label{fig:statistics}
\end{figure*}

By filtering out the MPIDs that have a ``negativity'' tag in the NAC calculations, we are left with \TPDBNbrKyotoMaterialsWithNACNoNegative\ and    \TPDBNbrMPMaterialsWithNACNoNegative\ ``high-quality'' phonon band structures in the \webphonondbshort\ and \webMP, respectively. Fig.~\ref{fig:statistics} illustrates a summary of the statistics of materials hosting non-atomic band sets, \ie either strong topology, fragile topology, OABR or OOABR. Strikingly, about half of the materials host at least one non-atomic cumulative band set: \TPDBPercentMPMaterialsWoNACNonTrivialNonAtomicNoNegative\% of the materials from the \webMP\ and \TPDBPercentKyotoMaterialsWoNACNonTrivialNonAtomicNoNegative\% of those from \webphonondbshort.  In \siref{app:statistics}, we provide an in-depth discussion of statistics, with or without NAC, and with or without the materials with negative frequencies.

The percentage of materials hosting non-atomic phonon band sets, while substantial, is not as important as the same percentage for electronic bands with spin-orbit coupling (\TQCDBNbrTopoBandPercent) but is actually higher than the percentage for electronic bands without spin-orbit coupling (\TQCDBNbrNoSOCTopoBandPercent\ as reported in Ref.~\cite{Vergniory2021}). Also, the main contribution to this percentage strongly differs in electrons and phonons. As can be observed in Fig.~\ref{fig:statistics}, there is an abundance of materials with trivial but non-atomic cumulative phonon band sets, \ie either OABR or OOABR. A qualitative explanation can be found by considering optical vibrational modes. 
Indeed, if we consider the simplest case of a one dimensional system with inversion symmetry, the optical (or acoustic) mode contributed by the atoms at $2c$ (\ie the general Wyckoff position) has to be obstructed (OABR/OOABR) at one of the inversion centers ($1a$ and $1b$) no matter if the center is occupied by an atom or not. It is thus somehow expected that OABR or OOABR are common in phonon materials. As a side note, we observe an almost absence of fragile cumulative topology for phonons.

\subsection{Interpretation of the topological indices}
For each (isolated and cumulative) topological set of bands, we have calculated its topological indices \cite{bradlyn_topological_2017,po_symmetry-based_2017,watanabe2018structure,song_diagnosis_2018,SlagerSymmetry,MTQC}. As proved in Ref. \cite{song_diagnosis_2018}, although the band structures along all the high-symmetry paths satisfy all the compatibility relations, the non-zero topological indices defined in spinless space groups (namely without spin-orbit coupling) are necessarily indicating gapless band nodes (points or lines).
Compared with the symmetry-enforced band nodes which break the compatibility relations, symmetry-indicated band nodes occur in general at generic momenta in the Brillouin zone. It also shows that all the non-trivial topological indices in (non-) centrosymmetric space groups indicate nodal lines (Weyl nodes). Remarkably, in 33 out of the 41 centrosymmetric space groups which have topological indices well defined in TQC, there is always at least one set of indices that indicates the presence of nodal lines with $Z_2$-monopole charge. For each set of topological bands in the \webTQCphonon, we have tabulated the essential topological indices that were defined and physically interpreted in Ref.~\cite{song_diagnosis_2018}. In the high-throughput calculations, discarding the MPIDs tagged as having ``negativities'', all the symmetry-indicated spinless topologies including Weyl nodes, nodal lines with and without $Z_2$-monopole charge were found in \TPDBNbrKyotoMaterialsWoNACTopologicalNoNegative(\TPDBPercentKyotoMaterialsWoNACTopologicalNoNegative\%) and \TPDBNbrMPMaterialsWoNACTopologicalNoNegative(\TPDBPercentMPMaterialsWoNACTopologicalNoNegative\%) MPID entries on \webphonondbshort\ and \webMP, respectively.

\section{Remarkable topological phonon materials}

\subsection{``Ideal'' phonon materials with non-atomic cumulative topologies}

The \webTQCphonon\ allows the search of the most promising candidates, that we will refer to as ``ideal'' phonon materials, exhibiting any of the non-atomic properties and additional criteria, detailed in 
\siref{app:bestmaterials}, where the main results of the search are also presented. Basically, in addition to hosting either a non-atomic band gap or ``ideal'' symmetry-enforced band nodes, we imposed the following restrictions:
\begin{enumerate}
    \item The material is not tagged as having ``negativities'' in the database, \ie its minimum frequency is above -5.0meV, so that the crystal structure is considered as relatively stable. 
    \item The maximal direct band gap along all the high-symmetry paths between the non-atomic cumulative set of bands and the first band above it (or between the bands forming a symmetry-enforced band node) is larger than 1meV, which is the typical experimental resolution to distinguish a phonon band gap. 
    \item The indirect gap along all the high-symmetry paths is positive for non-atomic cumulative band sets and and non-negative for symmetry-enforced band nodes, respectively.
\end{enumerate}

Applying these three filtering additional criteria, we obtained \TPDBNbrIdealTopoMaterials(\TPDBPercentIdealTopoMaterials\%), \TPDBNbrIdealOABRMaterials(\TPDBPercentIdealOABRMaterials\%) and \TPDBNbrIdealOOABRMaterials(\TPDBPercentIdealOOABRMaterials\%) ``ideal'' phonon materials with strong, OABR and OOABR topologies, respectively. In addition, we also select \TPDBNbrIdealSMMaterials(\TPDBPercentIdealSMMaterials\%) phonon materials hosting ``ideal'' symmetry-enforced band nodes. Note that we find no ``ideal'' phonon material with fragile cumulative topology satisfying the above criteria which is unsurprising due to the almost absence of fragile cumulative topology as discussed previously. The material lists and statistics for each type of ``ideal'' materials on both \webphonondbshort\ and \webMP\ are detailed in \siref{app:bestmateriallist}.
Moreover in \siref{app:bestmaterialbandstr}, we provide the phonon spectrum plot near the topological band gap and band nodes for ``ideal'' materials with strong topologies and symmetry-enforced band nodes.

\subsection{Prototypical topological phonon materials}
\begin{figure*}
    \centering
    \includegraphics[width=7.0in]{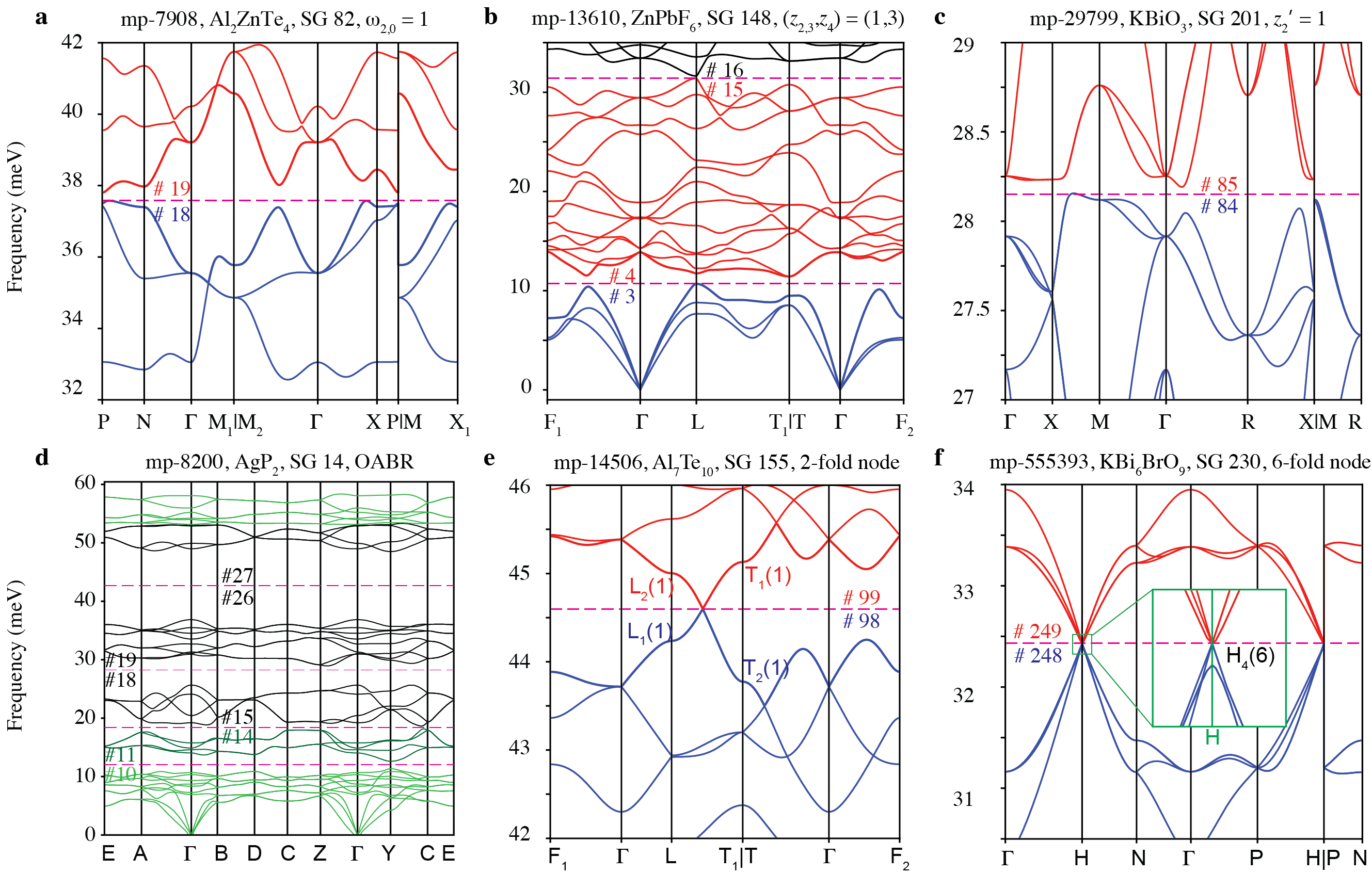}
    \caption{Phonon band dispersion of our relation of six prototypical materials. (a)\ce{Al2ZnTe4}. An ideal topological band gap is between the bands \#18 and \#19, where blue and red bands are below and above the topological gap, respectively. Topological index of the blue bands is $\omega_{2,0}=1$, indicating four Weyl points on the $k_z=0$ slice. (b)\ce{ZnPbF6}. There are two ideal topological band gaps: one is located between bands \#3 and \#4 and the another one between the bands \#15 and \#16. Symmetry data vectors of bands below the two gaps have the same set of topological indices $(z_{2,3},z_4)=(1,3)$, indicating nodal lines of $\pi$-Berry phase. (c)\ce{KBiO3}. The topological band gap between the bands \#84 and \#85 has a topological index $z_2'=1$, indicating three pairs of $Z_2$-monopole charge nodal lines. (d) \ce{AgP2}. There are three isolated band sets (plotted with light, dark and light green lines, respectively) diagnosed as OABR topology. The calculation of the (cumulative) RSIs up to a given band shows that there are four cumulative band sets hosting both an OABR topology and an indirect band gap (indicated by the pink dashed lines). (e)\ce{Al7Te10}. The first 98 sets of irreps at maximal {\bf q}-vectors do not form a fully connected set of bands. The state of band \#98 at the $L$ point labelled as $L_1(1)$ (in parentheses the dimension of the irrep) is connected to the irrep of \#99 at the $T$ point ($T_1(1)$) and not to the irrep of \#98 ($T_2(1)$) due to the compatibility relations along the intermediate $L-Y-T$ path ($L_1$($L_2$) and $T_2$($T_1$) have different symmetry eigenvalues of the common 2-fold rotation). There is thus a symmetry-enforced band crossing between the bands \#98 and \#99, giving a 2-fold Weyl node. (f)\ce{KBi6BrO9}. There is a 6-fold degeneracy point at $\Gamma$, composed by the bands \#247$\sim$252. The 6-dimensional irrep ($H_4$) is protected by inversion symmetry, 3-fold rotation symmetry along the (111) direction, 4-fold rotation symmetry combined with a fractional translation $\{C_{4z}|(\frac{1}{4},\frac{3}{4},\frac{1}{4})\}$ and a 2-fold screw  symmetry $\{C_{2y}|(\frac{1}{2},0,0)\}$ \cite{bradlyn2016beyond}.}
    \label{fig:mainfig2}
\end{figure*}

We here highlight six representative ``ideal'' phonon materials with symmetry-indicated topological bands, OABR bands or symmetry-enforced band nodes. The six materials are \ce{Al2ZnTe6} [\mpidweb{7908} SG 82 (\sgsymb{82})], \ce{ZnPbF6} [\mpidweb{13610} SG 148 (\sgsymb{142})], \ce{KBiO3} [\mpidweb{29799} SG 201 (\sgsymb{201})], \ce{AgP2} [\mpidweb{8200} SG 14 (\sgsymb{14})], \ce{Al7Te10} [\mpidweb{14506} SG 155 (\sgsymb{155})], and \ce{KBi6BrO9} [\mpidweb{555393} SG 230(\sgsymb{230})]. As shown in Fig. \ref{fig:mainfig2}, the first three materials have symmetry-indicated topological bands, \ce{AgP2} has three OABR-topology isolated band sets, and the last two materials have symmetry-enforced band nodes. Details about the topological properties and surface state calculations of these materials are provided in \siref{app:examples}. In the following, we briefly discuss the non-atomic bands in \ce{Al2ZnTe6}, \ce{KBiO3} and \ce{AgP2}.

In the phonon band structure of non-centrosymmetric \ce{Al2ZnTe6} (Fig. \ref{fig:mainfig2}a), the cumulative band set of indices $\#1\sim18$ is topological, indicated by a $Z_2$ index $\omega_{2,0}=1$. In Ref. \cite{song_diagnosis_2018}, $\omega_{2,0}\times \pi$ is interpreted as the Berry phase of a loop
enclosing a quarter of the BZ at $k_z = 0$ plane (which is invariant under a 4-fold rotation). Hence, $\omega_{2,0}=1$ indicates 4 mod 8 Weyl points in between bands \#18 and \#19 on the $k_z = 0$ slice. 
More details about the topological index, Weyl points, and surface state (phonon ``Fermi''-arcs) calculations are provided in \siref{app:ssAlZnTe}.

\ce{KBiO3} crystallize in the centrosymmetric space group \sgsymb{201}. As shown in Fig. \ref{fig:mainfig2}c, the cumulative band set of indices $\#1\sim84$ has a non-zero topological index $z_2'=1$, which indicates three pairs of nodal lines of $Z_2$-monopole charge because of $C_3$ symmetry, according to Ref. \cite{song_diagnosis_2018}. 
Unlike the $\pi$-Berry phase nodal line (which is also indicated by non-zero topological indices), $Z_2$-monopole charge nodal line has to be created (or annihilated) in pairs and is characterized by the second Stiefel-Whitney class, which is indicated by a $Z_2$ indicator $w_2$ and can be read off from the Wilson loop spectrum \cite{fang2015,bjyang2018}. 
In \siref{app:KBiO}, we calculate the Wilson-loop evolution for this cumulative band set on a 2D closed manifold wrapping a nodal line, where an odd number of crossing points at WCC$=\pi$ line corresponds to a $w_2=1$ phase and indicates the $Z_2$-monopole charge.

In the phonon spectrum of \ce{AgP2} (from the Kyoto database) shown in Fig. \ref{fig:mainfig2}d, there are six cumulative band sets and three isolated band sets (the green bands) that have an OABR topology, among which four cumulative band sets have an indirect band gap (indicated by the pink dashed lines).
The non-zero RSIs of these three isolated band sets are $\delta_1(a)=1$ for band set $\#1\sim10$, $\{\delta_1(a)=-1,\delta_4(d)=1\}$ for band set $\#11\sim14$, and $\delta_4(d)=-1$ for band set $\#31\sim36$, where a nozero RSI $\delta(w)$ implies unavoidable Wannier functions at the Wyckoff position $w$ (\ie $a$ or $d$) and neither $a$ nor $d$ are occupied by any atom. The RSIs of each cumulative band set can be obtained simply by summing the RSIs of all the isolated band sets within it.

\begin{figure*}
    \centering
    \includegraphics[width=7.0in]{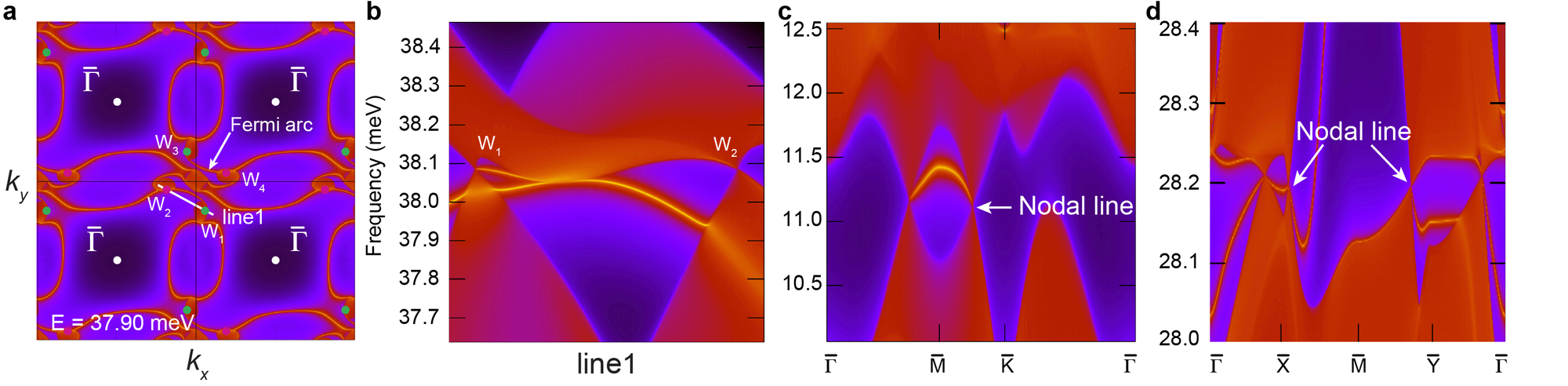}
    \caption{Surface state calculations for three prototypical topological phonon materials. (a-b)Surface state calculation for \ce{Al2ZnTe4} on the (001) surface. (a)Energy contour ($E=37.9$meV) of the surface states in a $2\times 2$ surface BZ, where the locations of Weyl points with chirality +1(-1) are indicated by the green(magenta) dots. Surface state dispersion along the path that connects the two Weyl points $W_1$ and $W_2$ is shown in (b). The topological phonon ``Fermi''-arc states can be distinguished from the bulk states both in (a) and (b). (c)Drum-head-like surface states of \ce{ZnPbF6} on the (001) surface. The nodal line bulk states are formed by the band crossing between bands \#3 and \#4, which is indicated by the nontrivial topological indices $(z_{2,3},z_4)=(1,3)$ in SG 148. (d)Drum-head-like surface states connecting the topological nodal lines of $Z_2$-monopole charge on the (001) surface of \ce{KBiO3}. The nodal lines are formed by the band crossing between bands \#84 and \#85, indicated by the nontrivial topological index $w_2=1$ in SG 201.}
    \label{fig:mainfig4}
\end{figure*}

In Fig. \ref{fig:mainfig4}, we performed surface state calculations with Green's function method \cite{sancho1985highly} for the three symmetry-indicated topological phonon bands in the first row of Fig. \ref{fig:mainfig2}, namely, \ce{Al2ZnTe4} with two pairs of Weyl nodes as shown in Figs. \ref{fig:mainfig4}(a-b), \ce{ZnPbF6} with topological nodal line of $\pi$-Berry phase (Fig. \ref{fig:mainfig4}(c)), and \ce{KBiO3} (Fig. \ref{fig:mainfig4}(d)) with topological nodal line of $Z_2$ monopole charge. In the energy contour and dispersion plots, the ``Fermi''-arc surface states connecting one pair of Weyl nodes and the drum-head-like surface states connecting the nodal lines can be distinguished. In \siref{app:examples}, we give for these three materials additional surface state calculations with different surface terminations. 

\section{Discussion}

We have performed the first complete topological classification for more than ten thousand phonon materials in a high-throughput search and we have made the results available at our public website \webTQCphonon.
We identify the band representations and topology of each phonon band set by calculating the topological indices and RSIs. 
As it happens in the electronic structures of solid state materials, non-trivial band topologies are ubiquitous in phonon systems: \TPDBPercentMaterialsWoNACNonTrivialNonAtomicNoNegative\% of the curated materials have some kind of non-trivial topology. 
Remarkably, we predict more than one thousand phonon materials that have at least one ``ideal'' non-atomic phonon band set. These ideal materials constitute an invaluable platform for further in-depth theoretical and experimental studies about phonon-band topologies.
Although we do not find any phonon materials hosting ``ideal'' band sets with both cumulative fragile topology and finite indirect band gap, there exist several materials with perfect isolated fragile band sets, such as the $C_2\cdot T$-symmetry-protected fragile band sets in \ce{HfIN} [\mpidweb{567441} SG 166(\sgsymb{166})] discussed in \siref{app:HfIN}.
As phonons are bosonic excitations, topological band gaps (or nodes) are not limited to a specific energy level and the nontrivial topology at any frequency level could be of interest and detectable in experiments. Based on the formalism and results of this work, future extension of the catalogue of topological phonons will encompass all the solid state materials contained in the Inorganic Crystal Structure Database\cite{ICSD}.

\acknowledgments
We thank Atsushi Togo, Shyam Dwaraknath and Kristin A. Persson for the helpful discussion about the phonon material databases. Y.X. thanks Shuyuan Zhang, Xun Jia and Chaoxing Liu for the helpful discussion about experimental detection of phonon topologies. We acknowledge the computational resources Cobra/Draco in the Max Planck Computing and Data Facility (MPCDF). This work is part of a project that has received funding from the European Research Council (ERC) under the European Union's Horizon 2020 research and innovation programme (grant agreement no. 101020833). 
 N.R. was supported by the DOE Grant No. DE-SC0016239 and by the Princeton Global Network Funds. B.A.B. and N.R. were also supported by a Simons Investigator grant (No. 404513), the Office of Naval Research (ONR Grant No. N00014-20-1-2303), the Schmidt Fund for Innovative Research, the BSF Israel US foundation (Grant No. 2018226), the Gordon and Betty Moore Foundation through Grant No. GBMF8685 towards the Princeton theory program and Grant No. GBMF11070 towards the EPiQS Initiative, and NSF-MRSEC (Grant No. DMR-2011750). 
L.E. was supported by the Government of the Basque Country (Project  IT1458-22) and the Spanish Ministry of Science and Innovation (PID2019-106644GB-I00). 
M.G.V. acknowledges the Spanish Ministerio de Ciencia e Innovaci\'on (grant PID2019-109905GB-C21) and the Deutsche Forschungsgemeinschaft (DFG, German Research Foundation) GA 3314/1-1 - FOR 5249 (QUAST). The work of J.L.M. has been supported by Spanish Science Ministry grants PGC2018-094626-B-C21 and PID2021-123703NB-C21 (MCIU/AEI/FEDER, EU), and Basque Government grants IT979-16 and IT1628-22. Z.D.S. also acknowledges National Natural Science Foundation of China (General Program No. 12274005), and National Key Research and Development Program of China (No. 2021YFA1401900). D.-S. Ma acknowledges the National Natural Science Foundation of China (Grants No.~12204074) and the China National Postdoctoral Program for Innovative Talent (Grant No. BX20220367).

{\bf Competing interests}
The authors declare no competing interests.

{\bf Data availability}
All data is available in the Supplementary Information and through our public website .

\onecolumngrid
\renewcommand{\thesection}{Appendix \arabic{section}}
\renewcommand{\thesubsection}{\arabic{section}.\arabic{subsection}}

\appendix
\clearpage
\begin{center}
{\bf Supplementary materials of "Catalogue of topological phonon materials"}
\end{center}

\tableofcontents

\clearpage

\addtocontents{toc}{\protect\setcounter{tocdepth}{3}}
\addtocontents{lot}{\protect\setcounter{lotdepth}{3}}

\section{Introduction of the Appendices}\label{app:introduction}
In the present work, we have performed the catalogue of phonon bands in both topological and obstructed atomic  band representation (OABR) phases. In addition to the main text, we provide in these Supplementary Appendices an in-depth discussion of our methodology and high-throughput results. 
In \siref{app:concepts}, we briefly introduce the theory of topological quantum chemistry (TQC) \cite{bradlyn_topological_2017, MTQC}, and the concept of obstructed atomic insulator (OAI) \cite{song2020,xu2021three}, defined in electronic structures, and its analogue OABR for phonons.
In \siref{app:phonondatabases}, we give an overview of the two phonon material databases, \webMP\ and \webphonondb (PhononDB@kyoto-u), on which we relied for our catalogue of topological and OABR phonon bands.
In \siref{app:highthroughput}, we give details of the methods used in the calculation of the irreducible representations (irreps) for phonon bands at the high-symmetry momenta. Then, we detail the workflow of the high-throughput calculations using TQC theory and the three-dimensional real space invariants (3D RSIs) \cite{song2020,xu2021three}.
In \siref{app:surfacestates}, for each phonon band structure, we performed the surface state calculations.
In \siref{app:examples}, we highlight several prototypical materials hosting non-atomic phonon bands and analyze their band topologies and surface states in detail.
In \siref{app:TQCphonondb}, based on the high-throughput calculations, we present our database of topological phonon materials, hosted in the website \webTQCphonon\ and the statistics for non-trivial phonon band sets. We also explain how dispersions  with negative energies are handled in the database.
In \siref{app:bestmaterials}, we provide the lists that contain the selected ``ideal'' phonon materials hosting non-atomic cumulative topologies, including strong topology, OABR and  orbital-selected obstructed atomic band representation (OOABR). We also give the lists of ``ideal'' phonon materials hosting symmetry-enforced band nodes.
Finally in \siref{app:phononmaterials}, we provide the complete lists of materials for both \webphonondb\ and \webMP\ that were considered in the high-throughput calculations. For each MPID entry, we provide its statistics of the non-atomic band sets and whether it exhibits ``negativity'' issues in its phonon dispersion.

\section{Concepts}\label{app:concepts}
In this section, we will briefly review the topological quantum chemistry (TQC) theory \cite{bradlyn_topological_2017, MTQC}, the symmetry-eigenvalue indicated topological phases \cite{po_symmetry-based_2017,watanabe2018structure,song_diagnosis_2018,song_quantitative_2018,SlagerSymmetry}, the definition of RSI \cite{song2020,xu2021three}, and the concept of OAI \cite{bradlyn_topological_2017,xu2021filling,xu2021three}. Note that we will here stick to the name OAI used in TQC while we use the acronym OABR for phonons where the notion of insulator is not meaningful.

\subsection{Topological quantum chemistry}\label{app:TQC}
In band theory, the symmetry properties of a band structure are characterized by the irreducible (co-)representations (irreps) at all the maximal $k$-vectors, whose little groups are maximal subgroups of the (magnetic) space group. Each maximal $k$-vector in the Brillouin zone (BZ) is kept invariant (mod translations of the reciprocal lattice) under a subset of operations of the space group that form the little group $G_{k}$ of $k$, and whose irreps are denoted as $\rho_{k}^i$.

With $N_{e}$ the number of valence electrons of the crystal material in one unit cell, the first $N_{e}$ bands at every $k$ point represents the set of \emph{occupied} bands.
These bands are characterized by the set of multiplicities $m(\rho_{k}^i)$ of every irrep $\rho_{k}^i$ of $G_{k}$ at every maximal $k$-vector, $\{m(\rho_{k_j}^i)|i=1,2,...,N(k_j), j=1,2,...,N_k\}$, where $N(k_j)$ is the number of irreps of $G_{k_j}$ and $N_k$ is the number of maximal $k$-vectors in the BZ of the considered space group.
For convenience, we introduce the symmetry data vector \cite{bradlyn_topological_2017,cano_building_2018,elcoro_double_2017,vergniory_graph_2017,SlagerSymmetry},
\begin{equation}
B = [
m(\rho_{k_1}^1), m(\rho_{k_1}^2), ..., m(\rho_{k_1}^{N(k_1)}),
m(\rho_{k_2}^1), m(\rho_{k_2}^2), ..., m(\rho_{k_2}^{N(k_2)}),
...,
m(\rho_{k_{N_k}}^1), m(\rho_{k_{N_k}}^2), ..., m(\rho_{k_{N_k}}^{N(k_{N_k})}) ]^T.\label{eq:B-vector}
\end{equation}
with $\sum_{i=1}^{N_k}N(k_i)$ components, that give the multiplicity of the corresponding irrep.

In the analysis of the electronic band structures done in the framework of TQC \cite{bradlyn_topological_2017, MTQC} and the equivalent method of symmetry-based indicators \cite{po_symmetry-based_2017,watanabe2018structure,song_quantitative_2018}, a band structure is gapped if its symmetry data vector satisfies all the compatibility relations between any two maximal $k$-vectors, \ie along the lines or planes that connect every pair of maximal $k$-vectors, the multiplicities of the irreps of the little group of the intermediate line or plane, given by the compatibility relations at both end points, are exactly the same. Otherwise, the band structure is \emph{necessarily} gapless and referred to as \emph{enforced semi-metal}. In the TQC analysis, the gapped band structures have been divided into three different types of topological insulating phases: strong topological insulator, fragile topological insulator, and topologically trivial insulator.

Following the terminology of Zak \cite{zak1980symmetry,zak1981band, zak1982band}, the electronic band structure of an atomic insulator is a band representation (BR) and the generators of the BRs are elementary BRs (EBRs), which are the induced band representations from all the possible atomic orbitals symmetrically centered at the maximal Wyckoff positions \cite{zak1982band,bradlyn_topological_2017,MTQC}. A BR can always be decomposed into a linear combination of EBRs (LCEBR) with non-negative-integer coefficients. The EBRs are thus the basis of the BR of topologically trivial insulators that can be induced from the atomic orbitals and are then Wannierizable. 
A gapped band structure can be symmetry-indicated as topologically trivial insulator if its symmetry data vector (in Eq. \ref{eq:B-vector}) is a linear combination of the symmetry data vectors of the EBRs. If not, it is topologically nontrivial and the material has strong or fragile topology. The symmetry data vector of fragile topology can not be written as a sum of EBRs but as a difference of EBRs with integer coefficients.
For strong topology, the symmetry data vector can not be expressed as a linear combination of EBRs with integer coefficients. Strong topology can be further distinguished on the basis of whether its symmetry data vector could be written as sums of parts of split EBRs \cite{bradlyn_topological_2017}: If it could be written as a sum of split (disconnected) EBRs, its topology is denoted SEBRs; If it could not be written as a sum of split EBRs but were still not a sum or a difference of EBRs, it is denoted more generically ‘no linear combination’ (NLC) of EBRs \cite{vergniory_complete_2019,Vergniory2021}.

\subsection{Obstructed atomic insulator and Real space invariant}\label{app:OAI}

In real crystalline materials, the BR of an insulator tagged as topologically trivial in the TQC analysis is always induced from local orbitals centered at different Wyckoff positions that are occupied or not occupied by atoms. If a BR cannot be induced \emph{only} from the atomic orbitals centered at the occupied Wyckoff positions, it is in the obstructed atomic insulator (OAI) phase \cite{bradlyn_topological_2017,xu2021three,xu2021filling}.
For the OAIs, there are \emph{necessarily} localized Wannier functions pinned at some empty sites.

Given the symmetry data vector $B$ (Eq.~\ref{eq:B-vector}) of the BR of a trivial insulator, the material is identified as an OAI if $B$ \emph{does not} satisfy the following condition,
\begin{equation}
B=\sum_{j=1}^{M}  \sum_{i=1}^{N_{rep, \alpha_j}} N_{i,j}(\rho_{\alpha_j}^i\uparrow\mathcal{G}), \;\;\; N_{i,j} \ge 0, \in \mathbf{Z}. \label{eq:OAI}
\end{equation} 
where $\{\alpha_j, j=1,2,...,M\}$ is the set of occupied Wyckoff positions, $\rho_{\alpha_j}^i\uparrow\mathcal{G}$ is the induced BR from an irrep $\rho_{\alpha_j}^i$ at the occupied position $\alpha_j$, $N_{rep, \alpha_j}$ is the multiplicity of irreps of the site-symmetry group of $\alpha_j$ and $N_{i,j}$ are non-negative integers. 

In the electron band structure, if the BR of a topologically trivial insulator satisfies the condition given in Eq.~\ref{eq:OAI}, it still can be diagnosed as an orbital-selected obstructed atomic insulator (OOAI) if Eq.~\ref{eq:OAI} \emph{necessarily} contains an irrep $\rho_{\alpha_j}^i$ (with $N_{i,j} > 0$) which \emph{cannot} be induced from the irrep or irreps that correspond to electronic orbitals in the outer-shell of the atoms that sit at ${\alpha_j}$. 
However, in phonon band structures, a phonon BR satisfying Eq.~\ref{eq:OAI} is an OOAI if Eq. \ref{eq:OAI} \emph{necessarily} contains an irrep $\rho_{\alpha_j}^i$ that \emph{is not} contained in the vector representation (\ie the representation of the $p$ orbitals) or, equivalently, $B$ is not a linear combination of BRs included in the mechanical representations of the occupied Wyckoff positions.

The topologically trivial insulators that are not diagnosed as either OAIs or OOAIs are referred to as atomic insulators (AIs) although a further even finer classification is possible. In Ref. \cite{xu2021three}, it was argued that both OAI and OOAI phases of a topologically trivial insulator can be identified by calculating the three-dimensional real space invariants (3D RSIs). The RSI indices are sets of indicators that show the possibility (or not) of moving charge centers (atomic orbitals) along different Wyckoff positions in a crystal system in an adiabatic process, in which the symmetry group does not change \cite{song2020}. In Ref. \cite{xu2021three}, we have generalized the RSI indices to 3D crystalline structures of the 1,651 Shubnikov space groups (SSGs) with and without SOC. 
We derived and tabulated their expressions in terms of the multiplicities of the momentum-space irreps on the \webBCS\ tools \href{http://www.cryst.ehu.es/cryst/RSI}{RSIsg} and  \href{http://www.cryst.ehu.es/cryst/MagRSI}{RSImag}, and each RSI is related with a set of Wyckoff positions $\{\alpha\}$.
For a given band structure, the RSIs are obtained by substituting its symmetry data vector into the RSIs' formula. A non-zero value of a given RSI indicates that there must be a set of orbitals located at the Wyckoff positions $\{\alpha\}$ related with this particular RSI or at Wyckoff positions of higher symmetry connected to $\{\alpha\}$.
If all the positions $\{\alpha\}$ and all the positions of higher symmetry connected to them are empty, we denote the topologically trivial insulator as an OAI and the Wyckoff positions $\{\alpha\}$ are named as \emph{obstructed Wannier charge centers} (OWCCs). 
Although the band topology in terms of EBRs of these systems are trivial, some Wannier charge centers of the occupied bands are out of the atoms. 
However, even when some Wyckoff positions in the subset $\{\alpha\}$ are occupied, we can distinguish two cases: if the irreps that enter into the definition of a non-zero-integer RSI do not correspond to orbitals in the outer-shell of the specific atoms sitting at $\{\alpha\}$, this RSI indicates an OOAI; otherwise, this RSI indicates the most trivial atomic insulator (AI).

\section{Phonon material databases}\label{app:phonondatabases}
In this section, we introduce the two materials databases \webMP~and \webphonondb(PhononDB@kyoto-u) that we rely on for the high-throughput search of topological, OABR  and OOABR phonon bands.

\subsection{Phonon materials on \webMP}\label{app:materialproject}

The phonon materials database on \webMP~\cite{petretto2018high} includes \TPDBNbrMPMaterials~MPIDs, which are inorganic solid semiconductors and insulators. The database provides the phonon band structures and derived quantities calculated by the {\it ab initio} software package {\it ABINIT} \cite{gonze2002first,gonze2009abinit,gonze2016recent} based on density functional perturbation theory (DFPT) \cite{RevModPhys.73.515}, used to calculate a perturbation of a finite momenta ($\bf q\neq 0$) within a primitive cell. In DFPT calculations, the second order derivatives of the total energies with respect to atomic displacements on a regular grid in the Brillouin zone, \ie the dynamical matrix, was stored in the ABINIT derivative database (DDB) file format. In the present work, we mainly use the material quantities stored in the DDB file to generate the force constant matrix for each material. 

\subsection{Phonon materials on \webphonondbshort}\label{app:phonondbkyoto}

\webphonondbshort~is a phonon database that provides phonon calculations and thermal property analyses for \TPDBNbrKyotoMaterials~MPIDs. Note that we only consider entries whose MPID are still valid in \webMP\ and whose space group is correctly reported.
All the crystal structures used in \webphonondbshort\ were originally obtained from \webMP\ and each crystal structure was relaxed and symmetrized before starting the phonon calculation.

The {\it ab initio} calculations of \webphonondbshort\ were performed using Density Functional Theory (DFT)~\cite{Hohenberg-PR64,Kohn-PR65} and its implementation in the {\it Vienna Ab-initio Simulation Package} (VASP)~\cite{vasp1,PhysRevB.48.13115}. The interaction between the ion cores and the valence electrons were treated using the projector augmented-wave method~\cite{paw1}, and the generalized gradient approximation (GGA) with the Perdew-Burke-Ernzerhof (PBE) type exchange-correlation potential was adopted \cite{PhysRevLett.77.3865}. As opposed to the phonon calculations (with DFPT method) on \webMP, phonon calculations on \webphonondbshort\ were performed with the finite difference method by constructing a commensurate supercell, which can only evaluate the perturbation dispersion around the $\Gamma$ point ($\bf q=0$). Using this method, the momenta located away from the $\Gamma$ point in the first Brillouin zone are mapped into the $\Gamma$ point of the first Brillouin zone of the supercell. In this work, we mainly use the sets of force, \ie the FORCE\_SETS files obtained from the finite difference method, to generate the force constant matrix for each material on \webphonondbshort. 

\section{High-throughput calculations of phonon irreps and topological classifications of phonon bands}\label{app:highthroughput}
In this section, we detail the method used in the calculation of the irreps of phonon bands and the followed workflow for the diagnosis of topological and OABR phonon bands.

\subsection{Irreps of phonon bands}\label{app:phononirrep}
The phonon spectrum (or phonon band structure) and the symmetry properties of phonon modes of a crystalline material are determined by the phonon Hamiltonian, namely the dynamical matrix in reciprocal space or the force constant matrix in real space. As explained in section \ref{app:materialproject}, the material databases \webMP\ and \webphonondbshort\ provide the force constant matrix obtained from {\it ab initio} calculations. Taking these data as starting point, we have calculated the phonon band structure and irreps at all the high-symmetry $q$ vectors, \ie the maximal momenta in the Brillouin zone as introduced in \siref{app:concepts}. The details of our calculations are explained in the following sections.

\subsubsection{Phonon frequencies and lattice vibration modes}\label{app:phononcal}
Given a phonon material in a space group $\mathcal{G}$, its force constant matrix is defined as the second-order derivative of the total energy with respect to the atomic displacements \cite{RevModPhys.73.515},
\begin{equation}
    \mathcal{F}_{\alpha\mu,\beta\nu}({\bf R}-{\bf R'})=\frac{\partial^2E}{\partial u_{\alpha\mu}({\bf R})\partial u_{\beta\nu}({\bf R'})} \mid_{u=0}\label{eq:fcmat}
\end{equation}
where {\bf R} and {\bf R}$^\prime$ represent translations of the Bravais lattice and $u_{\alpha\mu}(\bf R)$ ($u_{\beta\nu}(\bf R')$) refers to the atomic displacement of the atom labelled as $\mu$ ($\nu$) in the unit cell {\bf R} ({\bf R}$^\prime$) along the $\alpha$($\beta$) direction. The dynamical matrix is the mass-reduced discrete Fourier transform of the force constant matrix,

\begin{equation}
    \mathcal{D}_{\alpha\mu,\beta\nu}({\bf q})=\frac{1}{N_R\sqrt{M_{\mu}M_{\nu}}}\sum_{\bf R} e^{i \bf q\cdot R} \mathcal{F}_{\alpha\mu,\beta\nu}({\bf R})\label{eq:dmat}
\end{equation}
where $M_{\mu}$ is the mass of atom $\mu$ and whose basis is defined as
\begin{equation}
    \ket{u_{\alpha\mu}({\bf q})}= \frac{1}{\sqrt{N_R}}\sum_{\bf R}e^{-i{\bf q \cdot {\bf R}}} \ket{u_{\alpha\mu}({\bf R})} =\frac{1}{\sqrt{N_R}}\sum_{\bf R}e^{-i{\bf q \cdot {\bf R}}} \ket{u_{\alpha}({\bf R}+\tau_{\mu})},
\end{equation}
and $\tau_{\mu}$ is the position of atom $\mu$ in the unit cell ${\bf R}$.

For a given ${\bf q}$-vector, the eigenvalues of the dynamical matrix $\mathcal{D}({\bf q})$ give the squares of phonon frequencies $\omega^2_n({\bf q})$ ($n=1,2,...,3N$ and N is the number of atoms in one unit cell),
\begin{equation}
    \mathcal{D}({\bf q}) \ket{\phi_n^{\bf q}}= \omega^2_n({\bf q}) \ket{\phi_n^{\bf q}}\label{eq:phonon}
\end{equation}
 Since the dynamical matrix $\mathcal{D}({\bf q})$ is Hermitian, the eigenvalues $\{\omega^{2}_n({\bf q})\}$ are real. The eigenvector $\ket{\phi_n^{\bf q}}$ characterizes the lattice vibrational mode of frequency $\omega_n({\bf q})$ at $\bf q$.

Using the force constant matrices as provided on the two databases, the eigenvalue problem in Eq. \ref{eq:phonon} can be solved with the {\it Analysis of Derivative DataBase} (\href{https://github.com/abinit/abinit/blob/master/src/98_main/anaddb.F90}{Anaddb}) code (for \webMP) and the {\it phononpy} (version 2.14) package \cite{phonopy} (for \webphonondbshort). In the high-throughput calculations, we have also computed the phonon band structure along high-symmetry paths and density of states for each entry of the two databases.

\subsubsection{Non-analytical correction to the dynamical matrix}\label{app:NAC}

In polar semiconductors, whose positive and negative ions lie on separate planes, the long range macroscopic electric field induced by long wavelength (with $\bf q \rightarrow 0$) longitudinal optical (LO) phonons is non-negligible \cite{RevModPhys.73.515}. The coupling between LO phonons and electric fields result in the difference in frequency between LO and transverse optical (TO) phonons at the Brillouin zone centre, namely the LO-TO splitting. 

To quantitatively characterize the LO-TO splitting of phonon frequencies, we need an additional non-analytical term correction (NAC) in the dynamical matrix. The origin of this term is the long range dipole-dipole interaction  given by \cite{gonze1997dynamical,wang2010mixed},
\begin{equation}
    \mathcal{D}^{NA}_{\alpha\mu,\beta\nu}({\bf q \rightarrow 0})= \frac{4\pi e^2}{\Omega\sqrt{M_{\mu}M_{\nu}}} \frac{(\bf q \cdot Z^{*}_{\mu})_{\alpha}(\bf q \cdot Z^{*}_{\nu})_{\beta}}{\bf q \cdot \epsilon_{\infty} \cdot q} \label{eq:NAC}
\end{equation}

where $\Omega$ is the volume of one unit cell, and $\epsilon_{\infty}$ is the high frequency static dielectric tensor.  $Z^{*}_{\mu}$ is the Born effective charge tensor of atom $\mu$ in one unit cell, which describes the polarization ($\Vec{\mathcal{P}}$) induced by the displacement of atom $\mu$ ($\Vec{\mathcal{\tau}}_{\mu}$) under the condition of zero electric field ($\Vec{E}=0$) and defined as,
\begin{equation}
    Z^{*}_{\mu,\alpha\beta} = \Omega \frac{\partial\mathcal{P}_{\beta}}{\partial\tau_{\mu,\alpha}}\mid_{\Vec{E}=0}
\end{equation}
where $\alpha,\beta=x,y,z$.

Hence, the total dynamical matrix for $\bf q \rightarrow 0$ is expressed as,
\begin{equation}
    \mathcal{D}^{Total}({\bf q \rightarrow 0}) = \mathcal{D}({\bf q}) + \mathcal{D}^{NA}(\bf q)
\end{equation}
where $\mathcal{D}(\bf q)$ is the analytical term in Eq. \ref{eq:dmat}. Note that the NAC term is 0 at $\bf q=0$.

After the introduction of the NAC term in our high-throughput calculations, we have also calculated the phonon band structures, phonon density of states, and identified the corresponding band irreps and band topology for all the polar semiconductors in the two databases. 

\subsubsection{Symmetry eigenvalues of phonon bands}\label{app:trace}

Given a space group $\mathcal{G}$, a general symmetry operator $g \in \mathcal{G}$ is defined as $g = \{\mathcal{R}|{\tau}\}$ where $\mathcal{R}$ is a symmetry operation of the point group and $\tau$ represents a translation. For symmorphic space groups expressed in a primitive basis and a proper election of the origin, $\tau$ has integer components in all symmetry operations. However, in non-symmorphic space groups, the translation $\tau$ of some symmetry operations takes fractional values. In reciprocal space, the little group of a {\bf q}-point is $\mathcal{G}_{\bf q} \subset \mathcal{G}$, whose elements $\tilde{g}=\{\mathcal{\tilde{R}}|\tilde{\tau}\}$ satisfy the following condition, 
\begin{equation}
   \mathcal{\tilde{R}} {\bf q} \cdot = {\bf q} +{\bf K}\label{eq:lg}
\end{equation}  
where ${\bf K}$ belongs to the reciprocal lattice.

In real materials, the dynamical matrix is written in the basis of $\ket{u_{\alpha\mu}({\bf q})}$. When a general symmetry operator $\tilde{g}=\{\mathcal{\tilde{R}}\mid\tilde{\tau}\}$ acts on $\ket{u_{\alpha\mu}({\bf q})}$, we have,

\begin{equation}
    \begin{aligned}
    \{\mathcal{\tilde{R}}|\tilde{\tau}\} \ket{u_{\alpha\mu}({\bf q})}
    &=\frac{1}{\sqrt{N_R}}\sum_R e^{-i{\bf q}\cdot {\bf R}} \ket{u_{\tilde{\mathcal{R}}\alpha}(\mathcal{\tilde{R}}{\bf R}+\tilde{\mathcal{R}}\tau_{\mu}+\tilde{\tau})} \\
    &=\frac{1}{\sqrt{N_R}}\sum_R e^{-i(\tilde{\mathcal{R}}{\bf q})\cdot \tilde{\mathcal{R}}{\bf R}} \ket{u_{\tilde{\mathcal{R}}\alpha}(\mathcal{\tilde{R}}{\bf R}+\tilde{\mathcal{R}}\tau_{\mu}+\tilde{\tau})} \\
    &=\frac{1}{\sqrt{N_{R^\prime}}}\sum_{R^\prime,\mu\prime} e^{-i{\bf q} [{\bf R^\prime}+\tau_{\mu^\prime}-\tilde{\mathcal{R}}\tau_{\mu}-\tilde{\tau}]} \ket{u_{\tilde{\mathcal{R}}\alpha}({\bf R^\prime}+\tau_{\mu^\prime})} \\
    &=\sum_{\mu\prime,\alpha\prime}\mathcal{O}_{\alpha^{\prime}\alpha}e^{-i{\bf q}[\tau_{\mu^\prime}-\tilde{\mathcal{R}}\tau_{\mu}-\tilde{\tau}]}  \underline{\frac{1}{\sqrt{N_{R^\prime}}}\sum_{R^\prime}e^{-i{\bf q}{\bf R^\prime}} \ket{u_{\alpha\prime}({\bf R^\prime}+\tau_{\mu\prime})}} \\
    &= \sum_{\mu^\prime,\alpha^\prime} \mathcal{O}_{\alpha^{\prime}\alpha} \mathcal{A}_{\mu^{\prime}\mu} \underline{\ket{u_{\alpha^{\prime}\mu^{\prime}}({\bf q})}}\\
    \end{aligned}
\end{equation}
where we have taken $\mathcal{\tilde{R}}{\bf R}+\tilde{\mathcal{R}}\tau_{\mu}+\tilde{\tau}={\bf R^\prime}+\tau_{\mu\prime}$, and $\ket{u_{\alpha^{\prime}\mu^{\prime}}({\bf q})}=\frac{1}{\sqrt{N_{R^\prime}}}\sum_{R^\prime}e^{-i{\bf q}{\bf R^\prime}} \ket{u_{\alpha\prime}({\bf R^\prime}+\tau_{\mu\prime})}$. $\mathcal{O}$ and $\mathcal{A}$ are the transformation matrices of the vibrational directions and atomic positions under the operation $\tilde{g}$, respectively. 
As derived above, the element of $\mathcal{A}$ is expressed as,
\begin{equation}
    \mathcal{A}_{\mu^\prime\mu}=
    \left\{
    \begin{array}{ll}
         e^{-i {\bf q}\cdot [\tau_{\mu\prime}-(\tilde{\mathcal{R}}\tau_{\mu}+\tilde{\tau})]}, & \text{when~} \tau_{\mu\prime}=(\tilde{\mathcal{R}}\tau_{\mu}+\tilde{\tau}) \text{~mod~} {\bf R}  \\
         0, & \text{otherwise.} \\
    \end{array}
    \right.
\end{equation}

The matrix $\mathcal{O}$ is actually the representation of $\mathcal{\tilde{R}}$ under the basis of Cartesian coordinates, \ie $[\mathcal{\tilde{R}}]_{c}$. However, in the standard setting of the space group $\mathcal{G}$, the point group operations $\mathcal{\tilde{R}}$ are expressed in the basis of the lattice of translations given by three vectors ${\bf a_{i=1,2,3}}$ which are, in general, not orthogonal. If we represent the symmetry operation in the basis of the lattice as $[\mathcal{\tilde{R}}]_{\ell}$, the transformation between the two matrices is given by,
\begin{equation}
    \mathcal{O}=[\mathcal{\tilde{R}}]_{c} =  U[\mathcal{\tilde{R}}]_{\ell}U^{-1}
\end{equation}
where the transformation matrix is
\begin{equation}
    U=    \left[
    \begin{array}{l}
         {\bf a_1} \\
         {\bf a_2} \\
         {\bf a_3} \\
    \end{array}
    \right]^T,
\end{equation}
and ${\bf a_{i=1,2,3}}$ are the three lattice vectors expressed in Cartesian coordinates.

Given a phonon mode eigenfunction $\ket{\phi_{n}({\bf q})}$, whose little group at ${\bf q}$ is $\mathcal{G}_{\bf q}$, the expectation value (trace) of the symmetry operator $\tilde{g}\in\mathcal{G}_{\bf q}$ can be calculated as $\braket{\phi_n({\bf q})\mid\mathcal{O}\bigotimes\mathcal{A}\mid\phi_n({\bf q})}$.

Based on the above derivation, for each phonon material in \webMP\ and \webphonondbshort, we have evaluated the symmetry eigenvalues for all the phonon modes at all the maximal ${\bf q}$-vectors both with and without the NAC term. The symmetry eigenvalues are collected and stored in the \emph{trace.txt} format, which was originally used in Refs. \cite{vergniory_complete_2019,xu2020high} and on the \webBCS\ to identify the band irreps (or equivalently, the symmetry data vector) and band topologies of electronic bands with and without spin-orbit coupling.

\subsection{Topology of phonon bands}\label{app:topphonon}

By feeding the \emph{trace.txt} files into the tool of \href{https://www.cryst.ehu.es/cryst/checktopologicalphonos.html}{Check Topological Phonons} hosted in the \webBCS\ and described in the next section, we have successfully identified the irreps of phonon bands at all the maximal high-symmetry $\bf q$ points for \TPDBNbrMPMaterials\ phonon materials on \webMP\ and \TPDBNbrKyotoMaterials\ phonon materials on \webphonondbshort. 

In polar semiconductors, the NAC term in Eq. \ref{eq:NAC} makes the phonon dispersion discontinuous when approaching to the $\Gamma$ point along different directions ($\omega({\bf q}_1 \rightarrow 0)\neq \omega({\bf q}_2 \rightarrow 0)$), which is referred to as the LO-TO splitting \cite{RevModPhys.73.515}. For example, the phonon spectrum of \ce{ZnGeN2} [\mpidweb{2979} SG 33 (\sgsymb{33})] is discontinuous when approaching to the $\Gamma$ point along $Y\Gamma$ and $Z\Gamma$ directions. 
In particular cases, the energy order of two bands (of different band representations) can be inverted by the LO-TO splitting at a finite momentum close to the $\Gamma$ point and the order in energy of the two irreps can be different from the order at the $\Gamma$ point (it can happen due to of the discontinuity). In the TQC high-throughput method\cite{bradlyn_topological_2017,vergniory_complete_2019,MTQC,xu2020high,Vergniory2021}, as we evaluate the compatibility relations and band topologies using only the maximal {\bf q}-vectors, the potential band inversion originated by the LO-TO splitting can be missed and the 
topological classification might fail. 
In this work, using the symmetry data vectors of phonon bands, we adopt the TQC method \cite{bradlyn_topological_2017,vergniory_complete_2019,MTQC,xu2020high,Vergniory2021} to perform the high-throughput topological classification for phonon bands with and without NAC, where the classification with NAC is just for reference. 

{\bf Compatibility relations}
Unlike electronic band structures, no Fermi level or chemical potential is defined in phonon band structures. Indeed, phonons are bosonic excitations and the number of phonon branches is always three times the number of atoms in a primitive unit cell. The concept of \emph{occupied valence bands} or empty bands (conduction bands in electronic terminology)makes no sense. In this work, we focus not only on the topology of phonon bands below a specific energy band gap, but also on the topology of each set of bands that are separated from the other sets of bands along all the high-symmetry paths by an energy gap, which is identified by solving the compatibility relations. We refer to this set as an \emph{isolated set of bands}. The symmetry data vector of such a set of bands satisfies all the compatibility relations in the sense explained in section \ref{sec:workflow} of the main text, and there are no symmetry-protected band crossings along the high-symmetry lines or planes in the BZ. If the compatibility relations are not satisfied at least in a high-symmetry line or plane, this set of bands must be connected with the bands above or below and the symmetry-protected band crossing occur at the high-symmetry points, lines or planes in the BZ. 
In our analysis we consider every \emph{minimal} set of isolated bands. An isolated set of bands is minimal when it cannot be split into different sub-bands which internally also satisfy the compatibility relations. In other words, the isolated set of bands cannot be decomposed into sets of isolated bands.

{\bf Topological classification}
In the high-throughput calculations, for each phonon band structure we first identify all the isolated sets and their corresponding symmetry data vectors $\{B_i\mid i=1,2,...,N_B\}$ (where $N_B$ is the number of isolated sets and $\{B_i\}$ are sorted in order of energy). As introduced in TQC \cite{bradlyn_topological_2017,MTQC} and \siref{app:concepts}, to identify the symmetry-based topology of an isolated set of bands, the symmetry data vector of an isolated set $B_i$ is expressed as a linear combination of \underline{the symmetry data vectors of EBRs} (\ie $[EBR]$) of the space group, \begin{equation}
    B_i = [EBR] \cdot [X]
\end{equation}
where $[X]$ is the vector of coefficients. 
Then, the topology of $B_i$ can be diagnosed as strong, fragile or trivial topology based on the following criteria,
\begin{itemize}
    \item[1.] Strong topology. $B_i$ cannot be expressed as a linear combination of the symmetry data vectors of EBRs with integer coefficients, \ie $[X]$ contains at least one non-integer component.
    \item[2.] Fragile topology. $B_i$ can be expressed as a linear combination of the symmetry data vectors of EBRs with integer coefficients and $[X]$ contains necessarily at least one negative component. Note that, in general, the EBRs form an over-complete basis of band representations and, then, the $[X]$ coefficient vector is not unique in general. The topology compound is diagnosed as fragile when, in every possible linear combination with integer components [X], at least one coefficient is negative.
    \item[3.] Trivial topology. $B_i$ can be expressed as a linear combination of the symmetry data vectors of EBRs with non-negative integer coefficients.
\end{itemize}

{\bf Topological indices} For each isolated set of bands, we have also calculated its topological indices or symmetry-based indicators \cite{bradlyn_topological_2017,watanabe2018structure,song_diagnosis_2018,SlagerSymmetry,MTQC} if they are defined in the related space group.  
For an isolated set of phonon bands that is diagnosed as strong topological, it has at least one nonzero topological index. 
As interpreted in Ref. \cite{song_diagnosis_2018}, all the non-zero topological indices defined in the spinless space groups (without spin-orbit coupling) necessarily correspond to topological semimetals, \ie the set of topological bands are not compatible with a gaped band structure in three dimensions and there are topological band crossing between this \emph{isolated} set and the isolated set of bands above or below. 
Compared with the symmetry-protected band crossings, the topological band crossings indicated by non-zero topological indices occur, in general, at generic momenta in the Brillouin zone.

In addition, for each isolated set, we have also calculated the \emph{cumulative} symmetry data vector and the corresponding topological indices for all the phonon bands up to it. Given an isolated set whose symmetry data vector is $B_i$, its related cumulative symmetry data vector is defined as $\tilde{\mathcal{B}}_i=\sum_{j\leq i}B_j$. Then, the topological band crossing can be identified by the cumulative topological indices using the following criteria,
\begin{itemize}
    \item If at least one of the cumulative topological indices of $\tilde{\mathcal{B}}_i$ is non-zero (non-trivial), there is a topological band crossing between the isolated sets $B_i$ and the band set above it ($B_{i+1}$).
    \item Similarly, if at least one of the cumulative topological indices of $\tilde{\mathcal{B}}_{i-1}$ is non-zero, there is a topological band crossing between the isolated sets $B_{i-1}$ and the band set above it ($B_{i}$).
    \item Otherwise, there is no topological band crossing between the isolated set $B_i$ and the other sets.
\end{itemize}

\subsection{Check Topological Phonons tool}\label{app:checktop}
The tool \href{https://www.cryst.ehu.es/cryst/checktopologicalphonos.html}{Check Topological Phonons}, hosted in the \webBCS, can perform the identification of the irreps at high-symmetry points in the first Brillouin zone, the check of the fulfillment of the compatibility relations and the topological classification of the phonon bands. This tool is almost identical to the previously implemented tool \href{https://www.cryst.ehu.es/cryst/checktopologicalmat.html}{Check Topological Mat.} for electronic structures, but adapted to phonons. Details of the algorithm used and workflow of the program can be found in refs. \cite{vergniory_complete_2019,Vergniory2021}. The new tool introduces extra checks on the complete set of phonon bands given in the \emph{trace.txt}. Unlike the electronic systems, the whole set of phonon bands is necessarily a linear combination of EBRs and it must be identified as trivial. The information given in the output of the program is also adapted to the analysis of phonons.

\subsection{3D RSIs of phonon band structure and the obstructed atomic phonon bands}\label{app:oaiphonon}

For a given phonon band structure of trivial topology, its BR can be expressed as a sum of EBRs, \ie the BRs induced from the irreps at the maximal Wyckoff positions, with non-negative integer coefficients ~\cite{bradlyn_topological_2017,po2017symmetry,MTQC}. 
Like the spinless electronic bands in crystalline materials, a BR of phonon bands is wannierizable and the corresponding Wannier functions (orbitals in electronic systems) are exponentially localized at a Wyckoff position of the crystal lattice. As introduced in \siref{app:OAI} and in Ref. \cite{xu2021three}, depending on the location and symmetry properties of the Wannier functions, a BR of phonon bands can be diagnosed as obstructed atomic BR (OABR), orbital-selected obstructed atomic BR (OOABR) or atomic BR. 
In the following sections , we describe the algorithms used to classify each topologically trivial isolated set of phonon bands and each cumulative set of bands of a trivial cumulative topology into the three categories OABR, OOABR or atomic BR, based on the calculation of the RSIs.

\subsubsection{OABR phonon bands}

Given a topologically trivial (isolated or cumulative) set of phonon bands, we can directly calculate its sigle-valued RSIs by substituting the corresponding symmetry data vector into the RSIs' formula, as tabulated on the \webBCS. If all the RSIs of the set are zero, the BR of this set of phonon bands is diagnosed as an atomic BR. 
Otherwise, if there are non-zero RSIs, we adopt the following strategies to check if the set of bands is an OABR.

\begin{enumerate}
    \item For a given material, we first identify its space group and the Wyckoff positions (WPs) $\{\alpha^{occupied}\}$ that are occupied by atoms.
    \item Then, for each non-zero RSI index $\delta(\{\alpha_{rsi}\})$ which is defined at the WPs $\{\alpha_{rsi}\}$, if at least one WP in $\{\alpha_{rsi}\}$ is occupied, the non-zero RSI index is compatible with an atomic BR. On the contrary, if all the WPs in $\{\alpha_{rsi}\}$ are empty, we go to the next step.
    \item For each of the non-zero RSIs which are defined at unoccupied WPs, as identified in the last step, we check whether these WPs $\{\alpha_{rsi}\}$ are maximal or non-maximal. 
    \begin{enumerate}
        \item If all the WPs in $\{\alpha_{rsi}\}$ are maximal, this RSI indicates an OABR. 
        \item If $\{\alpha_{rsi}\}$ include some non-maximal WPs and none of these non-maximal WPs (lines/planes) is connected to an occupied WP of higher symmetry, this RSI indicates an OABR.  Otherwise, this non-zero RSI index is compatible with an atomic BR.
    \end{enumerate}
    \item If at least one RSI index indicates an OABR in the previous step, we refer this set of phonon bands to as an OABR.
\end{enumerate}

Using the above method, we have performed the high-throughput screening of OABR phonon bands.

\subsubsection{OOABR phonon bands}\label{app:methodOOAI}

As introduced in \siref{app:concepts}, 
if the irreps (indicated by the non-zero RSIs) of a set of topologically trivial phonon bands are localized at the atomic positions, but the multiplicity of at least one of these irreps in the decomposition of the vector representation (the representation of $p$ orbitals) is zero, the set of phonon bands is an OOABR. 

In the classification of OOABR phonon bands, we first identify the set of irreps $\{\rho_i\}$ with non-zero multiplicity in the decomposition of the vector representation into irreps of the site-symmetry group of each WP. Then, we select the RSIs with non-zero value and that are referred to as \emph{trivial RSIs} (they do not identify the material as OABR). If none of the irreps included in the definition of one of these RSIs belongs to the subset $\{\rho_i\}$, the material is tagged as OOABR. In the momentum space this means that, in the linear combination of the symmetry data vector, at least one BR with non-zero multiplicity is not included in the decomposition of the mechanical representations of the occupied WPs.
Using this method, we have performed the high-throughput screening of OOABR phonon bands.

\section{Surface state calculations}\label{app:surfacestates}

In this Appendix, we detail the calculations of phonon surface states. The computational method that we used divides into three steps. First, for each MPID entry we obtain the FORCE\_CONSTANTS file based on the output files from VASP \cite{vasp1,VASP1996} and {\it phonopy} \cite{phonopy}, \ie the {\it FORCE\_SETS},  {\it phonopy.conf}, and {\it POSCAR-unitcell} files. Then, we transform the format of {\it FORCE\_CONSTANTS} files into the format of {\it wannier90\_hr.dat}, which is the tight-binding Hamiltonian of electronic bands that are generated by the Wannier90 package. We generate the {\it phonopyTB\_hr.dat} file using a python script {\it phonon\_hr.py} \cite{WU2017}. 
Finally, based on the {\it phonopyTB\_hr.dat} file, we compute the slab-structure surface states for each phonon material using the WannierTools \cite{WU2017} package. The Miller indices of the open-boundary surfaces and high-symmetry k-paths used in the slab calculations are detailed in the following subsections.
 
\subsection{Miller indices of the open-boundary surfaces}

In three-dimensional space groups, there are 14 different Bravais lattices, as tabulated in the first column of Table \ref{tab:surfacestat}.
For each type of Bravais lattice, the working lattice unit cell can be chosen as primitive or as conventional. A primitive unit cell contains a single lattice point and it is thus the unit cell with the smallest volume. In some cases, it is not easy to infer at first sight the full symmetry of the crystal lattice from the primitive lattice, and the conventional unit cell is preferred.
A conventional cell is a unit cell with the full symmetry of its corresponding space group and for some Bravais lattices it is also a primitive unit cell. In the other cases the conventional unit cell is centered ant it contains more than one lattice point. In the present work, we define the Miller indices of a surface in the conventional unit cell.

For a conventional unit cell of lattice vectors $\{\bf a_1, a_2, a_3\}$ and reciprocal lattice vectors $\{\bf b_1, b_2, b_3\}$, the surface of Miller indices $(hkl)$ is defined as a 2D plane of normal vector $\vec{b}_{hkl}=h\vec{b_1}+k\vec{b_2}+l\vec{b_3}$.  
The 2D surface of normal vector $\vec{b}_{hkl}$ is defined by two vectors $\vec{R}_1=\sum_{i=1}^{3}R_{1,i}\vec{a}_i$ and $\vec{R}_2=\sum_{i=1}^{3}R_{2,i}\vec{a}_i$ that are parallel to the 2D surface and perpendicular to $\vec{b}_{hkl}$. 

Since all our {\it phonopyTB\_hr.dat} files are written in a primitive unit cell of lattice vectors $\{\bf p_1, p_2, p_3\}$, we first transform the two vectors $\vec{R}_{1,2}$ into the primitive convention. For a given type of Bravais lattice, the transformations between lattice vectors of primitive and conventional unit cell are defined by a matrix $M$,
\begin{equation}
    ({\bf p_1, p_2, p_3})^T=M\cdot ({\bf a_1, a_2, a_3})^T
\end{equation}
where the $3\times3$ $M$ matrices for all the 14 Bravais lattices are tabulated in the second column of Table \ref{tab:surfacestat}. Then, the two vectors $\vec{R}_{1,2}$, as defined above, can be expressed in the primitive convention, 
\begin{align*}
        \vec{R}_{i=1,2}&=(R_{i,1},R_{i,2},R_{i,3})({\bf a_1, a_2, a_3})^T \\
        &=(R_{i,1},R_{i,2},R_{i,3}) M^{-1} ({\bf p_1, p_2, p_3})^T \\
        &=(\Tilde{R}_{i,1},\Tilde{R}_{i,2},\Tilde{R}_{i,3}) ({\bf p_1, p_2, p_3})^T.
\end{align*}
In the surface state calculations with the WannierTools package, we adopt the $R_{1,2}$ vectors in its primitive convention $\left(\begin{array}{ccc} \Tilde{R}_{1,1}&\Tilde{R}_{1,2}&\Tilde{R}_{1,3} \\ \Tilde{R}_{2,1}&\Tilde{R}_{2,2}&\Tilde{R}_{2,3}\end{array}\right)$.
For each type of Bravais lattice, we have defined three non-equivalent open-boundary surfaces, whose Miller indices and the corresponding $R_{1,2}$ vectors are tabulated in the $4^{th}$ to $8^{th}$ columns of Table \ref{tab:surfacestat}.

\subsection{High-symmetry paths of the surface Brillouin Zone}

For all the 2D open-boundary surfaces that are defined in Table \ref{tab:surfacestat}, there are five types of 2D Brillouin zones which correspond to five types of 2D Bravais lattices. In the following, for each case we define its high-symmetry paths that are used in the surface states calculation (the coordinates are expressed in the reciprocal basis of the primitive basis vectors):

\begin{enumerate}
    \item For 2D primitive Monoclinic and primitive Orthorhombic (MO) Bravais lattices, the k-path is:
    
    $\mathrm{\Gamma\left(0,0\right)\rightarrow X\left(\frac{1}{2},0\right)\rightarrow M\left(\frac{1}{2},\frac{1}{2}\right)\rightarrow Y\left(0,\frac{1}{2}\right)\rightarrow\Gamma\left(0,0\right)\rightarrow M\left(\frac{1}{2},\frac{1}{2}\right)}$. We label this case as $k-MO$ in Table \ref{tab:surfacestat}.
    \item For 2D primitive Tetragonal and centered Orthorhombic (TO) Bravais lattices, the k-path is:
    
    $\mathrm{\Gamma\left(0,0\right)\rightarrow X\left(\frac{1}{2},0\right)\rightarrow M\left(\frac{1}{2},\frac{1}{2}\right)\rightarrow\Gamma\left(0,0\right)}$. We label this case as $k-TO$ in Table \ref{tab:surfacestat}.  
    \item For 2D Hexagonal (H) Bravais lattice, the  k-path is:
    
    $\mathrm{\Gamma\left(0,0\right)\rightarrow K\left(\frac{1}{3},\frac{1}{3}\right)\rightarrow M\left(\frac{1}{2},0\right)\rightarrow\Gamma\left(0,0\right)}$. We label this case as $k-H$ in Table \ref{tab:surfacestat}.
    
\end{enumerate}

The surface state calculations were performed by constructing a slab-structure finite-size unit cell, whose thickness is larger than 100 $\mathrm{\mathring{A}}$ along the open-boundary direction, which is large enough to avoid the coupling between the top and bottom surfaces.

\begin{table}
\caption[Definition of open-boundary surfaces and the corresponding high-symmetry paths for surface states calculation]{The definition of open-boundary surfaces and the corresponding high-symmetry paths for surface states calculation. In the first column we identify the 14 Bravais lattices in 3D. The second column gives the transformation matrix between the lattice vectors of primitive and conventional unit cells. The third column tabulates the schematic diagram for all the 14 Bravais lattices. The $R_{1,2}$ matrices of primitive convention that are used to define the lattice vector of open-boundary surfaces and the high-symmetry k-paths on the surface Brillouin zone are tabulated in the 4-8th columns. }\label{tab:surfacestat}
\begin{ruledtabular} %
\begin{tabular}{p{1.8cm}<{ \centering}p{1.5cm}<{ \centering}p{2.4cm}<{ \centering}p{1.5cm}<{ \centering}p{1.5cm}<{ \centering}p{1.5cm}<{ \centering}p{2cm}<{ \centering}p{2cm}<{ \centering}}
Bravais lattice  & Lattice Vector   & Lattice & (100) Surface &(010) Surface & (001) Surface & (110) Surface & (111) Surface\tabularnewline
\hline 
Primitive Triclinic & 
$\left(\begin{array}{ccc} 1 & 0 & 0\\ 0 & 1 & 0\\ 0 & 0 & 1 \end{array}\right)$ & 
\begin{minipage}{0.1\textwidth} \includegraphics[width=30mm]{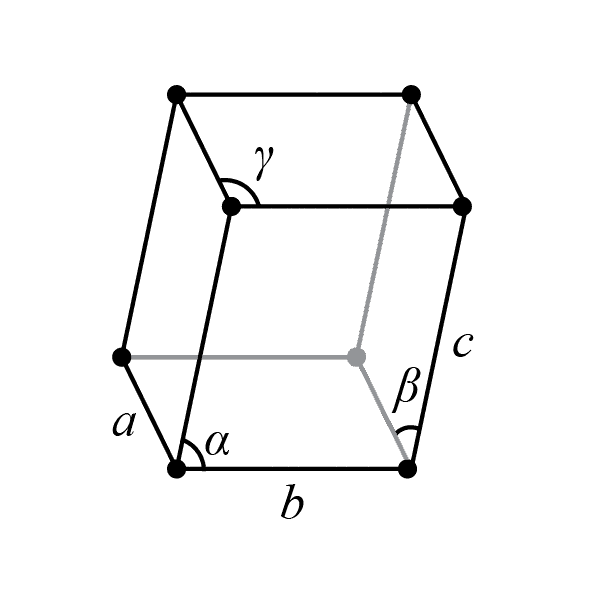}\end{minipage} & 
$\begin{array}{c} k-MO  \\ \left(\begin{array}{ccc} 0 & 1 & 0\\ 0 & 0 & 1 \end{array}\right) \end{array} $&  
$\begin{array}{c} k-MO  \\ \left(\begin{array}{ccc} 0 & 0 & 1\\ 1 & 0 & 0 \end{array}\right) \end{array}$ &
$\begin{array}{c} k-MO  \\ \left(\begin{array}{ccc} 1 & 0 & 0\\ 0 & 1 & 0 \end{array}\right) \end{array}$ & - & -
\tabularnewline

\hline 
Primitive Monoclinic & 
$\left(\begin{array}{ccc} 1 & 0 & 0\\ 0 & 1 & 0\\ 0 & 0 & 1 \end{array}\right)$ & 
\begin{minipage}{0.1\textwidth} \includegraphics[width=30mm]{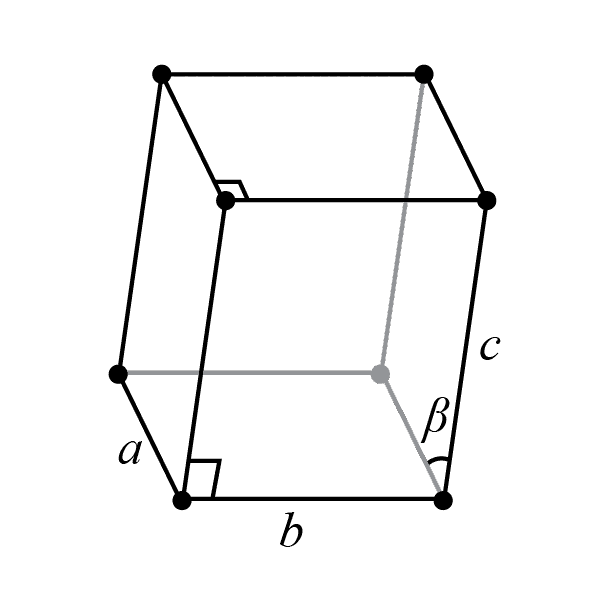}\end{minipage} & 
$\begin{array}{c} k-MO \\ \left(\begin{array}{ccc} 0 & 1 & 0\\ 0 & 0 & 1 \end{array}\right)\end{array}$ &  
$\begin{array}{c} k-MO \\ \left(\begin{array}{ccc} 0 & 0 & 1\\ 1 & 0 & 0 \end{array}\right)\end{array}$ &
$\begin{array}{c} k-MO \\ \left(\begin{array}{ccc} 1 & 0 & 0\\ 0 & 1 & 0 \end{array}\right)\end{array}$ & - & -
\tabularnewline
\hline 
Base-centered Monoclinic & 
$\left(\begin{array}{ccc} \frac{1}{2} & \frac{1}{2} & 0\\ -\frac{1}{2} & \frac{1}{2} & 0\\ 0 & 0 & 1 \end{array}\right)$ & 
\begin{minipage}{0.1\textwidth} \includegraphics[width=30mm]{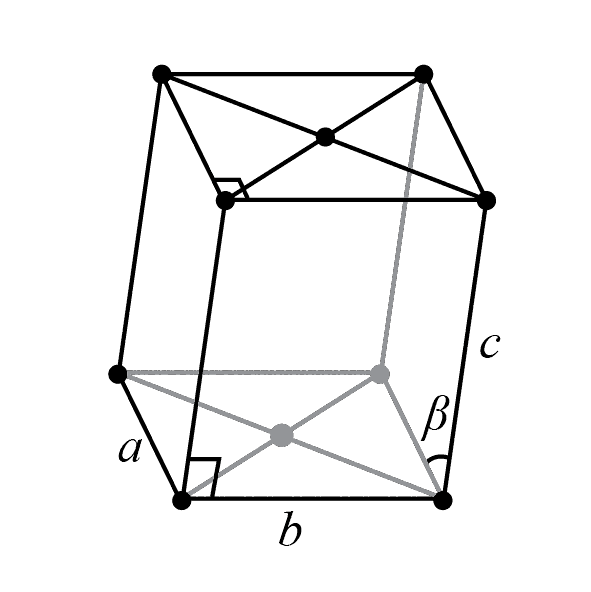}\end{minipage} & 
$\begin{array}{c} k-MO \\ \left(\begin{array}{ccc} 1 & 1 & 0\\ 0 & 0 & 1 \end{array}\right)\end{array}$ &  
$\begin{array}{c} k-MO \\ \left(\begin{array}{ccc} 1 & -1 & 0\\ 0 & 0 & 1 \end{array}\right)\end{array}$ &
$\begin{array}{c} k-MO \\ \left(\begin{array}{ccc} 1 & 0 & 0\\ 0 & 1 & 0 \end{array}\right)\end{array}$ & - & -
\tabularnewline
\hline 
Primitive  Orthorhombic & 
$\left(\begin{array}{ccc} 1 & 0 & 0\\ 0 & 1 & 0\\ 0 & 0 & 1 \end{array}\right)$ & 
\begin{minipage}{0.1\textwidth} \includegraphics[width=30mm]{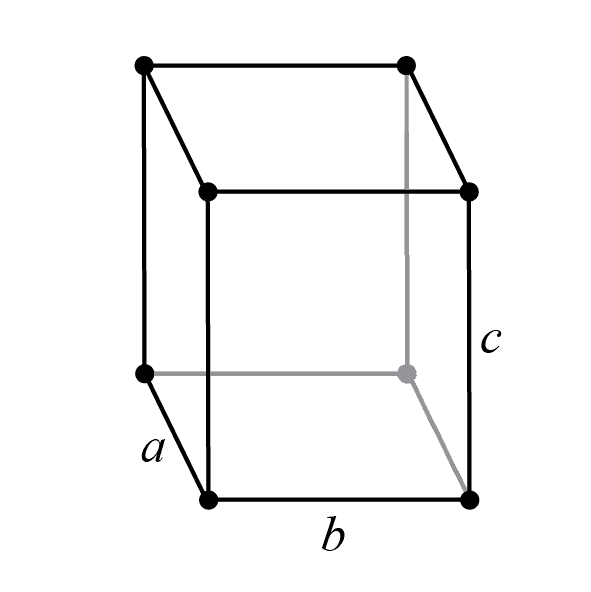}\end{minipage} & 
$\begin{array}{c} k-MO \\ \left(\begin{array}{ccc} 0 & 1 & 0\\ 0 & 0 & 1 \end{array}\right)\end{array}$ &  
$\begin{array}{c} k-MO \\ \left(\begin{array}{ccc} 0 & 0 & 1\\ 1 & 0 & 0 \end{array}\right)\end{array}$ &
$\begin{array}{c} k-MO \\ \left(\begin{array}{ccc} 1 & 0 & 0\\ 0 & 1 & 0 \end{array}\right)\end{array}$ & - & -
\tabularnewline
\hline 
Base-centered   Orthorhombic & 
$\left(\begin{array}{ccc} \frac{1}{2} & \frac{1}{2} & 0\\ -\frac{1}{2} & \frac{1}{2} & 0\\ 0 & 0 & 1 \end{array}\right)$ & 
\begin{minipage}{0.1\textwidth} \includegraphics[width=30mm]{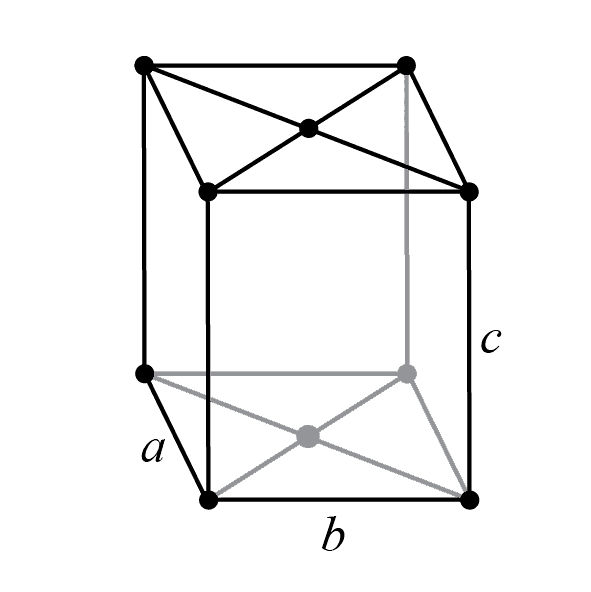}\end{minipage} & 
$\begin{array}{c} k-MO \\ \left(\begin{array}{ccc} 1 & 1 & 0\\ 0 & 0 & 1 \end{array}\right)\end{array}$ &  
$\begin{array}{c} k-MO \\ \left(\begin{array}{ccc} 1 & -1 & 0\\ 0 & 0 & 1 \end{array}\right)\end{array}$ &
$\begin{array}{c} k-MO \\ \left(\begin{array}{ccc} 1 & 0 & 0\\ 0 & 1 & 0 \end{array}\right)\end{array}$ & - & -
\tabularnewline
\hline 
Body-centered   Orthorhombic & 
$\left(\begin{array}{ccc} -\frac{1}{2} & \frac{1}{2} & \frac{1}{2}\\ \frac{1}{2} & -\frac{1}{2} & \frac{1}{2}\\ \frac{1}{2} & \frac{1}{2} & -\frac{1}{2} \end{array}\right)$ & 
\begin{minipage}{0.1\textwidth} \includegraphics[width=30mm]{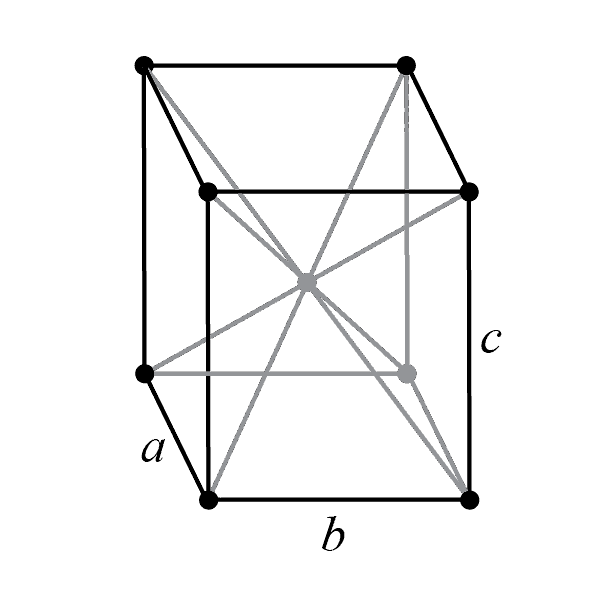}\end{minipage} & 
$\begin{array}{c} k-MO \\ \left(\begin{array}{ccc} 1 & 1 & 0\\ 1 & 0 & 1 \end{array}\right)\end{array}$ &  
$\begin{array}{c} k-MO \\ \left(\begin{array}{ccc} 0 & 1 & 1\\ 1 & 1 & 0 \end{array}\right)\end{array}$ &
$\begin{array}{c} k-MO \\ \left(\begin{array}{ccc} 0 & 1 & 1\\ 1 & 0 & 1 \end{array}\right)\end{array}$ & - & -
\tabularnewline
\hline 
Face-centered   Orthorhombic & 
$\left(\begin{array}{ccc} 0 & \frac{1}{2} & \frac{1}{2}\\ \frac{1}{2} & 0 & \frac{1}{2}\\ \frac{1}{2} & \frac{1}{2} & 0 \end{array}\right)$ & 
\begin{minipage}{0.1\textwidth} \includegraphics[width=30mm]{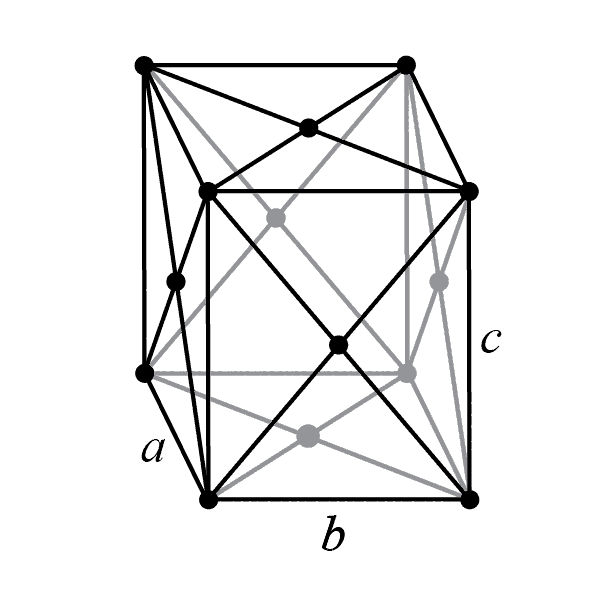}\end{minipage} & 
$\begin{array}{c} k-MO \\ \left(\begin{array}{ccc} 1 & 0 & 0\\ 0 & 1 & -1 \end{array}\right)\end{array}$ &  
$\begin{array}{c} k-MO \\ \left(\begin{array}{ccc} 0 & 1 & 0\\ 1 & 0 & -1 \end{array}\right)\end{array}$ &
$\begin{array}{c} k-MO \\ \left(\begin{array}{ccc} 0 & 0 & 1\\ 1 & -1 & 0 \end{array}\right)\end{array}$ & - & -
\tabularnewline
\end{tabular}
\end{ruledtabular}
\end{table}

\begin{table}
\caption{Continuation of Table \ref{tab:surfacestat}}\label{tab1}
\begin{ruledtabular} %
\begin{tabular}{p{1.8cm}<{ \centering}p{1.5cm}<{ \centering}p{2.4cm}<{ \centering}p{1.5cm}<{ \centering}p{1.5cm}<{ \centering}p{1.5cm}<{ \centering}p{2cm}<{ \centering}p{2cm}<{ \centering}}
Bravais lattice  & Lattice Vector   & Lattice & (100) Surface &(010) Surface & (001) Surface & (110) Surface & (111) Surface\tabularnewline
\hline 
Primitive  Tetragonal & 
$\left(\begin{array}{ccc} 1 & 0 & 0\\ 0 & 1 & 0\\ 0 & 0 & 1 \end{array}\right)$ & 
\begin{minipage}{0.1\textwidth} \includegraphics[width=30mm]{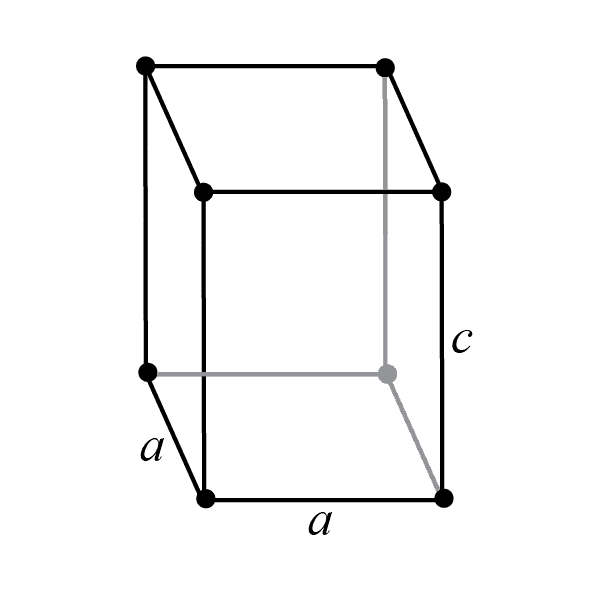}\end{minipage} & 
$\begin{array}{c} k-MO \\ \left(\begin{array}{ccc} 0 & 1 & 0\\ 0 & 0 & 1 \end{array}\right)\end{array}$ & - & 
$\begin{array}{c} k-TO \\ \left(\begin{array}{ccc} 1 & 0 & 0\\ 0 & 1 & 0 \end{array}\right)\end{array}$ &
$\begin{array}{c} k-MO \\ \left(\begin{array}{ccc} 1 & -1 & 0\\ 0 & 0 & 1 \end{array}\right)\end{array}$ & - 
\tabularnewline
\hline 
Body-centered  Tetragonal & 
$\left(\begin{array}{ccc} -\frac{1}{2} & \frac{1}{2} & \frac{1}{2}\\ \frac{1}{2} & -\frac{1}{2} & \frac{1}{2}\\ \frac{1}{2} & \frac{1}{2} & -\frac{1}{2} \end{array}\right)$ & 
\begin{minipage}{0.1\textwidth} \includegraphics[width=30mm]{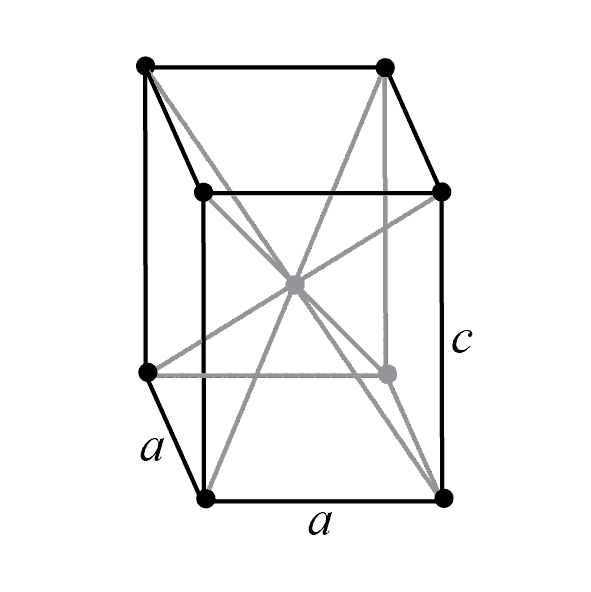}\end{minipage} & 
$\begin{array}{c} k-MO \\ \left(\begin{array}{ccc} 1 & 1 & 0\\ 1 & 0 & 1 \end{array}\right)\end{array}$ & - & 
$\begin{array}{c} k-TO \\ \left(\begin{array}{ccc} 0 & 1 & 1\\ 1 & 0 & 1 \end{array}\right)\end{array}$ &
$\begin{array}{c} k-MO \\ \left(\begin{array}{ccc} 0 & 1 & 0\\ 1 & 0 & 0 \end{array}\right)\end{array}$ & - 
\tabularnewline
\hline 
Rhombohedral & 
$\left(\begin{array}{ccc} \frac{2}{3} & -\frac{1}{3} & -\frac{1}{3}\\ \frac{1}{3} & \frac{1}{3} & -\frac{2}{3}\\ \frac{1}{3} & \frac{1}{3} & \frac{1}{3} \end{array}\right)$ & 
\begin{minipage}{0.1\textwidth} \includegraphics[width=30mm]{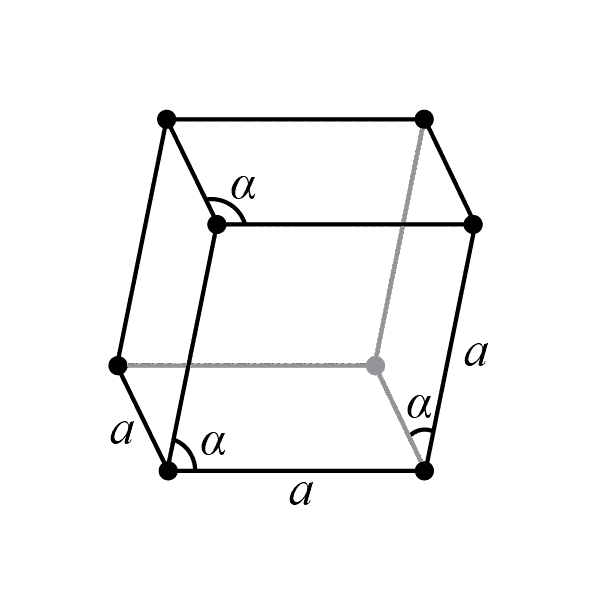}\end{minipage} & 
$\begin{array}{c} k-MO \\ \left(\begin{array}{ccc} -1 & 1 & 0\\ 1 & 1 & 1 \end{array}\right)\end{array}$ & - & 
$\begin{array}{c} k-TO \\ \left(\begin{array}{ccc} -1 & 0 & 1\\ 1 & -1 & 0 \end{array}\right)\end{array}$ &
$\begin{array}{c} k-MO \\ \left(\begin{array}{ccc} 0 & 1 & 1\\ 1 & 0 & 0 \end{array}\right)\end{array}$ & - 
\tabularnewline
\hline 
Hexagonal & 
$\left(\begin{array}{ccc} 1 & 0 & 0\\ 0 & 1 & 0\\ 0 & 0 & 1 \end{array}\right)$ & 
\begin{minipage}{0.1\textwidth} \includegraphics[width=30mm]{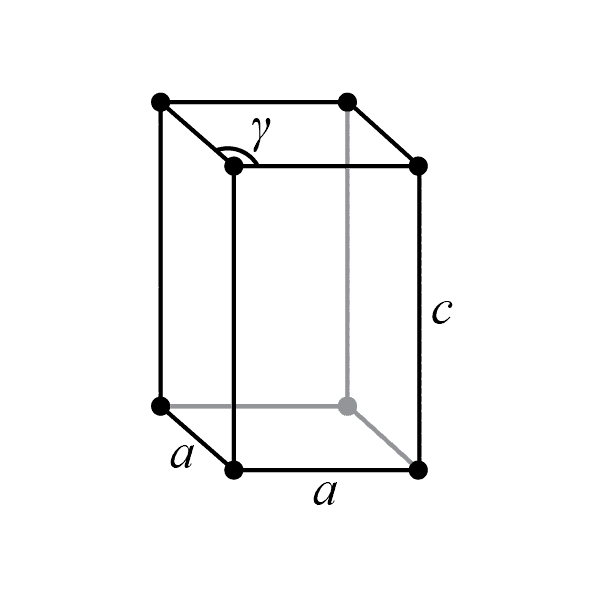}\end{minipage} & 
$\begin{array}{c} k-MO \\ \left(\begin{array}{ccc} 0 & 1 & 0\\ 0 & 0 & 1 \end{array}\right)\end{array}$ & - & 
$\begin{array}{c} k-H \\ \left(\begin{array}{ccc} 1 & 0 & 0\\ 0 & 1 & 0 \end{array}\right)\end{array}$ &
$\begin{array}{c} k-MO \\ \left(\begin{array}{ccc} 1 & -1 & 0\\ 1 & 0 & 0 \end{array}\right)\end{array}$ & - 
\tabularnewline
\hline 
Primitive   Cubic & 
$\left(\begin{array}{ccc} 1 & 0 & 0\\ 0 & 1 & 0\\ 0 & 0 & 1 \end{array}\right)$ & 
\begin{minipage}{0.1\textwidth} \includegraphics[width=30mm]{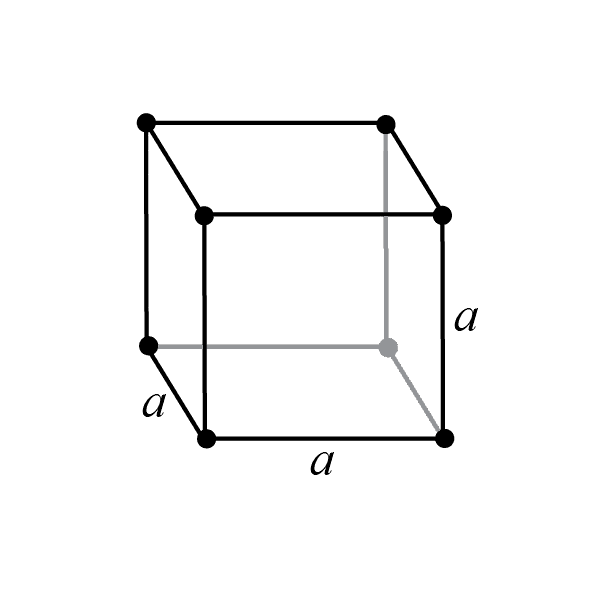}\end{minipage} & 
- & - & 
$\begin{array}{c} k-TO \\ \left(\begin{array}{ccc} 1 & 0 & 0\\ 0 & 1 & 0 \end{array}\right)\end{array}$ &
$\begin{array}{c} k-MO \\ \left(\begin{array}{ccc} 1 & -1 & 0\\ 0 & 0 & 1 \end{array}\right)\end{array}$ & 
$\begin{array}{c} k-H \\ \left(\begin{array}{ccc} 1 & -1 & 0\\ 0 & 1 & -1 \end{array}\right)\end{array}$  
\tabularnewline
\hline 
Body-centered   Cubic & 
$\left(\begin{array}{ccc} -\frac{1}{2} & \frac{1}{2} & \frac{1}{2}\\ \frac{1}{2} & -\frac{1}{2} & \frac{1}{2}\\ \frac{1}{2} & \frac{1}{2} & -\frac{1}{2} \end{array}\right)$ & 
\begin{minipage}{0.1\textwidth} \includegraphics[width=30mm]{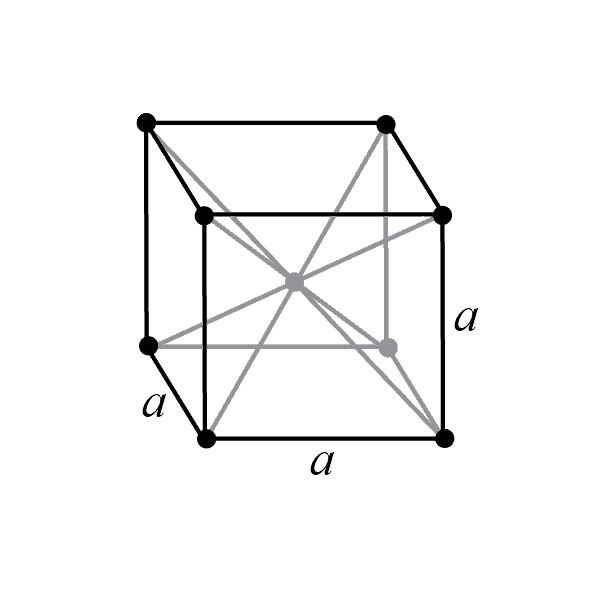}\end{minipage} & 
- & - & 
$\begin{array}{c} k-TO \\ \left(\begin{array}{ccc} 0 & 1 & 1\\ 1 & 0 & 1 \end{array}\right)\end{array}$ &
$\begin{array}{c} k-MO \\ \left(\begin{array}{ccc} 0 & 1 & 0\\ 1 & 0 & 0 \end{array}\right)\end{array}$ & 
$\begin{array}{c} k-H \\ \left(\begin{array}{ccc} 1 & -1 & 0\\ 0 & 1 & -1 \end{array}\right)\end{array}$  
\tabularnewline
\hline 
Face-centered   Cubic & 
$\left(\begin{array}{ccc} 0 & \frac{1}{2} & \frac{1}{2}\\ \frac{1}{2} & 0 & \frac{1}{2}\\ \frac{1}{2} & \frac{1}{2} & 0 \end{array}\right)$ & 
\begin{minipage}{0.1\textwidth} \includegraphics[width=30mm]{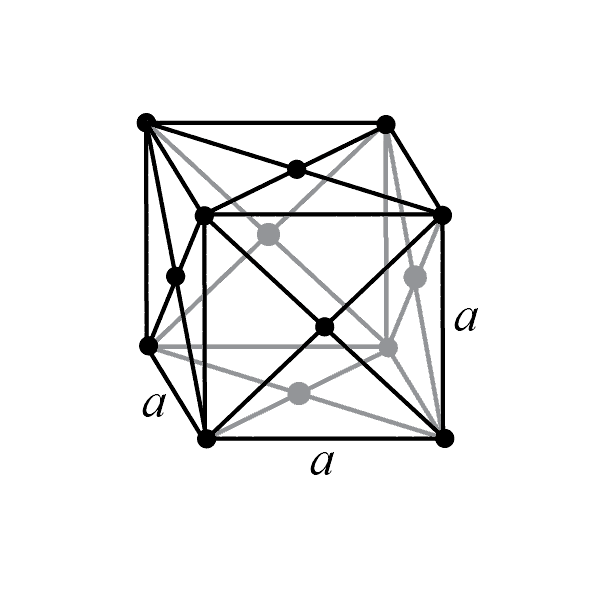}\end{minipage} & 
- & - & 
$\begin{array}{c} k-TO \\ \left(\begin{array}{ccc} 1 & -1 & 0\\ 0 & 0 & 1 \end{array}\right)\end{array}$ &
$\begin{array}{c} k-MO \\ \left(\begin{array}{ccc} 1 & -1 & 0\\ 1 & 1 & -1 \end{array}\right)\end{array}$ & 
$\begin{array}{c} k-H \\ \left(\begin{array}{ccc} 1 & -1 & 0\\ 0 & 1 & -1 \end{array}\right)\end{array}$  
\tabularnewline
\end{tabular}
\end{ruledtabular}
\end{table}

\section{Material examples of non-atomic phonon bands}\label{app:examples}

\subsection{Symmetry-indicated Weyl points in \ce{Al2ZnTe4}}\label{app:ssAlZnTe}

The non-centrosymmetric material \ce{Al2ZnTe4} of \mpidweb{7908} has a body-centered tetragonal lattice with SG 82 (\sgsymb{82}). Its full phonon band dispersion along high-symmetry paths, the band representations and band topologies indicated by the symmetry-based indicators are detailed on the \webTQCphonon(\mpidweb{7908}). 

The calculation based on Kyoto database shows that, the phonon bands (21 in total) of \ce{Al2ZnTe4} are divided into 9 isolated band sets (correspond to 9 cumulative band sets).
The calculation of topological indices show that all the cumulative band sets are diagnosed as topologically trivial (8 sets) and topological (1 set). Among the 8 topologically trivial sets, 3 sets are diagnosed as OABR by the single-valued (without spin-orbit coupling) RSIs in SG 82. In the following, we analyze in detail one of the topological band sets whose band indices are $\#1\sim18$.

\begin{figure}
    \centering
    \includegraphics[width=7.0in]{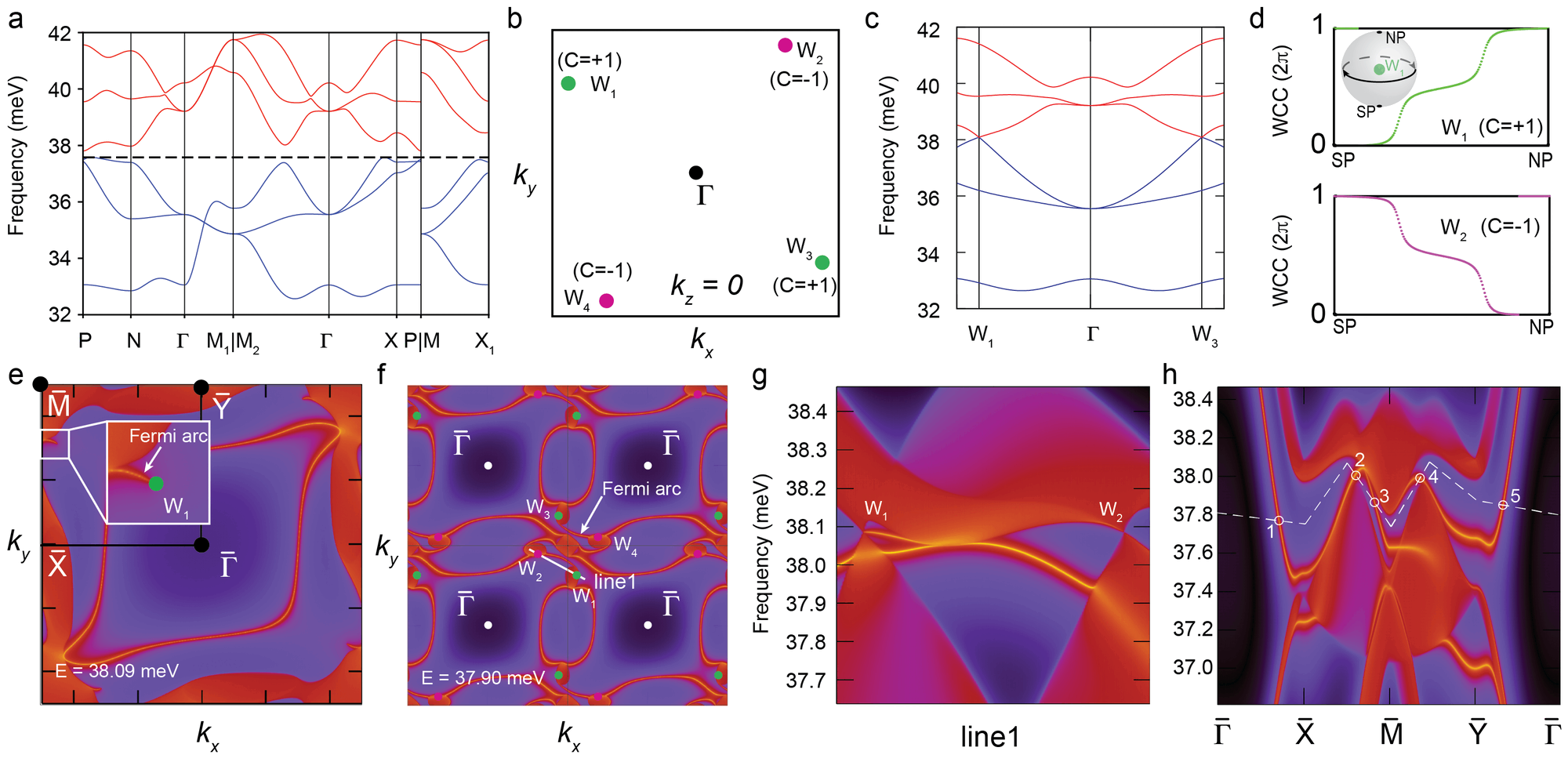}
    \caption{(a) Phonon band dispersion in an energy window around the topological band gap between the bands of indices 18 and 19, where the blue and red lines are the respective bands below and above the topological gap. (b) The distribution of Weyl points ($W_{1,2,3,4}$) on the $k_z=0$ slice of the bulk BZ. The green ($W_{1,3}$) and magenta ($W_{2,4}$) points represent the Weyl points of chirality +1 (C=+1) and -1 (C=-1), respectively. (c) Band dispersion along the line connecting Weyl points $W_1$ and $W_2$. (d) Identification of Weyl points $W_1$ and $W_2$ with Wilson-loop method. The Wilson-loop calculations are performed on a closed sphere enclosing $W_1$ (top panel) and $W_2$ (bottom panel) in the bulk BZ, where each point represents the Berry phase calculated by the integration of Berry connection along a weft of the sphere. The evolution of Wilson-loop is from the south pole (SP) to the north pole (NP) of the sphere. (e-h) Surface state calculations on the (001) surface of \ce{Al2ZnTe4}. (e) Energy contour ($E=38.09 meV$) of the surface states in the first surface BZ. Inset: Fermi arc connecting the Weyl point $W_1$. (f) Energy contour ($E=37.9 meV$) of the surface states in a $2\times 2$ surface BZ. (g) Surface state dispersion along the ``line1'' in (f), which connects the Weyl points $W_1$ and $W_2$. (h) Surface state dispersion along the high-symmetry lines $\bar{\Gamma}$-$\bar{X}$-$\bar{M}$-$\bar{Y}$-$\bar{\Gamma}$ enclosing a quarter of the surface BZ as shown in (e). By defining a curved energy level in the bulk energy gap, as represented by the dashed white lines, there is an odd number crossing points between between the surface states and the dashed lines, indicating an odd number of Weyl points in a quarter of the surface BZ.}
    \label{fig:7908ss}
\end{figure}

In the high-throughput calculations, symmetry properties of the phonon bands $\#1\sim18$ are characterized by a symmetry data vector which is,
\begin{equation}
    \scriptsize{B=3\Gamma_1+5\Gamma_2+5\Gamma_3\Gamma_4+4M_1+4M_2+5M_3M_4+18N_1+4P_1+4P_2+5P_3+5P_4+4PA_1+4PA_2+5PA_3+5PA_4+8X_1+10X_2}
    \label{eq:7908sdv}
\end{equation}
The $B$-vector in Eq. \ref{eq:7908sdv} satisfies all the compatibility relations in SG 82 but is not compatible with any atomic orbitals in the lattice of SG 82; the bands have to be topological. The topological band gap is shown in the band plotted in Fig. \ref{fig:7908ss}(a).
As defined in Refs. \cite{bradlyn_topological_2017,song_diagnosis_2018,po_symmetry-based_2017}, the symmetry-based indicator (or topological index) of SG 82 is
\begin{equation}
    \omega_{2,0} = N(\Gamma_{\xi=-1}) + N(M_{\xi=-1}) + N(X_{\eta=-1})~{\rm mod}~2= m(\Gamma_2) + m(M_2) + m(X_2)~{\rm mod}~2,
    \label{eq:SI82}
\end{equation}
where $\xi$ and $\eta$ are the symmetry eigenvalues of $S_4$ and $C_{2z}$, respectively. $N(K_{\xi(\eta)})$ is the number of bands of symmetry eigenvalue $\xi(\eta)$ at the $K$ point. $m(R)$ is the multiplicity of irrep $R$ of the related set of bands. By substituting the $B$-vector (Eq. \ref{eq:7908sdv}) into Eq. \ref{eq:SI82}, we have $\omega_{2,0}=1$. As interpreted in Ref. \cite{song_diagnosis_2018}, $\omega_{2,0}\times \pi$ is the Berry phase of a loop enclosing a quarter of the BZ at the $k_z=0$ plane. When $\omega_{2,0}=1$, it indicates 4 mod 8 Weyl points on the $k_z=0$ slice. Since the $S_4$ operation is an improper rotation, two Weyl points related by $S_4$ have opposite monopole charges (chirality). The identification and distribution of these four Weyl points on the $k_z=0$ slice are shown in Fig. \ref{fig:7908ss}(b-d). Surface state calculations and phonon ``Fermi arc'' states connecting the Weyl points are also analyzed and plotted in Fig. \ref{fig:7908ss}(e-h).

\subsection{Symmetry-indicated nodal line in ZnPbF$_6$}\label{app;ZnPbF6}

The centrosymmetric material \ce{ZnPbF6} of \mpidweb{13610} has a rhombohedral lattice with SG 148 (\sgsymb{148}). Its full phonon band dispersion along high-symmetry paths, the band representations and band topologies indicated by the symmetry-based indicators are detailed on the \webTQCphonon(\mpidweb{13610}). 

In summary, the phonon bands (24 in total) of \ce{ZnPbF6} are divided into 12 isolated band sets (correspond to 12 cumulative band sets). By calculating the topological indices defined in this space group, 7 of the cumulative band sets are diagnosed as topological with strong topology (9 sets) and one cumulative band set is classified as OOABR. In the following, we analyze in detail two of the topological band sets whose band indices are $\#1\sim3$ and $\#1\sim15$, respectively.

In the high-throughput calculations, symmetry properties of the phonon bands $\#1\sim3$ and $\#1\sim15$ are characterized by the symmetry data vectors $B_1$ and $B_2$, respectively, which are,
\begin{equation}
    B_1=\Gamma_1^{-} +\Gamma_2^{-}\Gamma_3^{-}+3 F_1^{-}+2L_1^{+}+L_1^{-}+T_1^{+}+T_2^{+}T_3^{+}
    \label{eq:13610sdv1}
\end{equation}
\begin{equation}
    B_2=2\Gamma_1^{+}+3\Gamma_1^{-} +2\Gamma_2^{+}\Gamma_3^{+}+3\Gamma_2^{-}\Gamma_3^{-}+ 6F_1^{+}+9F_1^{-}+8L_1^{+}+7L_1^{-}+3T_1^{+}+2T_1^{-}+3T_2^{+}T_3^{+}+2T_2^{-}T_3^{-}
    \label{eq:13610sdv2}
\end{equation}
The $B$-vectors in both Eq. \ref{eq:13610sdv1} and \ref{eq:13610sdv2} satisfy all the compatibility relations in SG 148 but are not compatible with any sum of atomic orbitals in the lattice of SG 148 so that it has strong topology. 

\begin{figure}
    \centering
    \includegraphics[width=6.9in]{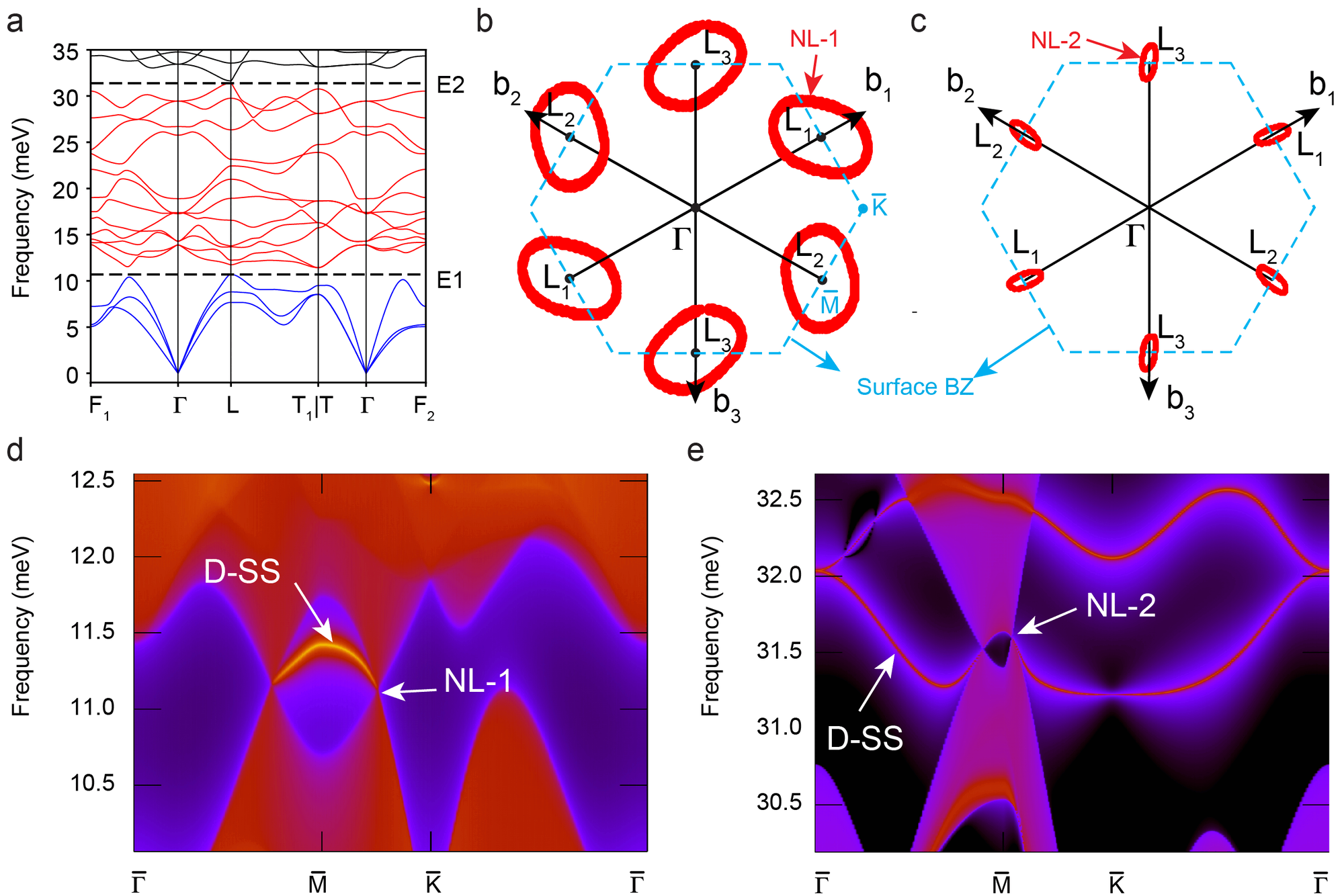}
    \caption{(a) Phonon band dispersion around the topological band gaps at $E1$ and $E2$. (b) Top view of the distribution of nodal lines at $E=E1$ (NL-1) centering at the $L$ points. (c) The same with (b) but for the nodal lines at $E=E2$ (NL-2). The six nodal lines in (b) or (c) are symmetry related by $C_3$ and inversion operations.
    (d) and (e) are the surface state calculations on the (001) surface whose surface BZ is plotted in (b). Due to the $\pi$-Berry phase associated with any loop that links with a nodal line, there are drumhead-like surface states connecting with the NL-1 (NL-2).}
    \label{fig:13610ss}
\end{figure}

The topological band gaps for the two B-vectors are shown in the band plotted in Fig. \ref{fig:13610ss}(a), \ie the dashed lines at E1 and E2.
As defined in Refs. \cite{bradlyn_topological_2017,song_diagnosis_2018,po_symmetry-based_2017}, there are two symmetry-based indicators (or topological indices) in SG 148 (without SOC). They are a $Z_2$ index on the $k_3=\pi$ plane $z_{2,3}$ and a $Z_4$ index $z_{4}$ as defined in the following equations,
\begin{equation}
\begin{aligned}
    z_{2,3} =& \sum_{K\in TRIM,k_3=\pi}^{} \frac{N_{-}(K)-N_{+}(K)}{2} ~{\rm mod}~2 \\
    =& \frac{M(L_1^{-})+M(T_1^{-})+2M(T_2^{-}T_3^{-})+2M(F_1^{-})-M(L_1^{+})-M(T_1^{+})-2M(T_2^{+}T_3^{+})-2M(F_1^{+})}{2}~{\rm mod}~2
\end{aligned}\label{eq:SI148z2}
\end{equation}

\begin{equation}
\begin{aligned}
    z_{4} =& \sum_{K\in TRIM}^{} \frac{N_{-}(K)-N_{+}(K)}{2} ~{\rm mod}~4 \\
    =& [\frac{M(\Gamma_1^{-})+2M(\Gamma_2^{-}\Gamma_3^{-})+3M(L_1^{-})+M(T_1^{-})+2M(T_2^{-}T_3^{-})+3M(F_1^{-})}{2} \\
    &-\frac{M(\Gamma_1^{+})+2M(\Gamma_2^{+}\Gamma_3^{+})+3M(L_1^{+})+M(T_1^{+})+2M(T_2^{+}T_3^{+})+3M(F_1^{+})}{2}]~{\rm mod}~4
\end{aligned}\label{eq:SI148z4}
\end{equation}
where $N(K_{\xi})$ is the number of bands of parity $\xi$ at the $K$ point. $m(R)$ is the multiplicity of irrep $R$ of the related set of bands. For $z_{2,3}$ in Eq. \ref{eq:SI148z2}, $K$ belongs to the four time-reversal invariant momenta (TRIM) on the $k_3=\pi$ plane. 
For $z_{4}$ in Eq. \ref{eq:SI148z4}, $K$ belongs to the eight time-reversal invariant momenta (TRIM) of the 3D BZ. 

By substituting the two $B$-vectors (Eqs. \ref{eq:13610sdv1} and \ref{eq:13610sdv2}) into Eqs. \ref{eq:SI148z2}-\ref{eq:SI148z4}, the indicators are $(z_{2,3},z_4)=(1,3)$ for both of the two sets of bands, \ie $\#1\sim3$ and $\#1\sim15$. As interpreted in Ref. \cite{song_diagnosis_2018}, the indicators $(z_{2,3},z_4)=(1,3)$ indicate an odd number of nodal lines centering at the $L$ point or equivalently at the $T$ point. Note that there are three equivalent $L$ points and one $T$ point in the first BZ.

For both of the two sets of bands, we find the nodal lines indicated by $(z_{2,3},z_4)=(1,3)$ are centering at the $L$ points, as shown in Fig. \ref{fig:13610ss}(b-c). Due to the $\pi$-Berry phase associated with any
loop that links with a nodal line \cite{Kim2015,Chen2015,song_diagnosis_2018}, drumhead-like surface states exist connecting with each nodal line on the surface.
Surface state calculations along (001) direction for the two topological sets are plotted in Fig. \ref{fig:13610ss}(d-e).

\subsection{Symmetry-indicated nodal line of $Z_2$-monopole charge in KBiO$_3$}\label{app:KBiO}
The centrosymmetric material \ce{KBiO3} of \mpidweb{29799} has a simple cubic lattice with SG 201 (\sgsymb{201}). Its phonon band dispersion along high-symmetry paths, the band representations and band topologies indicated by the symmetry-based indicators are detailed on the \webTQCphonon(\mpidweb{29799}). 

In summary, the phonon bands (180 in total) of \ce{KBiO3} are divided into 13 isolated band sets (correspond to 13 cumulative band sets). By calculating the topological indices defined in this space group, all the cumulative band sets are diagnosed as topologically trivial (8 sets) and topological with strong topologies (5 sets). Among the 8 topologically trivial sets of bands, 3 are diagnosed as OABR by the RSIs of SG 201.
In the following, we analyze in detail one of the topological cumulative band sets whose band indices are $\#1\sim84$.

In the high-throughput calculations, symmetry properties of the phonon bands $\#79\sim84$ are characterized by the symmetry data vectors $B$, which is
\begin{equation}
    \scriptsize{B=2\Gamma_1^{+}+4\Gamma_1^{-} +3\Gamma_2^{+}\Gamma_3^{+}+3\Gamma_2^{-}\Gamma_3^{-}+10\Gamma_4^{+}+12\Gamma_4^{-}+21M_1+21M_2+4R_1^{+}+3R_1^{-} +4R_2^{+}R_3^{+}+3R_2^{-}R_3^{-}+12R_4^{+}+9R_4^{-}+20X_1+22X_2}
    \label{eq:29799sdv}
\end{equation}

\begin{figure}
    \centering
    \includegraphics[width=7.0in]{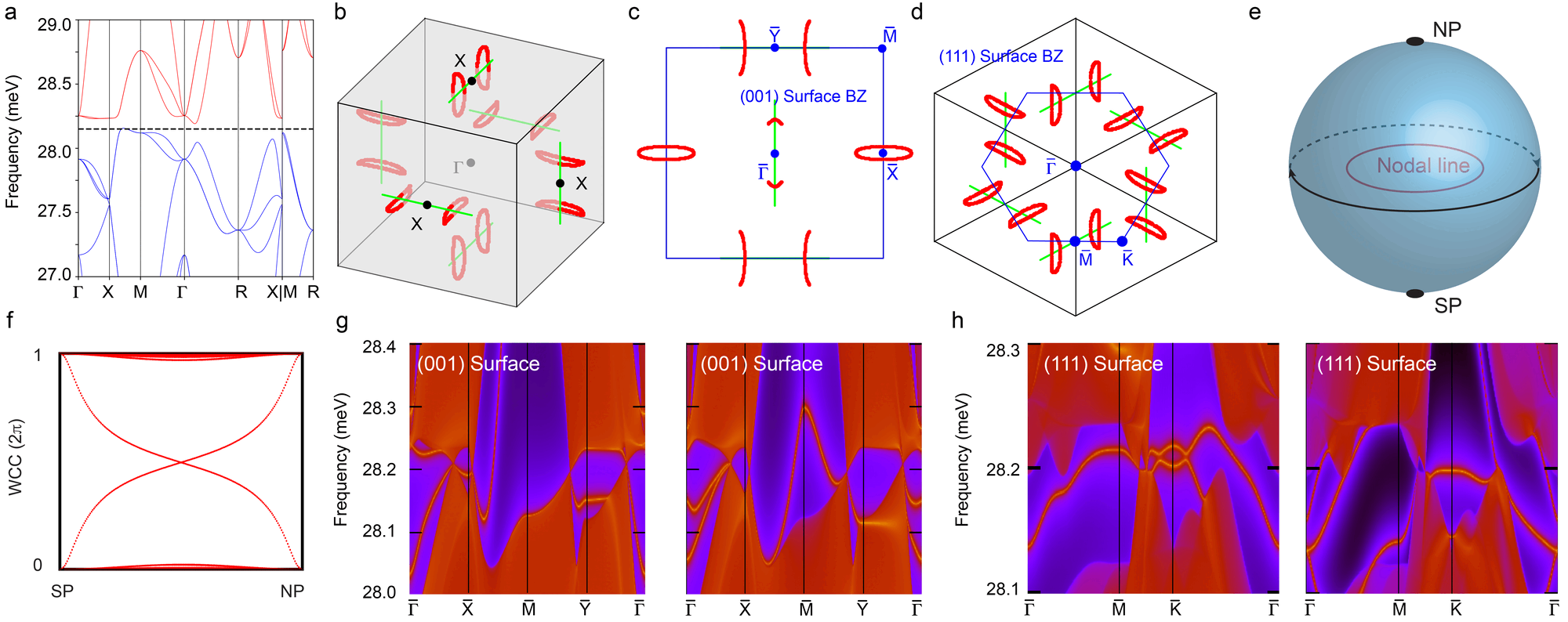}
    \caption{(a) Phonon band dispersion around the topological band gaps. (b) Distribution of the 3 pairs of topological nodal rings (in red) of $Z_2$-monopole charge in the 3D bulk BZ. Top view of the nodal rings along the (c) (001) and (d) (111) directions. (e) A sphere enclosing one of the nodal rings. 
    (f) Wilson-loop calculation performed on the sphere wrapping a nodal ring in (e). The evolution of Wilson-loop is from the south pole (SP) to the north pole (NP) of the sphere. 
    (g) Surface state calculations on the (001) surface with two different surface termination. (h) Surface state calculations on the (111) surface with two different surface termination. }
    \label{fig:29799ss}
\end{figure}

The topological band gap for the B-vector (Eq. \ref{eq:29799sdv}) is shown in the band plotted in Fig. \ref{fig:29799ss}(a), \ie the dashed line at 28.16 meV.
As defined in Refs. \cite{bradlyn_topological_2017,song_diagnosis_2018,po_symmetry-based_2017}, the symmetry-based indicator of SG 201 is $z_2^{\prime}$ which is simply defined by the half of $z_4$ in Eq. \ref{eq:SI148z4}, \ie $z_2^{\prime}=\frac{z_4}{2}$. For the $B$ vector in Eq. \ref{eq:29799sdv}, its indicator is $z_2^{\prime}=1$, showing 6 mod 12 nodal rings in the first BZ \cite{song_diagnosis_2018}, which are related by inversion and three-fold rotation symmetries.  Unlike the $\pi$-Berry phase nodal lines in \ce{ZnPbF6} (SG 148, \mpidweb{13610}), each nodal ring indicated by the topological index $z_2^{\prime}=1$ has a $Z_2$-monopole charge \cite{Chen2015,song_diagnosis_2018,Kim2015}. 
A topological Nodal ring with $Z_2$-monopole charge can only be created and annihilated in pairs, as the total charge of the BZ must be zero. The addition of a finite perturbation can make a nodal line with $Z_2$-monopole charge shrink to an accidental
nodal point, but cannot gap it \cite{Chen2015}. Due to the three-fold rotation symmetry, there are 3 pairs of nodal rings for the $B$ vector in Eq. \ref{eq:29799sdv}. As shown in Fig. \ref{fig:29799ss}(b-d), each pair of topological nodal rings (red rings) are centering around the 2-fold rotational axis (green lines) on the $k_x=\pi$ plane, and the two nodal rings of each pair are related by the inversion symmetry at the $X$ point.

In the presence of inversion symmetry (time reversal is assumed in the phonon spectrum), the monopole charge carried by a nodal line is characterized by the second Stiefel-Whitney class, which is indicated by a $Z_2$ indicator $w_2$ and can be read off from the Wilson loop spectrum \cite{fang2015,bjyang2018}. As shown in Fig. \ref{fig:29799ss}(f), we calculated the Wilson-loop evolution for phonon bands $1\sim84$ on a 2D closed manifold wrapping a nodal line (Fig. \ref{fig:29799ss}(d)). The single crossing point at $WCC=\pi$ line corresponds to $w_2=1$ phase and indicates the nodal line of $Z_2$-monopole charge.

In Figs. \ref{fig:29799ss}(g-h), drum-head-like surface phonon modes connecting to the nodal lines were simulated on the (001) and (111) planes.

\subsection{Symmetry-enforced nodal points and lines}\label{app:bandnodes}
Except for the Weyl nodes and nodal lines that are indicated by the nontrivial symmetry indicators, the symmetry-enforced band degeneracy is very common in phonon band structures. In general, if a symmetry data vector does not satisfy any compatibility relations of the related space group, the related bands of this symmetry data vector necessarily have degenerate point/line/plane with the bands above or below. 
As introduced in Section \ref{app:topphonon}, for the phonon band structure of each MPID entry, all the isolated sets of bands that satisfy the compatibility relations were identified. In general, there is no symmetry-enforced band degeneracy between any two isolated sets. And for each isolated set (which is of more than one band), every band has symmetry enforced degenerate point/line/plane with other bands within the same set.

\begin{figure}
    \centering
    \includegraphics[width=7.0in]{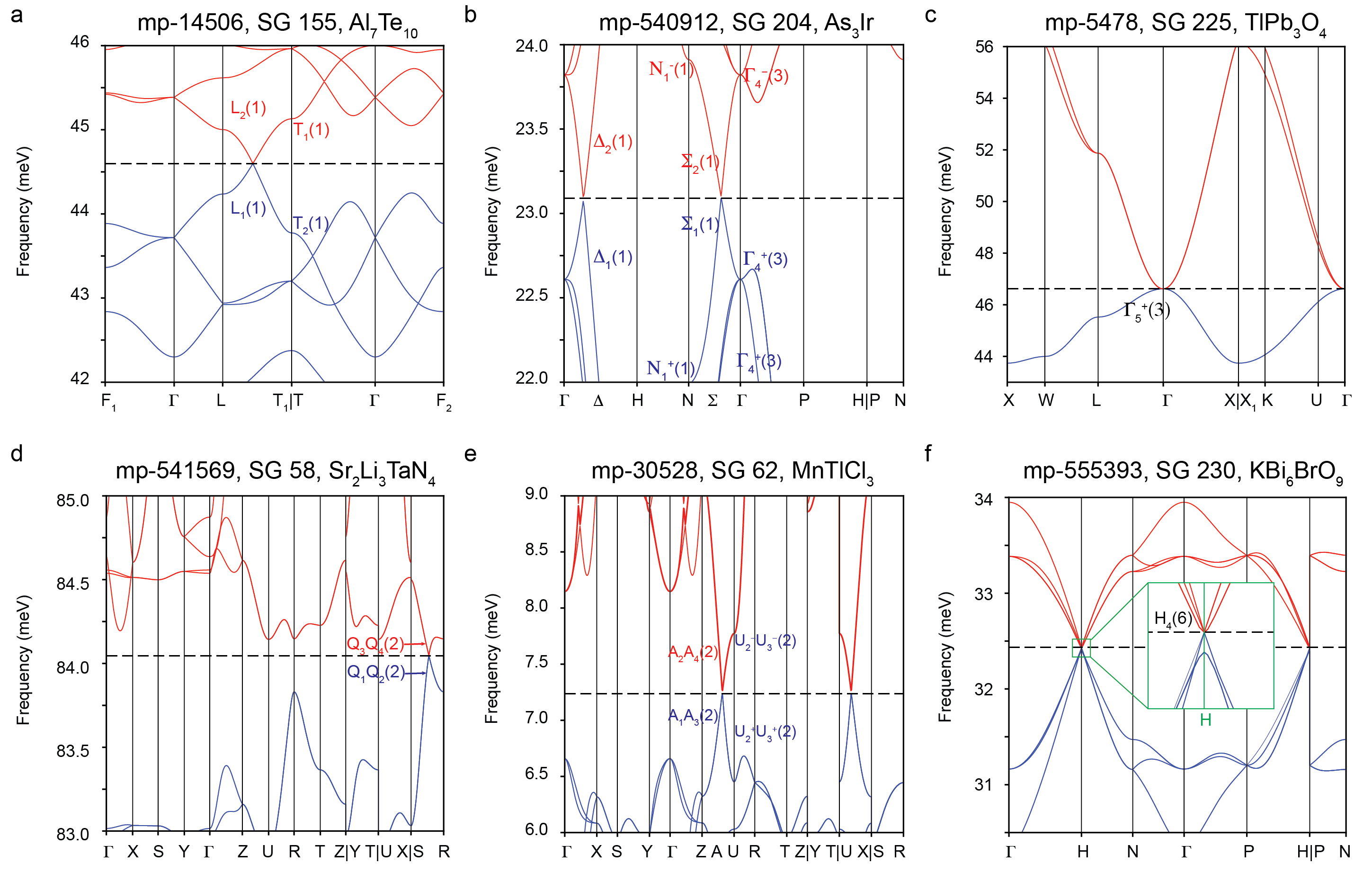}
    \caption{Band structure and symmetry analysis of the symmetry-enforced band crossings for the six prototypical phonon materials presented in Table \ref{tab:ES}. (a) Weyl node protected by the 2-fold rotation symmetry along $L$-$T_1$ path, \ie the (110) direction of the crystalline lattice. 
    The node is formed by two one-dimensional(1D) bands of different $C_2^{110}$ eigenvalues, corresponding to the irreps $L_1$($T_1$) and $L_2$($T_2$) at $L$ ($T$) point. (b) Weyl nodal line protected by the mirror symmetry on $k_z=0$ plane. The node line is formed by two one-dimensional (1D) bands of different $M_z$ eigenvalues, corresponding to the irreps $\Delta_1$ and $\Delta_2$ at $\Delta$ point (or $\Sigma_1$ and $\Sigma_2$ at $\Sigma$ point).
    (c) 3-fold degeneracy at the $\Gamma$ point. The three-dimensional irrep ($\Gamma_5^+$) is protected by the 3-fold rotation along (111) direction, 2-fold rotation $C_{2x}$ and $C_{2z}$ symmetries \cite{bradlyn2016beyond}. (d) Dirac node of 4-fold degeneracy on the $S$-$R$ path. The node is formed by the crossing of two 2-dimensional (2D) bands of different $C_{2z}$ eigenvalues, corresponding to the irreps  $Q_1Q_2$ and $Q_3Q_4$ at the $Q$ line. (e) Dirac nodal line protected by the mirror symmetry $M_z$ on the $k_z=0$ plane. The node line is formed by the crossing between two 2D bands of different $M_{z}$ eigenvalues, corresponding to the irreps $A_1A_3$ and $A_2A_4$ at the $A$ line. (f) 6-fold degeneracy point at $H$. The 6-dimensional irrep ($H_4(6)$) is protected by the inversion symmetry, 3-fold rotation symmetry along the (111) direction, 4-fold rotation symmetry combined with a fractional translation $\{C_{4z}|(\frac{1}{4},\frac{3}{4},\frac{1}{4})\}$ and a 2-fold screw axis $\{C_{2y}|(\frac{1}{2},0,0)\}$ \cite{bradlyn2016beyond}. Note that the dimension for each band irrep has been indicated in parentheses.}
    \label{fig:ES}
\end{figure}

In this section, we select 6 prototypical phonon materials which host different types of symmetry enforced band degeneracy. 
The 6 materials and their symmetry-enforced degeneracy points/lines are tabulated in Table \ref{tab:ES}. In Fig. \ref{fig:ES}, we have plotted and analyzed the band dispersion around the degenerate points/lines for each of these 6 materials.

\begin{table}[]
    \centering
    \caption{The 6 prototypical phonon band structures hosting symmetry enforced band degeneracy. In the first three columns, we tabulate the chemical formula (Chem. Form.), MPID number, and space group (SG) for each material. In the last two columns, we tabulate the related band indices where the band degeneracy occurs and the type of symmetry-enforced band crossing.}
    \label{tab:ES}
    \begin{tabular}{c|c|c|c|c}
        \hline
        Chem. Form. & MPID & SG & Band Indices & Type \\
        \hline
        \ce{Al7Te10} & \mpidweb{14506} & SG 155 (\sgsymb{155}) & \#98$\sim$99 & 2-fold degeneracy (Weyl) node \\
        \hline
        \ce{As3Ir} & \mpidweb{540912} & SG 204 (\sgsymb{204}) & \#21$\sim$22 & 2-fold degeneracy (Weyl) nodal line \\
        \hline
        \ce{TlPb3O4} & \mpidweb{5478} & SG 225 (\sgsymb{225}) & \#25$\sim$27 & 3-fold degeneracy node \\
        \hline
        \ce{Sr2Li3TaN4} & \mpidweb{541569} & SG 58 (\sgsymb{58}) & \#109$\sim$112 & 4-fold degeneracy (Dirac) node \\
        \hline
        \ce{MnTlCl3} & \mpidweb{30528} & SG 62 (\sgsymb{62}) & \#11$\sim$14 & 4-fold degeneracy (Dirac) nodal line \\
        \hline
        \ce{KBi6BrO9} & \mpidweb{555393} & SG 230 (\sgsymb{230}) & \#247$\sim$252 & 6-fold degeneracy node \\
        \hline
    \end{tabular}
\end{table}

\subsection{Fragile phonon bands in \ce{HfIN}}\label{app:HfIN}

\begin{figure}
    \centering
    \includegraphics[width=7.0in]{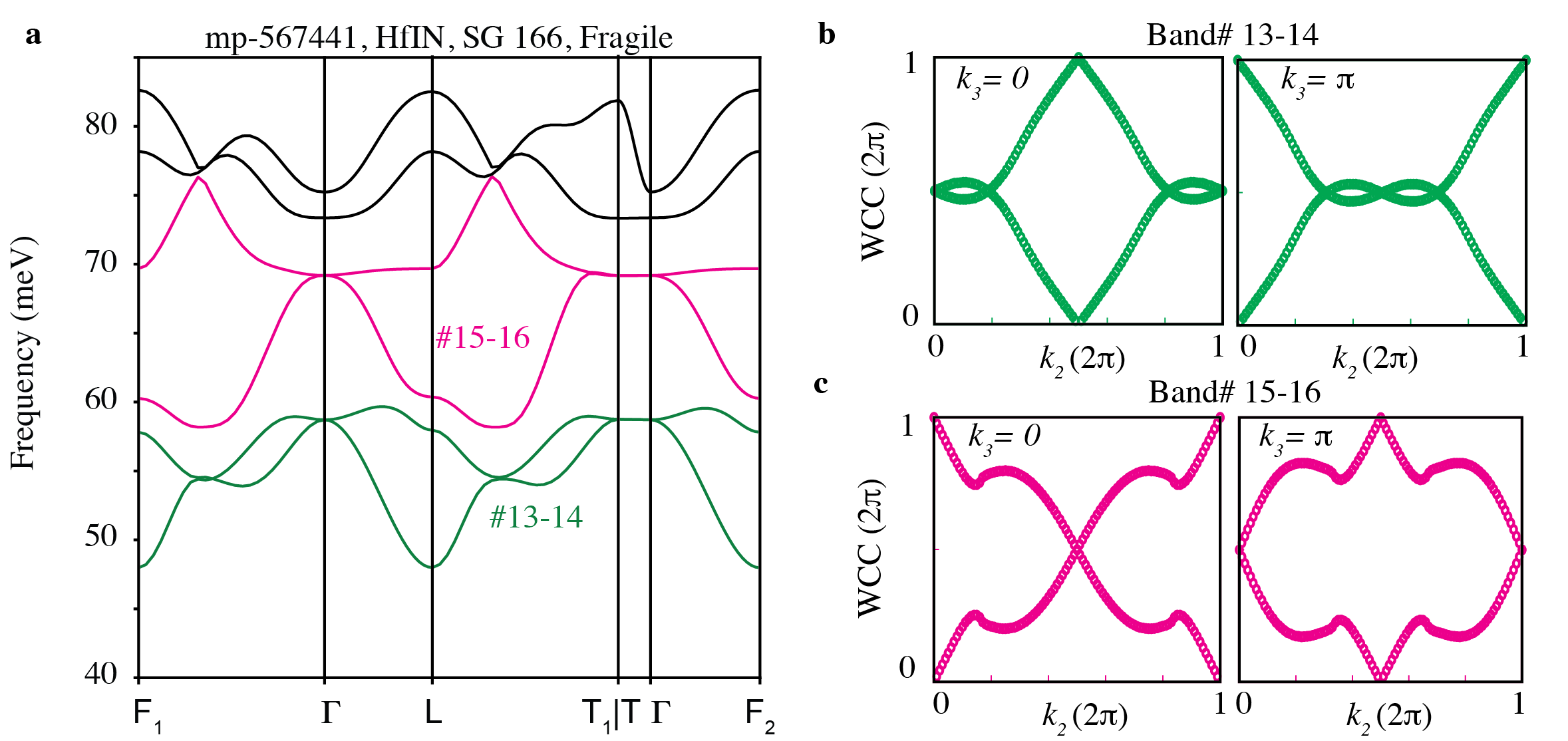}
    \caption{Phonon band spectrum and Wilson loop spectrum of the fragile phonon bands in \ce{HfIN}. (a) The band dispersion of the fragile bands in \ce{HfIN}. There are two fragile isolated band sets corresponding to band indices $\#13\sim14$(green) and $\#15\sim16$(pink), respectively.(b-c) Wilson-loop spectrum of the fragile sets $\#13\sim14$ (b) and $\#15\sim16$ (c) in the phonon band structure of \ce{HfIN}. For each fragile set, the winding numbers of WCC are equal to one at both of the $C\cdot T$-invariant planes $k_3=0,\pi$.}
    \label{fig:567441}
\end{figure}

As mentioned in the main text, we did not find any phonon material with ``ideal'' fragile band set (according to the criteria that we have selected). There are still a few candidates that are close to this definition and we here present one of them. The centrosymmetric material \ce{HfIN} of \mpidweb{567441} has a rhombohedral lattice with SG 166 (\sgsymb{166}).
In Fig. \ref{fig:567441}(a), the phonon spectrum of \ce{HfIN} has two fragile isolated band sets: one has band indices $\#13\sim14$ and another one has indices $\#15\sim16$. Their respective symmetry data vectors are $B_{f1}=(\Gamma_3^-, T_3^+, F_1^+ + F_2^+, L_1^- + L_2^-)$ and $B_{f2}=(\Gamma_3^+, T_3^-, F_1^- + F_2^-, L_1^+ + L_2^+)$. Although neither of them can be expressed as a summation of elementary band representations (EBRs), both of them can be expressed as a difference of two EBRs, \ie
\begin{equation}
\begin{aligned}
    &B_{f1}=A_u \uparrow G_d \ominus A_{1u} \uparrow G_b, \\
   &B_{f2}=A_g \uparrow G_d \ominus A_{1g} \uparrow G_b,
\end{aligned}
\end{equation}
where $\rho \uparrow G_{\alpha}$ represents the EBR induced by irrep $\rho$ at the Wyckoff position $\alpha$ (of site symmetry $G_{\alpha}$). Hence, they are symmetry-indicated fragile bands \cite{song_fragile_2019}. In addition, the two-band fragile topology of each of the two sets of bands is characterized by the winding number of Wilson-loop spectrum \cite{song2019all,ahn2019failure,PhysRevB.99.195455}. 
In Figs. \ref{fig:567441}(b-c), 
we show the Wannier charge center (WCC) evolution of the two sets of bands on both $k_3=0$ and $k_3=\pi$ planes, both of which are two-dimensional and $C_2\cdot T$ (a composite operation of two-fold rotation and time-reversal) invariant planes. The winding number of each panel is equal to 1, indicating that this three-dimensional fragile set comes from the stacking of two-dimensional fragile bands.

\section{Topological phonon database}\label{app:TQCphonondb}
\subsection{Overview of the \webTQCphonon}\label{app:webphonon}

The website \webTQCphonon\ offers a convenient and publicly available way to browse and search through the high-throughput phonon calculations presented in this article. The style and interface are similar to those of the three websites: the \webTQC\cite{vergniory_complete_2019,Vergniory2021}, the \webMTQC\cite{xu2020high} or the \webflatband\cite{Regnault2022}. In this Appendix, we provide a quick overview and guideline for the \webTQCphonon .

\begin{figure*}
    \centering
    \begin{flushleft}a\\
    \end{flushleft}
    \includegraphics[width=.9\textwidth]{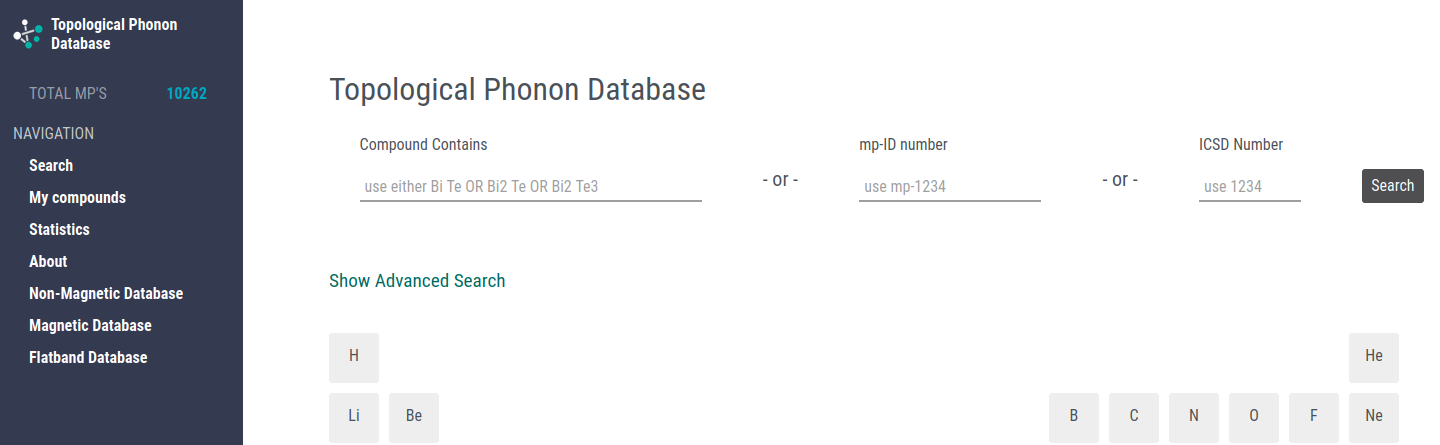}\\
    \begin{flushleft}b\\
    \end{flushleft}
    \includegraphics[width=.9\textwidth]{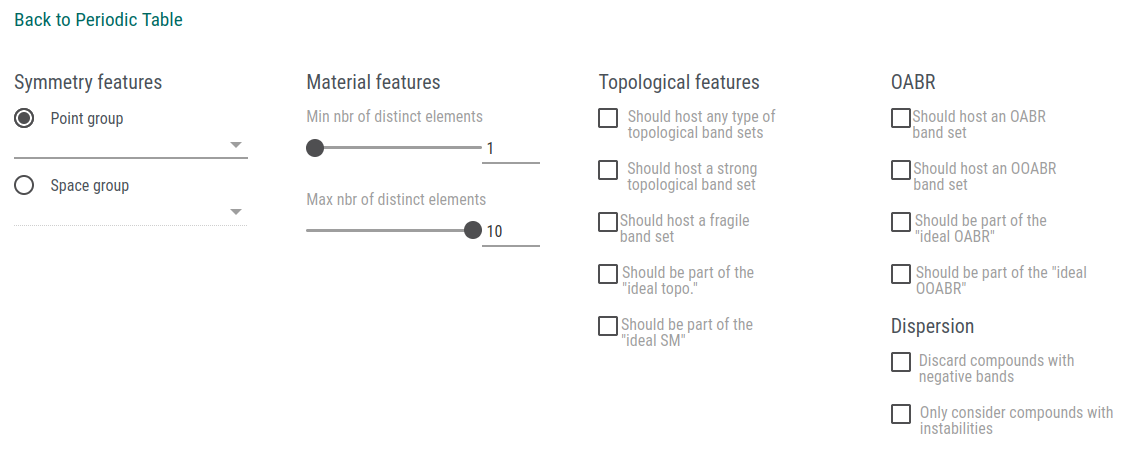}
    \caption{The two search modes of the \webTQCphonon. (a) Basic and default search mode. (b) Advanced search mode}
    \label{app:fig:search}
\end{figure*}

The search engine has two modes. The basic mode is the default one, where the compounds can be searched either by atom type, exact stoichiometric formula, material project ID (MPID) or ICSD number as shown in Fig.~\ref{app:fig:search}a. Note that the exact stoichiometric search can be enforced when a single atom is involved by typing, e.g., ${\rm Bi}1$ rather than ${\rm Bi}$. In the advanced search mode, additional filters can be applied, as depicted in Fig.~\ref{app:fig:search}b. These filters include the space or point groups, compounds hosting certain type of band sets such as fragile band sets, the absence of negative energies or the presence of instabilities (as defined in \siref{app:negativenergies}).

\begin{figure*}
    \centering
 \begin{tabular}{ll}
   a & b \\
    \includegraphics[width=.35\textwidth]{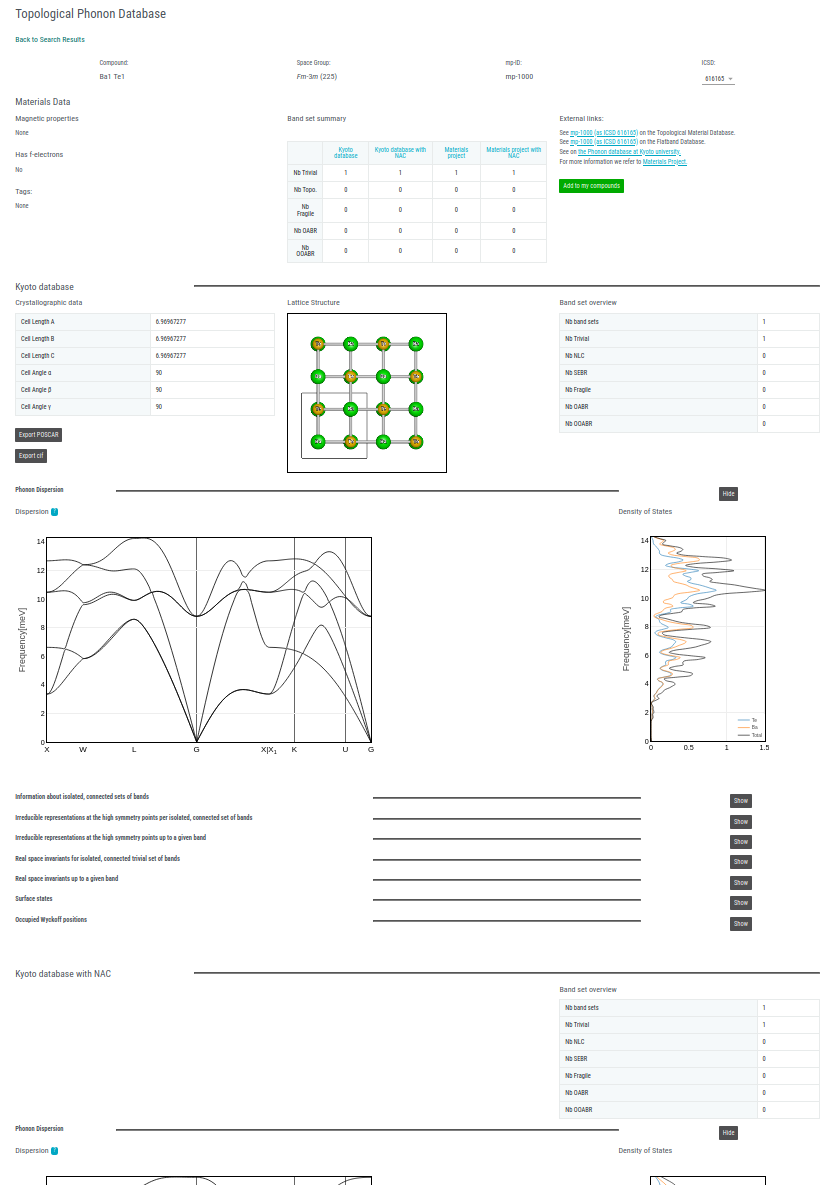} & \includegraphics[width=.60\textwidth]{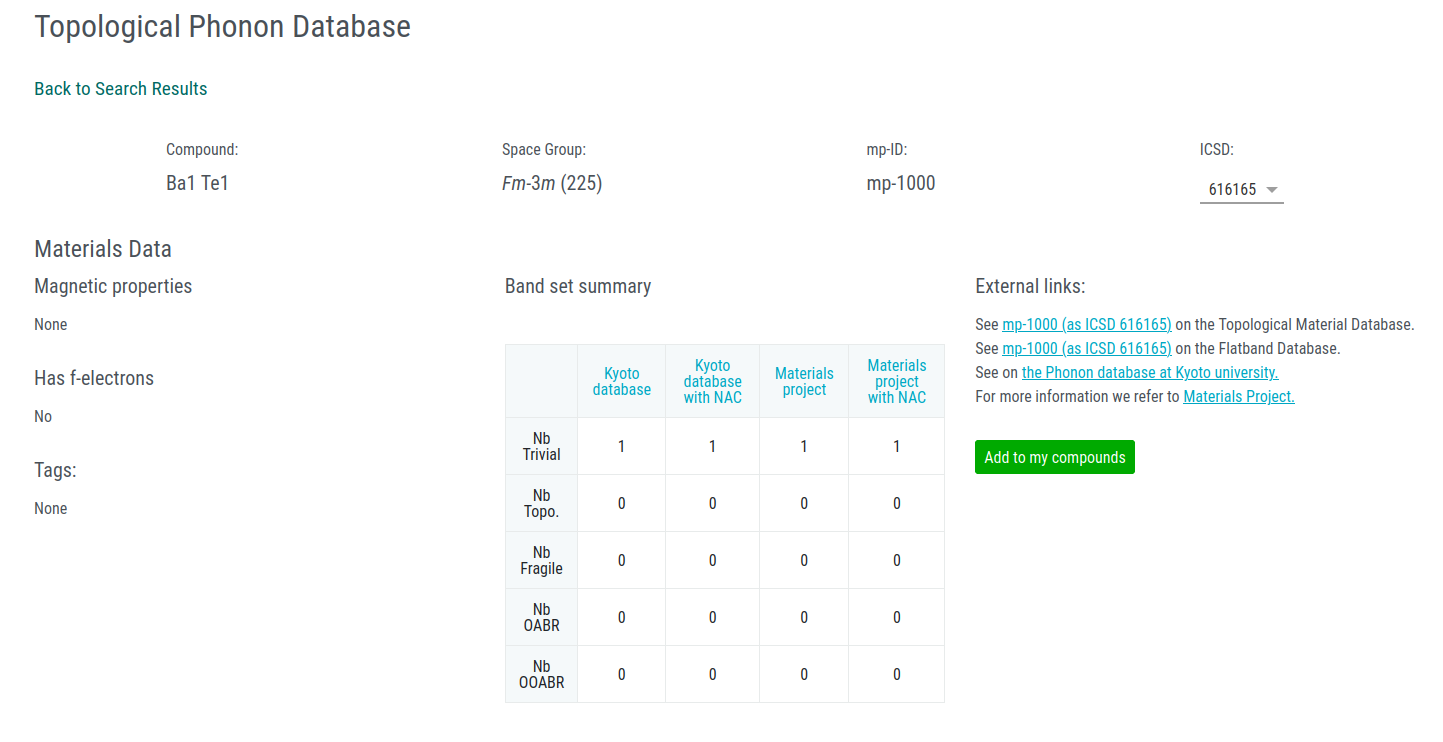} \\ \end{tabular}
    \caption{Information for each compound as provided on the \webTQCphonon. (a) A partial overview of the webpage displaying the compound information layout, including a summary and a detail result section for each type of calculations and crystallographic data source. (b) Compound summary. In particular, the table "Band set summary'' gives an overview of the topological content for the phonons depending on the data source and with or without NAC. Each column title is a link to the corresponding detail result section.}
    \label{app:fig:compoundsummary}
\end{figure*}

Once a compound has been selected, its webpage usually contains several types of information as shown in  Fig.~\ref{app:fig:compoundsummary}a. On top of the page, we provide a brief summary about the compound (see Fig.~\ref{app:fig:compoundsummary}b) including the structural chemical formula, the space group, the MPID and any potentially related ICSD. Direct links (when available) to several other websites are listed including \webMP, \webphonondb\ and \webTQC. For a given compound, its phonon-band topological characterization might change according to the data source, \webMP\ or \webphonondb, and if the calculations were performed with or without NAC. For that reason, a table summarizes the topological content for the phonons, as shown in Fig.~\ref{app:fig:compoundsummary}, for the at most four cases depending on the data source availability for the compound. The column title provides a link within the page for the calculation details.

\begin{figure*}
    \centering
 \begin{tabular}{ll}
   a & b \\
    \includegraphics[width=.40\textwidth]{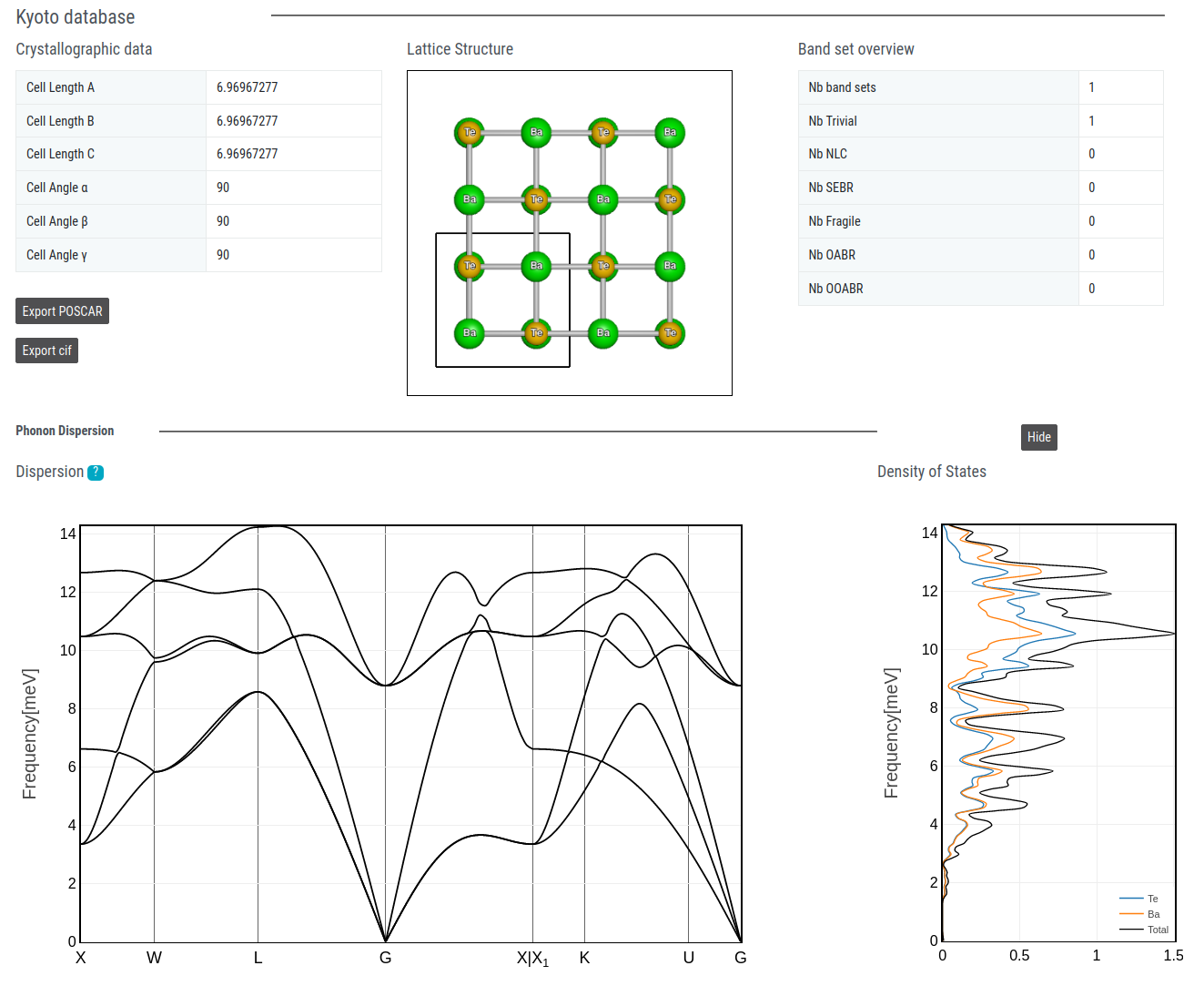} & \includegraphics[width=.55\textwidth]{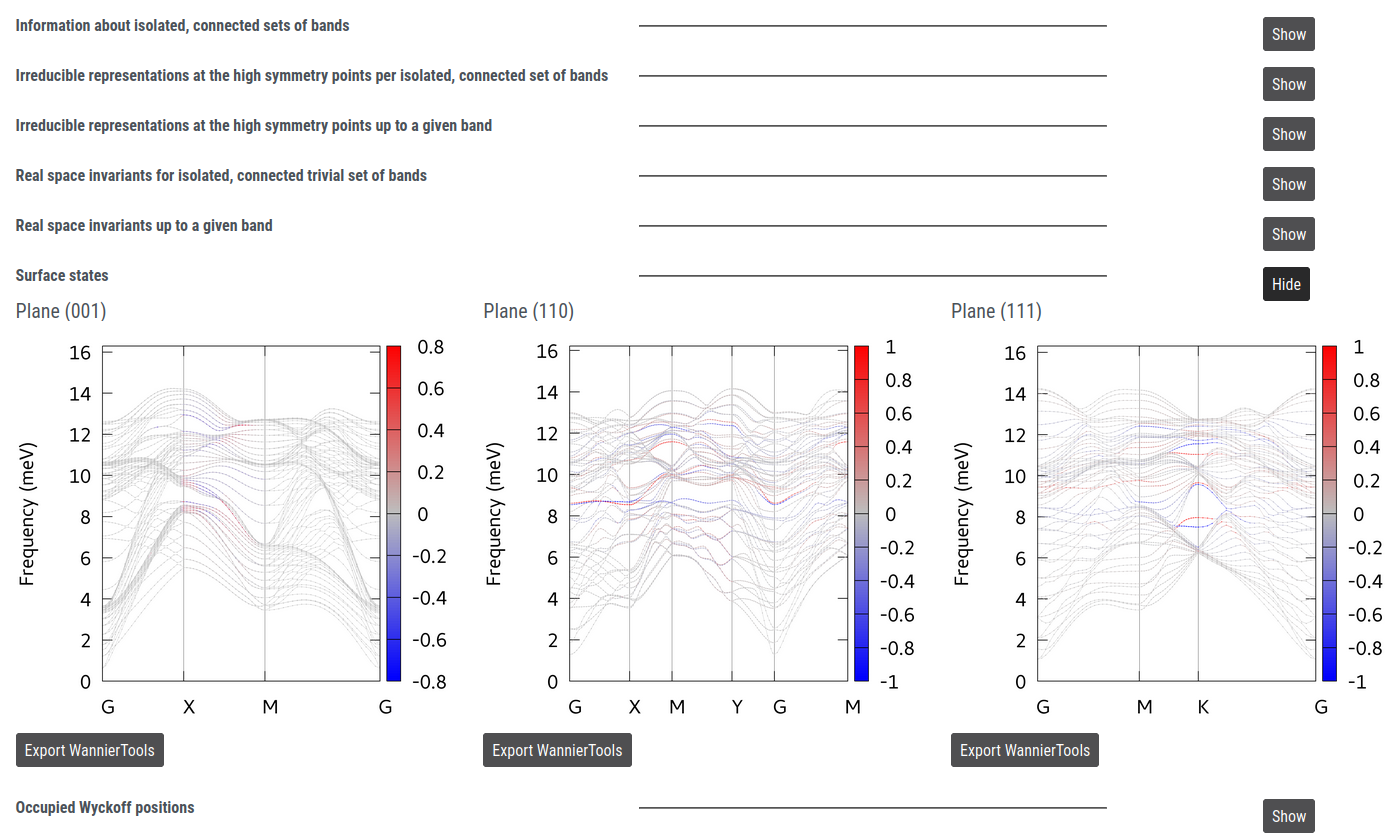} \\ \end{tabular}
    \caption{The detail results provided for each type of data set, without or with NAC for a compound on the \webTQCphonon. Here we use $\rm Ba Te$ in SG 225 (\mpidweb{1000}) as an example. (a) The top part of the detail results contains the crystallographic information, the band set overview, the phonon dispersion and the total and projected density of states. (b) The lower part of detail results has several items (hidden by default) explained in \siref{app:webphonon}. As an example, here we show how the surface states are displayed.}
    \label{app:fig:webdetailresults}
\end{figure*}

For each type of data set and each type of calculation (without or with NAC), the website provides a complete review of the numerical results. The first part contain the crystallographic data (the POSCAR and cif files can be downloaded from here), the band set overview, the phonon dispersion (the bands are in blue for strong topology, and in red for fragile topology) and the projected and total density of states (see Fig.~\ref{app:fig:webdetailresults}a). Note that the crystallographic data is only provided for the results without NAC since this data is identical for the results with NAC. The other results are hidden by default (see Fig.~\ref{app:fig:webdetailresults}b) and cover the following items

\begin{itemize}
\item Information about isolated, connected sets of bands.
\item The irreducible representations at the high symmetry points per isolated, connected set of bands.
\item The irreducible representations at the high symmetry points up to a given band.
\item The real space invariants for isolated, connected trivial set of bands.
\item The real space invariants up to a given band.
\item The surface states (except for the calculations performed with NAC and data coming from \webMP). For each surface state, the WannierTools\cite{WU2017} input file can be downloaded.
\item The list of occupied Wyckoff positions and the corresponding atoms.
\end{itemize}

For convenience, the bottom of each page provides the Wyckoff positions and the K-vector types for the material space group, following the convention of the \webBCSfull.

Finally, the navigation menu contains several links. It contains links to our related material websites, \webTQC\cite{vergniory_complete_2019,Vergniory2021}, the \webMTQC\cite{xu2020high} and the \webflatband\cite{Regnault2022}, and link to the list of materials hosting the most promising nontrivial phonon bands presented in \siref{app:bestmaterials}. In addition, the page "Statistics" offers all the raw data used to generate the different tables of this article in a CSV file format.

\subsection{Dispersions with negative energies}\label{app:negativenergies}

In general, the eigenvalues of the dynamical matrix of a stable crystalline structure are non-negative, while in {\it ab initio} calculations the eigenvalues could be negative and hence the phonon frequencies(the square root of eigenvalues) are imaginary. The reasons of a negativity are multiple. When the negativity is weak, it usually comes from numerical error or weak violation of translational invariance. It can be removed by imposing the acoustic sum rule or rotational sum rule. A larger negativity may come from the supercell finite size (or k-point mesh) used in the calculation or an instability of the crystalline structure that is considered at 0K. 

In the present high-throughput calculation, the 'negativity' problem occurs for most of the materials. As tabulated in Table \ref{tab:negativebandthreshold}, by considering a given energy threshold, we have counted all the materials of a minimum frequency lower than this threshold. Although more than \TPDBPercentKyotoMPMaterialsWithNACStrictNegative\% of the phonon material entries have strictly speaking negative frequencies, most of the negative values are tiny and can be taken as a numerical error. In this work, we use a negativity threshold to avoid such cases: if a phonon spectrum (with NAC) has a minimum frequency of less than $-5$ meV, we treat the corresponding MPID entry as having `negativity' problem. 

\begin{figure}[ht]
\centering
\begin{tabular}{cc}
a\hspace{0.4cm} \scriptsize{\mpidweb{2097}, SG 129 (\sgsymb{129}), \ce{SnO}} & b\hspace{0.4cm} \scriptsize{\mpidweb{29364}, SG 12 (\sgsymb{12}), \ce{Li5SbO5}} \\
\includegraphics[width=0.5\textwidth,angle=0]{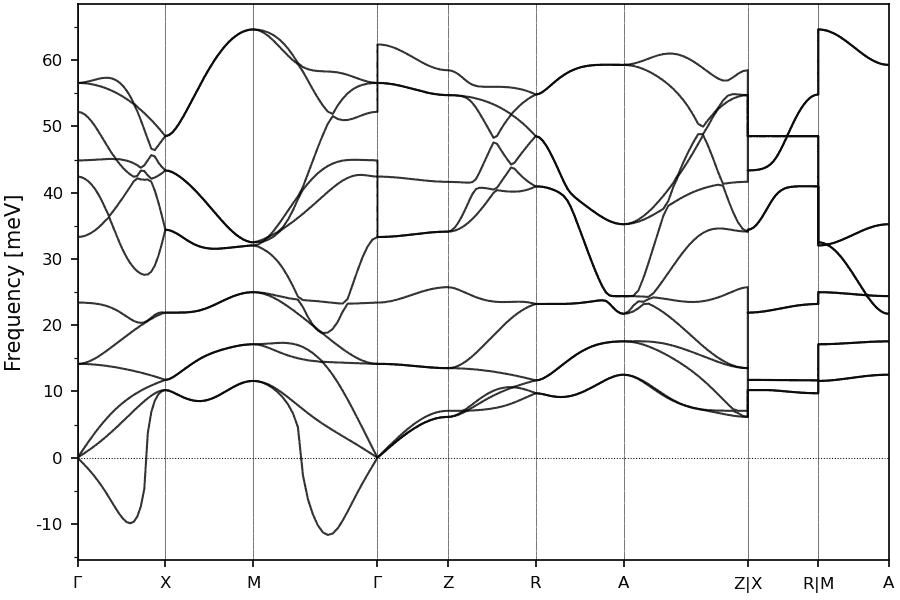} & \includegraphics[width=0.5\textwidth,angle=0]{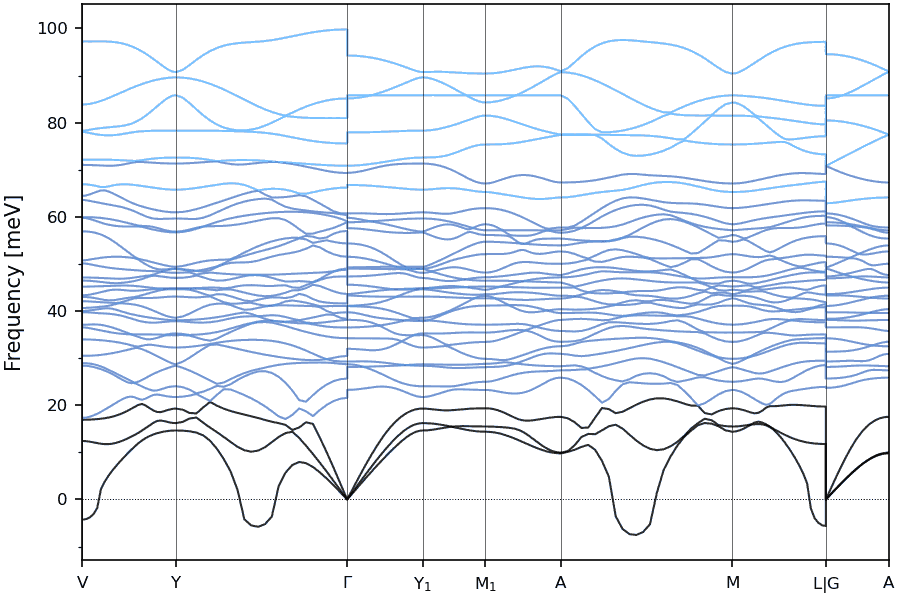} \\
c\hspace{0.4cm} \scriptsize{\mpidweb{776466}, SG 8 (\sgsymb{8}), \ce{Ba2Ti6N2O11}} & d\hspace{0.4cm} \scriptsize{\mpidweb{540784}, SG 12 (\sgsymb{12}), \ce{Rb2Ti6O13}} \\
\includegraphics[width=0.5\textwidth,angle=0]{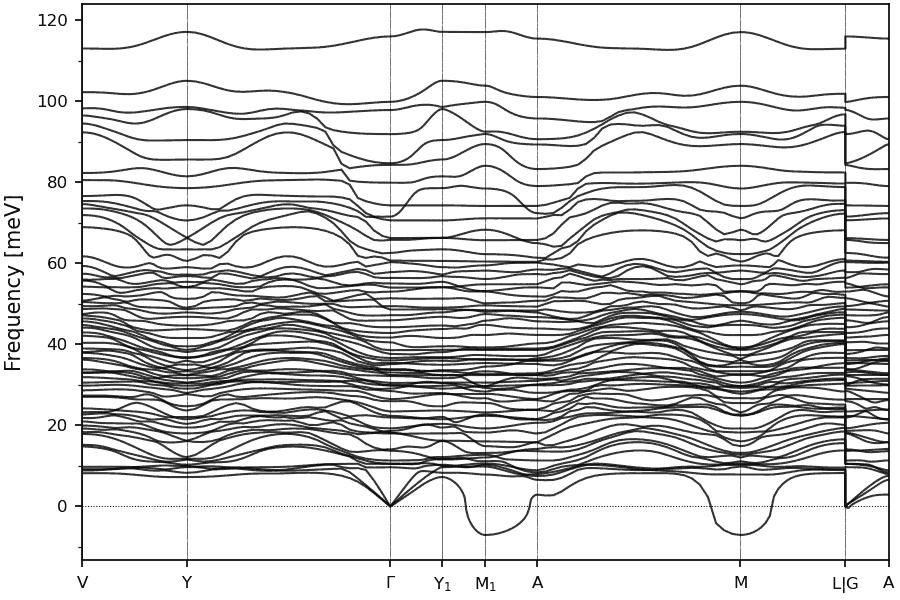} & \includegraphics[width=0.5\textwidth,angle=0]{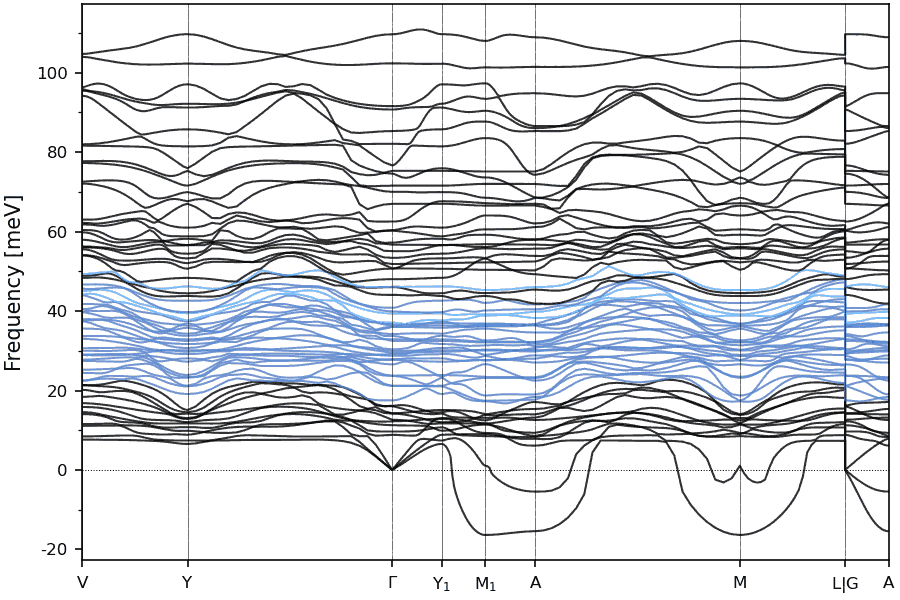} \\
e\hspace{0.4cm} \scriptsize{\mpidweb{766422}, SG 62 (\sgsymb{62}), \ce{Li3 Bi S4}} &
f\hspace{0.4cm} \scriptsize{\mpidweb{757917}, SG 145 (\sgsymb{145}), \ce{BeGa2O4}} \\
\includegraphics[width=0.5\textwidth,angle=0]{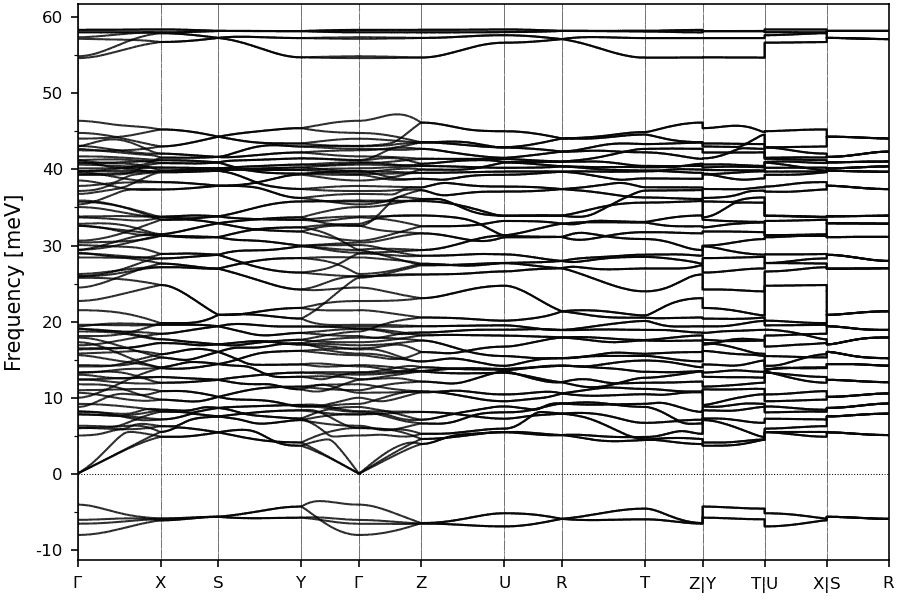} &
\includegraphics[width=0.5\textwidth,angle=0]{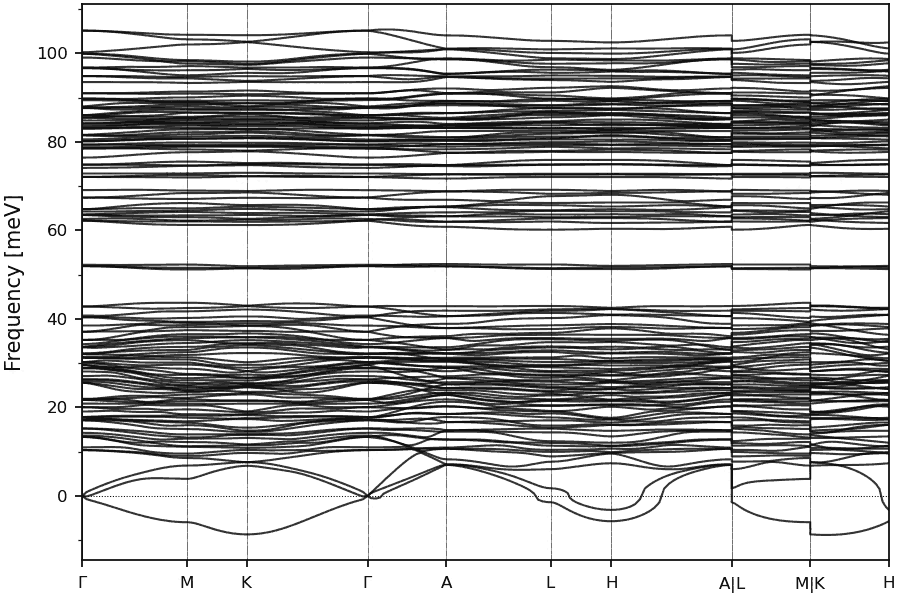} \\ \end{tabular}
\caption{Six materials from the Kyoto database exemplifying the different types of negativities in a phonon spectrum. {\it Top row}: \mpidweb{2097} (a) and \mpidweb{29364} (b) have interpolation errors. {\it Middle row}: \mpidweb{776466} (c) and \mpidweb{540784} (d)  have instabilities at 0K around the $M$ point for both structures. {\it Bottom row} \mpidweb{766422} (e) and \mpidweb{757917} (f) have bands that are completely for (e) or substantially for (f) negative over the whole Brillouin zone and are tagged as ``discarded".}
\label{fig:negativityexamples}
\end{figure}

By manually analyzing the feature of the phonon spectrum, we divide the negativities for all the \TPDBNbrKyotoMaterialsNegativeBands\ entries on the Kyoto phonon database and \TPDBNbrMPMaterialsNegativeBands\ entries on the Material Project phonon database into the following three categories:
\begin{enumerate}
    \item[1] {\it Interpolation error}. The negativity is weak and it usually occurs on a single band around a few high-symmetry points or along some high-symmetry lines. For example, the negativities in \mpidweb{2097} (Fig.~\ref{fig:negativityexamples}a) and \mpidweb{29364} (Fig.~\ref{fig:negativityexamples}b) are in this case.
    \item[2] {\it Instability}. It has one or two bands of strong negativity and the negativity occurs around a few high-symmetry points or along high-symmetry lines. For example, the negativities in \mpidweb{776466} (Fig.~\ref{fig:negativityexamples}c) and \mpidweb{540784} (Fig.~\ref{fig:negativityexamples}d) belong to this category.
    \item[3] {\it Discarded}. There are so many negative bands (more than two) or the negativity is so important over most part of the Brillouin zone that it is hard to clarify if there is an interpolation error or instability. The calculated phonon spectrum of this type of negativity is usually not reliable and can be discarded. For example, the phonon spectrum of materials \mpidweb{766422} (Fig.~\ref{fig:negativityexamples}e) and \mpidweb{757917} (Fig.~\ref{fig:negativityexamples}f) have bands that are completely (or substantially) negative over the whole Brillouin zone and hence can be discarded.
\end{enumerate}

We emphasize that the above criteria are empirical and qualitative, so that the categories are not strict and should be seen as a guideline. In Tabs.~\ref{tab:negativebandmp} and~\ref{tab:negativebandkyoto}, for each space group in the Material Project and Kyoto databases, we provide the number of material entries that are tagged as interpolation error, instability or discarded. Every material entry with a`negativity' category has been tagged on the \webTQCphonon\ and in the full material lists in \siref{app:phononmaterials}, namely Tabs.~\ref{tab:listphononmaterials} and~\ref{tab:listphononmaterialskyoto}.

\LTcapwidth=1.0\textwidth
\renewcommand\arraystretch{1.0}
\begin{longtable*}{|c|c|}
  \caption[Number of material entries with negative energies lower than a given energy threshold]{Number of materials (second column) having dispersions with negative energies lower than a given energy threshold (first column). The material entries treat separately data from the two database sources. As such, the total number that the percentage refers to, is the total number of dispersion that have been computed. Note that only the phonon dispersions obtained from the calculations with NAC were considered here.}
    \label{tab:negativebandthreshold}\\
    \hline
    \hline
    Threshold (mev) & \# materials  \\
        \hline
0 & 9105 {\tiny(79.1\%)} \\ 
-1 & 4171 {\tiny(36.2\%)} \\ 
-2 & 3465 {\tiny(30.1\%)} \\ 
-3 & 2942 {\tiny(25.6\%)} \\ 
-4 & 2513 {\tiny(21.8\%)} \\ 
-5 & 2130 {\tiny(18.5\%)} \\ 
-6 & 1807 {\tiny(15.7\%)} \\ 
-7 & 1519 {\tiny(13.2\%)} \\ 
-8 & 1257 {\tiny(10.9\%)} \\ 
-9 & 1073 {\tiny(9.3\%)} \\ 
-10 & 901 {\tiny(7.8\%)} \\ 
    \hline
\end{longtable*}

\LTcapwidth=1.0\textwidth
\renewcommand\arraystretch{1.0}
\begin{longtable*}{|c|c|c|c|c|}
  \caption[Number of materials with negative energies from the Materials Project phonon database]{Number of materials having dispersions with negative energies from the Materials Project phonon database. The material entries treat separately data from the two database sources. Note that only the phonon dispersions obtained from the calculations with NAC were considered here. For each space group (SG) given in the first column, we provide the number of materials (second column), the number of materials tagged as discarded (third column), the number of materials where the negative energies potentially come from interpolation errors (fourth column) and the number of materials where the negative energies potentially might emerge from instabilities (fifth column).}
    \label{tab:negativebandmp}\\
    \hline
    \hline
    {\scriptsize{SG}} & {\scriptsize{\# materials}}  & {\scriptsize{\begin{tabular}{c}\# discarded materials \end{tabular}}} & {\scriptsize{\begin{tabular}{c}\# materials with\\ interpolation errors \end{tabular}}} & {\scriptsize{\begin{tabular}{c}\# materials with\\ instabilities \end{tabular}}}\\
        \hline
1& 3 & ---  & ---  & ---  \\ 
2& 5 & ---  & ---  & ---  \\ 
4& 1 & ---  & ---  & ---  \\ 
5& 3 & ---  & 1 {\tiny(33.3\%)} & ---  \\ 
6& 1 & ---  & ---  & ---  \\ 
8& 7 & 1 {\tiny(14.3\%)} & ---  & 2 {\tiny(28.6\%)} \\ 
9& 9 & ---  & ---  & ---  \\ 
11& 24 & 2 {\tiny(8.3\%)} & ---  & 2 {\tiny(8.3\%)} \\ 
12& 57 & 6 {\tiny(10.5\%)} & 1 {\tiny(1.8\%)} & 1 {\tiny(1.8\%)} \\ 
13& 5 & ---  & ---  & ---  \\ 
14& 16 & ---  & ---  & ---  \\ 
15& 30 & ---  & ---  & ---  \\ 
19& 1 & ---  & ---  & ---  \\ 
23& 1 & ---  & ---  & ---  \\ 
24& 2 & ---  & ---  & 1 {\tiny(50.0\%)} \\ 
26& 1 & ---  & ---  & ---  \\ 
29& 3 & ---  & ---  & ---  \\ 
31& 3 & ---  & ---  & ---  \\ 
33& 1 & ---  & ---  & ---  \\ 
36& 27 & ---  & ---  & ---  \\ 
43& 4 & ---  & ---  & ---  \\ 
44& 1 & ---  & ---  & ---  \\ 
51& 2 & ---  & ---  & ---  \\ 
53& 1 & ---  & ---  & ---  \\ 
55& 2 & ---  & ---  & ---  \\ 
58& 14 & ---  & ---  & ---  \\ 
59& 2 & ---  & ---  & ---  \\ 
62& 18 & ---  & ---  & ---  \\ 
63& 64 & ---  & ---  & ---  \\ 
64& 2 & ---  & ---  & ---  \\ 
65& 6 & 1 {\tiny(16.7\%)} & ---  & ---  \\ 
67& 1 & ---  & ---  & ---  \\ 
69& 8 & 1 {\tiny(12.5\%)} & ---  & ---  \\ 
71& 18 & 1 {\tiny(5.6\%)} & ---  & ---  \\ 
72& 21 & ---  & ---  & ---  \\ 
74& 1 & ---  & ---  & 1 {\tiny(100\%)} \\ 
82& 37 & ---  & ---  & ---  \\ 
87& 5 & ---  & ---  & ---  \\ 
88& 8 & ---  & ---  & ---  \\ 
92& 2 & ---  & ---  & ---  \\ 
97& 2 & ---  & ---  & ---  \\ 
98& 1 & ---  & ---  & ---  \\ 
99& 3 & 2 {\tiny(66.7\%)} & ---  & ---  \\ 
107& 3 & 1 {\tiny(33.3\%)} & ---  & ---  \\ 
108& 1 & ---  & ---  & ---  \\ 
109& 2 & ---  & ---  & ---  \\ 
111& 2 & ---  & ---  & ---  \\ 
112& 1 & ---  & ---  & ---  \\ 
113& 3 & ---  & ---  & ---  \\ 
115& 1 & ---  & ---  & ---  \\ 
119& 2 & ---  & ---  & ---  \\ 
121& 16 & 1 {\tiny(6.2\%)} & ---  & ---  \\ 
122& 39 & ---  & ---  & 1 {\tiny(2.6\%)} \\ 
123& 13 & 1 {\tiny(7.7\%)} & ---  & 1 {\tiny(7.7\%)} \\ 
127& 2 & ---  & ---  & ---  \\ 
129& 56 & ---  & ---  & ---  \\ 
131& 2 & ---  & ---  & ---  \\ 
136& 15 & ---  & ---  & ---  \\ 
137& 1 & ---  & ---  & ---  \\ 
139& 72 & 7 {\tiny(9.7\%)} & 1 {\tiny(1.4\%)} & 4 {\tiny(5.6\%)} \\ 
140& 17 & 1 {\tiny(5.9\%)} & ---  & ---  \\ 
141& 16 & ---  & ---  & ---  \\ 
146& 1 & ---  & ---  & ---  \\ 
147& 2 & ---  & ---  & ---  \\ 
148& 46 & ---  & ---  & ---  \\ 
149& 2 & ---  & ---  & ---  \\ 
152& 6 & ---  & ---  & ---  \\ 
154& 2 & ---  & ---  & ---  \\ 
155& 7 & 2 {\tiny(28.6\%)} & ---  & ---  \\ 
156& 11 & ---  & ---  & ---  \\ 
160& 17 & 1 {\tiny(5.9\%)} & 1 {\tiny(5.9\%)} & ---  \\ 
161& 5 & ---  & ---  & ---  \\ 
162& 13 & ---  & ---  & ---  \\ 
164& 59 & ---  & ---  & ---  \\ 
166& 97 & 3 {\tiny(3.1\%)} & ---  & ---  \\ 
167& 20 & ---  & ---  & ---  \\ 
176& 2 & ---  & ---  & ---  \\ 
180& 1 & 1 {\tiny(100\%)} & ---  & ---  \\ 
186& 34 & ---  & ---  & ---  \\ 
187& 9 & ---  & ---  & 1 {\tiny(11.1\%)} \\ 
189& 14 & ---  & ---  & ---  \\ 
190& 1 & ---  & ---  & ---  \\ 
191& 1 & 1 {\tiny(100\%)} & ---  & ---  \\ 
194& 57 & ---  & ---  & ---  \\ 
198& 4 & ---  & ---  & ---  \\ 
199& 1 & ---  & ---  & ---  \\ 
205& 3 & ---  & ---  & ---  \\ 
206& 1 & ---  & ---  & ---  \\ 
215& 11 & ---  & ---  & ---  \\ 
216& 91 & 2 {\tiny(2.2\%)} & ---  & ---  \\ 
217& 3 & ---  & ---  & ---  \\ 
221& 61 & 8 {\tiny(13.1\%)} & ---  & 7 {\tiny(11.5\%)} \\ 
225& 240 & 46 {\tiny(19.2\%)} & ---  & 3 {\tiny(1.2\%)} \\ 
227& 7 & 5 {\tiny(71.4\%)} & ---  & ---  \\ 
\hline
total& 1516 & 94 {\tiny(6.2\%)} & 4 {\tiny(0.3\%)} & 24 {\tiny(1.6\%)}\\
    \hline
\end{longtable*}

\LTcapwidth=1.0\textwidth
\renewcommand\arraystretch{1.0}
\begin{longtable*}{|c|c|c|c|c|}
  \caption[Number of materials with negative energies from the Kyoto phonon database]{Number of materials having dispersions with negative energies from the Kyoto phonon database. The material entries treat separately data from the two database sources. Note that only the phonon dispersions obtained from the calculations with NAC were considered here. For each space group (SG) given in the first column, we provide the number of materials (second column), the number of materials tagged as discarded (third column), the number of materials where the negative energies potentially come from interpolation errors (fourth column) and the number of materials where the negative energies potentially might emerge from instabilities (fifth column).}
    \label{tab:negativebandkyoto}\\
    \hline
    \hline
    {\scriptsize{SG}} & {\scriptsize{\# materials}}  & {\scriptsize{\begin{tabular}{c}\# discarded materials \end{tabular}}} & {\scriptsize{\begin{tabular}{c}\# materials with\\ interpolation errors \end{tabular}}} & {\scriptsize{\begin{tabular}{c}\# materials with\\ instabilities \end{tabular}}}\\
        \hline
3& 11 & 3 {\tiny(27.3\%)} & ---  & ---  \\ 
4& 97 & 21 {\tiny(21.6\%)} & ---  & ---  \\ 
5& 55 & 9 {\tiny(16.4\%)} & 1 {\tiny(1.8\%)} & 3 {\tiny(5.5\%)} \\ 
6& 11 & 8 {\tiny(72.7\%)} & ---  & 1 {\tiny(9.1\%)} \\ 
7& 51 & 7 {\tiny(13.7\%)} & ---  & 2 {\tiny(3.9\%)} \\ 
8& 65 & 16 {\tiny(24.6\%)} & ---  & 15 {\tiny(23.1\%)} \\ 
9& 82 & 14 {\tiny(17.1\%)} & 1 {\tiny(1.2\%)} & 1 {\tiny(1.2\%)} \\ 
10& 15 & 3 {\tiny(20.0\%)} & ---  & ---  \\ 
11& 204 & 36 {\tiny(17.6\%)} & 2 {\tiny(1.0\%)} & 6 {\tiny(2.9\%)} \\ 
12& 365 & 69 {\tiny(18.9\%)} & 6 {\tiny(1.6\%)} & 22 {\tiny(6.0\%)} \\ 
13& 83 & 20 {\tiny(24.1\%)} & 1 {\tiny(1.2\%)} & 2 {\tiny(2.4\%)} \\ 
14& 1070 & 146 {\tiny(13.6\%)} & 10 {\tiny(0.9\%)} & 21 {\tiny(2.0\%)} \\ 
15& 599 & 95 {\tiny(15.9\%)} & 20 {\tiny(3.3\%)} & 15 {\tiny(2.5\%)} \\ 
17& 1 & ---  & ---  & ---  \\ 
18& 12 & 2 {\tiny(16.7\%)} & ---  & ---  \\ 
19& 118 & 7 {\tiny(5.9\%)} & 3 {\tiny(2.5\%)} & 3 {\tiny(2.5\%)} \\ 
20& 36 & 17 {\tiny(47.2\%)} & ---  & ---  \\ 
21& 1 & 1 {\tiny(100\%)} & ---  & ---  \\ 
22& 1 & 1 {\tiny(100\%)} & ---  & ---  \\ 
24& 7 & ---  & ---  & 1 {\tiny(14.3\%)} \\ 
25& 3 & 2 {\tiny(66.7\%)} & ---  & ---  \\ 
26& 22 & 4 {\tiny(18.2\%)} & ---  & 2 {\tiny(9.1\%)} \\ 
28& 1 & 1 {\tiny(100\%)} & ---  & ---  \\ 
29& 30 & 6 {\tiny(20.0\%)} & ---  & ---  \\ 
30& 2 & 1 {\tiny(50.0\%)} & ---  & ---  \\ 
31& 55 & 5 {\tiny(9.1\%)} & ---  & ---  \\ 
32& 6 & 3 {\tiny(50.0\%)} & ---  & 1 {\tiny(16.7\%)} \\ 
33& 128 & 32 {\tiny(25.0\%)} & 1 {\tiny(0.8\%)} & 1 {\tiny(0.8\%)} \\ 
34& 6 & 1 {\tiny(16.7\%)} & ---  & ---  \\ 
35& 3 & 2 {\tiny(66.7\%)} & ---  & ---  \\ 
36& 106 & 13 {\tiny(12.3\%)} & ---  & 2 {\tiny(1.9\%)} \\ 
37& 6 & 1 {\tiny(16.7\%)} & ---  & 2 {\tiny(33.3\%)} \\ 
38& 16 & 8 {\tiny(50.0\%)} & ---  & ---  \\ 
39& 7 & 4 {\tiny(57.1\%)} & ---  & ---  \\ 
40& 36 & 4 {\tiny(11.1\%)} & ---  & 3 {\tiny(8.3\%)} \\ 
41& 15 & 3 {\tiny(20.0\%)} & ---  & 1 {\tiny(6.7\%)} \\ 
42& 9 & 2 {\tiny(22.2\%)} & 1 {\tiny(11.1\%)} & 1 {\tiny(11.1\%)} \\ 
43& 30 & 5 {\tiny(16.7\%)} & 2 {\tiny(6.7\%)} & 1 {\tiny(3.3\%)} \\ 
44& 12 & 3 {\tiny(25.0\%)} & ---  & ---  \\ 
45& 7 & ---  & ---  & ---  \\ 
46& 12 & 5 {\tiny(41.7\%)} & ---  & ---  \\ 
47& 2 & 1 {\tiny(50.0\%)} & ---  & ---  \\ 
48& 2 & ---  & ---  & ---  \\ 
50& 1 & ---  & ---  & ---  \\ 
51& 14 & 5 {\tiny(35.7\%)} & ---  & ---  \\ 
52& 37 & 6 {\tiny(16.2\%)} & ---  & ---  \\ 
53& 12 & 1 {\tiny(8.3\%)} & ---  & ---  \\ 
54& 8 & 5 {\tiny(62.5\%)} & ---  & ---  \\ 
55& 47 & 9 {\tiny(19.1\%)} & ---  & ---  \\ 
56& 11 & 1 {\tiny(9.1\%)} & 1 {\tiny(9.1\%)} & ---  \\ 
57& 58 & 11 {\tiny(19.0\%)} & ---  & 1 {\tiny(1.7\%)} \\ 
58& 61 & 10 {\tiny(16.4\%)} & ---  & 1 {\tiny(1.6\%)} \\ 
59& 41 & 7 {\tiny(17.1\%)} & 2 {\tiny(4.9\%)} & ---  \\ 
60& 122 & 20 {\tiny(16.4\%)} & 5 {\tiny(4.1\%)} & 3 {\tiny(2.5\%)} \\ 
61& 128 & 15 {\tiny(11.7\%)} & 3 {\tiny(2.3\%)} & 2 {\tiny(1.6\%)} \\ 
62& 837 & 89 {\tiny(10.6\%)} & ---  & 5 {\tiny(0.6\%)} \\ 
63& 301 & 65 {\tiny(21.6\%)} & 2 {\tiny(0.7\%)} & 7 {\tiny(2.3\%)} \\ 
64& 91 & 10 {\tiny(11.0\%)} & ---  & 1 {\tiny(1.1\%)} \\ 
65& 18 & 6 {\tiny(33.3\%)} & ---  & ---  \\ 
66& 7 & 2 {\tiny(28.6\%)} & ---  & ---  \\ 
67& 8 & 1 {\tiny(12.5\%)} & ---  & ---  \\ 
68& 5 & 3 {\tiny(60.0\%)} & ---  & ---  \\ 
69& 13 & 1 {\tiny(7.7\%)} & ---  & ---  \\ 
70& 61 & 4 {\tiny(6.6\%)} & ---  & ---  \\ 
71& 43 & 9 {\tiny(20.9\%)} & 1 {\tiny(2.3\%)} & ---  \\ 
72& 61 & 3 {\tiny(4.9\%)} & ---  & 2 {\tiny(3.3\%)} \\ 
73& 7 & 4 {\tiny(57.1\%)} & ---  & ---  \\ 
74& 55 & 14 {\tiny(25.4\%)} & ---  & 5 {\tiny(9.1\%)} \\ 
76& 5 & ---  & ---  & ---  \\ 
77& 1 & ---  & ---  & ---  \\ 
79& 5 & 1 {\tiny(20.0\%)} & ---  & ---  \\ 
80& 1 & ---  & ---  & ---  \\ 
81& 2 & 1 {\tiny(50.0\%)} & ---  & ---  \\ 
82& 70 & 5 {\tiny(7.1\%)} & ---  & 3 {\tiny(4.3\%)} \\ 
83& 1 & ---  & ---  & ---  \\ 
84& 9 & 3 {\tiny(33.3\%)} & ---  & ---  \\ 
85& 14 & 5 {\tiny(35.7\%)} & ---  & ---  \\ 
86& 15 & 4 {\tiny(26.7\%)} & ---  & ---  \\ 
87& 40 & 4 {\tiny(10.0\%)} & 2 {\tiny(5.0\%)} & 4 {\tiny(10.0\%)} \\ 
88& 54 & 8 {\tiny(14.8\%)} & 1 {\tiny(1.9\%)} & ---  \\ 
91& 8 & 4 {\tiny(50.0\%)} & ---  & ---  \\ 
92& 28 & 11 {\tiny(39.3\%)} & ---  & ---  \\ 
95& 1 & ---  & ---  & ---  \\ 
96& 11 & 1 {\tiny(9.1\%)} & ---  & ---  \\ 
97& 5 & ---  & ---  & ---  \\ 
99& 4 & 3 {\tiny(75.0\%)} & ---  & ---  \\ 
100& 8 & 4 {\tiny(50.0\%)} & ---  & 1 {\tiny(12.5\%)} \\ 
102& 2 & 1 {\tiny(50.0\%)} & ---  & ---  \\ 
105& 6 & 4 {\tiny(66.7\%)} & ---  & ---  \\ 
107& 10 & 6 {\tiny(60.0\%)} & ---  & ---  \\ 
108& 6 & 4 {\tiny(66.7\%)} & ---  & 1 {\tiny(16.7\%)} \\ 
109& 6 & 1 {\tiny(16.7\%)} & ---  & ---  \\ 
110& 3 & 2 {\tiny(66.7\%)} & ---  & ---  \\ 
111& 5 & 2 {\tiny(40.0\%)} & ---  & ---  \\ 
112& 2 & ---  & ---  & ---  \\ 
113& 40 & 5 {\tiny(12.5\%)} & ---  & ---  \\ 
114& 13 & 2 {\tiny(15.4\%)} & ---  & ---  \\ 
115& 8 & 1 {\tiny(12.5\%)} & ---  & ---  \\ 
116& 3 & 1 {\tiny(33.3\%)} & ---  & ---  \\ 
117& 6 & 2 {\tiny(33.3\%)} & ---  & ---  \\ 
118& 4 & ---  & ---  & ---  \\ 
119& 8 & 1 {\tiny(12.5\%)} & 1 {\tiny(12.5\%)} & ---  \\ 
120& 2 & 2 {\tiny(100\%)} & ---  & ---  \\ 
121& 25 & 2 {\tiny(8.0\%)} & ---  & ---  \\ 
122& 115 & 7 {\tiny(6.1\%)} & ---  & 1 {\tiny(0.9\%)} \\ 
123& 61 & 15 {\tiny(24.6\%)} & ---  & 4 {\tiny(6.6\%)} \\ 
124& 3 & 1 {\tiny(33.3\%)} & ---  & ---  \\ 
125& 10 & 3 {\tiny(30.0\%)} & ---  & 2 {\tiny(20.0\%)} \\ 
126& 4 & ---  & ---  & ---  \\ 
127& 44 & 9 {\tiny(20.4\%)} & ---  & ---  \\ 
128& 14 & 2 {\tiny(14.3\%)} & ---  & ---  \\ 
129& 151 & 10 {\tiny(6.6\%)} & 1 {\tiny(0.7\%)} & ---  \\ 
130& 13 & 3 {\tiny(23.1\%)} & ---  & ---  \\ 
131& 3 & 2 {\tiny(66.7\%)} & ---  & ---  \\ 
132& 8 & 1 {\tiny(12.5\%)} & ---  & ---  \\ 
133& 3 & 1 {\tiny(33.3\%)} & ---  & ---  \\ 
135& 7 & 4 {\tiny(57.1\%)} & ---  & 1 {\tiny(14.3\%)} \\ 
136& 73 & 10 {\tiny(13.7\%)} & ---  & 1 {\tiny(1.4\%)} \\ 
137& 24 & 6 {\tiny(25.0\%)} & ---  & ---  \\ 
138& 8 & 2 {\tiny(25.0\%)} & ---  & ---  \\ 
139& 146 & 40 {\tiny(27.4\%)} & 4 {\tiny(2.7\%)} & 9 {\tiny(6.2\%)} \\ 
140& 97 & 15 {\tiny(15.5\%)} & ---  & 3 {\tiny(3.1\%)} \\ 
141& 106 & 11 {\tiny(10.4\%)} & 1 {\tiny(0.9\%)} & 1 {\tiny(0.9\%)} \\ 
142& 35 & 3 {\tiny(8.6\%)} & ---  & ---  \\ 
143& 4 & 2 {\tiny(50.0\%)} & ---  & ---  \\ 
144& 8 & ---  & ---  & ---  \\ 
145& 4 & 2 {\tiny(50.0\%)} & ---  & ---  \\ 
146& 47 & 17 {\tiny(36.2\%)} & 1 {\tiny(2.1\%)} & 3 {\tiny(6.4\%)} \\ 
147& 30 & 4 {\tiny(13.3\%)} & ---  & ---  \\ 
148& 196 & 35 {\tiny(17.9\%)} & ---  & 11 {\tiny(5.6\%)} \\ 
149& 1 & ---  & ---  & ---  \\ 
150& 42 & 9 {\tiny(21.4\%)} & ---  & ---  \\ 
151& 4 & 1 {\tiny(25.0\%)} & ---  & ---  \\ 
152& 37 & 14 {\tiny(37.8\%)} & 1 {\tiny(2.7\%)} & ---  \\ 
154& 11 & 1 {\tiny(9.1\%)} & ---  & ---  \\ 
155& 24 & 7 {\tiny(29.2\%)} & ---  & ---  \\ 
156& 171 & 5 {\tiny(2.9\%)} & ---  & ---  \\ 
157& 6 & 1 {\tiny(16.7\%)} & ---  & ---  \\ 
158& 1 & 1 {\tiny(100\%)} & ---  & ---  \\ 
159& 15 & 3 {\tiny(20.0\%)} & ---  & ---  \\ 
160& 83 & 18 {\tiny(21.7\%)} & 1 {\tiny(1.2\%)} & 2 {\tiny(2.4\%)} \\ 
161& 39 & 9 {\tiny(23.1\%)} & ---  & ---  \\ 
162& 22 & ---  & ---  & ---  \\ 
163& 31 & 2 {\tiny(6.5\%)} & ---  & ---  \\ 
164& 192 & 26 {\tiny(13.5\%)} & ---  & 3 {\tiny(1.6\%)} \\ 
165& 28 & 8 {\tiny(28.6\%)} & ---  & 2 {\tiny(7.1\%)} \\ 
166& 312 & 28 {\tiny(9.0\%)} & ---  & 5 {\tiny(1.6\%)} \\ 
167& 152 & 27 {\tiny(17.8\%)} & ---  & 4 {\tiny(2.6\%)} \\ 
169& 6 & ---  & ---  & ---  \\ 
172& 1 & 1 {\tiny(100\%)} & ---  & ---  \\ 
173& 86 & 8 {\tiny(9.3\%)} & 1 {\tiny(1.2\%)} & 1 {\tiny(1.2\%)} \\ 
174& 12 & 4 {\tiny(33.3\%)} & ---  & ---  \\ 
176& 88 & 32 {\tiny(36.4\%)} & ---  & 2 {\tiny(2.3\%)} \\ 
178& 3 & 1 {\tiny(33.3\%)} & ---  & ---  \\ 
179& 1 & 1 {\tiny(100\%)} & ---  & ---  \\ 
180& 11 & 3 {\tiny(27.3\%)} & ---  & ---  \\ 
181& 6 & 2 {\tiny(33.3\%)} & ---  & ---  \\ 
182& 8 & 4 {\tiny(50.0\%)} & ---  & ---  \\ 
185& 34 & 3 {\tiny(8.8\%)} & ---  & ---  \\ 
186& 115 & 15 {\tiny(13.0\%)} & ---  & ---  \\ 
187& 9 & ---  & ---  & ---  \\ 
188& 22 & 10 {\tiny(45.5\%)} & ---  & ---  \\ 
189& 35 & 14 {\tiny(40.0\%)} & ---  & 1 {\tiny(2.9\%)} \\ 
190& 15 & 5 {\tiny(33.3\%)} & ---  & ---  \\ 
191& 6 & 2 {\tiny(33.3\%)} & ---  & ---  \\ 
192& 4 & 1 {\tiny(25.0\%)} & ---  & ---  \\ 
193& 14 & 2 {\tiny(14.3\%)} & ---  & ---  \\ 
194& 176 & 39 {\tiny(22.2\%)} & ---  & 7 {\tiny(4.0\%)} \\ 
195& 1 & 1 {\tiny(100\%)} & ---  & ---  \\ 
196& 1 & ---  & ---  & ---  \\ 
197& 6 & 1 {\tiny(16.7\%)} & ---  & ---  \\ 
198& 70 & 13 {\tiny(18.6\%)} & ---  & ---  \\ 
199& 13 & 2 {\tiny(15.4\%)} & ---  & ---  \\ 
200& 1 & ---  & ---  & ---  \\ 
201& 16 & 3 {\tiny(18.8\%)} & ---  & ---  \\ 
202& 5 & 1 {\tiny(20.0\%)} & ---  & ---  \\ 
203& 5 & ---  & ---  & ---  \\ 
204& 5 & ---  & ---  & 1 {\tiny(20.0\%)} \\ 
205& 70 & 7 {\tiny(10.0\%)} & ---  & 1 {\tiny(1.4\%)} \\ 
206& 36 & ---  & ---  & ---  \\ 
212& 7 & 4 {\tiny(57.1\%)} & ---  & ---  \\ 
213& 7 & ---  & ---  & ---  \\ 
214& 2 & ---  & ---  & ---  \\ 
215& 19 & 3 {\tiny(15.8\%)} & ---  & ---  \\ 
216& 73 & 9 {\tiny(12.3\%)} & ---  & ---  \\ 
217& 50 & 7 {\tiny(14.0\%)} & ---  & 3 {\tiny(6.0\%)} \\ 
218& 51 & ---  & ---  & ---  \\ 
219& 2 & 1 {\tiny(50.0\%)} & ---  & ---  \\ 
220& 10 & ---  & ---  & ---  \\ 
221& 85 & 28 {\tiny(32.9\%)} & ---  & 12 {\tiny(14.1\%)} \\ 
223& 1 & 1 {\tiny(100\%)} & ---  & ---  \\ 
224& 5 & 2 {\tiny(40.0\%)} & ---  & ---  \\ 
225& 332 & 100 {\tiny(30.1\%)} & ---  & 1 {\tiny(0.3\%)} \\ 
227& 142 & 29 {\tiny(20.4\%)} & ---  & 2 {\tiny(1.4\%)} \\ 
229& 16 & 5 {\tiny(31.2\%)} & ---  & ---  \\ 
230& 45 & 6 {\tiny(13.3\%)} & ---  & ---  \\ 
\hline
total& 9991 & 1707 {\tiny(17.1\%)} & 76 {\tiny(0.8\%)} & 225 {\tiny(2.2\%)}\\
    \hline
\end{longtable*}

\subsection{Statistics for non-trivial phonon band sets}\label{app:statistics}

We now turn to the statistics for non-trivial phonon band sets. For a given material, the number of band sets might change depending whether the dispersion calculations were performed with or without NAC. Indeed, gaps might open or close depending on the type of calculations. For materials from the Kyoto database, \TPDBNbrKyotoMaterialsWoWNACBandChanges\ out of the \TPDBNbrKyotoMaterials\ entries (\TPDBPercentKyotoMaterialsWoWNACBandChanges\%) has different number of band sets with and without NAC. Similarly, \TPDBNbrMPMaterialsWoWNACBandChanges\ out of the \TPDBNbrMPMaterials\ materials (\TPDBPercentMPMaterialsWoWNACBandChanges\%) from the MP have a change of their band sets number. For that reason, for each table providing the statistics per space group of phonon band sets, Tab.~\ref{tab:statiticssgmp} for materials obtained from MP and Tab.~\ref{tab:statiticssgkyoto} for materials obtained from the Kyoto phonon database, we give the numbers of materials hosting strong topology, fragile topology, OABR or OOABR band sets both without and with NAC. As indicated at the end of these two tables, even if considering NAC does change the number of band sets for a small fraction of materials, the percentage of materials having at least one non atomic band set is almost unchanged. For the Kyoto database, this number goes down from \TPDBNbrKyotoMaterialsWoNACNonTrivialNonAtomic\ (\TPDBPercentKyotoMaterialsWoNACNonTrivialNonAtomic\%) to \TPDBNbrKyotoMaterialsWithNACNonTrivialNonAtomic\ (\TPDBPercentKyotoMaterialsWithNACNonTrivialNonAtomic\%) when using NAC and for the MP database, it changes from \TPDBNbrMPMaterialsWoNACNonTrivialNonAtomic\ (\TPDBPercentMPMaterialsWoNACNonTrivialNonAtomic\%) to \TPDBNbrMPMaterialsWithNACNonTrivialNonAtomic\ (\TPDBPercentMPMaterialsWithNACNonTrivialNonAtomic\%).

As we have discussed in \siref{app:negativenergies}, \TPDBNbrKyotoMaterialsNegativeBands\ materials from the Kyoto database and \TPDBNbrKyotoMaterialsInstabilities\ materials from the MP database have negative dispersions. As such, we provide the analogue of Tabs.~\ref{tab:statiticssgkyoto} and~\ref{tab:statiticssgmp} but where materials with negative dispersions have been filtered out: Tab.~\ref{tab:statiticssgkyotonnonnegative} for the Kyoto database and Tab.~\ref{tab:statiticssgmpnonnegative} for the MP database. We can observe that removing materials with negative energies barely changes the ratio of materials hosting at least one non-atomic band set. For example, if we focus on the \TPDBNbrKyotoMaterialsWithNACNoNegative\ materials from the Kyoto database without negative energies computed with NAC, \TPDBNbrKyotoMaterialsWithNACNonTrivialNoNegative\ entries, i.e., \TPDBPercentKyotoMaterialsWithNACNonTrivialNonAtomicNoNegative\%, have at least one non-atomic band set, as opposed to \TPDBPercentKyotoMaterialsWithNACNonTrivialNonAtomic\% mentioned previously when considering all the Kyoto database materials.

\LTcapwidth=1.0\textwidth
\renewcommand\arraystretch{1.0}
\begin{longtable*}{|c|c|c|c|c|c|c|c|}
  \caption[Statistics of non-trivial phonon band sets for the phonon materials from Materials Project]{Statistics of non-trivial phonon band sets for the phonon materials from Materials Project. The statistics are provided per space group (SG), shown in the first column. The second column gives the number of materials per SG. The third column provides the number and percentage of materials hosting at least one strong-topology band set. The fourth, fifth and sixth columns gives similar figures for fragile-topology, OABR and OOABR band sets. Note that a material can host several types of non-trivial or non-atomic band sets. Thus the percentages do not add up. In the seventh column, we show the number of materials hosting at least one non-atomic band set, i.e., topological, fragile, OABR or OOABR. Conversely, the last column gives the number of materials having solely atomic band sets. For each SG, the first line provides statistics for calculations performed without NAC while the second line gives statistics for calculations performed with NAC.}
    \label{tab:statiticssgmp}\\
    \hline
    \hline
    {\scriptsize{SG}} & {\scriptsize{\# materials}}  & {\scriptsize{\begin{tabular}{c}\# materials with\\ strong band sets \end{tabular}}} & {\scriptsize{\begin{tabular}{c}\# materials with\\ fragile band sets \end{tabular}}} & {\scriptsize{\begin{tabular}{c}\# materials with\\ OABR band sets \end{tabular}}} & {\scriptsize{\begin{tabular}{c}\# materials with\\ OOABR band sets \end{tabular}}}  & {\scriptsize{\begin{tabular}{c}\# materials with\\ non-atomic band sets \end{tabular}}} & {\scriptsize{\begin{tabular}{c}\# materials with\\ only atomic band sets \end{tabular}}}\\
        \hline
1& 3 & ---  & ---  & ---  & ---  & ---  & 3 {\tiny(100\%)} \\ 
 &  & ---  & ---  & ---  & ---  & ---  & 3 {\tiny(100\%)} \\ 
2& 5 & 5 {\tiny(100\%)} & ---  & 5 {\tiny(100\%)} & 1 {\tiny(20.0\%)} & 5 {\tiny(100\%)} & ---  \\
\hline
\end{longtable*}

\LTcapwidth=1.0\textwidth
\renewcommand\arraystretch{1.0}
\begin{longtable*}{|c|c|c|c|c|c|c|c|}
  \caption[Statistics of non-trivial phonon band sets for the phonon materials from the Kyoto phonon database]{Statistics of non-trivial phonon band sets for the phonon materials from the Kyoto phonon database. The statistics are provided per space group (SG), shown in the first column. The second column gives the number of materials per SG. The third column provides the number and percentage of materials hosting at least one strong-topology band set. The fourth, fifth and sixth columns gives similar figures for fragile-topology, OABR and OOABR band sets. Note that a material can host several types of non-trivial or non-atomic band sets. Thus the percentages do not add up. In the seventh column, we show the number of materials hosting at least one non-atomic band set, i.e., strong-topology, fragile-topology, OABR or OOABR. Conversely, the last column gives the number of materials having solely atomic band sets. For each SG, the first line provides statistics for calculations performed without NAC while the second line gives statistics for calculations performed with NAC.}
    \label{tab:statiticssgkyoto}\\
    \hline
    \hline
    {\scriptsize{SG}} & {\scriptsize{\# materials}}  & {\scriptsize{\begin{tabular}{c}\# materials with\\ strong band sets \end{tabular}}} & {\scriptsize{\begin{tabular}{c}\# materials with\\ fragile band sets \end{tabular}}} & {\scriptsize{\begin{tabular}{c}\# materials with\\ OABR band sets \end{tabular}}} & {\scriptsize{\begin{tabular}{c}\# materials with\\ OOABR band sets \end{tabular}}} & {\scriptsize{\begin{tabular}{c}\# materials with\\ non-atomic band sets \end{tabular}}} & {\scriptsize{\begin{tabular}{c}\# materials with\\ only atomic band sets \end{tabular}}} \\
        \hline
3& 11 & 11 {\tiny(100\%)} & ---  & 2 {\tiny(18.2\%)} & ---  & 11 {\tiny(100\%)} & ---  \\ 
 &  & 11 {\tiny(100\%)} & ---  & 2 {\tiny(18.2\%)} & ---  & 11 {\tiny(100\%)} & ---  \\
\hline
\end{longtable*}

\LTcapwidth=1.0\textwidth
\renewcommand\arraystretch{1.0}
\begin{longtable*}{|c|c|c|c|c|c|c|c|}
  \caption[Statistics of non-trivial phonon band sets for the phonon materials without negative energies from Materials Project]{Statistics of non-trivial phonon band sets for the phonon materials \emph{without negative energies} from Materials Project. The statistics are provided per space group (SG), shown in the first column. The second column gives the number of materials per SG. The third column provides the number and percentage of materials hosting at least one strong-topology band set. The fourth, fifth and sixth columns gives similar figures for fragile-topology, OABR and OOABR band sets. Note that a material can host several types of non-trivial or non-atomic band sets. Thus the percentages do not add up. In the seventh column, we show the number of materials hosting at least one non-atomic band set, i.e., strong-topology, fragile-topology, OABR or OOABR. Conversely, the last column gives the number of materials having solely atomic band sets. For each SG, the first line provides statistics for calculations performed without NAC while the second line gives statistics for calculations performed with NAC.}
    \label{tab:statiticssgmpnonnegative}\\
    \hline
    \hline
    {\scriptsize{SG}} & {\scriptsize{\# materials}}  & {\scriptsize{\begin{tabular}{c}\# materials with\\ strong band sets \end{tabular}}} & {\scriptsize{\begin{tabular}{c}\# materials with\\ fragile band sets \end{tabular}}} & {\scriptsize{\begin{tabular}{c}\# materials with\\ OABR band sets \end{tabular}}} & {\scriptsize{\begin{tabular}{c}\# materials with\\ OOABR band sets \end{tabular}}}  & {\scriptsize{\begin{tabular}{c}\# materials with\\ non-atomic band sets \end{tabular}}} & {\scriptsize{\begin{tabular}{c}\# materials with\\ only atomic band sets \end{tabular}}}\\
        \hline
1& 3 & ---  & ---  & ---  & ---  & ---  & 3 {\tiny(100\%)} \\ 
 & 3 & ---  & ---  & ---  & ---  & ---  & 3 {\tiny(100\%)} \\
\hline
\end{longtable*}

\LTcapwidth=1.0\textwidth
\renewcommand\arraystretch{1.0}
\begin{longtable*}{|c|c|c|c|c|c|c|c|}
  \caption[Statistics of non-trivial phonon band sets for the phonon materials without negative energies from the Kyoto phonon database]{Statistics of non-trivial phonon band sets for the phonon materials \emph{without negative energies} from the Kyoto phonon database. The statistics are provided per space group (SG), shown in the first column. The second column gives the number of materials per SG. The third column provides the number and percentage of materials hosting at least one strong-topology band set. The fourth, fifth and sixth columns gives similar figures for fragile-topology, OABR and OOABR band sets. Note that a material can host several types of non-trivial or non-atomic band sets. Thus the percentages do not add up. In the seventh column, we show the number of materials hosting at least one non-atomic band set, i.e., strong-topology, fragile-topology, OABR or OOABR. Conversely, the last column gives the number of materials having solely atomic band sets. For each SG, the first line provides statistics for calculations performed without NAC while the second line gives statistics for calculations performed with NAC.}
    \label{tab:statiticssgkyotonnonnegative}\\
    \hline
    \hline
    {\scriptsize{SG}} & {\scriptsize{\# materials}}  & {\scriptsize{\begin{tabular}{c}\# materials with\\ strong band sets \end{tabular}}} & {\scriptsize{\begin{tabular}{c}\# materials with\\ fragile band sets \end{tabular}}} & {\scriptsize{\begin{tabular}{c}\# materials with\\ OABR band sets \end{tabular}}} & {\scriptsize{\begin{tabular}{c}\# materials with\\ OsOABR band sets \end{tabular}}} & {\scriptsize{\begin{tabular}{c}\# materials with\\ non-atomic band sets \end{tabular}}} & {\scriptsize{\begin{tabular}{c}\# materials with\\ only atomic band sets \end{tabular}}} \\
        \hline
3& 4 & 4 {\tiny(100\%)} & ---  & 1 {\tiny(25.0\%)} & ---  & 4 {\tiny(100\%)} & ---  \\
\hline
\end{longtable*}

\section{``Ideal'' non-atomic phonon bands}\label{app:bestmaterials}

Among the \TPDBNbrKyotoMaterials\ MPID entries on \webphonondbshort\ and \TPDBNbrMPMaterials\ MPID entries on \webMP, we obtained \TPDBNbrKyotoMaterialsWoNACNonTrivialNonAtomicNoNegative(\TPDBPercentKyotoMaterialsWoNACNonTrivialNonAtomicNoNegative\%) and \TPDBNbrMPMaterialsWoNACNonTrivialNonAtomicNoNegative(\TPDBPercentMPMaterialsWoNACNonTrivialNonAtomicNoNegative\%) entries that have at least one type of non-atomic cumulative band topologies. In addition, all the symmetry-enforced band nodes were also identified by evaluating the compatibility relations.
In this Appendix, we perform a screening of the ``ideal'' phonon materials, which host either non-atomic band gap or ``ideal'' symmetry-enforced band nodes, with the following criteria (previously described in the main text):
\begin{enumerate}
    \item The material is not tagged as having ``negativities'' in the database, \ie its minimum frequency is above -5.0meV, such that its crystal structure is considered as relatively stable. 
    \item The maximal direct band gap along all the high-symmetry paths between the non-atomic cumulative set of bands and the first band above it (or between the bands forming a symmetry-enforced band node) is larger than 1meV, which is the typical experimental resolution to distinguish a phonon band gap. 
    \item The indirect gap along all the high-symmetry paths is positive for non-atomic cumulative band sets and and non-negative for symmetry-enforced band nodes, respectively.
\end{enumerate}

Using the above screening criteria, \TPDBNbrIdealTopoMaterials(\TPDBPercentIdealTopoMaterials\%), \TPDBNbrIdealOABRMaterials(\TPDBPercentIdealOABRMaterials\%) and \TPDBNbrIdealOOABRMaterials(\TPDBPercentIdealOOABRMaterials\%) ``ideal'' phonon materials with strong topology, OABR and OOABR are obtained, respectively. We also select \TPDBNbrIdealSMMaterials(\TPDBPercentIdealSMMaterials\%) phonon materials hosting ``ideal'' symmetry-enforced band nodes.
We find no ``ideal'' phonon material with fragile cumulative topology satisfying the above criteria. More specifically, we tabulate the statistics of ``ideal'' phonon materials for \webphonondbshort\ and \webMP\ in the Table \ref{tab:idealstat}. In Tables \ref{tab:besttopmp} to \ref{tab:bestooabrkyoto}, we give the complete lists of "ideal" phonon materials for each data source and each non-atomic type. In \siref{app:bestmaterialbandstr}, we also provide the phonon spectrum plot near the topological band gap and band nodes for ``ideal'' materials with strong topologies and symmetry-enforced band nodes, respectively.

\begin{table}[h]
    \centering
    \caption{Statistics for ``ideal'' phonon materials with non-atomic cumulative topologies or enforced band nodes on \webphonondbshort\ and \webMP. For each category, we provide the related percentage with respect to the total number of MPIDs in the corresponding database and a link to its material list.}
    \label{tab:idealstat}
\begin{tabular}{|c|c|c|c|c|}
    \hline\hline
        Data source &  Strong topology & Enforced nodes & OABR & OOABR \\
        \hline
        \webphonondbshort\ & \TPDBNbrIdealTopoKyotoMaterials(\TPDBPercentIdealTopoKyotoMaterials\%) Tab.\ref{tab:besttopkyoto}& \TPDBNbrIdealSMKyotoMaterials(\TPDBPercentIdealSMKyotoMaterials\%) Tab.\ref{tab:bestsmkyoto}& \TPDBNbrIdealOABRKyotoMaterials(\TPDBPercentIdealOABRKyotoMaterials\%) Tab.\ref{tab:bestoabrkyoto}& \TPDBNbrIdealOOABRKyotoMaterials(\TPDBPercentIdealOOABRKyotoMaterials\%) Tab.\ref{tab:bestooabrkyoto} \\
        \hline
        \webMP\ & \TPDBNbrIdealTopoMPMaterials(\TPDBPercentIdealTopoMPMaterials\%) Tab.\ref{tab:besttopmp}& \TPDBNbrIdealSMMPMaterials(\TPDBPercentIdealSMMPMaterials\%) Tab.\ref{tab:bestsmmp}& \TPDBNbrIdealOABRMPMaterials(\TPDBPercentIdealOABRMPMaterials\%) Tab.\ref{tab:bestoabrmp}& \TPDBNbrIdealOOABRMPMaterials(\TPDBPercentIdealOOABRMPMaterials\%) Tab.\ref{tab:bestooabrmp} \\
        \hline
    \end{tabular}
\end{table}

\subsection{``Ideal'' phonon materials with non-atomic cumulative band sets}\label{app:bestmateriallist}

\subsubsection{``Ideal'' topological phonon materials with cumulative strong topology}

\LTcapwidth=1.0\textwidth
\renewcommand\arraystretch{1.2}
\begin{longtable*}{|c|l|l|l|l|c|c|c|c|c|c|}
    \caption[The ``ideal'' phonon materials with cumulative strong topology on the \webMP]{The ``ideal'' topological phonon materials with cumulative strong topology on the \webMP. 
    For each MPID entry, the table provides its space group number and symbol (SG), MPID with a direct link to the Topological phonon website, chemical formula (chem. formula ), and a link to its band structure (BS) plot near the topological band gaps. Note that there can exist more than one ``ideal'' set of bands with cumulative strong topology in a phonon band structure. For each band set, we provide the band index of the top band (topBI), the maximal energy level (maxE) in unit of meV of the band set, the maximal direct band gap (maxDGap) and the indirect band gap (IGap) along the high-symmetry paths of the BZ, TQC topology of the cumulative set of bands (Topo.), and the related topological indices (Indices) that indicate the band topology.} 
    \label{tab:besttopmp}\\
    \hline
    \scriptsize{No.} & \scriptsize{SG} & \scriptsize{MPID} & \scriptsize{chem. formula} & \scriptsize{BS} & \scriptsize{topBI} & \scriptsize{E (meV)} & \scriptsize{maxDGap} & \scriptsize{IGap} & \scriptsize{Topo.} & \scriptsize{Indices} \\
    \hline
\multirow{1}{*}{\scriptsize{1}} & \multirow{1}{*}{\scriptsize{2 (\sgsymb{2})}} & \multirow{1}{*}{\scriptsize{\mpidweb{561710}}} & \multirow{1}{*}{\scriptsize{\ce{CsSnS$_3$}}} & \multirow{1}{*}{\scriptsize{Fig. \ref{fig:1_besttop_mp}}} & \scriptsize{17} & \scriptsize{18.741} & \scriptsize{0.048} & \scriptsize{0.005} & \scriptsize{NLC} & \scriptsize{($z_{2,1}$,$z_{2,2}$,$z_{2,3}$,$z_{4}$) = (0,1,0,0)} \\
\hline
\end{longtable*}

\LTcapwidth=1.0\textwidth
\renewcommand\arraystretch{1.2}
\begin{longtable*}{|c|l|l|l|l|c|c|c|c|c|c|}
    \caption[The ``ideal'' phonon materials with cumulative strong topology on the \webphonondb]{The ``ideal'' topological phonon materials with cumulative strong topology on the \webphonondb. 
    For each MPID entry, the table provides its space group number and symbol (SG), MPID with a direct link to the Topological phonon website, chemical formula (chem. formula ), and a link to its band structure (BS) plot near the topological band gaps. Note that there can exist more than one ``ideal'' set of bands with cumulative strong topology in a phonon band structure. For each band set, we provide the band index of the top band (topBI), the maximal energy level (maxE) in unit of meV of the band set, the maximal direct band gap (maxDGap) and the indirect band gap (IGap) along the high-symmetry paths of the BZ, TQC topology of the cumulative set of bands (Topo.), and the related topological indices (Indices) that indicate the band topology.} 
    \label{tab:besttopkyoto}\\
    \hline
    \scriptsize{No.} & \scriptsize{SG} & \scriptsize{MPID} & \scriptsize{chem. formula} & \scriptsize{BS} & \scriptsize{topBI} & \scriptsize{E (meV)} & \scriptsize{maxDGap} & \scriptsize{IGap} & \scriptsize{Topo.} & \scriptsize{Indices} \\
    \hline
\multirow{1}{*}{\scriptsize{1}} & \multirow{1}{*}{\scriptsize{3 (\sgsymb{3})}} & \multirow{1}{*}{\scriptsize{\mpidweb{571132}}} & \multirow{1}{*}{\scriptsize{\ce{LiMg$_10$AlH$_24$}}} & \multirow{1}{*}{\scriptsize{Fig. \ref{fig:1_besttop_kyoto}}} & \scriptsize{84} & \scriptsize{149.449} & \scriptsize{0.343} & \scriptsize{0.098} & \scriptsize{NLC} & \scriptsize{($\alpha_{2}$) = (1)} \\
\hline
\end{longtable*}

\subsubsection{``Ideal'' phonon materials with symmetry-enforced band nodes}

\LTcapwidth=1.0\textwidth
\renewcommand\arraystretch{1.2}
\begin{longtable*}{|c|l|l|l|l|c|c|c|}
    \caption[The ``ideal'' phonon materials with symmetry-enforced band nodes on the \webMP]{The ``ideal'' phonon materials with symmetry-enforced band nodes on the \webMP.
    For each MPID entry, the table provides its space group number and symbol (SG), MPID with a direct link to the Topological phonon website, chemical formula (chem. formula ), and a link to its band structure (BS) plot near the symmetry-enforced band nodes, number of bands below the symmetry-enforced band crossings (Nbr Bands), energy level (E) and the the maximal direct band gap (maxDGap) in unit of meV of the band nodes.} 
    \label{tab:bestsmmp}\\
    \hline
    \scriptsize{No.} & \scriptsize{SG} & \scriptsize{MPID} & \scriptsize{chem. formula} & \scriptsize{BandStr} & \scriptsize{Nbr Bands} & \scriptsize{E (meV)} & maxDGap\\
    \hline
\multirow{1}{*}{\scriptsize{1}} & \multirow{1}{*}{\scriptsize{11 (\sgsymb{11})}} & \multirow{1}{*}{\scriptsize{\mpidweb{998413}}} & \multirow{1}{*}{\scriptsize{\ce{CsSeBr$_3$}}} & \multirow{1}{*}{\scriptsize{Fig. \ref{fig:1_bestsm_mp}}} & \scriptsize{16} & \scriptsize{10.417} & \scriptsize{2.453} \\
\hline
\multirow{1}{*}{\scriptsize{2}} & \multirow{1}{*}{\scriptsize{12 (\sgsymb{12})}} & \multirow{1}{*}{\scriptsize{\mpidweb{10985}}} & \multirow{1}{*}{\scriptsize{\ce{RbCu$_3$S$_2$}}} & \multirow{1}{*}{\scriptsize{Fig. \ref{fig:1_bestsm_mp}}} & \scriptsize{35} & \scriptsize{39.528} & \scriptsize{1.427} \\
\hline
\multirow{1}{*}{\scriptsize{3}} & \multirow{1}{*}{\scriptsize{14 (\sgsymb{14})}} & \multirow{1}{*}{\scriptsize{\mpidweb{27904}}} & \multirow{1}{*}{\scriptsize{\ce{FeSbS}}} & \multirow{1}{*}{\scriptsize{Fig. \ref{fig:1_bestsm_mp}}} & \scriptsize{34} & \scriptsize{52.487} & \scriptsize{3.981} \\
\hline
\end{longtable*}

\LTcapwidth=1.0\textwidth
\renewcommand\arraystretch{1.2}
\begin{longtable*}{|c|l|l|l|l|c|c|c|}
    \caption[The ``ideal'' phonon materials with symmetry-enforced band nodes on the \webphonondb]{The ``ideal'' phonon materials with symmetry-enforced band nodes on the \webphonondb.
    For each MPID entry, the table provides its space group number and symbol (SG), MPID with a direct link to the Topological phonon website, chemical formula (chem. formula ), and a link to its band structure (BS) plot near the symmetry-enforced band nodes, number of bands below the symmetry-enforced band crossings (Nbr Bands), energy level (E) and the the maximal direct band gap (maxDGap) in unit of meV of the band nodes.} 
    \label{tab:bestsmkyoto}\\
    \hline
    \scriptsize{No.} & \scriptsize{SG} & \scriptsize{MPID} & \scriptsize{chem. formula} & \scriptsize{BandStr} & \scriptsize{Nbr Bands} & \scriptsize{E (meV)} & maxDGap\\
    \hline
\multirow{1}{*}{\scriptsize{1}} & \multirow{1}{*}{\scriptsize{4 (\sgsymb{4})}} & \multirow{1}{*}{\scriptsize{\mpidweb{5263}}} & \multirow{1}{*}{\scriptsize{\ce{Ca$_4$P$_2$O$_9$}}} & \multirow{1}{*}{\scriptsize{Fig. \ref{fig:1_bestsm_kyoto}}} & \scriptsize{118} & \scriptsize{56.278} & \scriptsize{2.105} \\
\hline
\multirow{1}{*}{\scriptsize{2}} & \multirow{1}{*}{\scriptsize{4 (\sgsymb{4})}} & \multirow{1}{*}{\scriptsize{\mpidweb{6903}}} & \multirow{1}{*}{\scriptsize{\ce{Na$_2$Ca$_4$ZrNbSi$_4$O$_17$F}}} & \multirow{1}{*}{\scriptsize{Fig. \ref{fig:1_bestsm_kyoto}}} & \scriptsize{154} & \scriptsize{103.989} & \scriptsize{4.624} \\
\hline
\multirow{1}{*}{\scriptsize{3}} & \multirow{1}{*}{\scriptsize{4 (\sgsymb{4})}} & \multirow{1}{*}{\scriptsize{\mpidweb{23544}}} & \multirow{1}{*}{\scriptsize{\ce{Bi$_4$Br$_2$O$_5$}}} & \multirow{1}{*}{\scriptsize{Fig. \ref{fig:1_bestsm_kyoto}}} & \scriptsize{72} & \scriptsize{19.5} & \scriptsize{5.03} \\
\hline
\end{longtable*}

\subsubsection{``Ideal'' phonon materials with cumulative OABR topology}

\LTcapwidth=1.0\textwidth
\renewcommand\arraystretch{1.2}
\begin{longtable*}{|c|l|l|l|l|c|c|c|c|p{5.5cm}|}
        \caption[The ``ideal'' phonon materials with cumulative OABR topology on the \webMP.]{The ``ideal'' phonon materials with cumulative OABR topology on the \webMP.
        For each MPID entry, the table provides its space group number and symbol (SG), MPID with a direct link to the Topological phonon website, chemical formula (chem. formula ). For each cumulative band set, we provide the band index of the top band (topBI), the maximal energy level (maxE) in unit of meV of the band set, the indirect band gap (IGap) and the maximal direct band gap (maxDGap) along the high-symmetry paths of the BZ, topology of the cumulative set of bands (Topo.), and the RSI indices that indicate the related topology. The defnition of RSIs can be found on the \webBCS, and the letter(s) in the parentheses is the Wyckoff position(s) where the RSI is defined.} 
    \label{tab:bestoabrmp}\\
    \hline
    \scriptsize{No.} & \scriptsize{SG} & \scriptsize{MPID} & \scriptsize{chem. formula} & \scriptsize{topBI} & \scriptsize{maxE(meV)} & \scriptsize{IGap} & \scriptsize{maxDgap} & \scriptsize{Topo.} & \scriptsize{RSI}\\
    \hline
\multirow{1}{*}{\scriptsize{1}} & \multirow{1}{*}{\scriptsize{2(\sgsymb{2}})} & \multirow{1}{*}{\scriptsize{\mpidweb{997036}}} & \multirow{1}{*}{\scriptsize{\ce{CdAuO$_2$}}} & \scriptsize{21} & \scriptsize{72.43} & \scriptsize{0.236} & \scriptsize{5.811} & \scriptsize{OABR} & \scriptsize{$\delta_7(g)=1$}  \\
\hline
\multirow{7}{*}{\scriptsize{2}} & \multirow{7}{*}{\scriptsize{2(\sgsymb{2}})} & \multirow{7}{*}{\scriptsize{\mpidweb{561710}}} & \multirow{7}{*}{\scriptsize{\ce{CsSnS$_3$}}} & \scriptsize{9} & \scriptsize{8.61} & \scriptsize{0.156} & \scriptsize{3.535} & \scriptsize{OABR} & \scriptsize{$\delta_3(c)=1$,$\delta_7(g)=1$,$\delta_8(h)=1$}  \\
\cline{5-10}
&&&& \scriptsize{15} & \scriptsize{15.587} & \scriptsize{0.132} & \scriptsize{2.346} & \scriptsize{OABR} & \scriptsize{$\delta_3(c)=1$} \\
\hline
\end{longtable*}

\LTcapwidth=1.0\textwidth
\renewcommand\arraystretch{1.2}
\begin{longtable*}{|c|l|l|l|l|c|c|c|c|p{5.5cm}|}
        \caption[The ``ideal'' phonon materials with cumulative OABR topology on the \webphonondb.]{The ``ideal'' phonon materials with cumulative OABR topology on the \webphonondb.
        For each MPID entry, the table provides its space group number and symbol (SG), MPID with a direct link to the Topological phonon website, chemical formula (chem. formula ). For each cumulative band set, we provide the band index of the top band (topBI), the maximal energy level (maxE) in unit of meV of the band set, the indirect band gap (IGap) and the maximal direct band gap (maxDGap) along the high-symmetry paths of the BZ, topology of the cumulative set of bands (Topo.), and the RSI indices that indicate the related topology. The defnition of RSIs can be found on the \webBCS, and the letter(s) in the parentheses is the Wyckoff position(s) where the RSI is defined.} 
    \label{tab:bestoabrkyoto}\\
    \hline
    \scriptsize{No.} & \scriptsize{SG} & \scriptsize{MPID} & \scriptsize{chem. formula} & \scriptsize{topBI} & \scriptsize{maxE(meV)} & \scriptsize{IGap} & \scriptsize{maxDgap} & \scriptsize{Topo.} & \scriptsize{RSI}\\
    \hline
\multirow{2}{*}{\scriptsize{1}} & \multirow{2}{*}{\scriptsize{10(\sgsymb{10}})} & \multirow{2}{*}{\scriptsize{\mpidweb{972120}}} & \multirow{2}{*}{\scriptsize{\ce{Sr(GaAs)$_2$}}} & \scriptsize{56} & \scriptsize{28.194} & \scriptsize{2.396} & \scriptsize{3.993} & \scriptsize{OABR} & \scriptsize{$\delta_1(a)=1$,$\delta_7(g)=1$}  \\
\cline{5-10}
&&&& \scriptsize{57} & \scriptsize{31.067} & \scriptsize{2.481} & \scriptsize{3.002} & \scriptsize{OABR} & \scriptsize{$\delta_7(g)=1$} \\
\hline
        \hline
\end{longtable*}

\subsubsection{``Ideal'' phonon materials with cumulative OOABR topology}

\LTcapwidth=1.0\textwidth
\renewcommand\arraystretch{1.2}
\begin{longtable*}{|c|l|l|l|l|c|c|c|c|p{5.5cm}|}
        \caption[The ``ideal'' phonon materials with cumulative OOABR topology on the \webMP.]{The ``ideal'' phonon materials with cumulative OOABR topology on the \webMP.
        For each MPID entry, the table provides its space group number and symbol (SG), MPID with a direct link to the Topological phonon website, chemical formula (chem. formula ). For each cumulative band set, we provide the band index of the top band (topBI), the maximal energy level (maxE) in unit of meV of the band set, the indirect band gap (IGap) and the maximal direct band gap (maxDGap) along the high-symmetry paths of the BZ, topology of the cumulative set of bands (Topo.), and the RSI indices that indicate the related topology. The defnition of RSIs can be found on the \webBCS, and the letter(s) in the parentheses is the Wyckoff position(s) where the RSI is defined.} 
    \label{tab:bestooabrmp}\\
    \hline
    \scriptsize{No.} & \scriptsize{SG} & \scriptsize{MPID} & \scriptsize{chem. formula} & \scriptsize{topBI} & \scriptsize{maxE(meV)} & \scriptsize{IGap} & \scriptsize{maxDgap} & \scriptsize{Topo.} & \scriptsize{RSI}\\
    \hline
\multirow{1}{*}{\scriptsize{1}} & \multirow{1}{*}{\scriptsize{12(\sgsymb{12}})} & \multirow{1}{*}{\scriptsize{\mpidweb{13055}}} & \multirow{1}{*}{\scriptsize{\ce{Ba$_2$LaTaO$_6$}}} & \scriptsize{29} & \scriptsize{85.64} & \scriptsize{6.989} & \scriptsize{30.625} & \scriptsize{OOABR} & \scriptsize{$\delta_4(d)=4$}  \\
\hline
\multirow{1}{*}{\scriptsize{2}} & \multirow{1}{*}{\scriptsize{12(\sgsymb{12}})} & \multirow{1}{*}{\scriptsize{\mpidweb{23127}}} & \multirow{1}{*}{\scriptsize{\ce{Ba$_2$BiSbO$_6$}}} & \scriptsize{29} & \scriptsize{79.8} & \scriptsize{0.265} & \scriptsize{17.342} & \scriptsize{OOABR} & \scriptsize{$\delta_1(a)=4$}  \\
\hline
\end{longtable*}

\LTcapwidth=1.0\textwidth
\renewcommand\arraystretch{1.2}
\begin{longtable*}{|c|l|l|l|l|c|c|c|c|p{5.5cm}|}
        \caption[The ``ideal'' phonon materials with cumulative OOABR topology on the \webphonondb.]{The ``ideal'' phonon materials with cumulative OOABR topology on the \webphonondb.
        For each MPID entry, the table provides its space group number and symbol (SG), MPID with a direct link to the Topological phonon website, chemical formula (chem. formula ). For each cumulative band set, we provide the band index of the top band (topBI), the maximal energy level (maxE) in unit of meV of the band set, the indirect band gap (IGap) and the maximal direct band gap (maxDGap) along the high-symmetry paths of the BZ, topology of the cumulative set of bands (Topo.), and the RSI indices that indicate the related topology. The defnition of RSIs can be found on the \webBCS, and the letter(s) in the parentheses is the Wyckoff position(s) where the RSI is defined.} 
    \label{tab:bestooabrkyoto}\\
    \hline
    \scriptsize{No.} & \scriptsize{SG} & \scriptsize{MPID} & \scriptsize{chem. formula} & \scriptsize{topBI} & \scriptsize{maxE(meV)} & \scriptsize{IGap} & \scriptsize{maxDgap} & \scriptsize{Topo.} & \scriptsize{RSI}\\
    \hline
\multirow{1}{*}{\scriptsize{1}} & \multirow{1}{*}{\scriptsize{12(\sgsymb{12}})} & \multirow{1}{*}{\scriptsize{\mpidweb{561973}}} & \multirow{1}{*}{\scriptsize{\ce{K(TeO$_3$)$_2$}}} & \scriptsize{105} & \scriptsize{94.913} & \scriptsize{1.875} & \scriptsize{4.314} & \scriptsize{OOABR} & \scriptsize{$\delta_4(d)=4$}  \\
\hline
\multirow{2}{*}{\scriptsize{2}} & \multirow{2}{*}{\scriptsize{12(\sgsymb{12}})} & \multirow{2}{*}{\scriptsize{\mpidweb{755013}}} & \multirow{2}{*}{\scriptsize{\ce{Li$_5$TaO$_5$}}} & \scriptsize{29} & \scriptsize{73.476} & \scriptsize{0.394} & \scriptsize{10.325} & \scriptsize{OOABR} & \scriptsize{$\delta_1(a)=4$}  \\
\cline{5-10}
&&&& \scriptsize{32} & \scriptsize{90.949} & \scriptsize{1.191} & \scriptsize{29.564} & \scriptsize{OOABR} & \scriptsize{$\delta_1(a)=4$} \\
\hline
        \hline
\end{longtable*}

\subsection{Band structures of the ``ideal'' topological phonon materials with cumulative strong topologies and symmetry-enforced band nodes}\label{app:bestmaterialbandstr}

For each band structure plot, we provide its material MPID with a direct link to the \webTQCphonon, the material SG and its chemical formula on the top of it. 
For each cumulative set of topological phonon bands that is selected as an ``ideal'' in Tables \ref{tab:besttopmp}-\ref{tab:bestsmkyoto}, its maximal frequency is indicated by a red dashed line.  
In each plot, the isolated topological band sets are indicated by different colors: blue dispersions are for strong topology, red dispersions are for fragile topology, green dispersions are for OABR, orange dispersions are for OOABR. Light and dark shades are a guide to distinguish separated band sets with identical properties.

\subsubsection{Band structures of the best topological phonon bands}

\begin{figure}[h]
\centering
\begin{tabular}{cc}
\scriptsize{\mpidweb{561710}, SG 2 (\sgsymb{2}), \ce{CsSnS$_3$}} & \scriptsize{\mpidweb{997036}, SG 2 (\sgsymb{2}), \ce{CdAuO$_2$}} \\
\includegraphics[width=0.5\textwidth,angle=0]{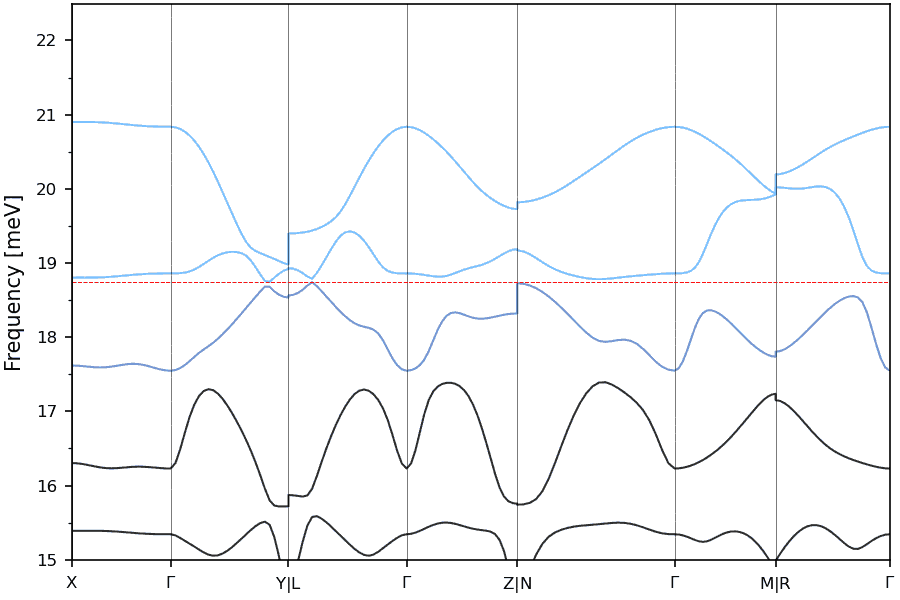} & \includegraphics[width=0.5\textwidth,angle=0]{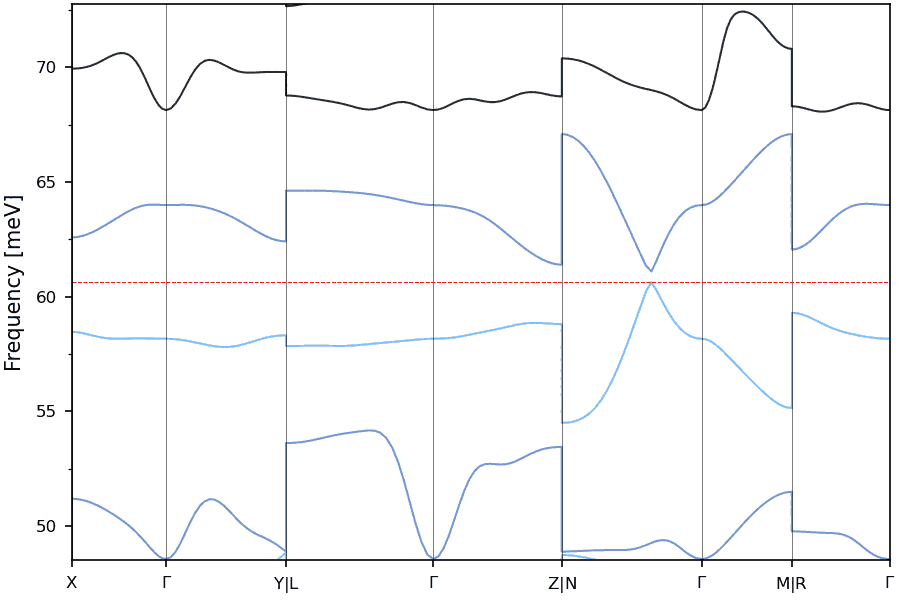} \\
\scriptsize{\mpidweb{7808}, SG 12 (\sgsymb{12}), \ce{BaP$_3$}} & \scriptsize{\mpidweb{568666}, SG 12 (\sgsymb{12}), \ce{RbAuI$_3$}} \\
\includegraphics[width=0.5\textwidth,angle=0]{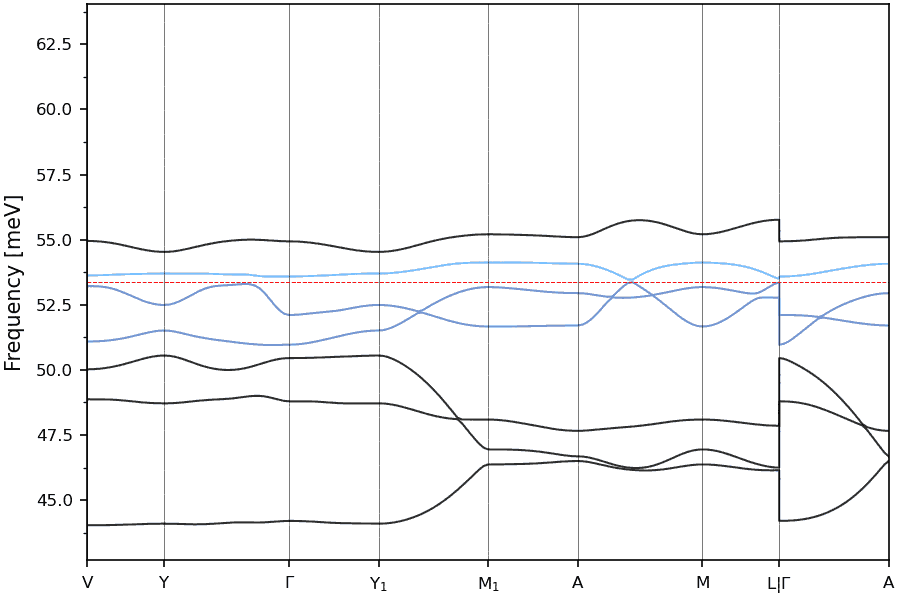} & \includegraphics[width=0.5\textwidth,angle=0]{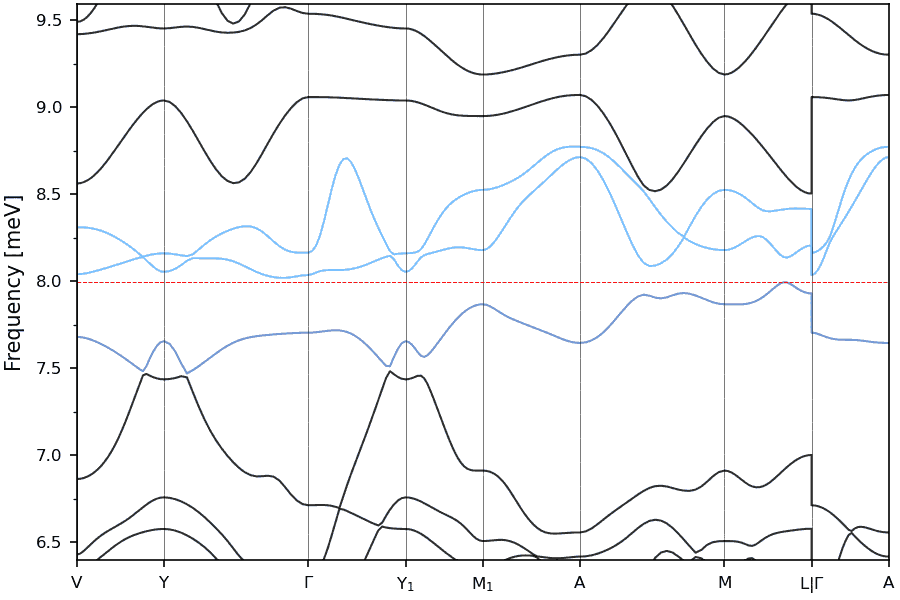} \\
\scriptsize{\mpidweb{28266}, SG 15 (\sgsymb{15}), \ce{P$_2$Pd}} & \scriptsize{\mpidweb{3277}, SG 82 (\sgsymb{82}), \ce{BAsO$_4$}} \\
\includegraphics[width=0.5\textwidth,angle=0]{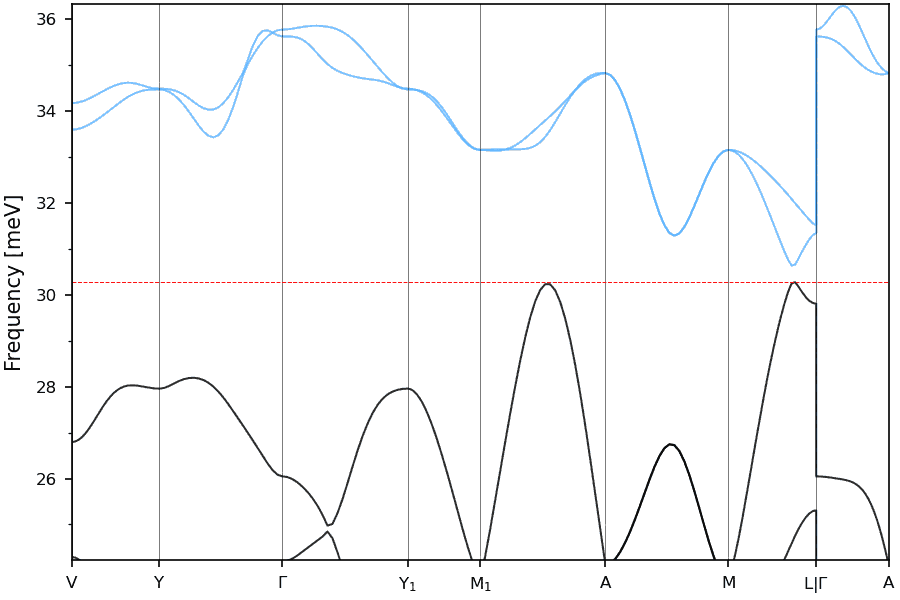} & \includegraphics[width=0.5\textwidth,angle=0]{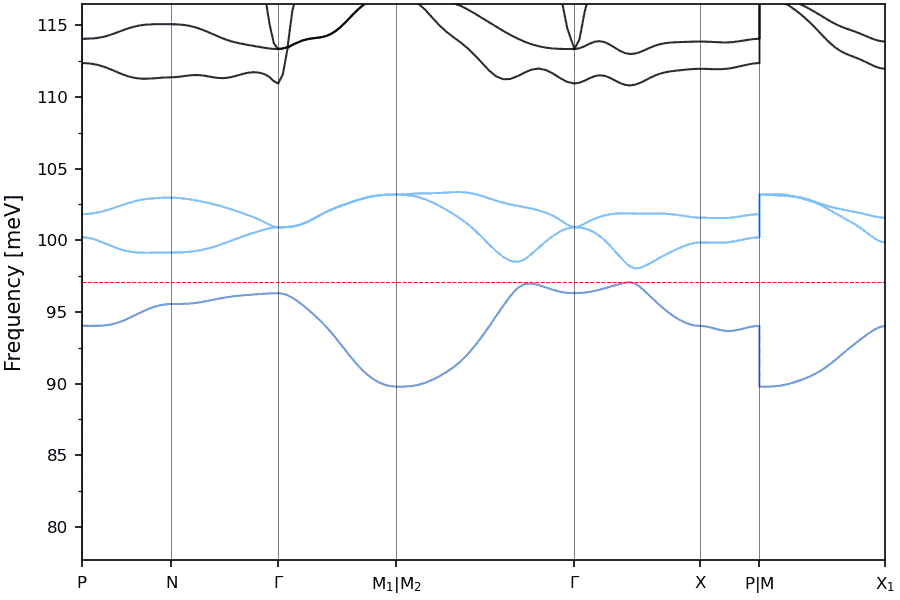} \\
\end{tabular}
\caption{Band structures of the ``ideal'' topological phonon bands with cumulative strong topology on the Material Project. (part 1/5)}
\label{fig:1_besttop_mp}
\end{figure}

\begin{figure}[h]
\centering
\begin{tabular}{cc}
\scriptsize{\mpidweb{4809}, SG 82 (\sgsymb{82}), \ce{Ga$_2$HgS$_4$}} & \scriptsize{\mpidweb{5046}, SG 82 (\sgsymb{82}), \ce{BeSO$_4$}} \\
\includegraphics[width=0.5\textwidth,angle=0]{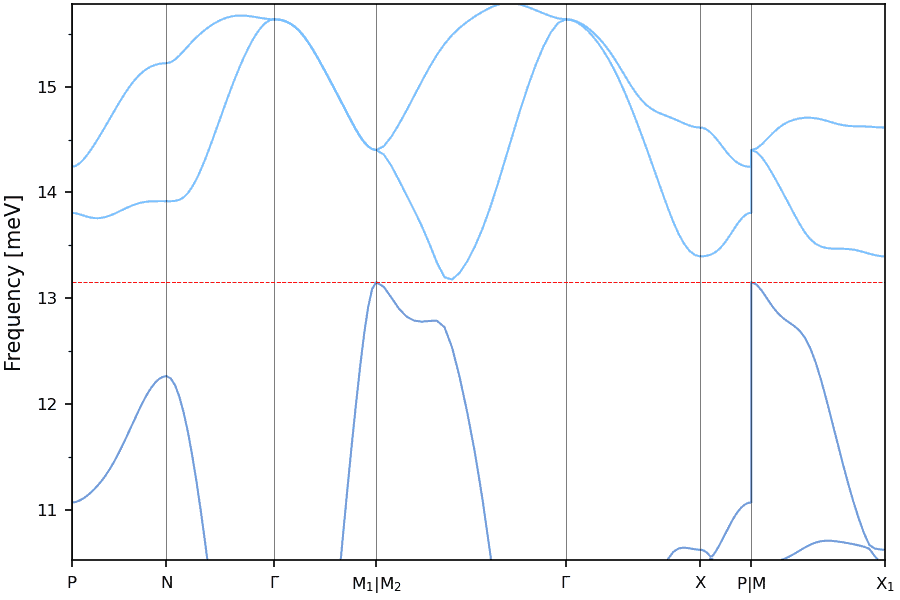} & \includegraphics[width=0.5\textwidth,angle=0]{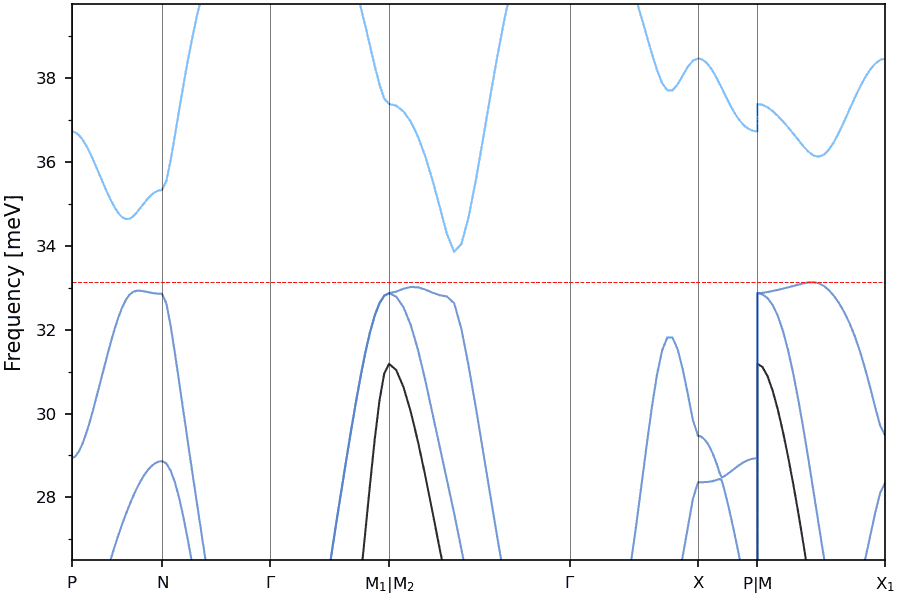} \\
\scriptsize{\mpidweb{7848}, SG 82 (\sgsymb{82}), \ce{AlPO$_4$}} & \scriptsize{\mpidweb{7849}, SG 82 (\sgsymb{82}), \ce{AlAsO$_4$}} \\
\includegraphics[width=0.5\textwidth,angle=0]{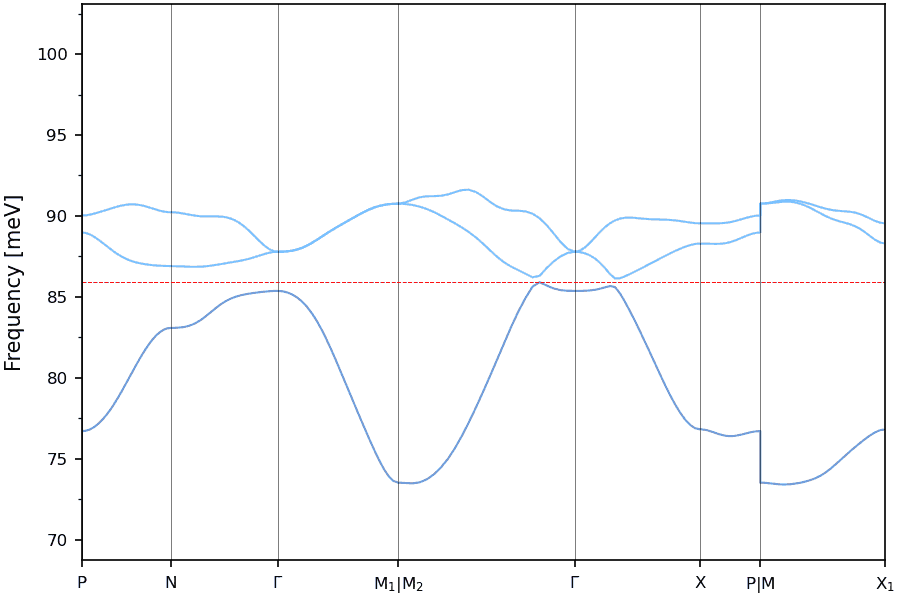} & \includegraphics[width=0.5\textwidth,angle=0]{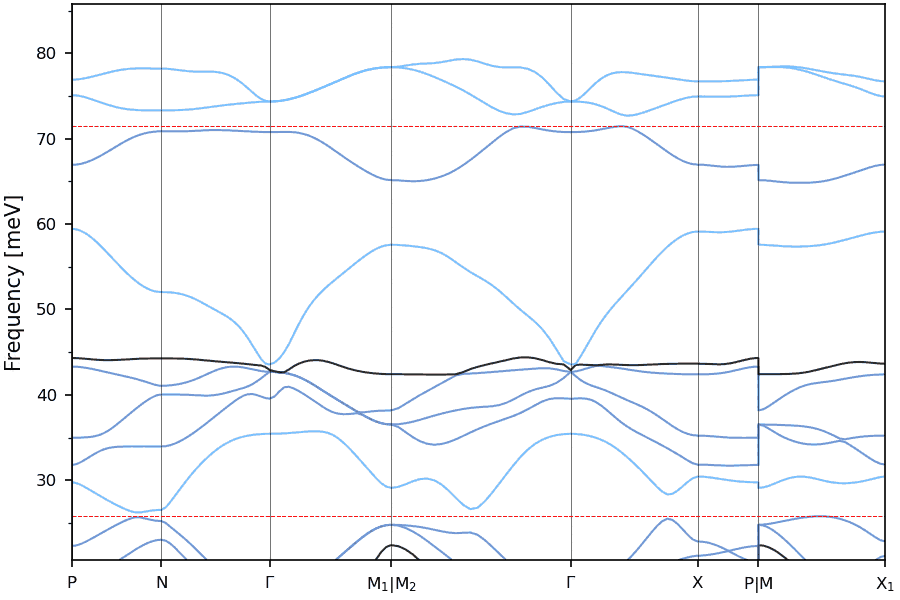} \\
\scriptsize{\mpidweb{7907}, SG 82 (\sgsymb{82}), \ce{Al$_2$ZnSe$_4$}} & \scriptsize{\mpidweb{7908}, SG 82 (\sgsymb{82}), \ce{Al$_2$ZnTe$_4$}} \\
\includegraphics[width=0.5\textwidth,angle=0]{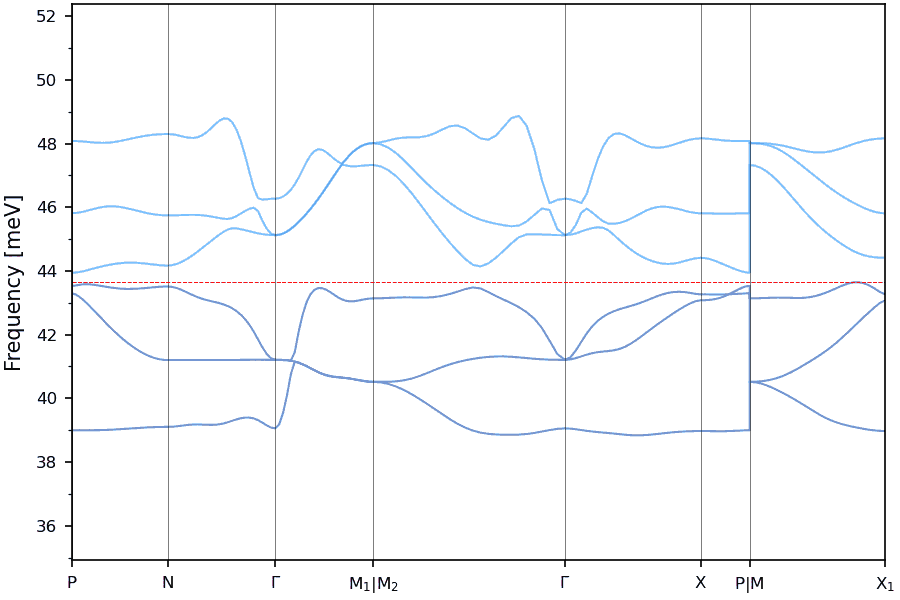} & \includegraphics[width=0.5\textwidth,angle=0]{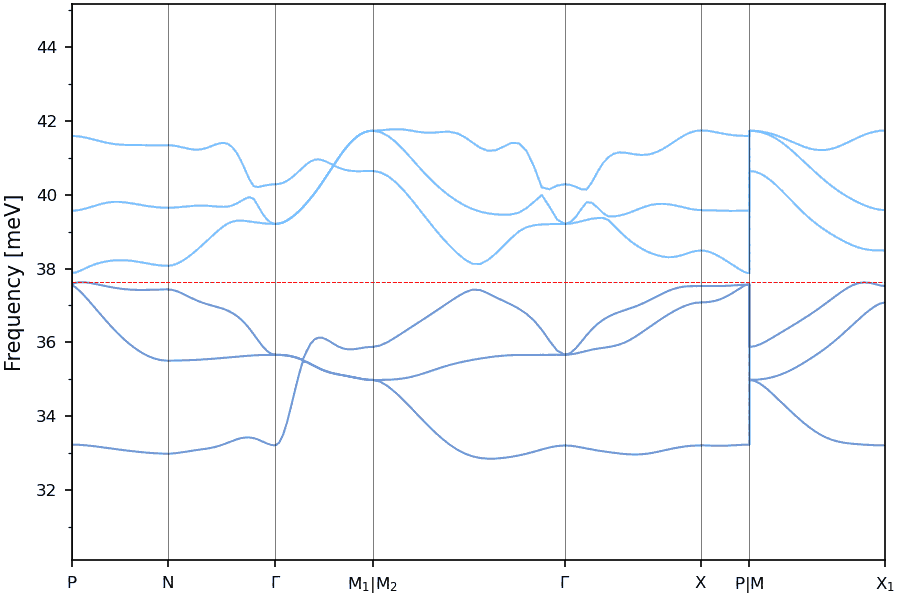} \\
\end{tabular}
\caption{Band structures of the ``ideal'' topological phonon bands with cumulative strong topology on the Material Project. (part 2/5)}
\label{fig:2_besttop_mp}
\end{figure}

\begin{figure}[h]
\centering
\begin{tabular}{cc}
\scriptsize{\mpidweb{8874}, SG 82 (\sgsymb{82}), \ce{LiSiBO$_4$}} & \scriptsize{\mpidweb{20731}, SG 82 (\sgsymb{82}), \ce{In$_2$HgSe$_4$}} \\
\includegraphics[width=0.5\textwidth,angle=0]{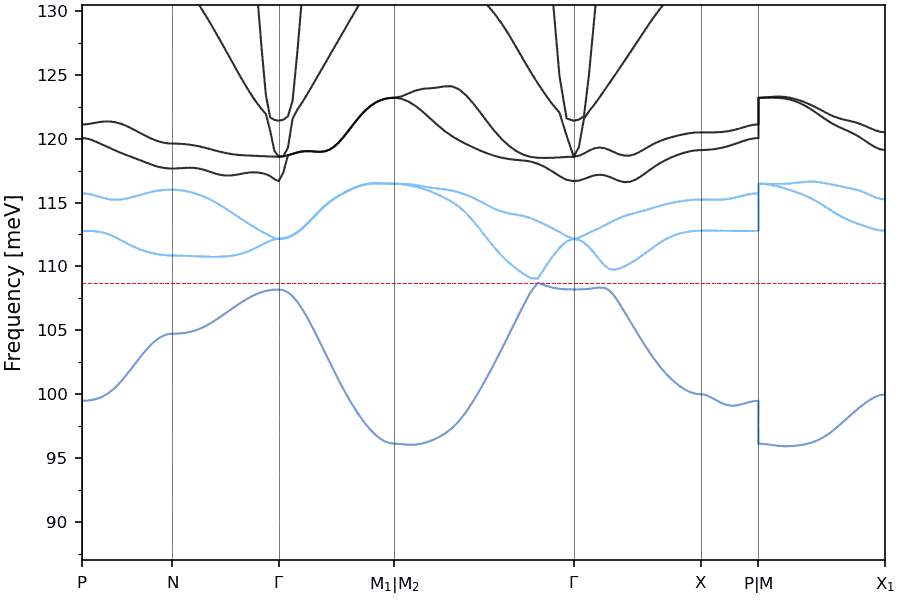} & \includegraphics[width=0.5\textwidth,angle=0]{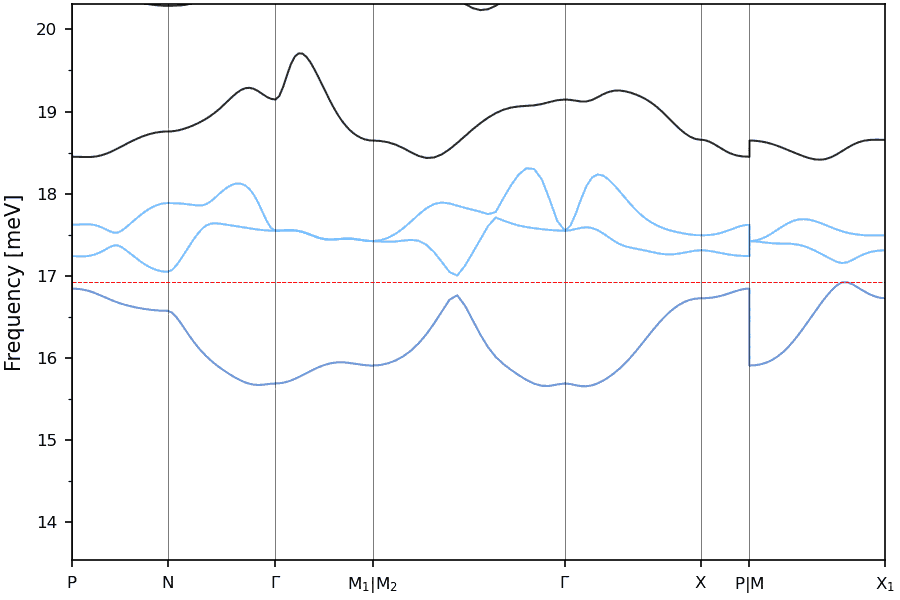} \\
\scriptsize{\mpidweb{9348}, SG 87 (\sgsymb{87}), \ce{KRb$_2$ScF$_6$}} & \scriptsize{\mpidweb{7921}, SG 148 (\sgsymb{148}), \ce{MgPdF$_6$}} \\
\includegraphics[width=0.5\textwidth,angle=0]{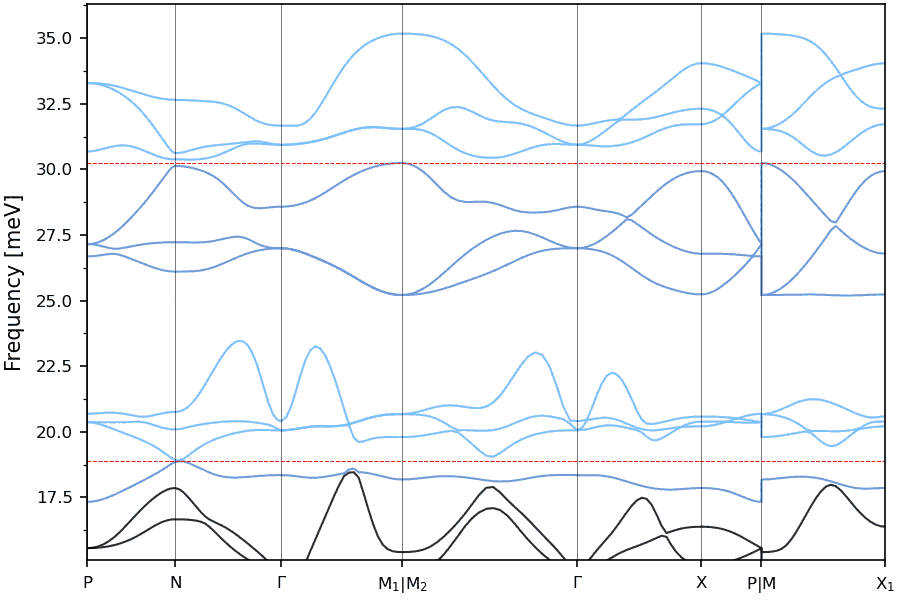} & \includegraphics[width=0.5\textwidth,angle=0]{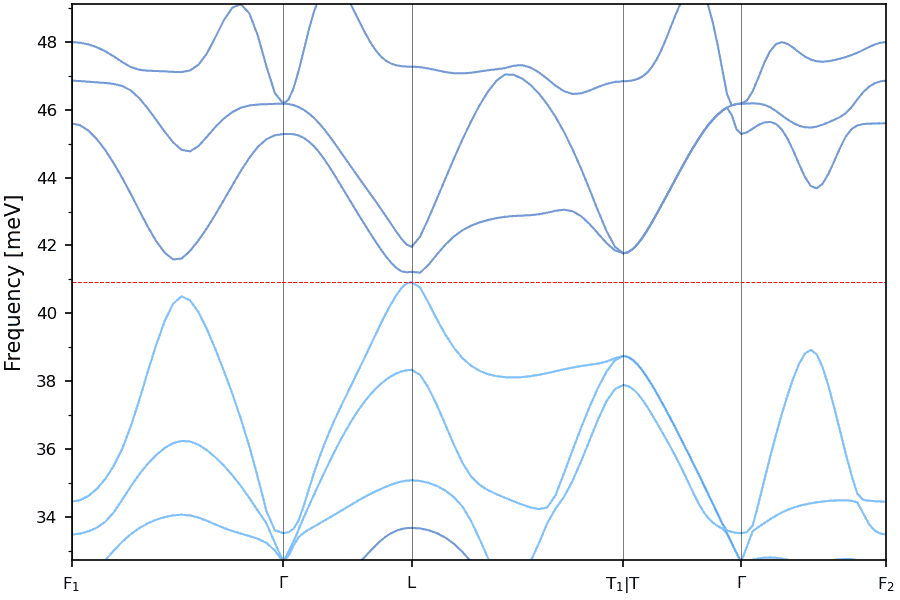} \\
\scriptsize{\mpidweb{8256}, SG 148 (\sgsymb{148}), \ce{ZnPtF$_6$}} & \scriptsize{\mpidweb{13610}, SG 148 (\sgsymb{148}), \ce{ZnPbF$_6$}} \\
\includegraphics[width=0.5\textwidth,angle=0]{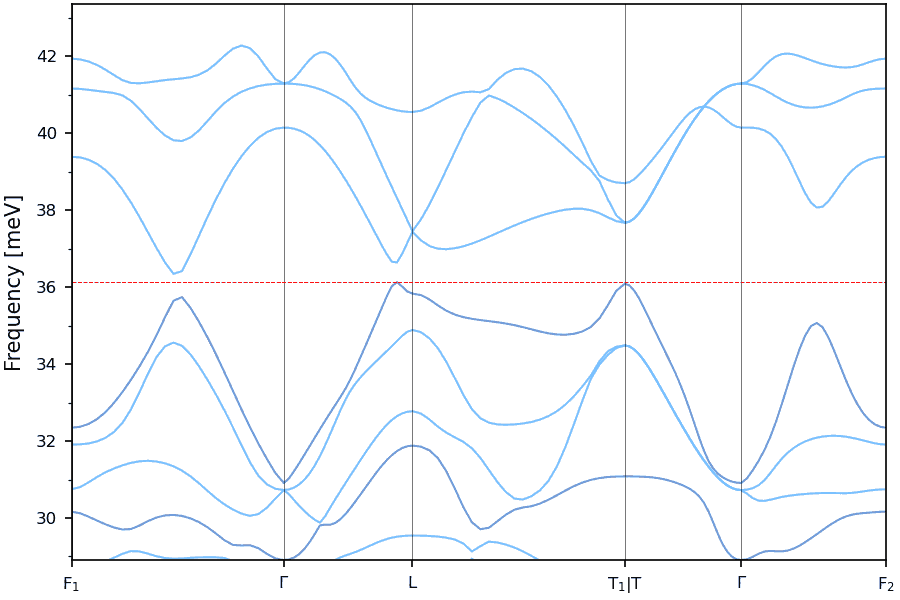} & \includegraphics[width=0.5\textwidth,angle=0]{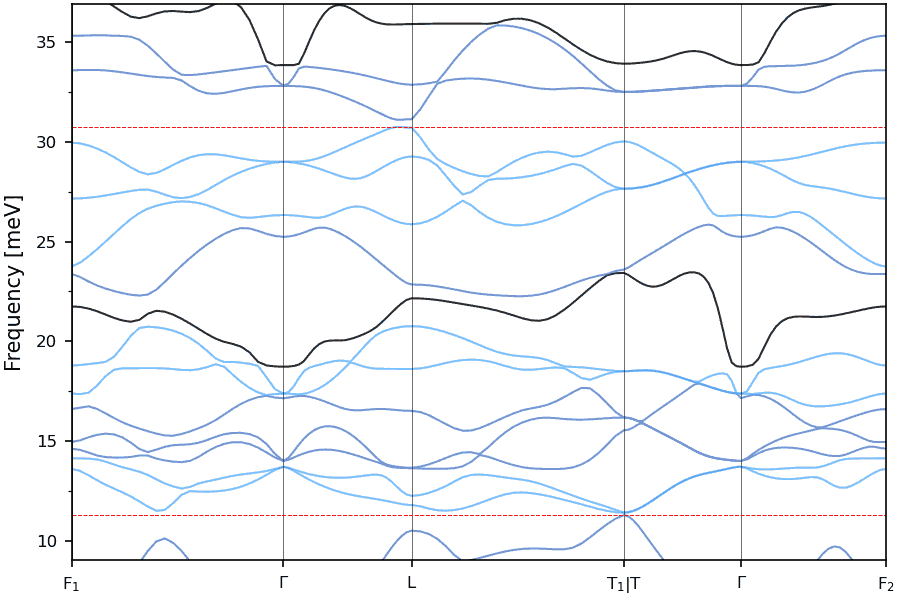} \\
\end{tabular}
\caption{Band structures of the ``ideal'' topological phonon bands with cumulative strong topology on the Material Project. (part 3/5)}
\label{fig:3_besttop_mp}
\end{figure}

\begin{figure}[h]
\centering
\begin{tabular}{cc}
\scriptsize{\mpidweb{13903}, SG 148 (\sgsymb{148}), \ce{ZnSnF$_6$}} & \scriptsize{\mpidweb{13907}, SG 148 (\sgsymb{148}), \ce{CdSnF$_6$}} \\
\includegraphics[width=0.5\textwidth,angle=0]{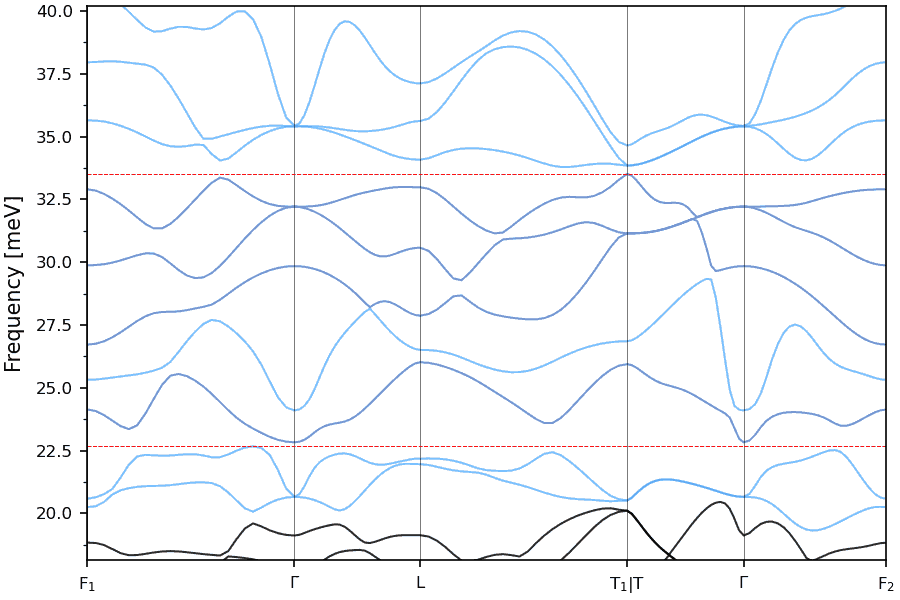} & \includegraphics[width=0.5\textwidth,angle=0]{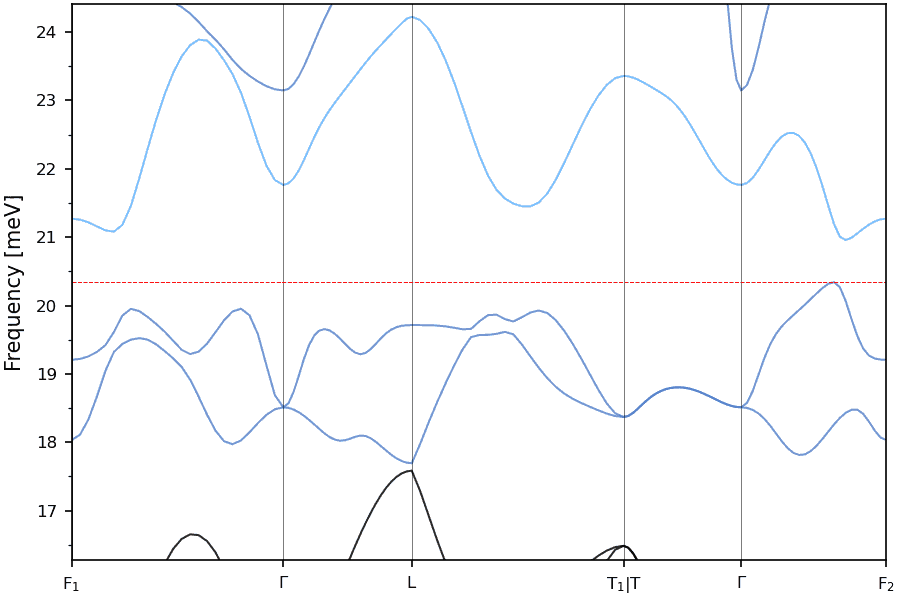} \\
\scriptsize{\mpidweb{13983}, SG 148 (\sgsymb{148}), \ce{ZnPdF$_6$}} & \scriptsize{\mpidweb{13984}, SG 148 (\sgsymb{148}), \ce{CdPdF$_6$}} \\
\includegraphics[width=0.5\textwidth,angle=0]{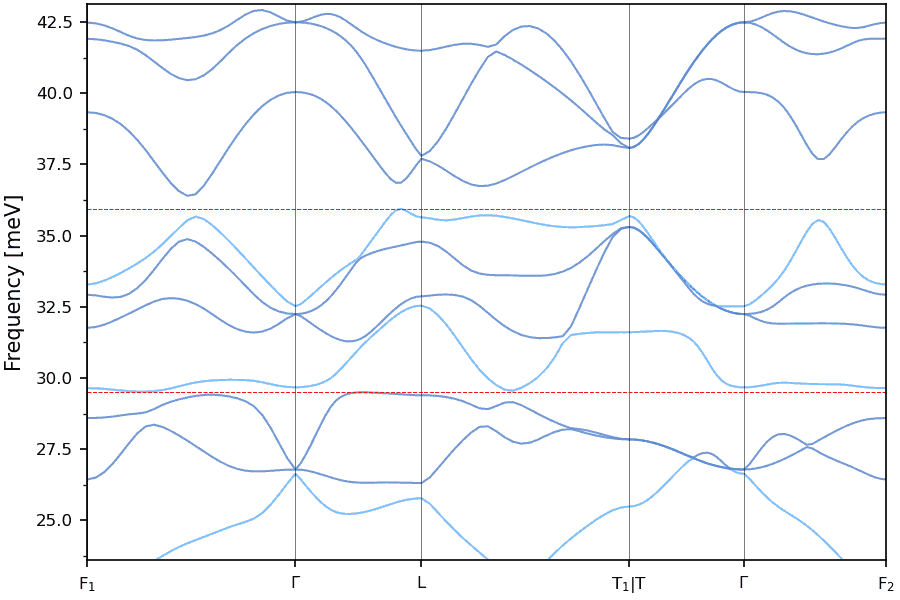} & \includegraphics[width=0.5\textwidth,angle=0]{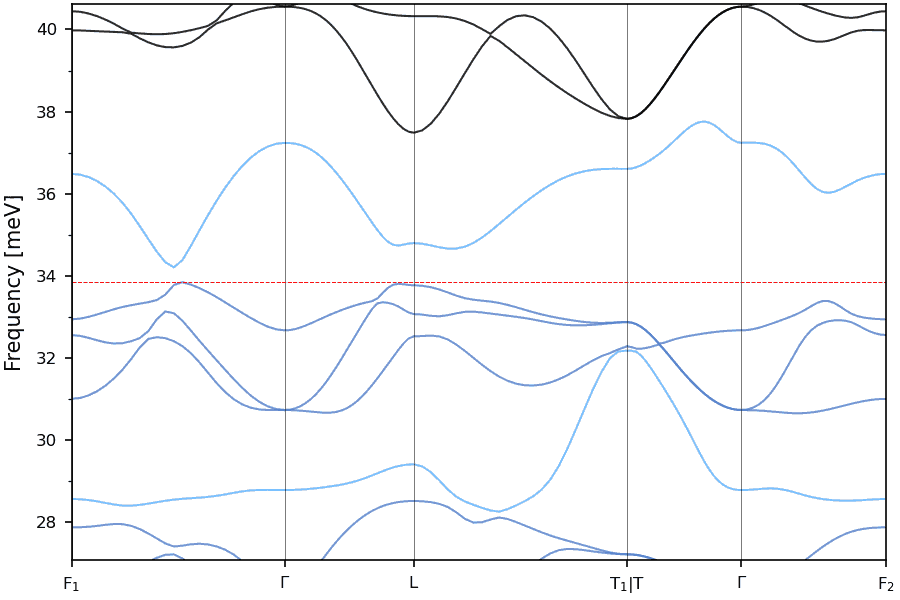} \\
\scriptsize{\mpidweb{19734}, SG 148 (\sgsymb{148}), \ce{MgPbF$_6$}} & \scriptsize{\mpidweb{23054}, SG 148 (\sgsymb{148}), \ce{NaBiO$_3$}} \\
\includegraphics[width=0.5\textwidth,angle=0]{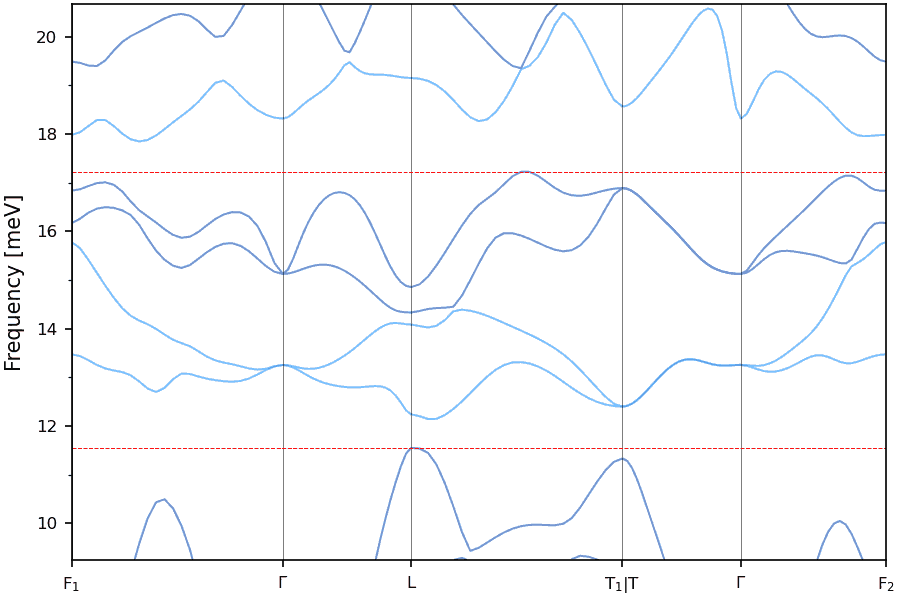} & \includegraphics[width=0.5\textwidth,angle=0]{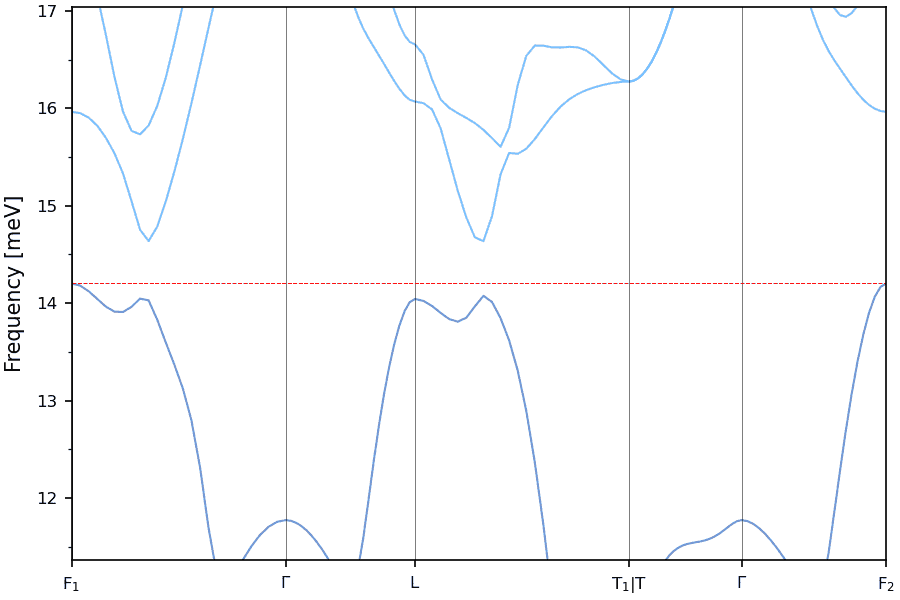} \\
\end{tabular}
\caption{Band structures of the ``ideal'' topological phonon bands with cumulative strong topology on the Material Project. (part 4/5)}
\label{fig:4_besttop_mp}
\end{figure}

\begin{figure}[h]
\centering
\begin{tabular}{cc}
\scriptsize{\mpidweb{36508}, SG 148 (\sgsymb{148}), \ce{SnHgF$_6$}} & \scriptsize{\mpidweb{37216}, SG 148 (\sgsymb{148}), \ce{CdPbF$_6$}} \\
\includegraphics[width=0.5\textwidth,angle=0]{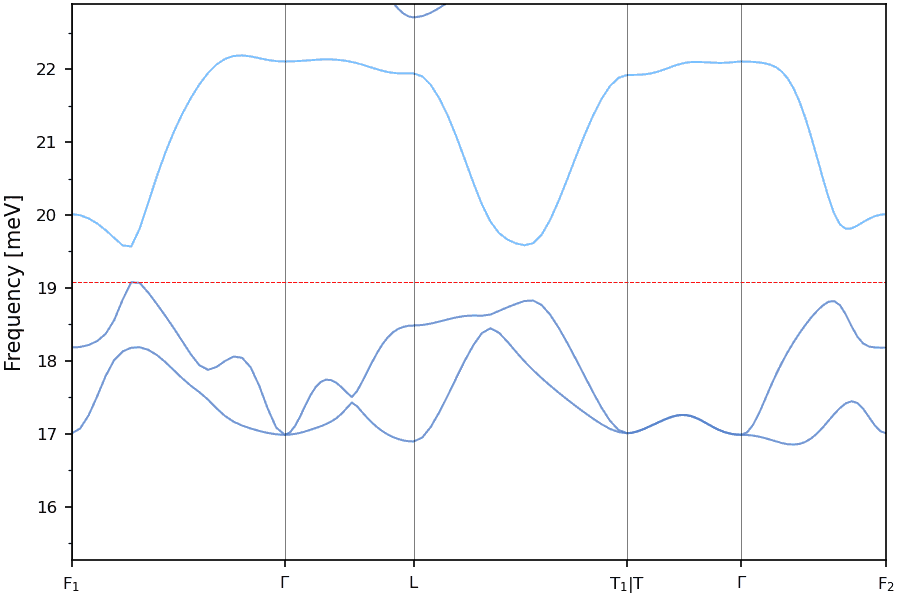} & \includegraphics[width=0.5\textwidth,angle=0]{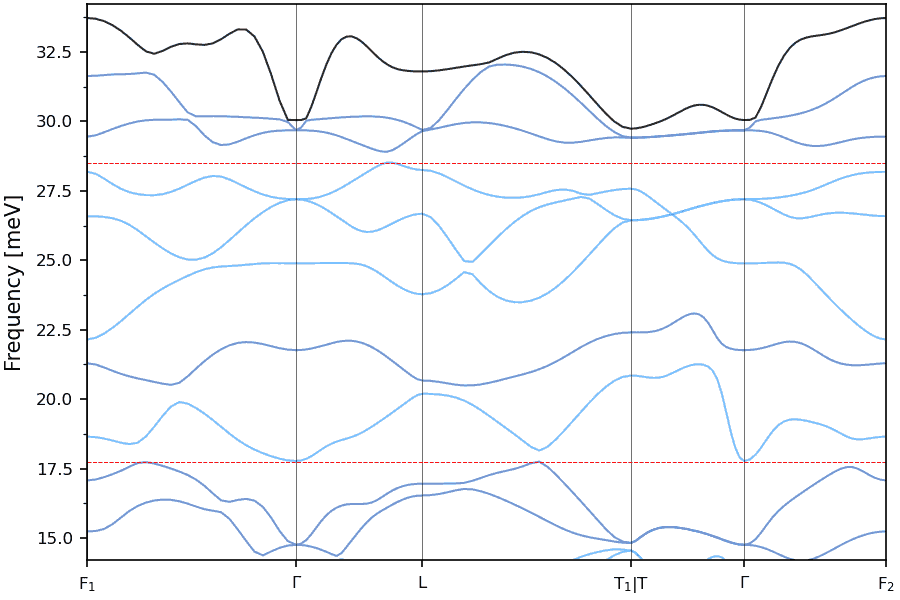} \\
\scriptsize{\mpidweb{547792}, SG 148 (\sgsymb{148}), \ce{KSbO$_3$}} & \scriptsize{\mpidweb{551269}, SG 148 (\sgsymb{148}), \ce{Ba$_2$LaSbO$_6$}} \\
\includegraphics[width=0.5\textwidth,angle=0]{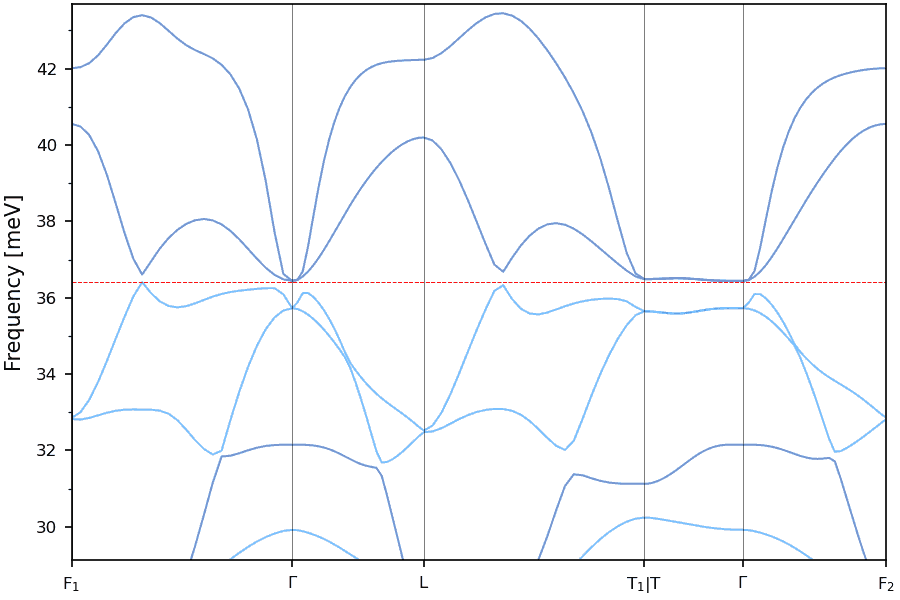} & \includegraphics[width=0.5\textwidth,angle=0]{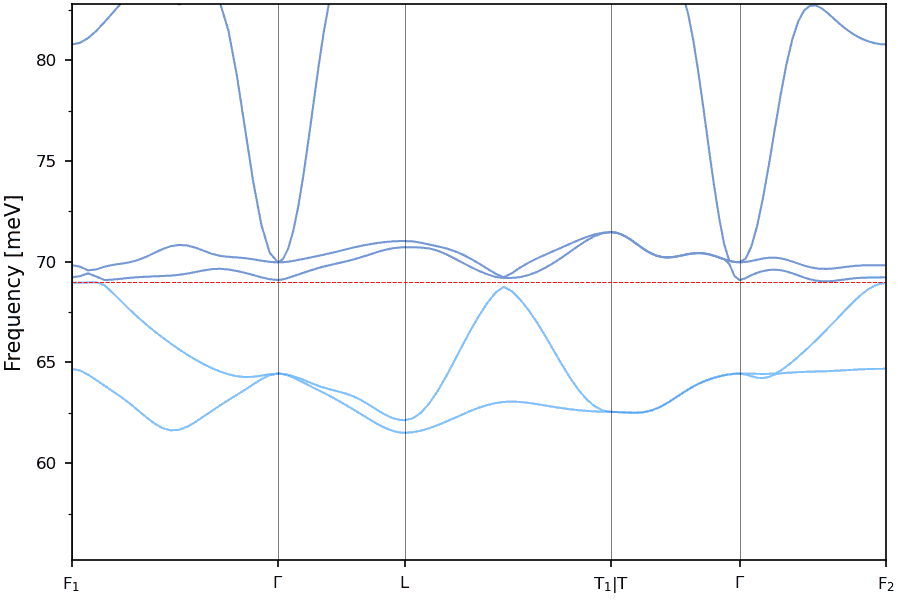} \\
\scriptsize{\mpidweb{754329}, SG 148 (\sgsymb{148}), \ce{CdSnO$_3$}} & \\
\includegraphics[width=0.5\textwidth,angle=0]{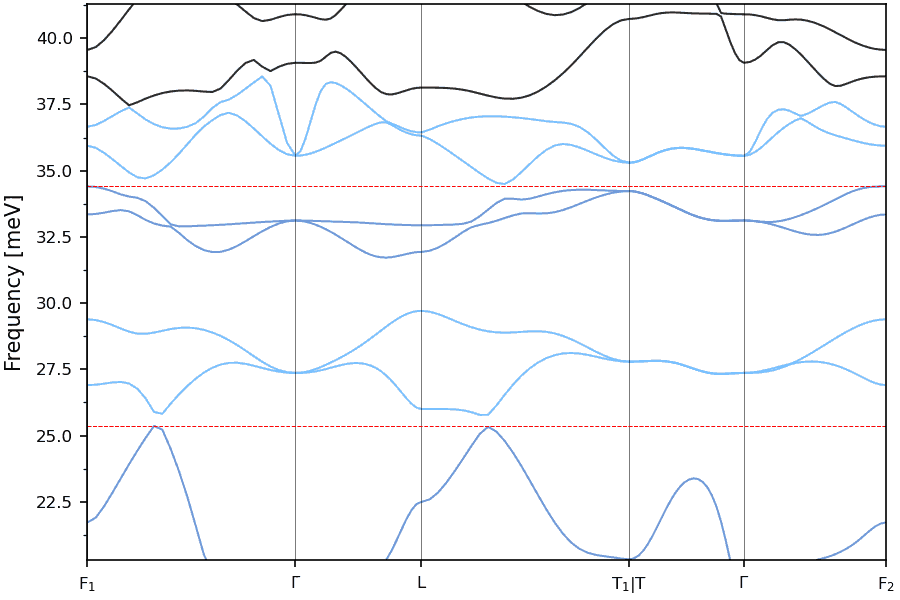} & \\
\end{tabular}
\caption{Band structures of the ``ideal'' topological phonon bands with cumulative strong topology on the Material Project. (part 5/5)}
\label{fig:5_besttop_mp}
\end{figure}
\clearpage
\begin{figure}[h]
\centering
\begin{tabular}{cc}
\scriptsize{\mpidweb{571132}, SG 3 (\sgsymb{3}), \ce{LiMg$_10$AlH$_24$}} & \scriptsize{\mpidweb{7808}, SG 12 (\sgsymb{12}), \ce{BaP$_3$}} \\
\includegraphics[width=0.5\textwidth,angle=0]{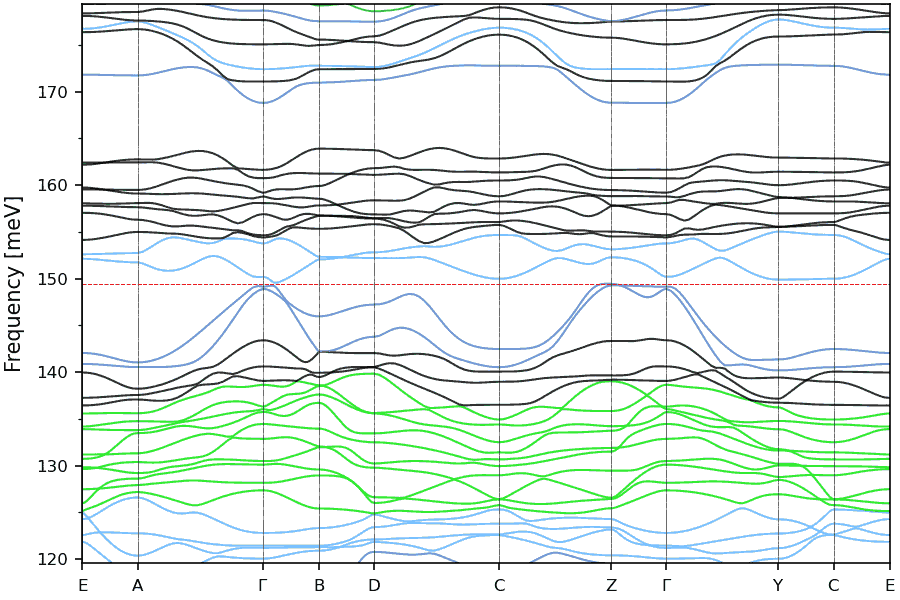} & \includegraphics[width=0.5\textwidth,angle=0]{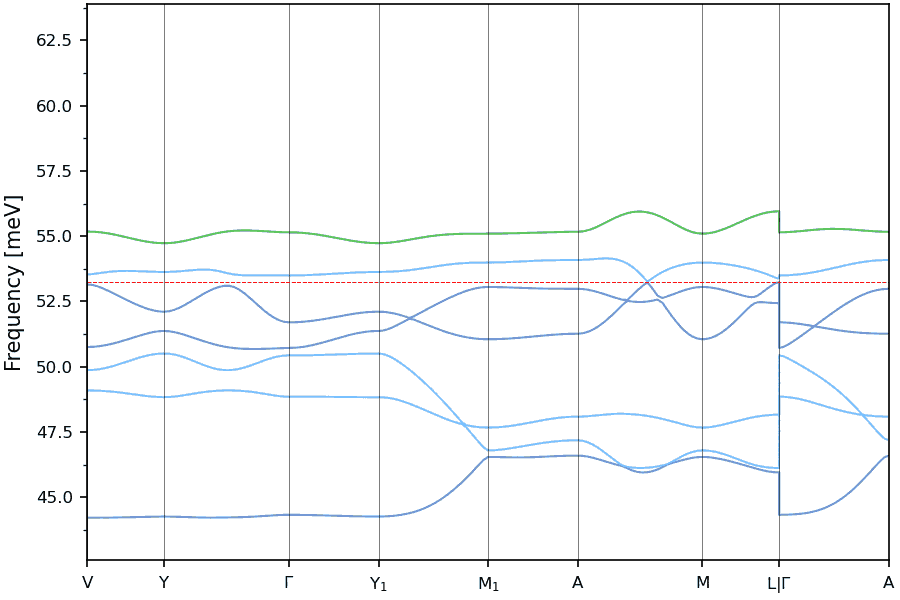} \\
\scriptsize{\mpidweb{568628}, SG 12 (\sgsymb{12}), \ce{CsHg$_5$Cl$_11$}} & \scriptsize{\mpidweb{569623}, SG 12 (\sgsymb{12}), \ce{Nb(SeBr)$_2$}} \\
\includegraphics[width=0.5\textwidth,angle=0]{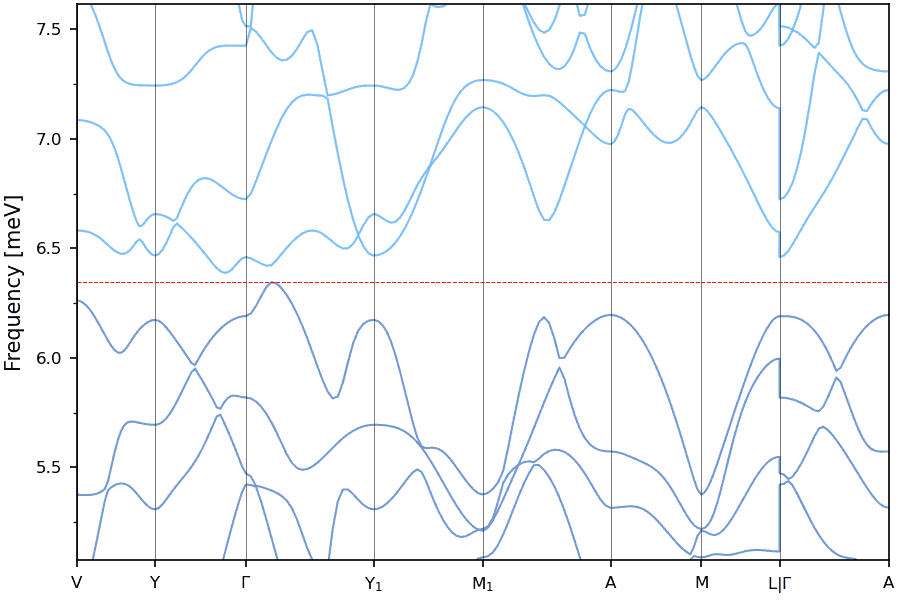} & \includegraphics[width=0.5\textwidth,angle=0]{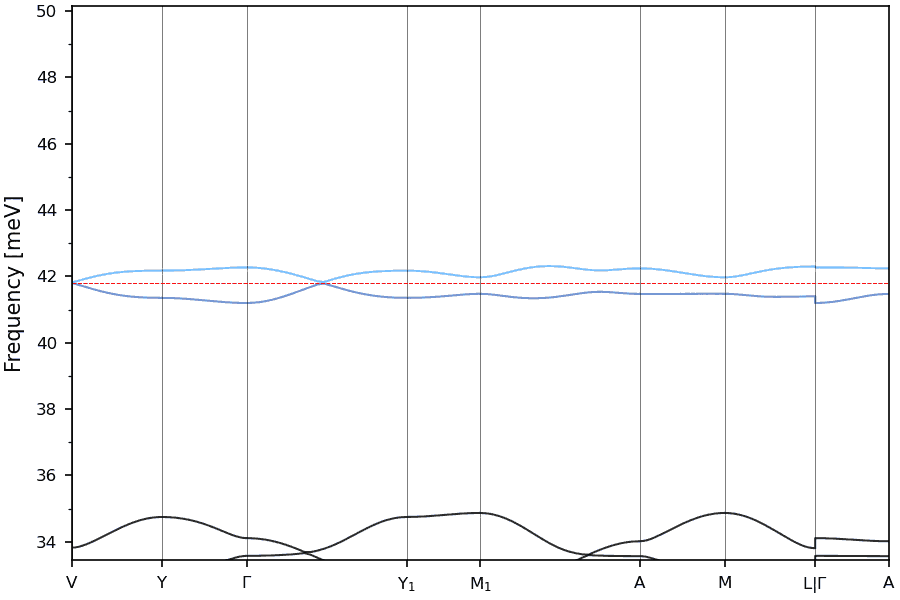} \\
\scriptsize{\mpidweb{571425}, SG 12 (\sgsymb{12}), \ce{Ag$_2$Hg$_7$(P$_4$Br$_3$)$_2$}} & \scriptsize{\mpidweb{733604}, SG 12 (\sgsymb{12}), \ce{Ba$_2$Mg$_3$H$_10$}} \\
\includegraphics[width=0.5\textwidth,angle=0]{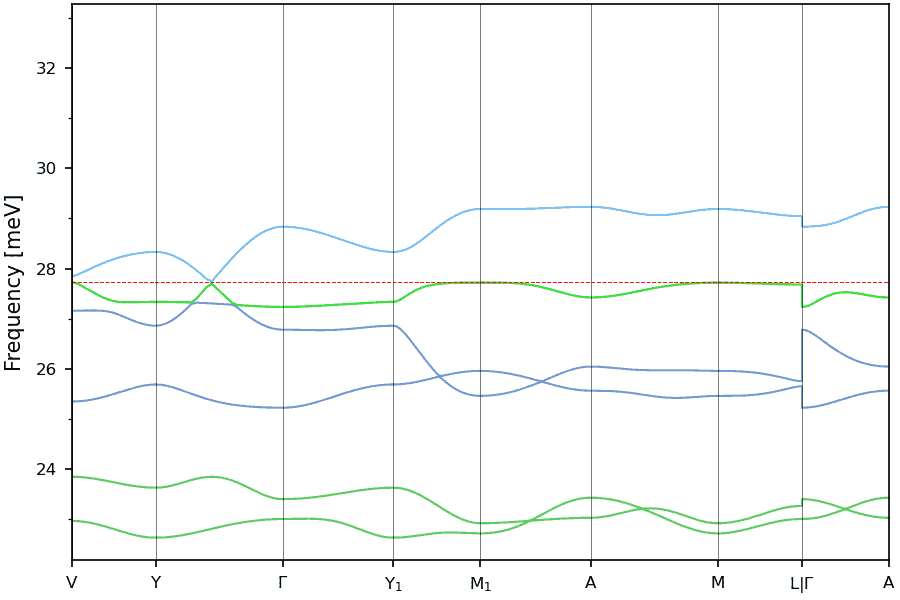} & \includegraphics[width=0.5\textwidth,angle=0]{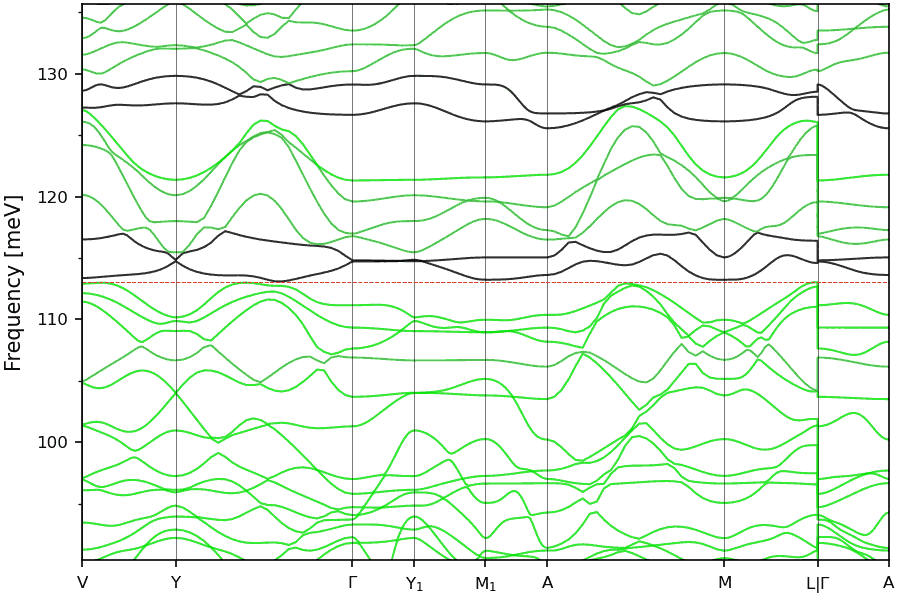} \\
\end{tabular}
\caption{Band structures of the ``ideal'' topological phonon bands with cumulative strong topology on the PhononDB@kyoto-u. (part 1/27)}
\label{fig:1_besttop_kyoto}
\end{figure}

\begin{figure}[h]
\centering
\begin{tabular}{cc}
\scriptsize{\mpidweb{753802}, SG 12 (\sgsymb{12}), \ce{Sc$_2$Ti$_2$O$_7$}} & \scriptsize{\mpidweb{755794}, SG 12 (\sgsymb{12}), \ce{Rb$_3$LuO$_3$}} \\
\includegraphics[width=0.5\textwidth,angle=0]{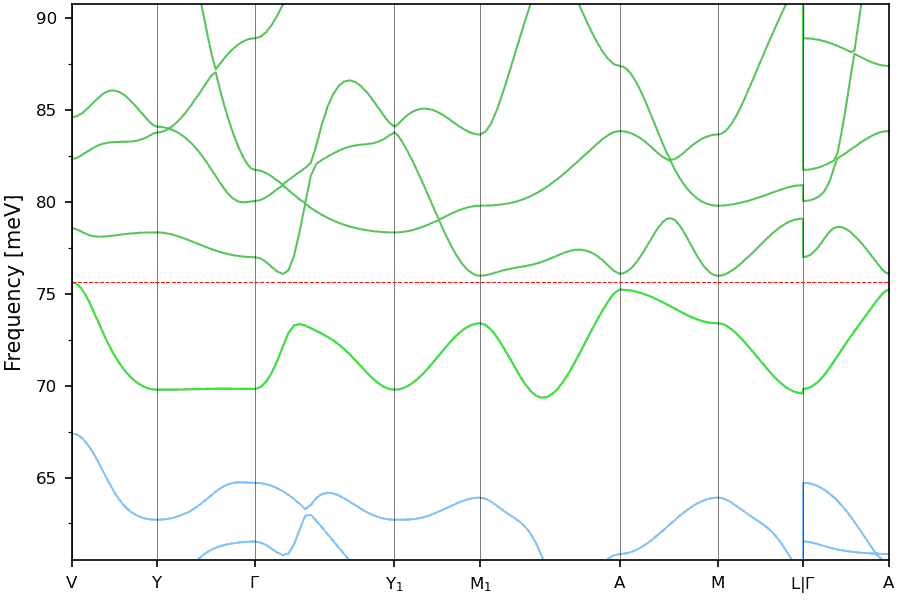} & \includegraphics[width=0.5\textwidth,angle=0]{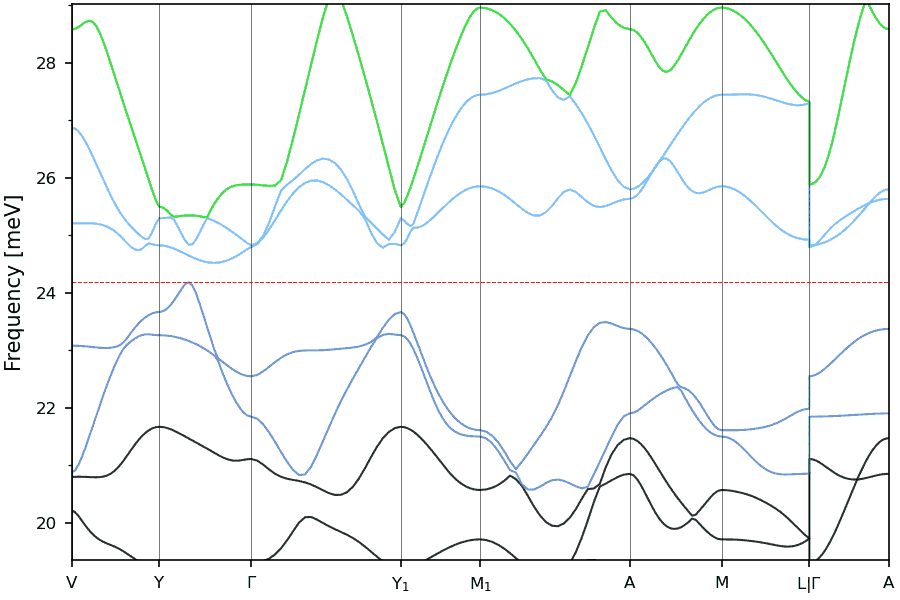} \\
\scriptsize{\mpidweb{6652}, SG 13 (\sgsymb{13}), \ce{KZn$_4$(BO$_3$)$_3$}} & \scriptsize{\mpidweb{540924}, SG 13 (\sgsymb{13}), \ce{NbTeI$_3$}} \\
\includegraphics[width=0.5\textwidth,angle=0]{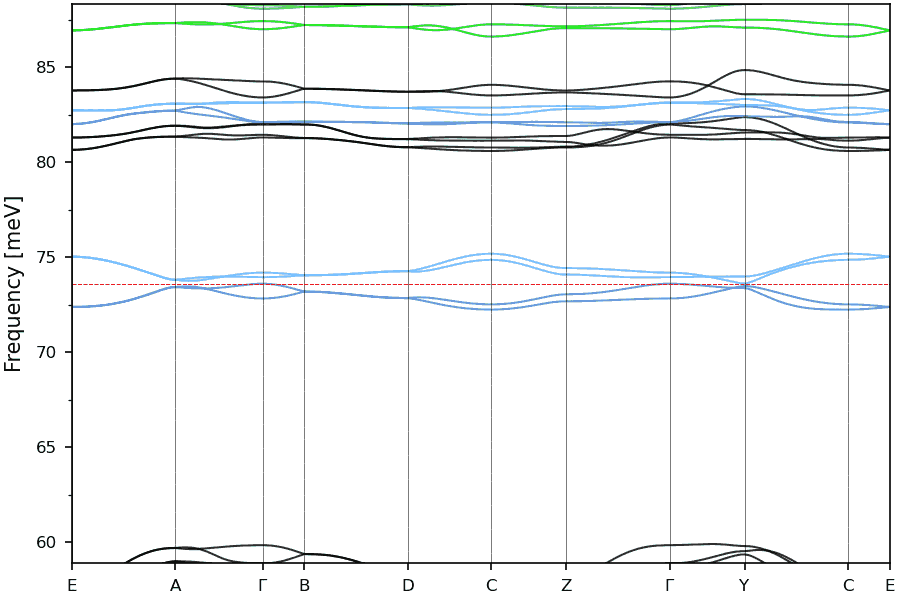} & \includegraphics[width=0.5\textwidth,angle=0]{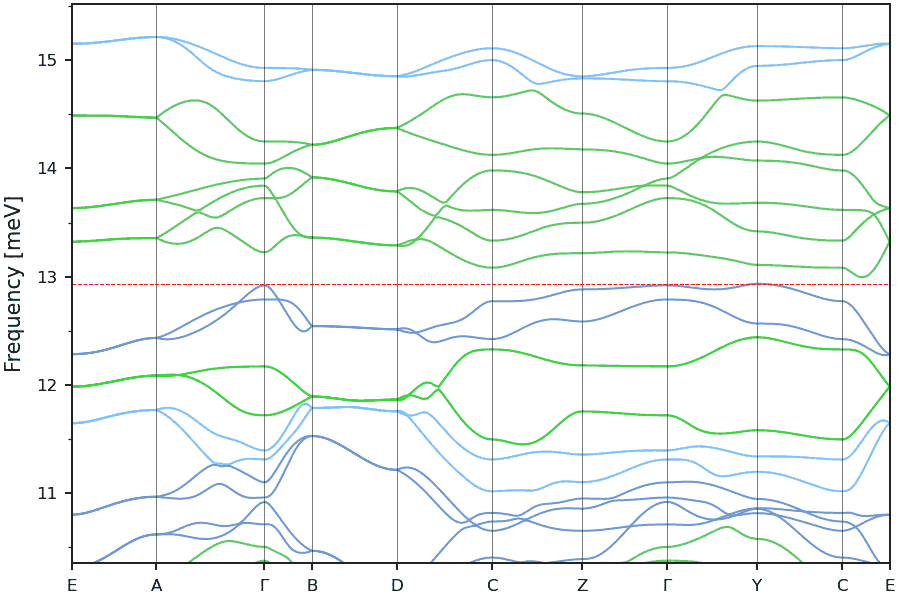} \\
\scriptsize{\mpidweb{542143}, SG 13 (\sgsymb{13}), \ce{RbZn$_4$(BO$_3$)$_3$}} & \scriptsize{\mpidweb{554736}, SG 13 (\sgsymb{13}), \ce{CaB$_2$(H$_6$O$_5$)$_2$}} \\
\includegraphics[width=0.5\textwidth,angle=0]{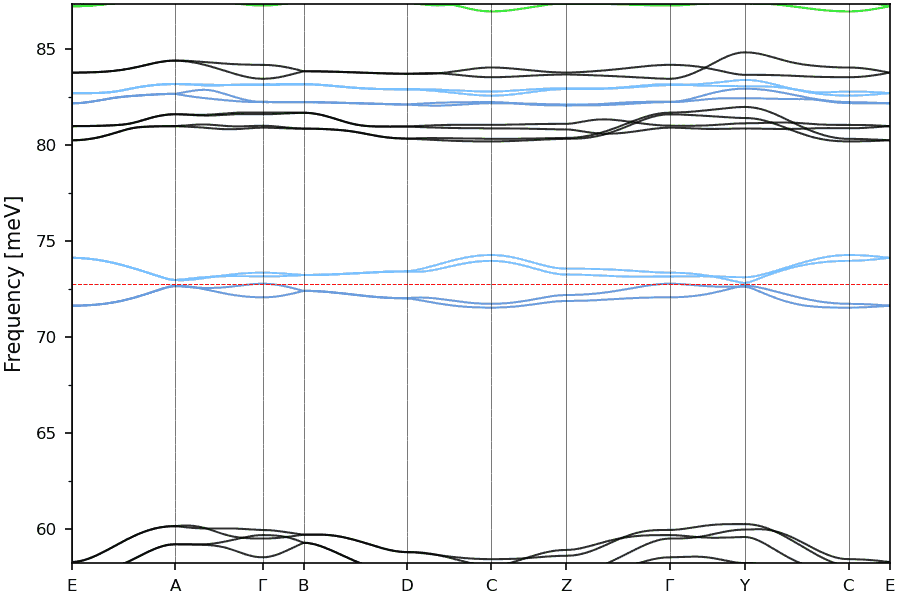} & \includegraphics[width=0.5\textwidth,angle=0]{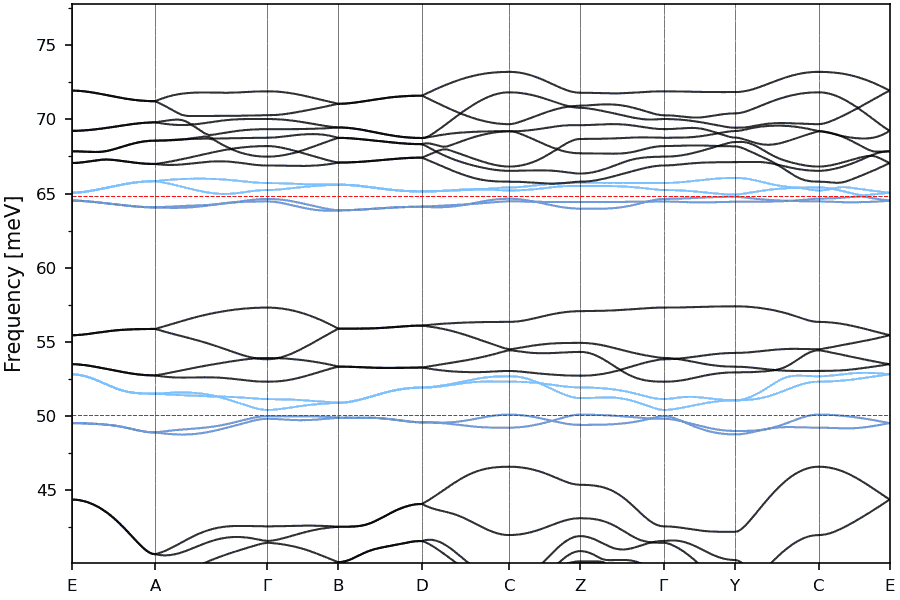} \\
\end{tabular}
\caption{Band structures of the ``ideal'' topological phonon bands with cumulative strong topology on the PhononDB@kyoto-u. (part 2/27)}
\label{fig:2_besttop_kyoto}
\end{figure}

\begin{figure}[h]
\centering
\begin{tabular}{cc}
\scriptsize{\mpidweb{555392}, SG 13 (\sgsymb{13}), \ce{K$_6$NaAu$_2$IO$_8$}} & \scriptsize{\mpidweb{570270}, SG 13 (\sgsymb{13}), \ce{NbSeCl$_3$}} \\
\includegraphics[width=0.5\textwidth,angle=0]{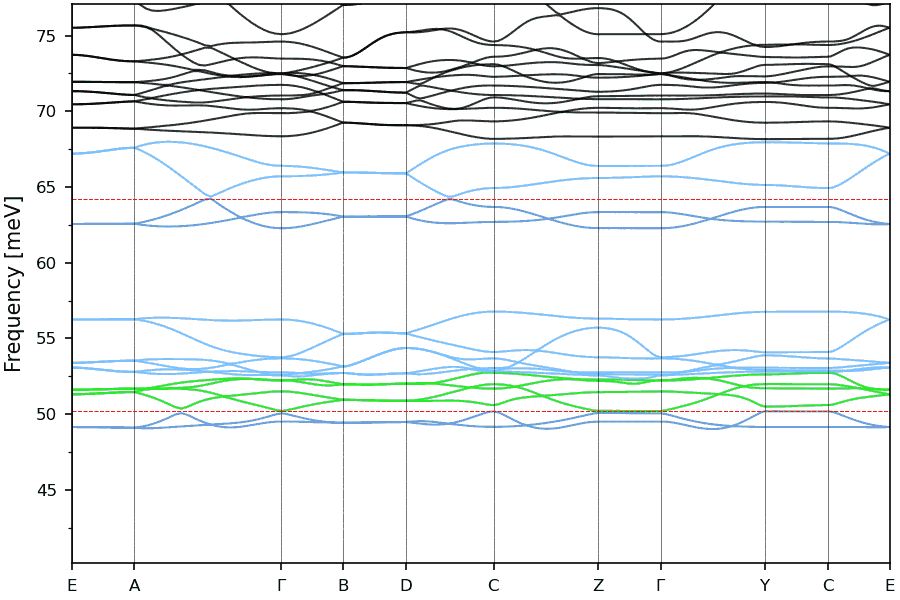} & \includegraphics[width=0.5\textwidth,angle=0]{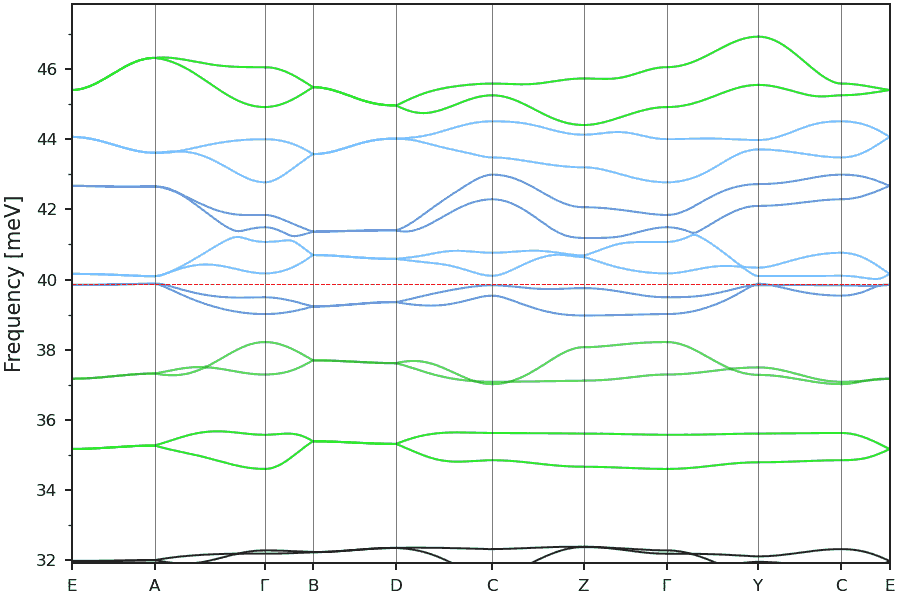} \\
\scriptsize{\mpidweb{867129}, SG 13 (\sgsymb{13}), \ce{RbZnH$_2$Se$_2$ClO$_6$}} & \scriptsize{\mpidweb{6331}, SG 14 (\sgsymb{14}), \ce{Na$_2$Y$_2$B$_2$O$_7$}} \\
\includegraphics[width=0.5\textwidth,angle=0]{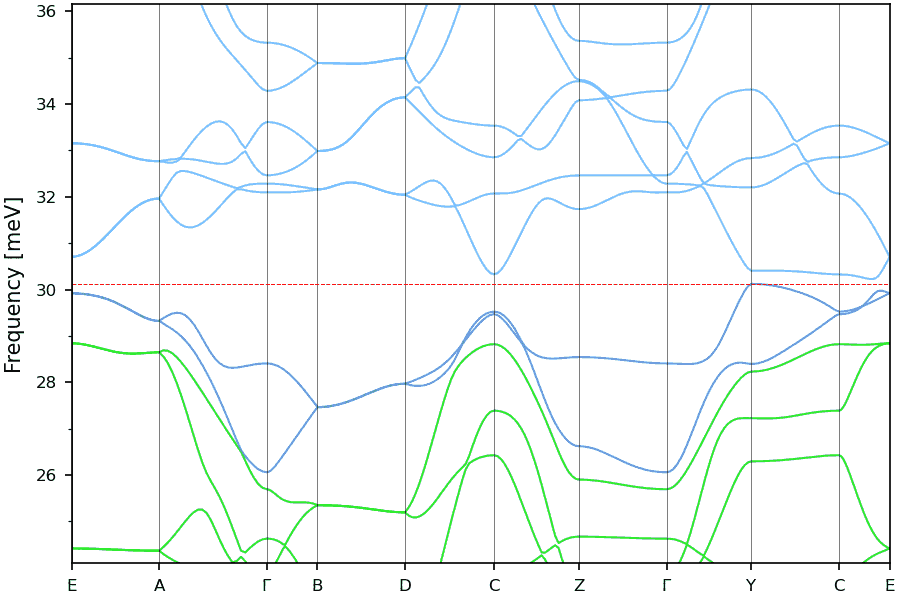} & \includegraphics[width=0.5\textwidth,angle=0]{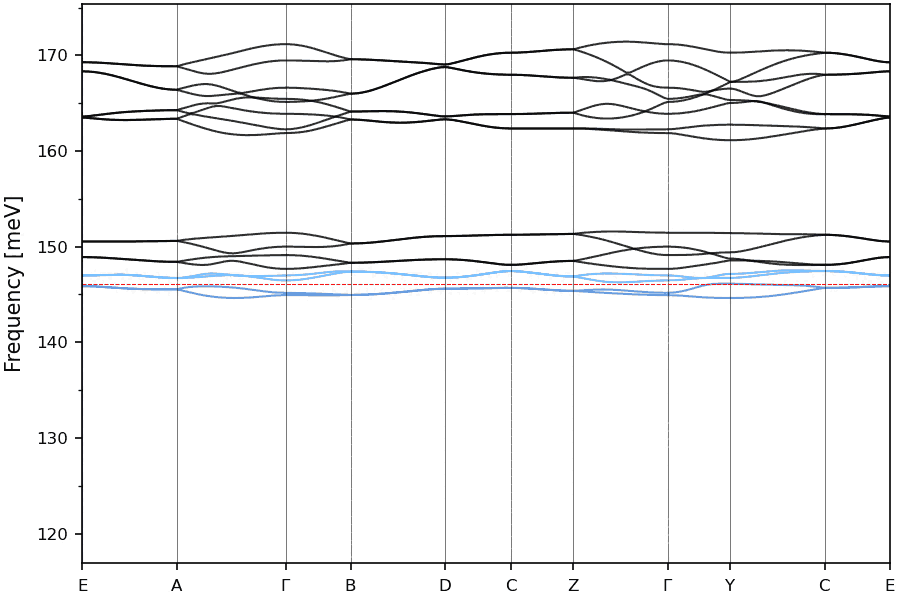} \\
\scriptsize{\mpidweb{7624}, SG 14 (\sgsymb{14}), \ce{Er$_2$Si$_2$O$_7$}} & \scriptsize{\mpidweb{11643}, SG 14 (\sgsymb{14}), \ce{K$_2$Mg(PSe$_3$)$_2$}} \\
\includegraphics[width=0.5\textwidth,angle=0]{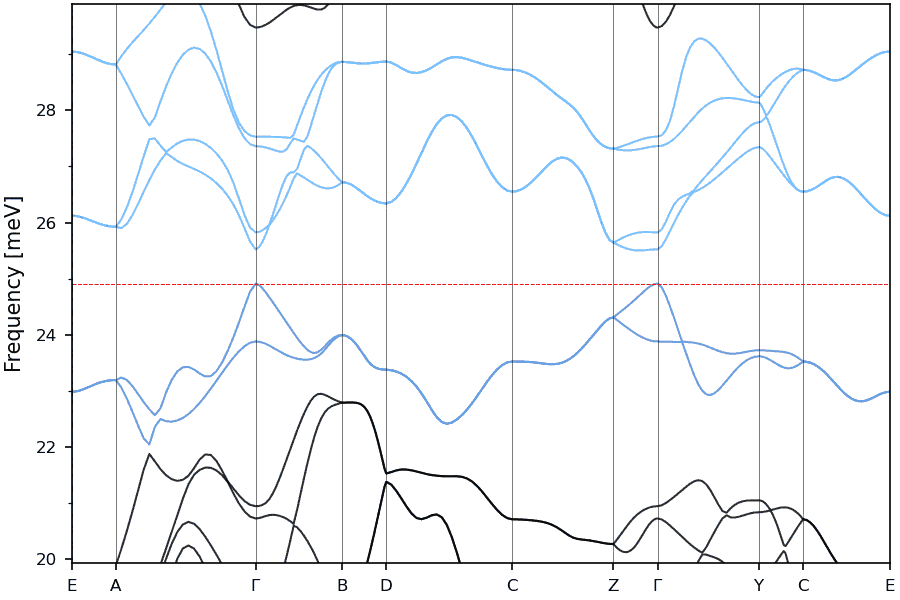} & \includegraphics[width=0.5\textwidth,angle=0]{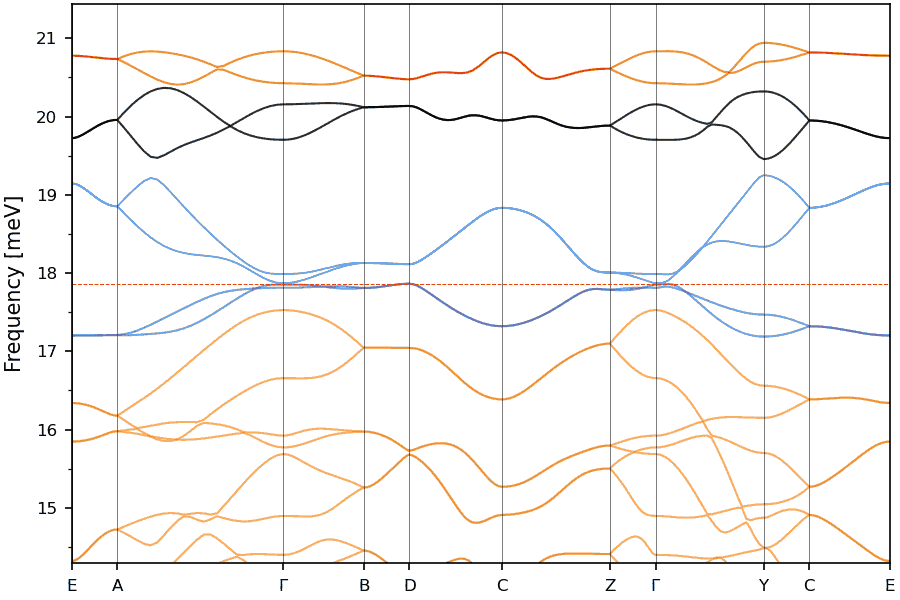} \\
\end{tabular}
\caption{Band structures of the ``ideal'' topological phonon bands with cumulative strong topology on the PhononDB@kyoto-u. (part 3/27)}
\label{fig:3_besttop_kyoto}
\end{figure}

\begin{figure}[h]
\centering
\begin{tabular}{cc}
\scriptsize{\mpidweb{12449}, SG 14 (\sgsymb{14}), \ce{LiPr$_2$OsO$_6$}} & \scriptsize{\mpidweb{16993}, SG 14 (\sgsymb{14}), \ce{Er$_2$SiO$_5$}} \\
\includegraphics[width=0.5\textwidth,angle=0]{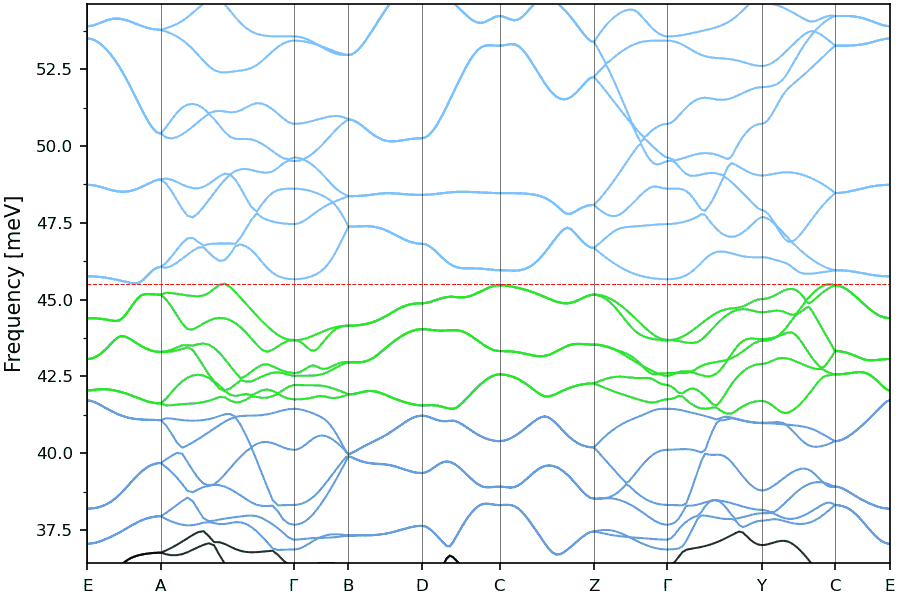} & \includegraphics[width=0.5\textwidth,angle=0]{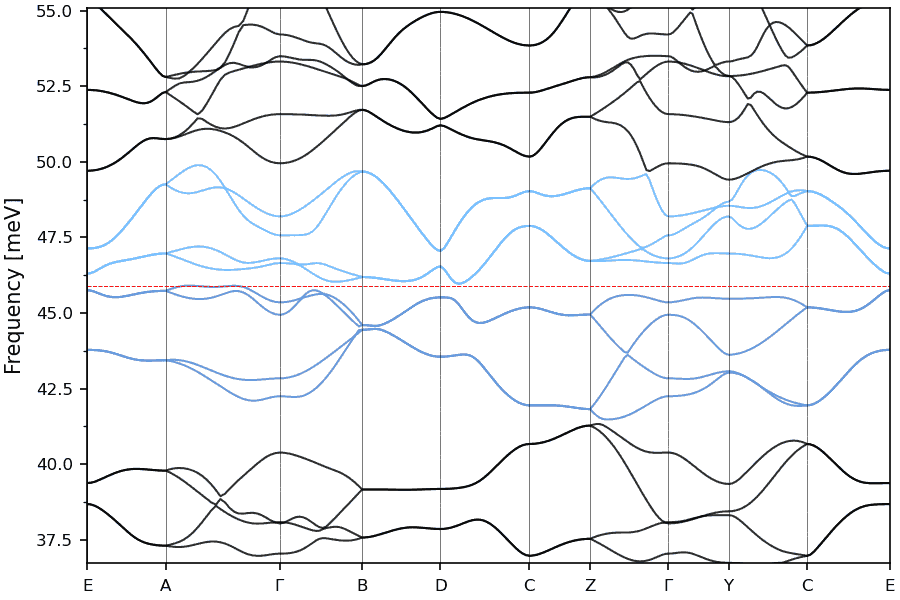} \\
\scriptsize{\mpidweb{17065}, SG 14 (\sgsymb{14}), \ce{Ho$_2$SiO$_5$}} & \scriptsize{\mpidweb{17774}, SG 14 (\sgsymb{14}), \ce{K$_2$Li$_3$GaO$_4$}} \\
\includegraphics[width=0.5\textwidth,angle=0]{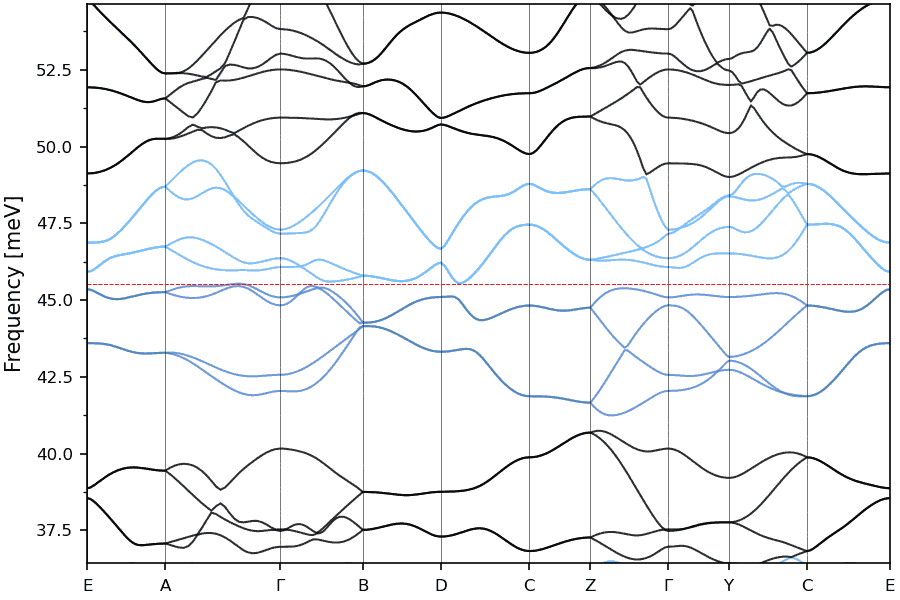} & \includegraphics[width=0.5\textwidth,angle=0]{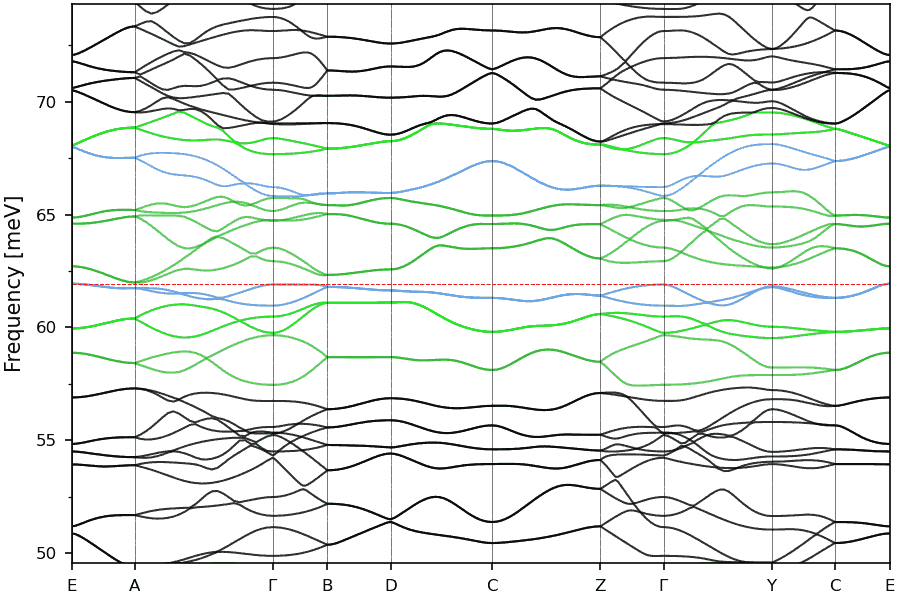} \\
\scriptsize{\mpidweb{17924}, SG 14 (\sgsymb{14}), \ce{ZnP$_2$PbO$_7$}} & \scriptsize{\mpidweb{26129}, SG 14 (\sgsymb{14}), \ce{LiBiP$_2$O$_7$}} \\
\includegraphics[width=0.5\textwidth,angle=0]{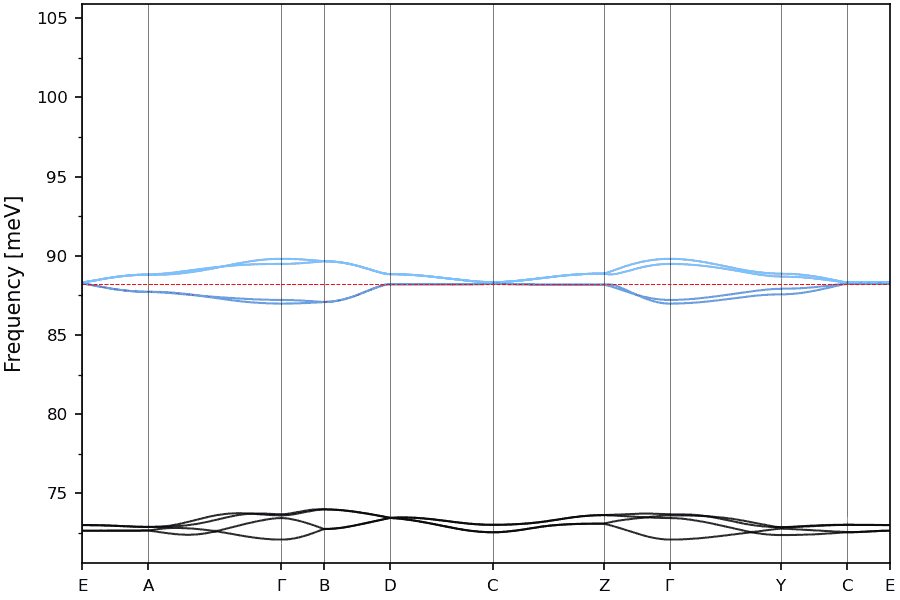} & \includegraphics[width=0.5\textwidth,angle=0]{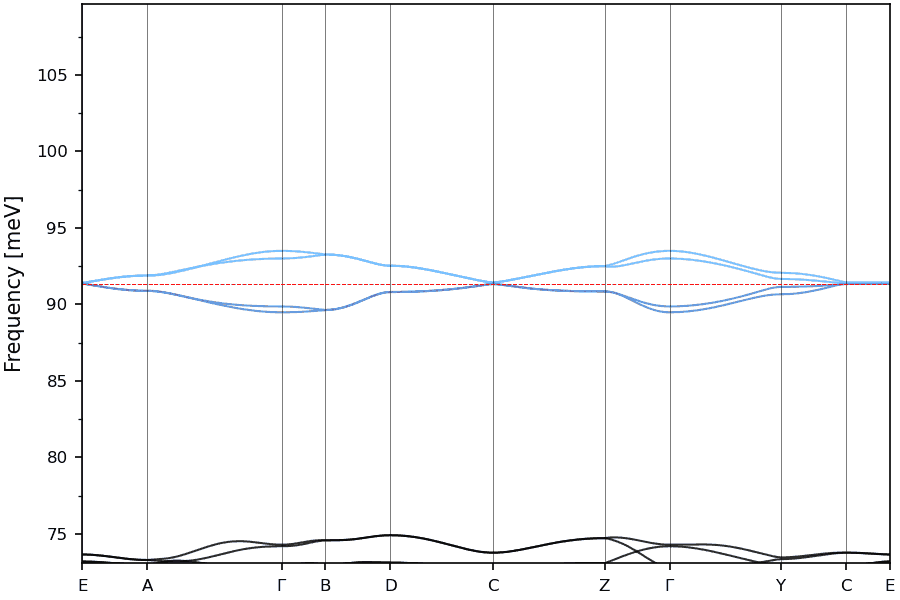} \\
\end{tabular}
\caption{Band structures of the ``ideal'' topological phonon bands with cumulative strong topology on the PhononDB@kyoto-u. (part 4/27)}
\label{fig:4_besttop_kyoto}
\end{figure}

\begin{figure}[h]
\centering
\begin{tabular}{cc}
\scriptsize{\mpidweb{27753}, SG 14 (\sgsymb{14}), \ce{In$_2$(PS$_3$)$_3$}} & \scriptsize{\mpidweb{28773}, SG 14 (\sgsymb{14}), \ce{Na$_4$Ga$_2$S$_5$}} \\
\includegraphics[width=0.5\textwidth,angle=0]{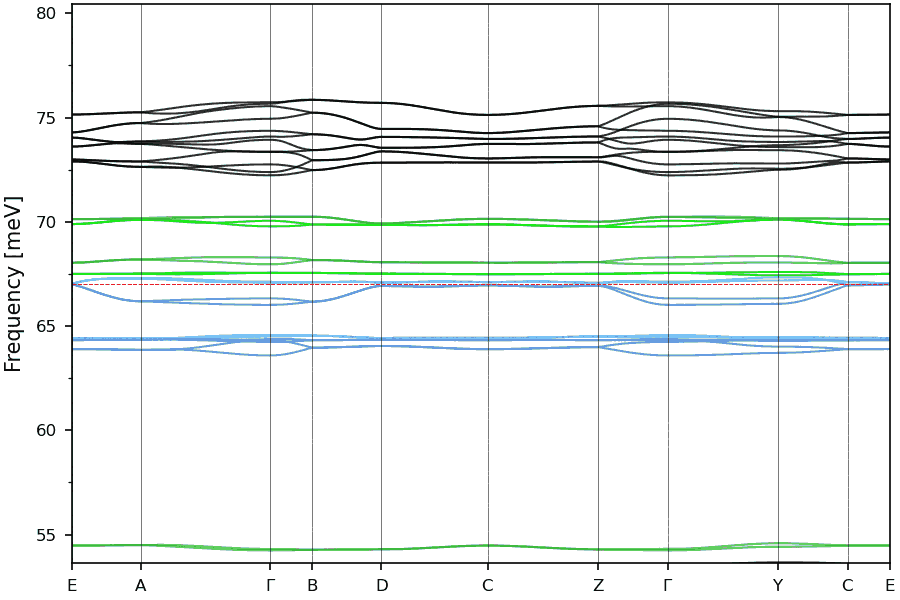} & \includegraphics[width=0.5\textwidth,angle=0]{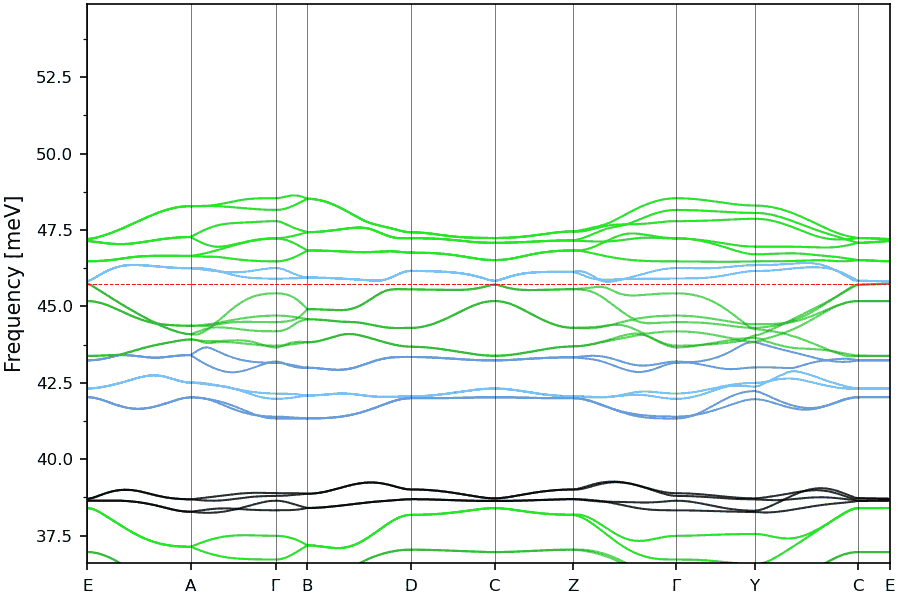} \\
\scriptsize{\mpidweb{29282}, SG 14 (\sgsymb{14}), \ce{P$_2$PtO$_7$}} & \scriptsize{\mpidweb{30926}, SG 14 (\sgsymb{14}), \ce{LaB$_5$O$_9$}} \\
\includegraphics[width=0.5\textwidth,angle=0]{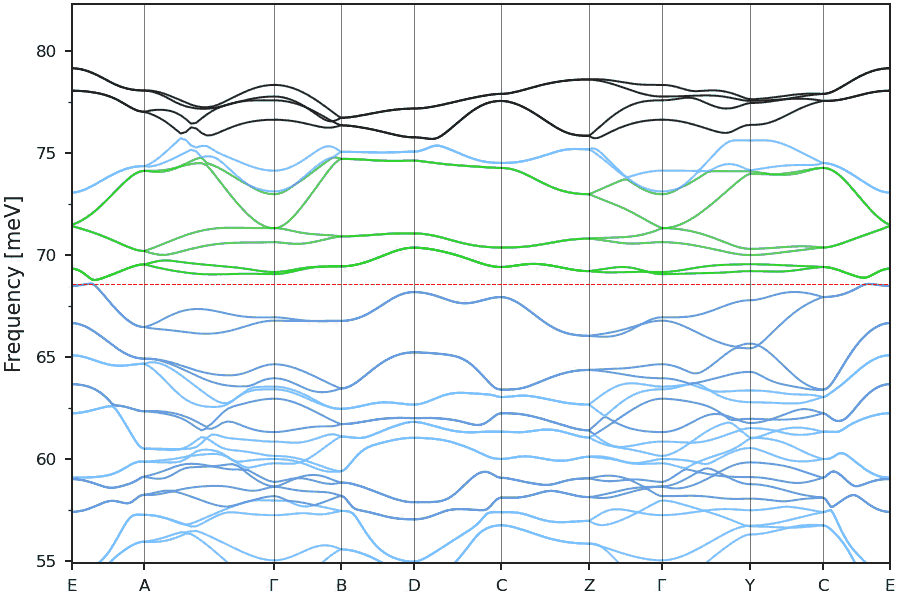} & \includegraphics[width=0.5\textwidth,angle=0]{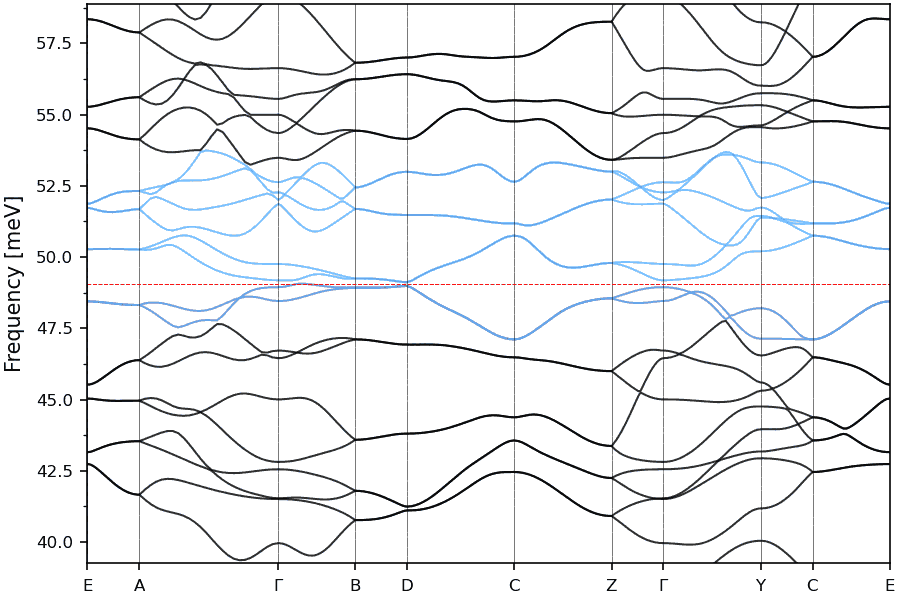} \\
\scriptsize{\mpidweb{542843}, SG 14 (\sgsymb{14}), \ce{MgB$_2$P$_2$H$_8$O$_13$}} & \scriptsize{\mpidweb{555104}, SG 14 (\sgsymb{14}), \ce{Na$_2$Er$_2$B$_2$O$_7$}} \\
\includegraphics[width=0.5\textwidth,angle=0]{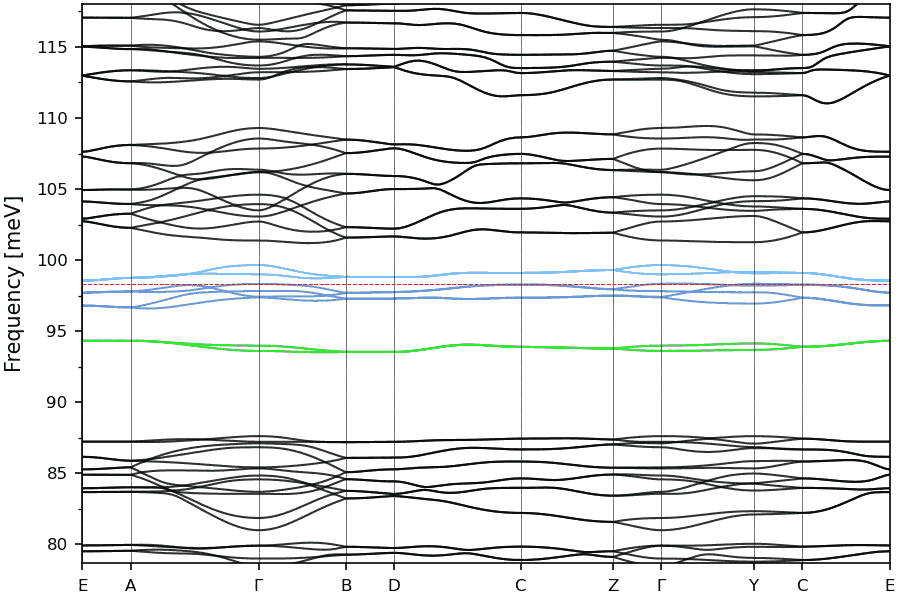} & \includegraphics[width=0.5\textwidth,angle=0]{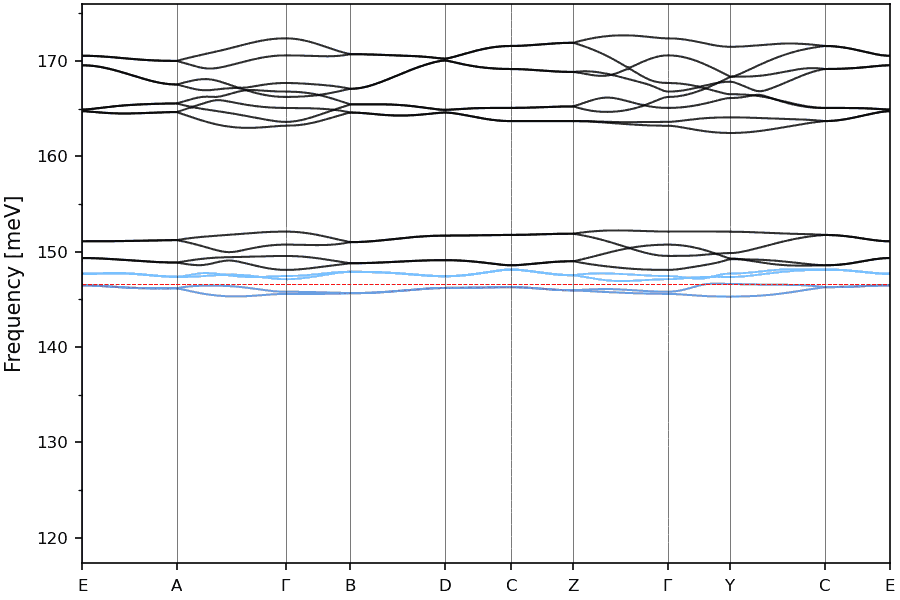} \\
\end{tabular}
\caption{Band structures of the ``ideal'' topological phonon bands with cumulative strong topology on the PhononDB@kyoto-u. (part 5/27)}
\label{fig:5_besttop_kyoto}
\end{figure}

\begin{figure}[h]
\centering
\begin{tabular}{cc}
\scriptsize{\mpidweb{556601}, SG 14 (\sgsymb{14}), \ce{BiBPbO$_4$}} & \scriptsize{\mpidweb{558616}, SG 14 (\sgsymb{14}), \ce{SrZnP$_2$O$_7$}} \\
\includegraphics[width=0.5\textwidth,angle=0]{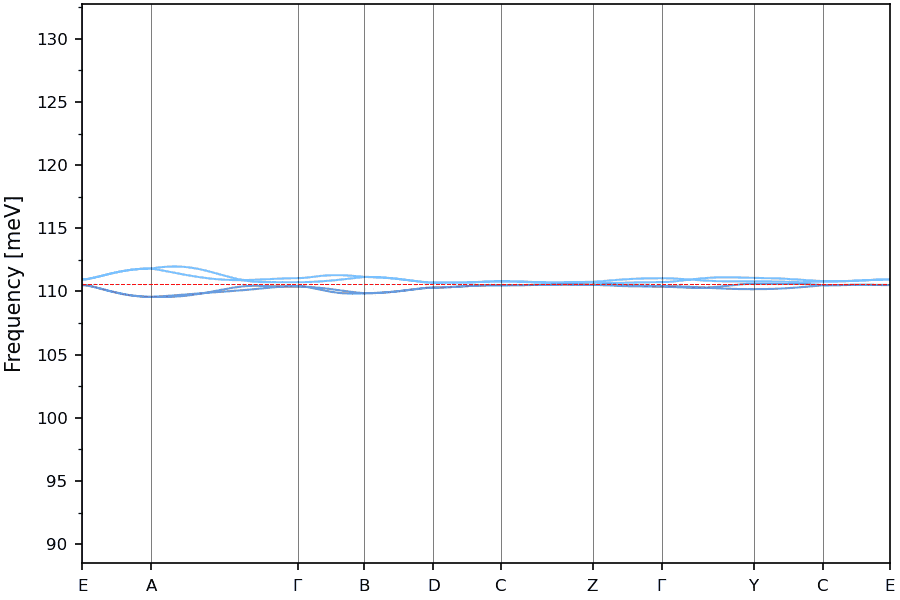} & \includegraphics[width=0.5\textwidth,angle=0]{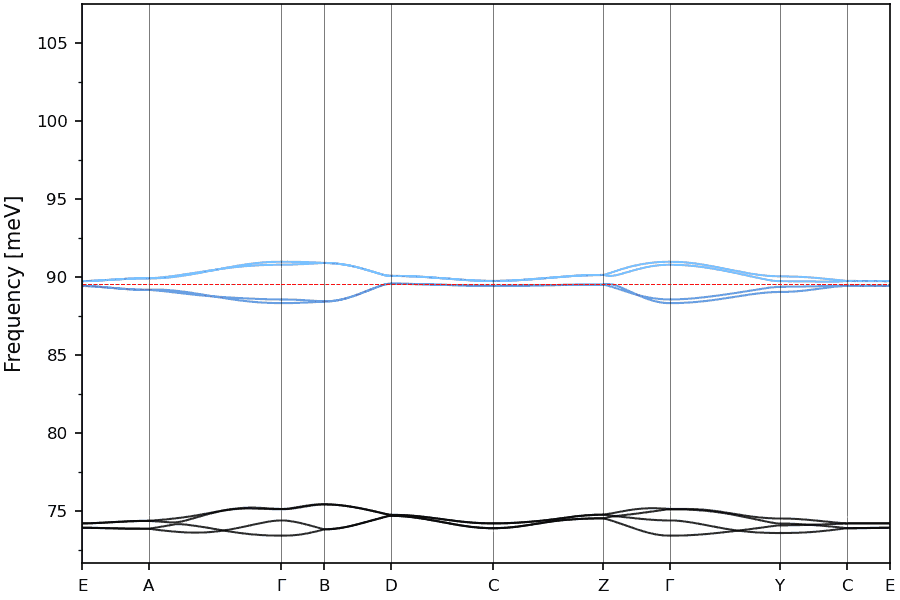} \\
\scriptsize{\mpidweb{559104}, SG 14 (\sgsymb{14}), \ce{Pr$_4$B$_10$O$_21$}} & \scriptsize{\mpidweb{571219}, SG 14 (\sgsymb{14}), \ce{Tl$_3$BiI$_6$}} \\
\includegraphics[width=0.5\textwidth,angle=0]{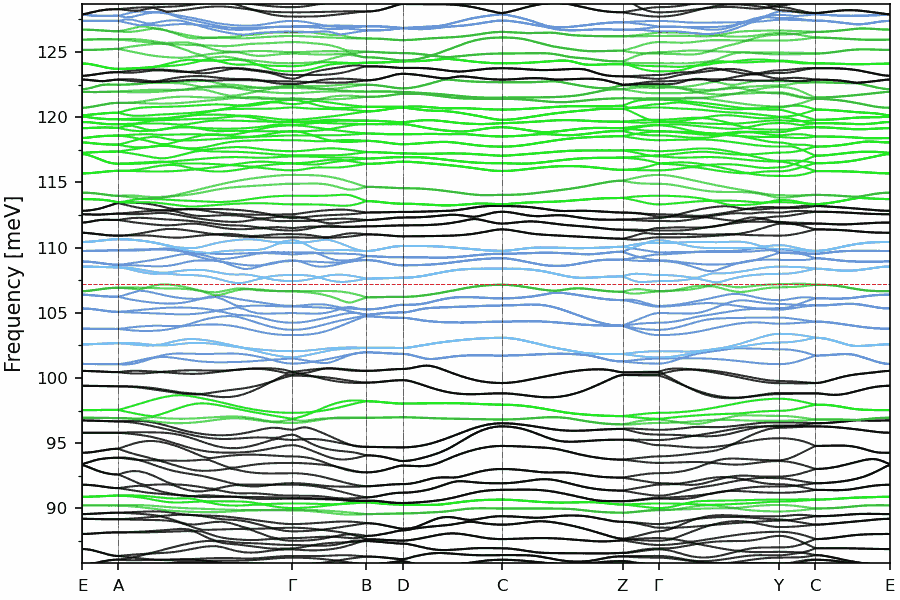} & \includegraphics[width=0.5\textwidth,angle=0]{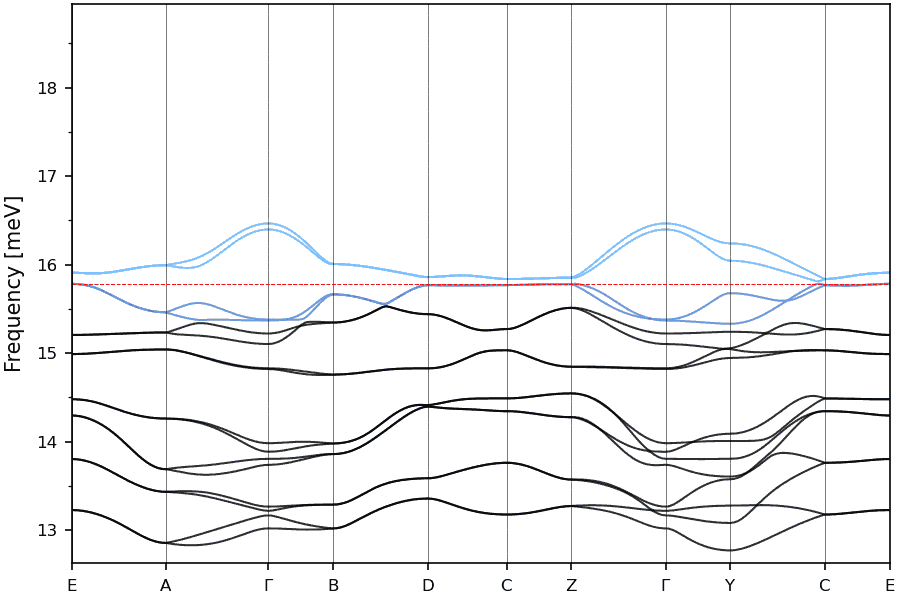} \\
\scriptsize{\mpidweb{581864}, SG 14 (\sgsymb{14}), \ce{CsAsSe$_2$}} & \scriptsize{\mpidweb{619651}, SG 14 (\sgsymb{14}), \ce{Cs$_3$AlP$_2$}} \\
\includegraphics[width=0.5\textwidth,angle=0]{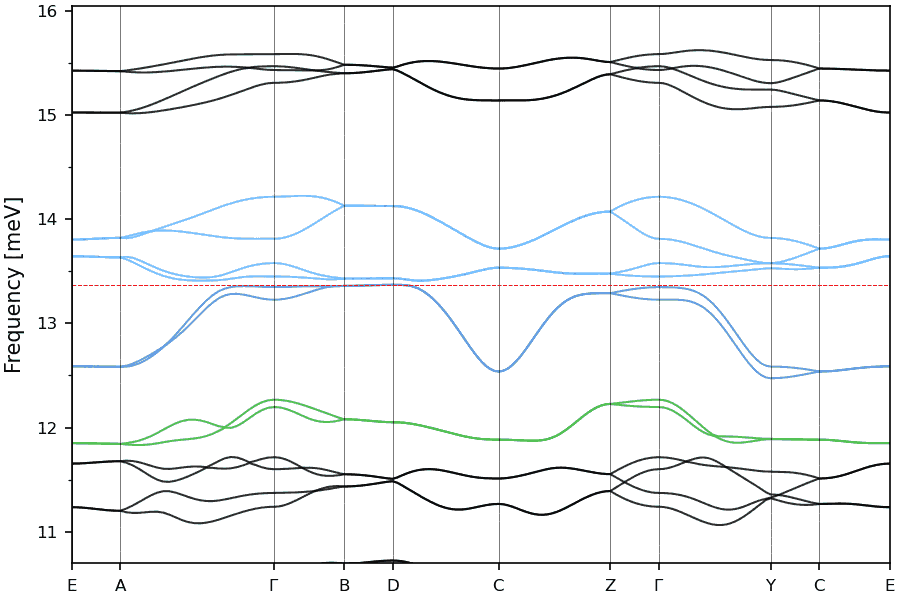} & \includegraphics[width=0.5\textwidth,angle=0]{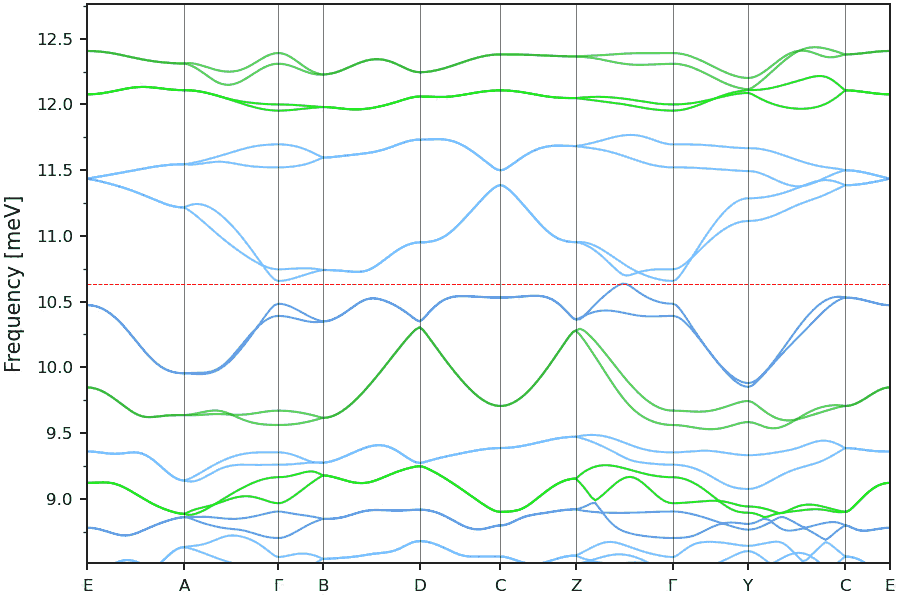} \\
\end{tabular}
\caption{Band structures of the ``ideal'' topological phonon bands with cumulative strong topology on the PhononDB@kyoto-u. (part 6/27)}
\label{fig:6_besttop_kyoto}
\end{figure}

\begin{figure}[h]
\centering
\begin{tabular}{cc}
\scriptsize{\mpidweb{696272}, SG 14 (\sgsymb{14}), \ce{HgHNO$_4$}} & \scriptsize{\mpidweb{733877}, SG 14 (\sgsymb{14}), \ce{CaB$_3$H$_13$O$_12$}} \\
\includegraphics[width=0.5\textwidth,angle=0]{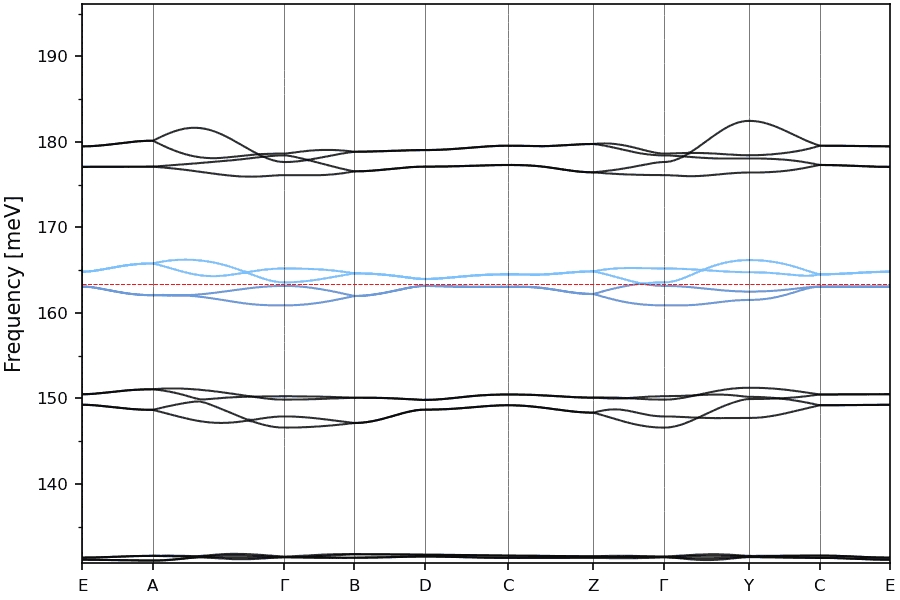} & \includegraphics[width=0.5\textwidth,angle=0]{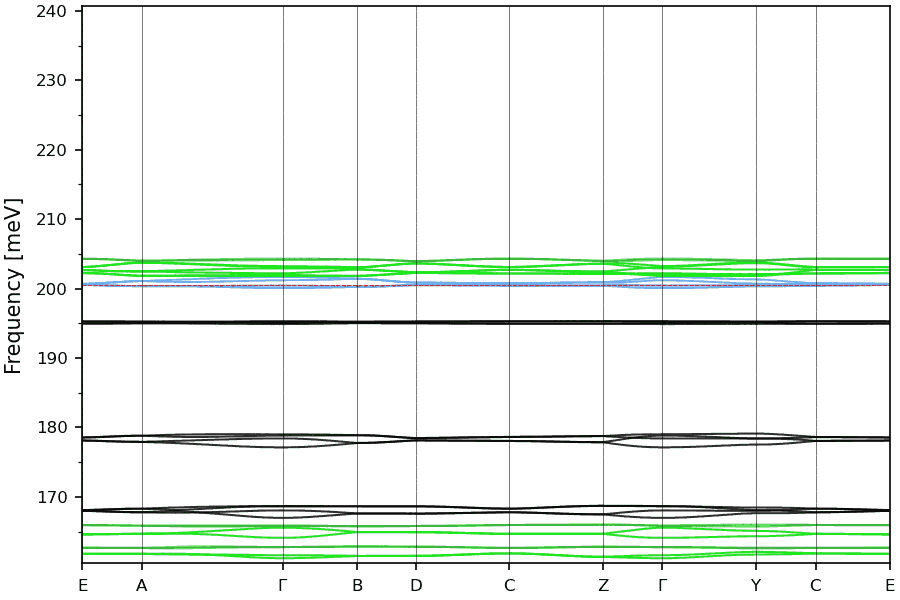} \\
\scriptsize{\mpidweb{765481}, SG 14 (\sgsymb{14}), \ce{Li$_2$AgF$_5$}} & \scriptsize{\mpidweb{771158}, SG 14 (\sgsymb{14}), \ce{Ba$_2$B$_2$O$_5$}} \\
\includegraphics[width=0.5\textwidth,angle=0]{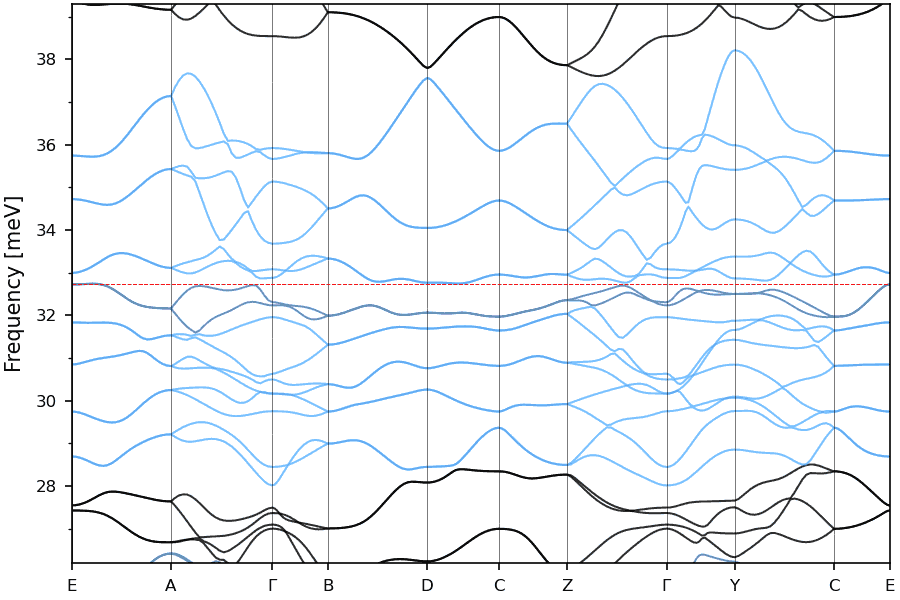} & \includegraphics[width=0.5\textwidth,angle=0]{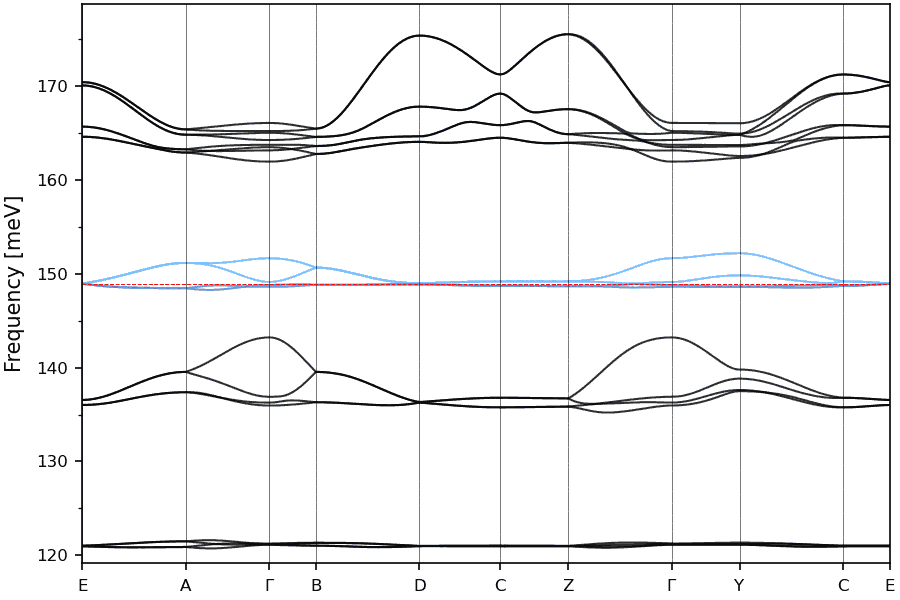} \\
\scriptsize{\mpidweb{772812}, SG 14 (\sgsymb{14}), \ce{Lu$_2$GeO$_5$}} & \scriptsize{\mpidweb{989614}, SG 14 (\sgsymb{14}), \ce{LaReN$_3$}} \\
\includegraphics[width=0.5\textwidth,angle=0]{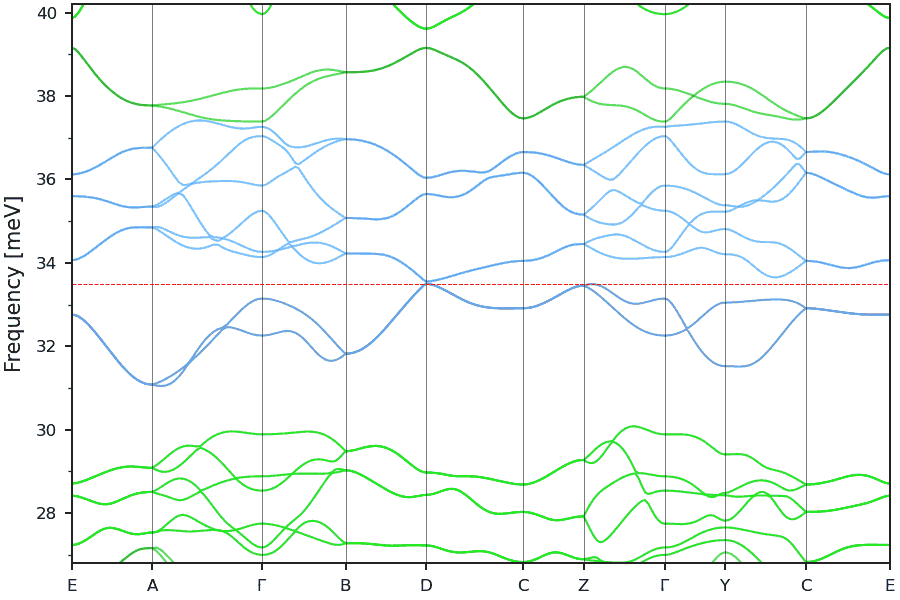} & \includegraphics[width=0.5\textwidth,angle=0]{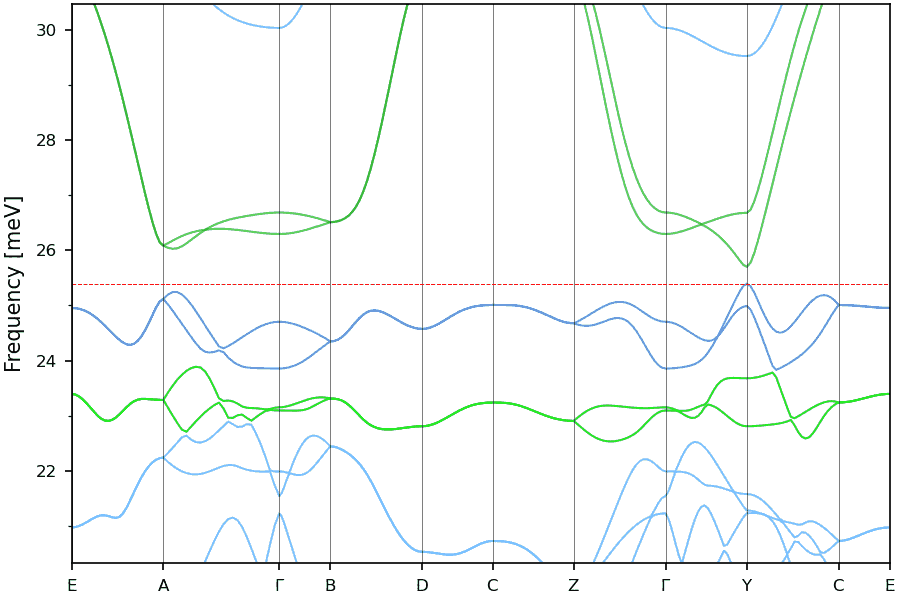} \\
\end{tabular}
\caption{Band structures of the ``ideal'' topological phonon bands with cumulative strong topology on the PhononDB@kyoto-u. (part 7/27)}
\label{fig:7_besttop_kyoto}
\end{figure}

\begin{figure}[h]
\centering
\begin{tabular}{cc}
\scriptsize{\mpidweb{3502}, SG 15 (\sgsymb{15}), \ce{Zn$_2$Te$_3$O$_8$}} & \scriptsize{\mpidweb{4344}, SG 15 (\sgsymb{15}), \ce{HoTaO$_4$}} \\
\includegraphics[width=0.5\textwidth,angle=0]{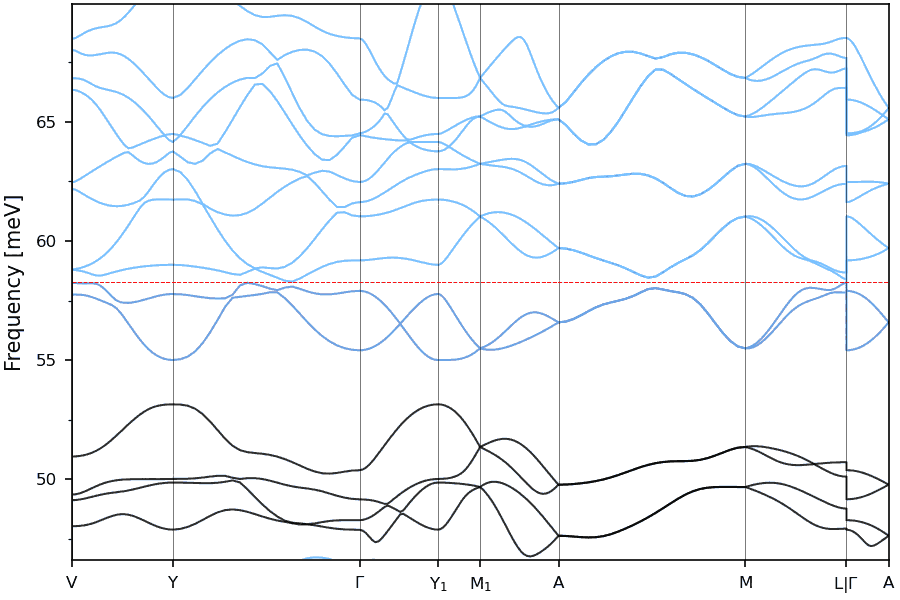} & \includegraphics[width=0.5\textwidth,angle=0]{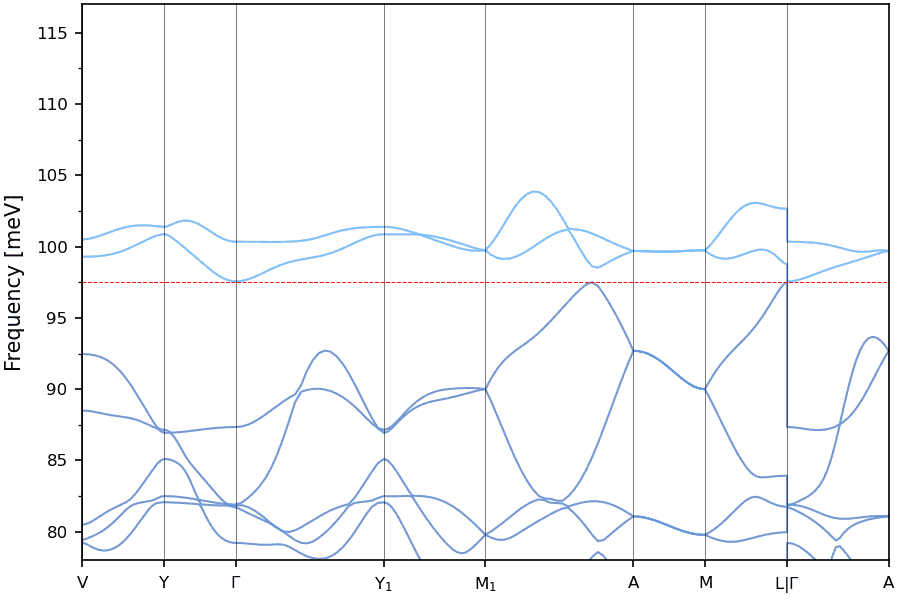} \\
\scriptsize{\mpidweb{5450}, SG 15 (\sgsymb{15}), \ce{K$_2$ZrF$_6$}} & \scriptsize{\mpidweb{6198}, SG 15 (\sgsymb{15}), \ce{SrGa$_2$(SiO$_4$)$_2$}} \\
\includegraphics[width=0.5\textwidth,angle=0]{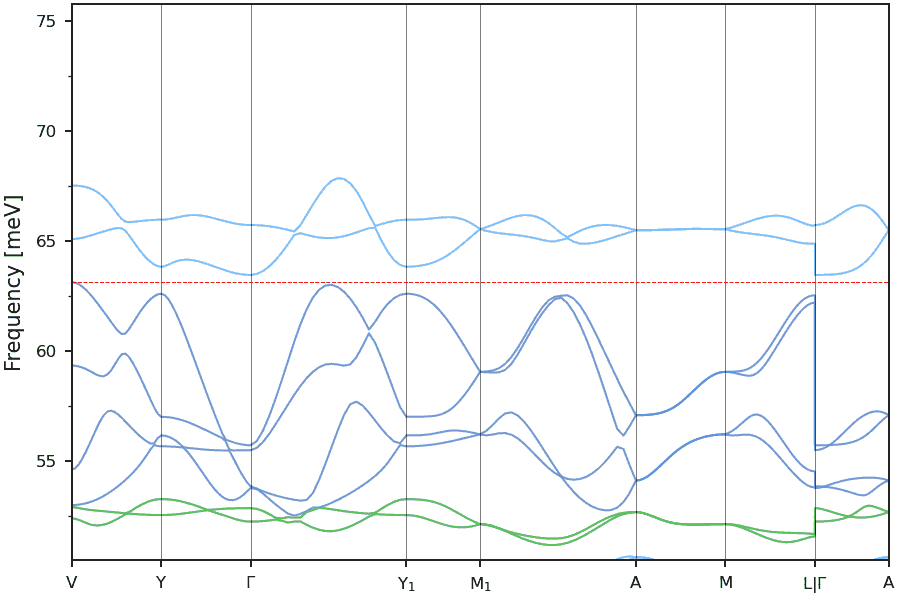} & \includegraphics[width=0.5\textwidth,angle=0]{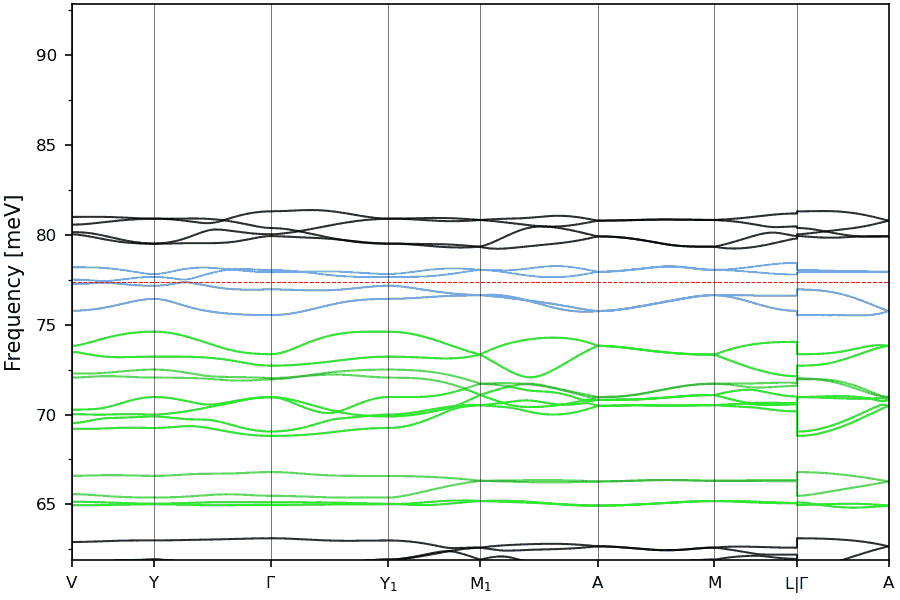} \\
\scriptsize{\mpidweb{6526}, SG 15 (\sgsymb{15}), \ce{NaGeSbO$_5$}} & \scriptsize{\mpidweb{6759}, SG 15 (\sgsymb{15}), \ce{BaGa$_2$(SiO$_4$)$_2$}} \\
\includegraphics[width=0.5\textwidth,angle=0]{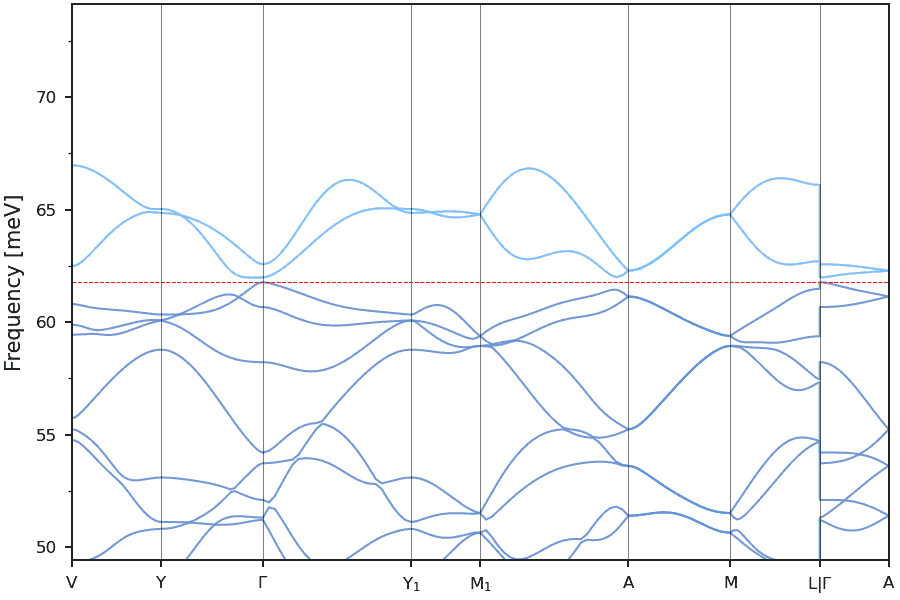} & \includegraphics[width=0.5\textwidth,angle=0]{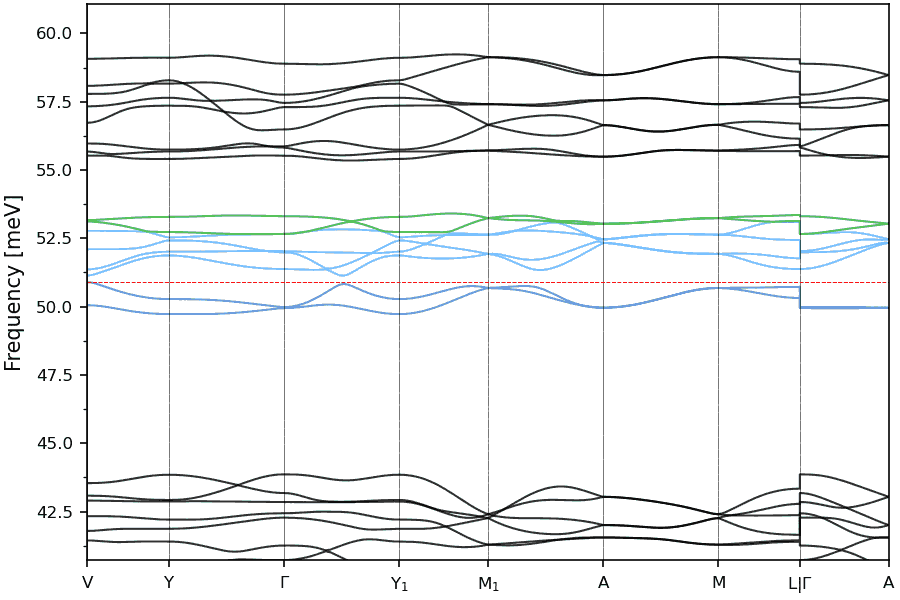} \\
\end{tabular}
\caption{Band structures of the ``ideal'' topological phonon bands with cumulative strong topology on the PhononDB@kyoto-u. (part 8/27)}
\label{fig:8_besttop_kyoto}
\end{figure}

\begin{figure}[h]
\centering
\begin{tabular}{cc}
\scriptsize{\mpidweb{7293}, SG 15 (\sgsymb{15}), \ce{HgPSe$_3$}} & \scriptsize{\mpidweb{8709}, SG 15 (\sgsymb{15}), \ce{KIn(GeO$_3$)$_2$}} \\
\includegraphics[width=0.5\textwidth,angle=0]{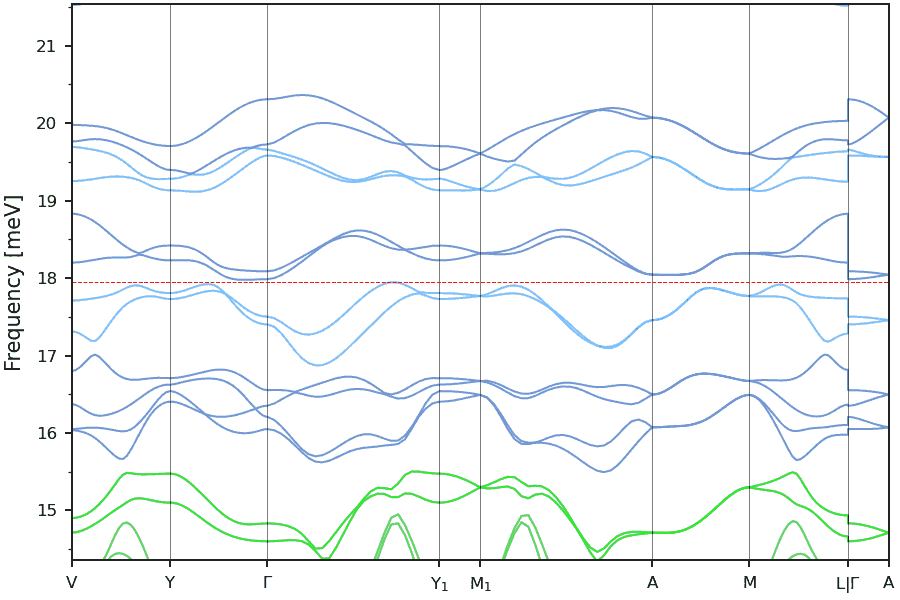} & \includegraphics[width=0.5\textwidth,angle=0]{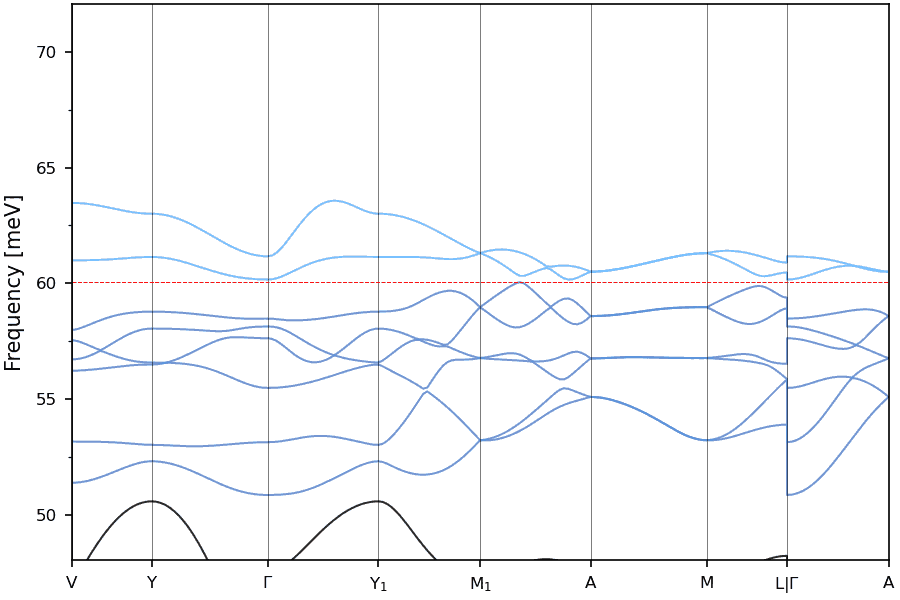} \\
\scriptsize{\mpidweb{17400}, SG 15 (\sgsymb{15}), \ce{Ba$_2$TaN$_3$}} & \scriptsize{\mpidweb{18255}, SG 15 (\sgsymb{15}), \ce{RbSm(PO$_3$)$_4$}} \\
\includegraphics[width=0.5\textwidth,angle=0]{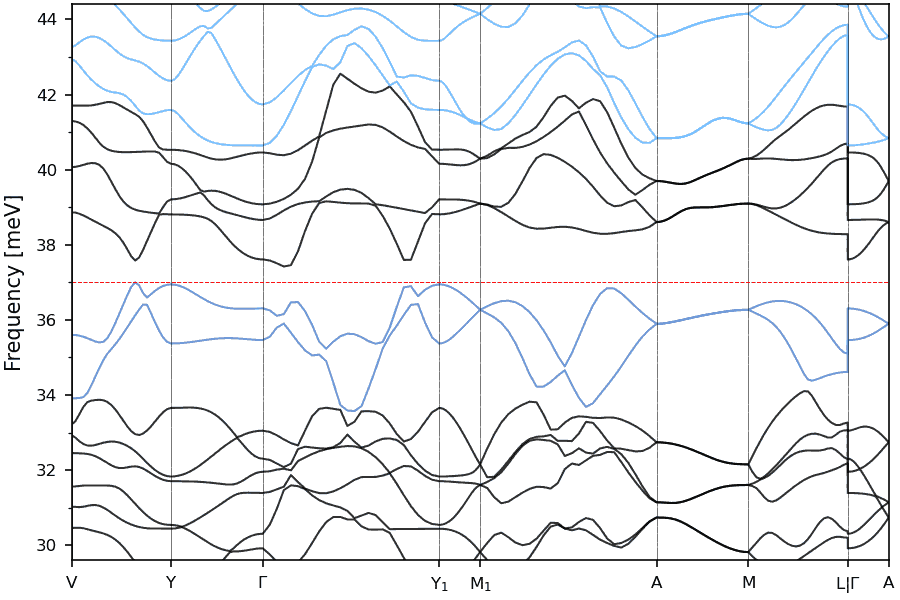} & \includegraphics[width=0.5\textwidth,angle=0]{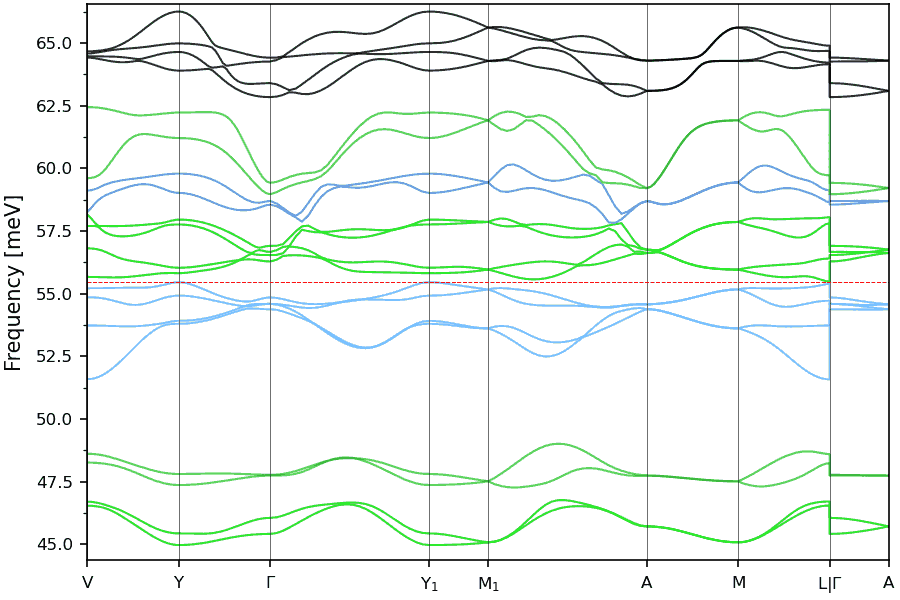} \\
\scriptsize{\mpidweb{18321}, SG 15 (\sgsymb{15}), \ce{Hg$_2$Te$_2$O$_7$}} & \scriptsize{\mpidweb{27436}, SG 15 (\sgsymb{15}), \ce{PaCl$_5$}} \\
\includegraphics[width=0.5\textwidth,angle=0]{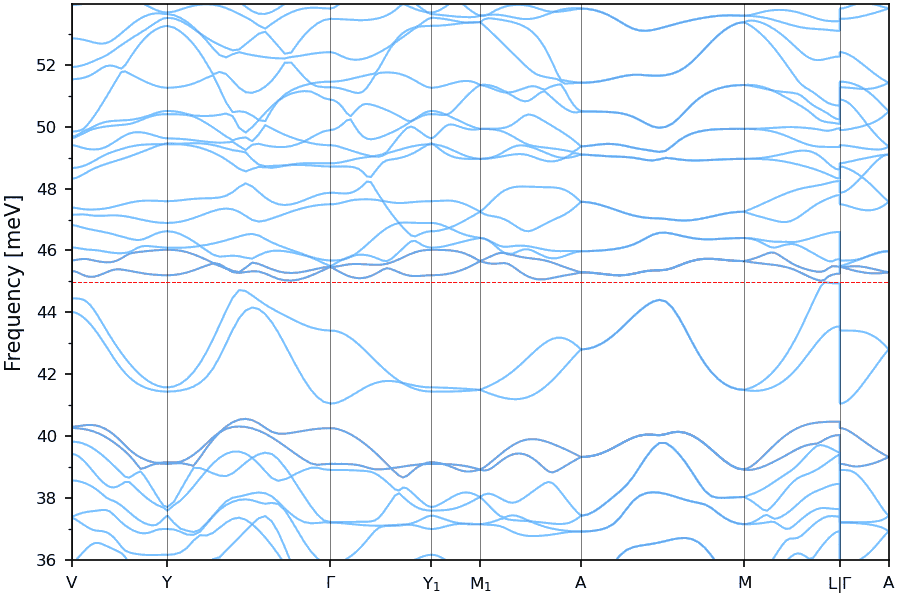} & \includegraphics[width=0.5\textwidth,angle=0]{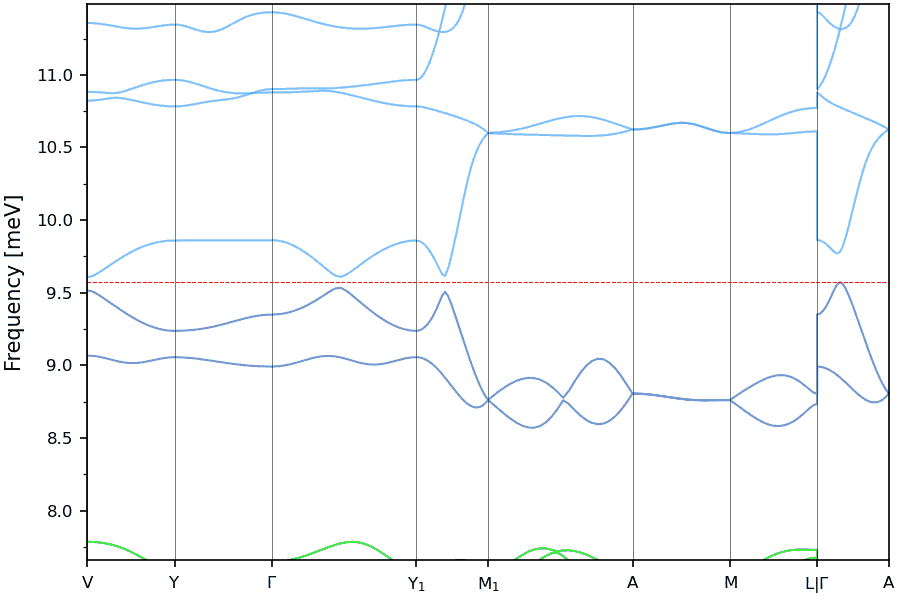} \\
\end{tabular}
\caption{Band structures of the ``ideal'' topological phonon bands with cumulative strong topology on the PhononDB@kyoto-u. (part 9/27)}
\label{fig:9_besttop_kyoto}
\end{figure}

\begin{figure}[h]
\centering
\begin{tabular}{cc}
\scriptsize{\mpidweb{28266}, SG 15 (\sgsymb{15}), \ce{P$_2$Pd}} & \scriptsize{\mpidweb{29302}, SG 15 (\sgsymb{15}), \ce{Ba$_3$(GaS$_3$)$_2$}} \\
\includegraphics[width=0.5\textwidth,angle=0]{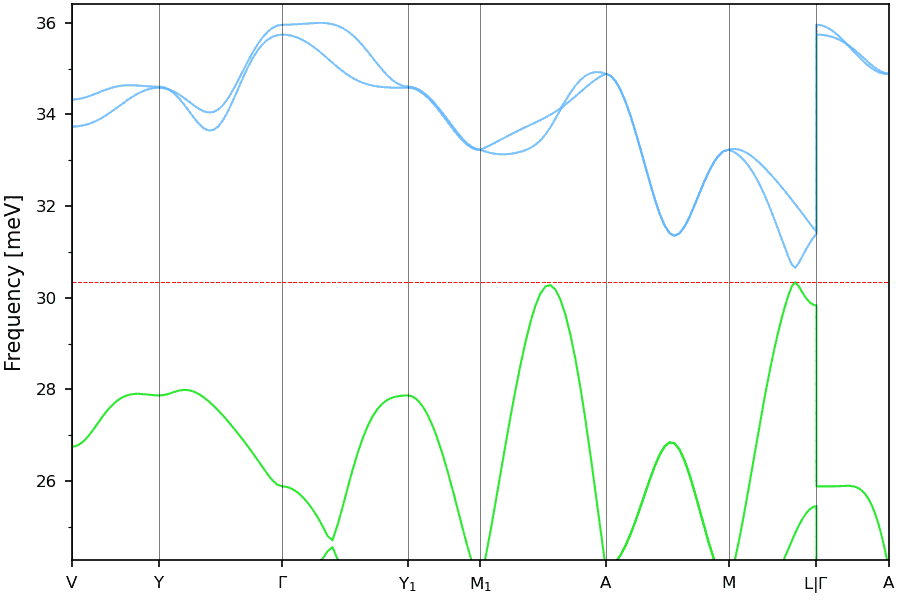} & \includegraphics[width=0.5\textwidth,angle=0]{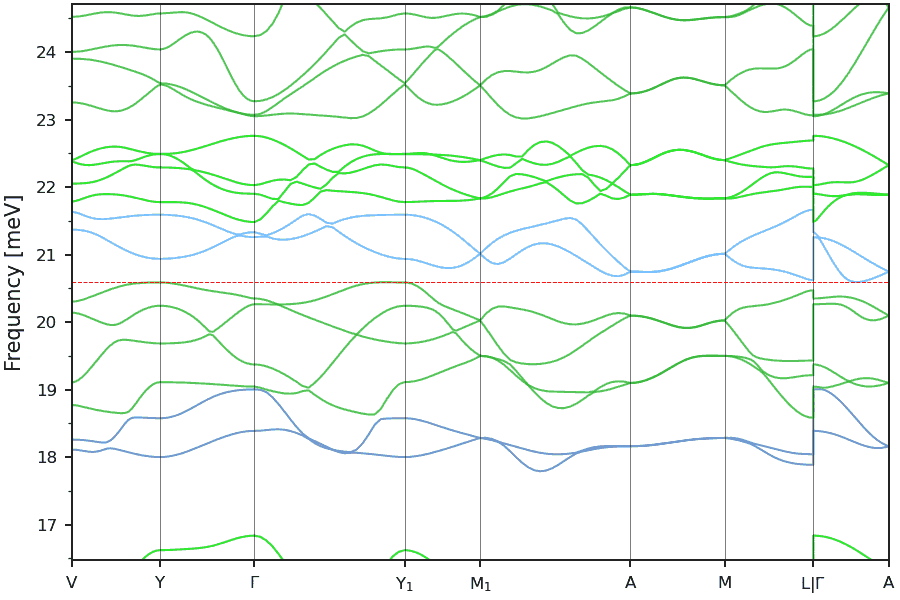} \\
\scriptsize{\mpidweb{30150}, SG 15 (\sgsymb{15}), \ce{GaBH$_6$}} & \scriptsize{\mpidweb{542459}, SG 15 (\sgsymb{15}), \ce{BaGa$_2$(GeO$_4$)$_2$}} \\
\includegraphics[width=0.5\textwidth,angle=0]{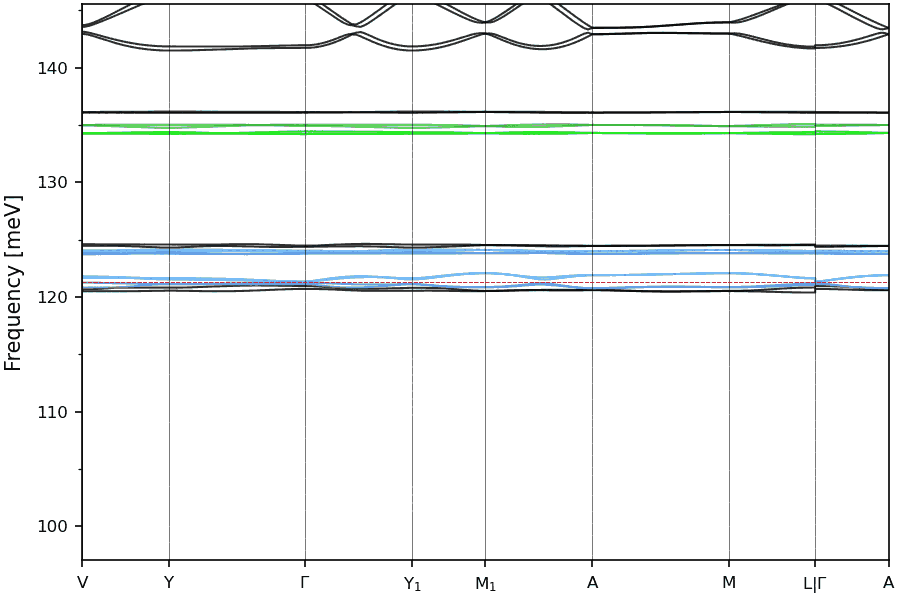} & \includegraphics[width=0.5\textwidth,angle=0]{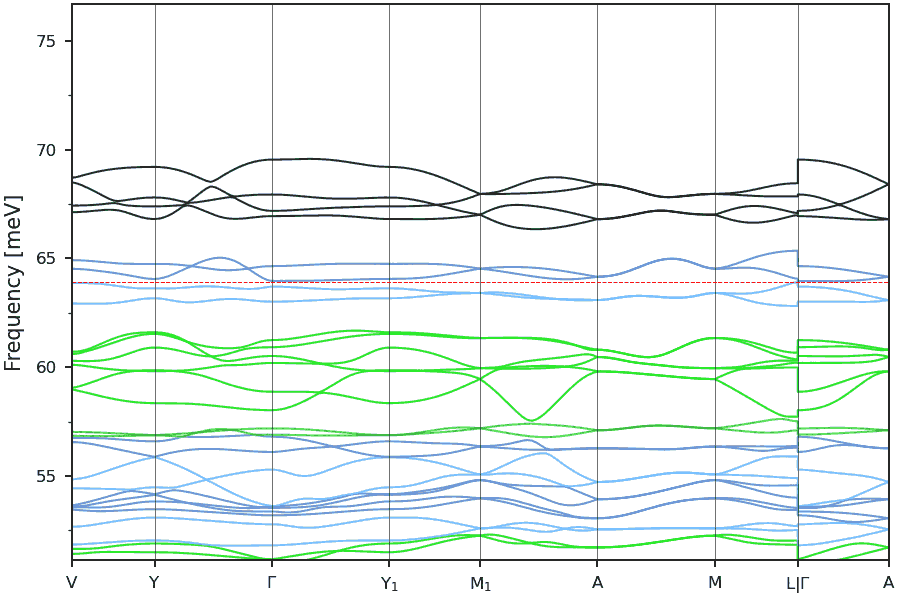} \\
\scriptsize{\mpidweb{555852}, SG 15 (\sgsymb{15}), \ce{Rb$_2$Ag$_2$GeS$_4$}} & \scriptsize{\mpidweb{557104}, SG 15 (\sgsymb{15}), \ce{LiAlB$_2$O$_5$}} \\
\includegraphics[width=0.5\textwidth,angle=0]{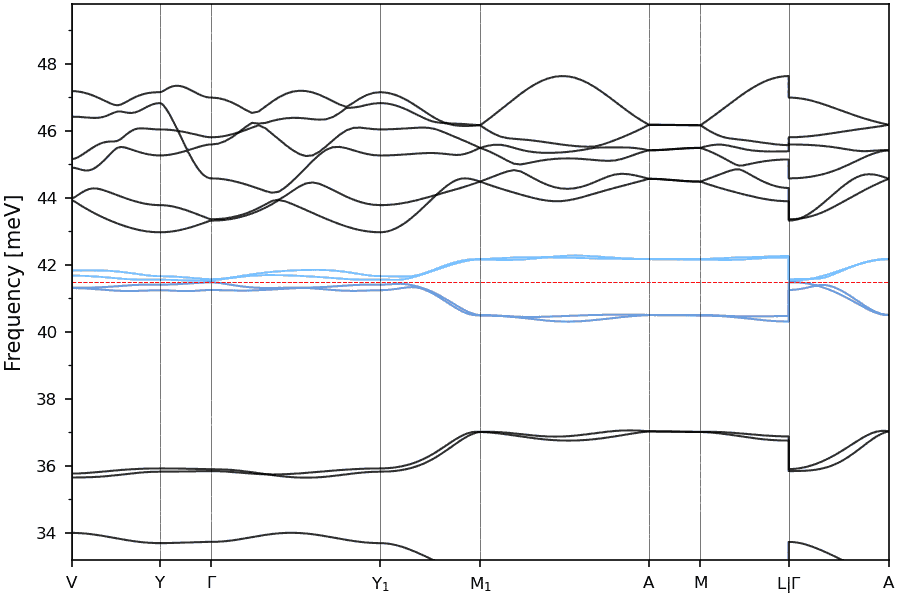} & \includegraphics[width=0.5\textwidth,angle=0]{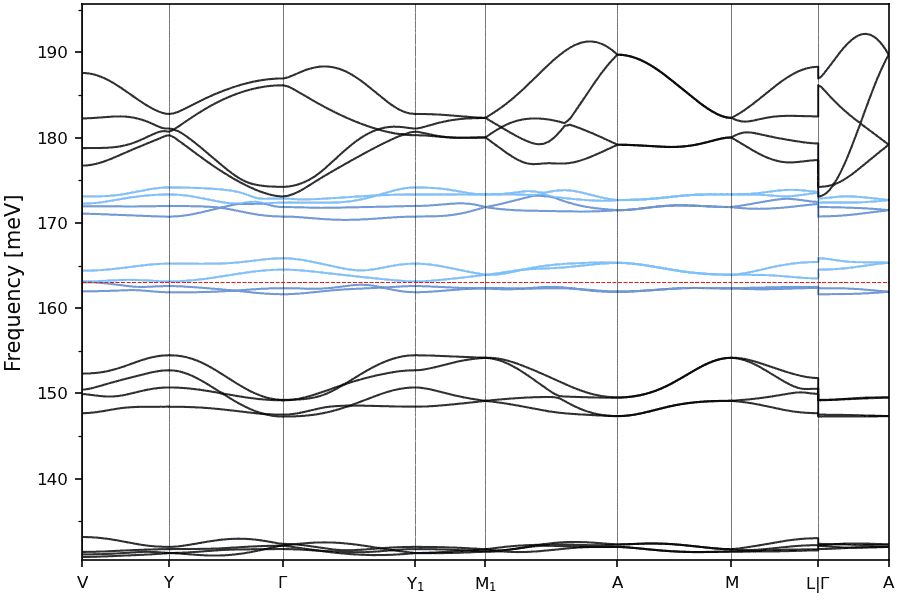} \\
\end{tabular}
\caption{Band structures of the ``ideal'' topological phonon bands with cumulative strong topology on the PhononDB@kyoto-u. (part 10/27)}
\label{fig:10_besttop_kyoto}
\end{figure}

\begin{figure}[h]
\centering
\begin{tabular}{cc}
\scriptsize{\mpidweb{559710}, SG 15 (\sgsymb{15}), \ce{NaInAs$_2$O$_7$}} & \scriptsize{\mpidweb{560493}, SG 15 (\sgsymb{15}), \ce{Ba$_2$V(CuS$_2$)$_3$}} \\
\includegraphics[width=0.5\textwidth,angle=0]{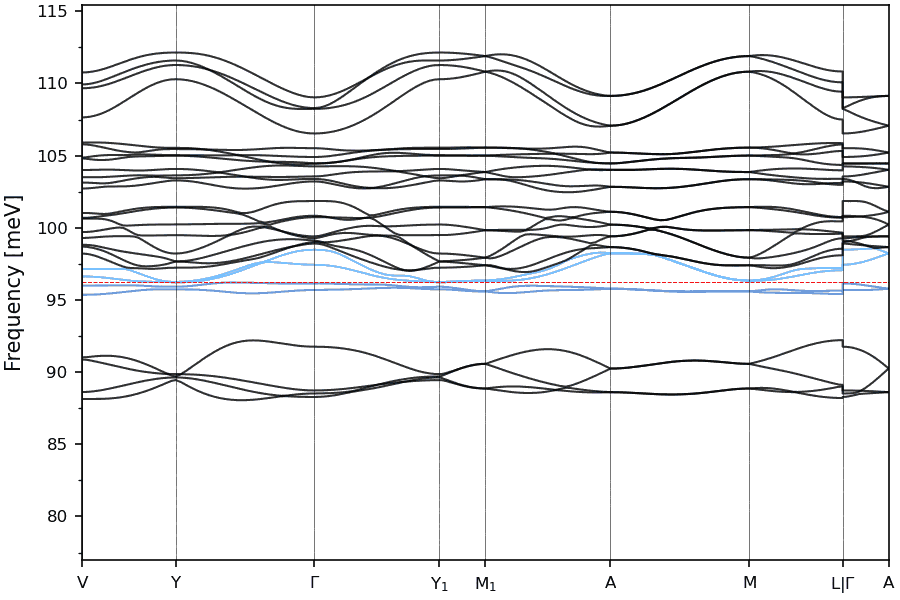} & \includegraphics[width=0.5\textwidth,angle=0]{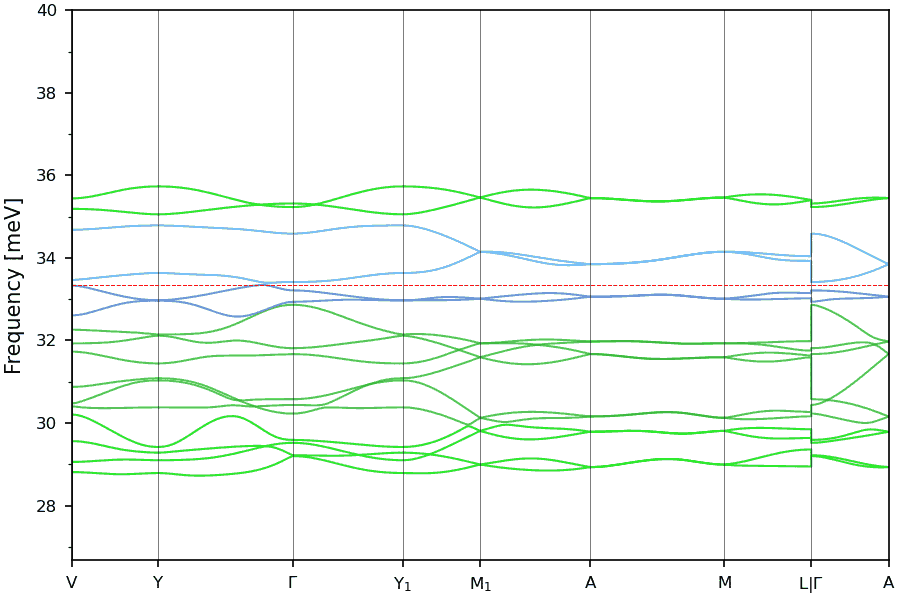} \\
\scriptsize{\mpidweb{560767}, SG 15 (\sgsymb{15}), \ce{KNaTiO$_3$}} & \scriptsize{\mpidweb{561257}, SG 15 (\sgsymb{15}), \ce{LiPr$_2$C$_2$N$_4$F$_3$}} \\
\includegraphics[width=0.5\textwidth,angle=0]{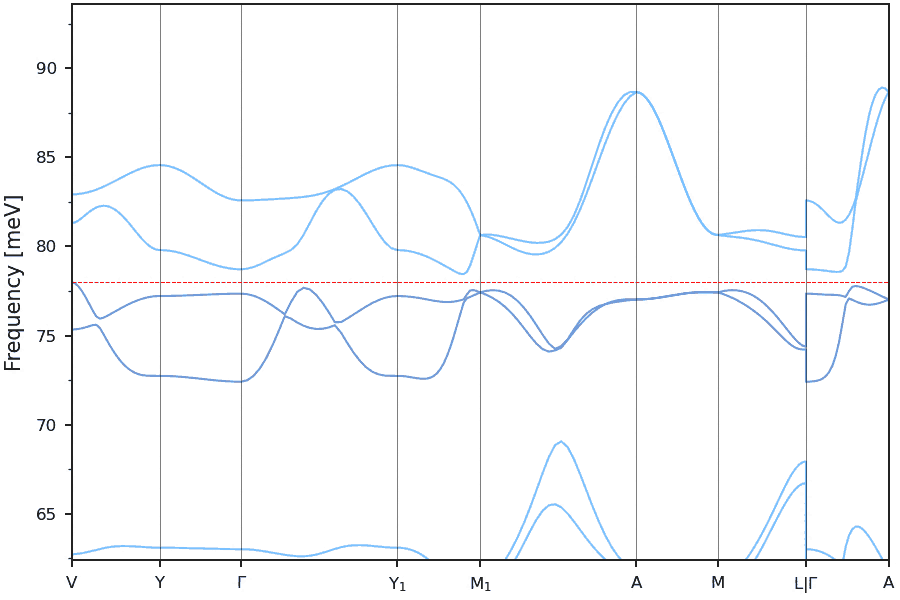} & \includegraphics[width=0.5\textwidth,angle=0]{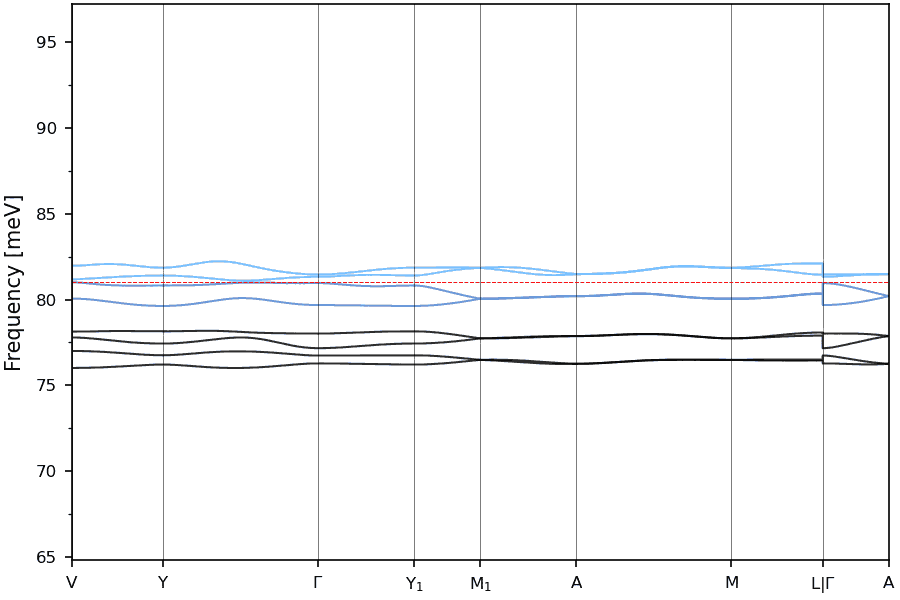} \\
\scriptsize{\mpidweb{561315}, SG 15 (\sgsymb{15}), \ce{Cs$_3$AlGe$_2$O$_7$}} & \scriptsize{\mpidweb{561467}, SG 15 (\sgsymb{15}), \ce{YAg$_3$(PS$_4$)$_2$}} \\
\includegraphics[width=0.5\textwidth,angle=0]{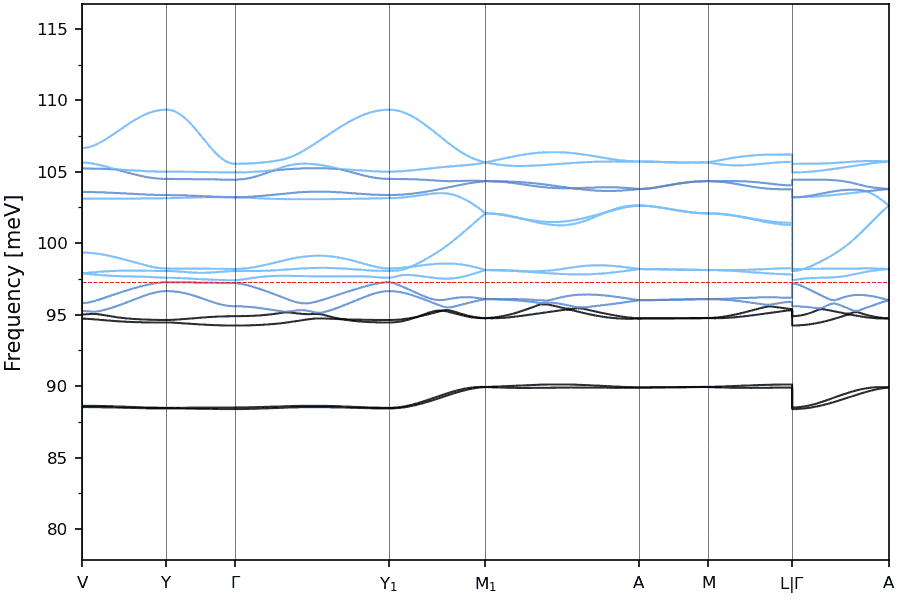} & \includegraphics[width=0.5\textwidth,angle=0]{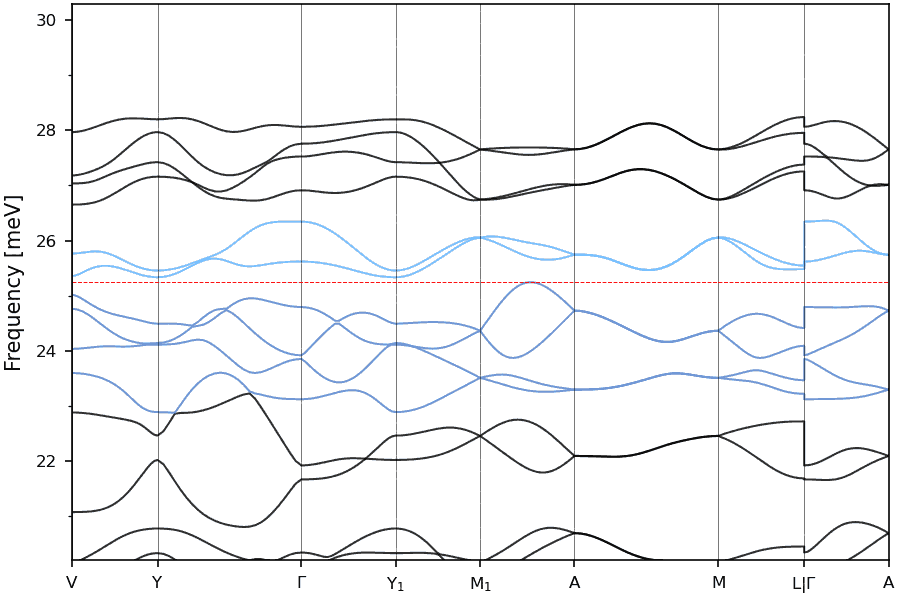} \\
\end{tabular}
\caption{Band structures of the ``ideal'' topological phonon bands with cumulative strong topology on the PhononDB@kyoto-u. (part 11/27)}
\label{fig:11_besttop_kyoto}
\end{figure}

\begin{figure}[h]
\centering
\begin{tabular}{cc}
\scriptsize{\mpidweb{676563}, SG 15 (\sgsymb{15}), \ce{Ta$_4$Cu$_2$O$_11$}} & \scriptsize{\mpidweb{774547}, SG 15 (\sgsymb{15}), \ce{Er$_2$GeO$_5$}} \\
\includegraphics[width=0.5\textwidth,angle=0]{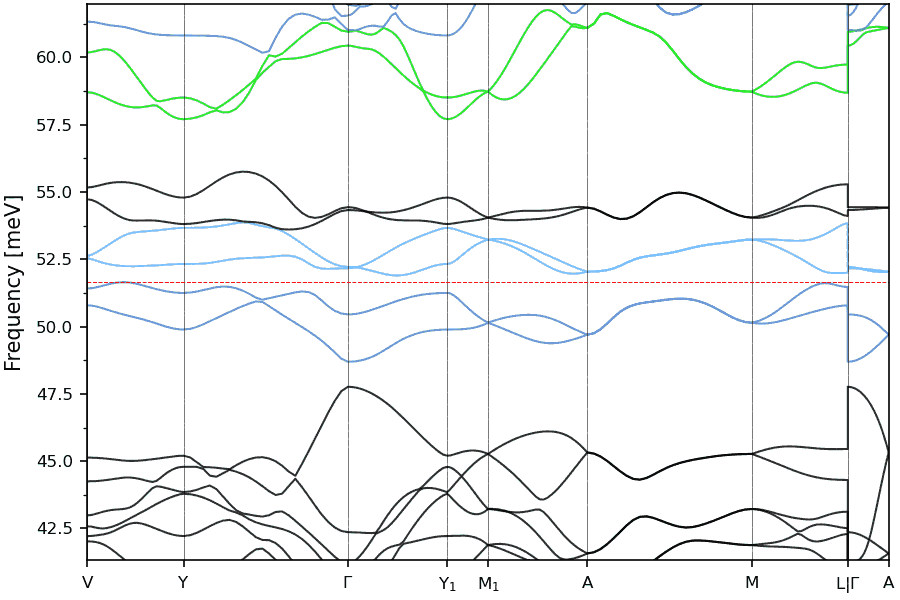} & \includegraphics[width=0.5\textwidth,angle=0]{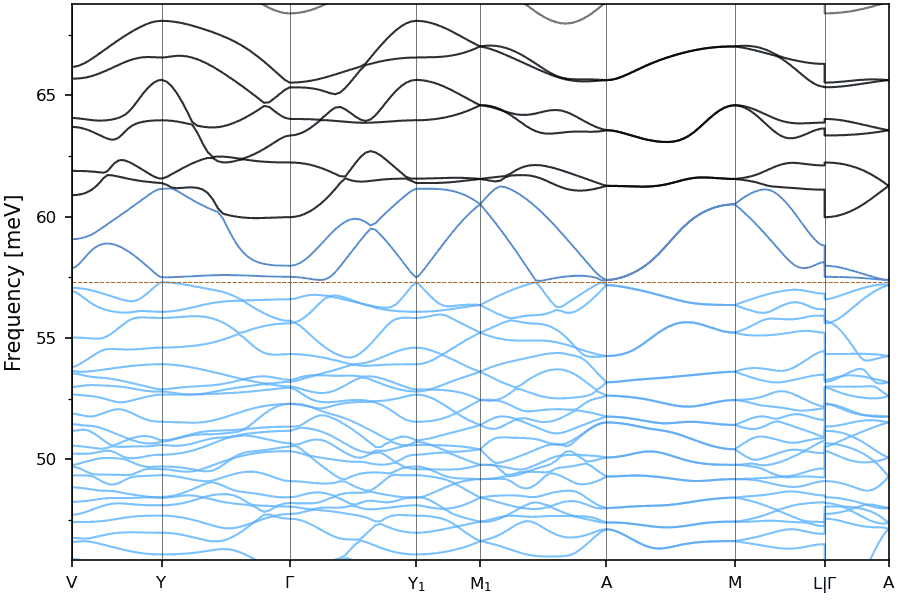} \\
\scriptsize{\mpidweb{866661}, SG 15 (\sgsymb{15}), \ce{K$_3$Dy(AsS$_4$)$_2$}} & \scriptsize{\mpidweb{985608}, SG 15 (\sgsymb{15}), \ce{Nd$_2$O$_5$}} \\
\includegraphics[width=0.5\textwidth,angle=0]{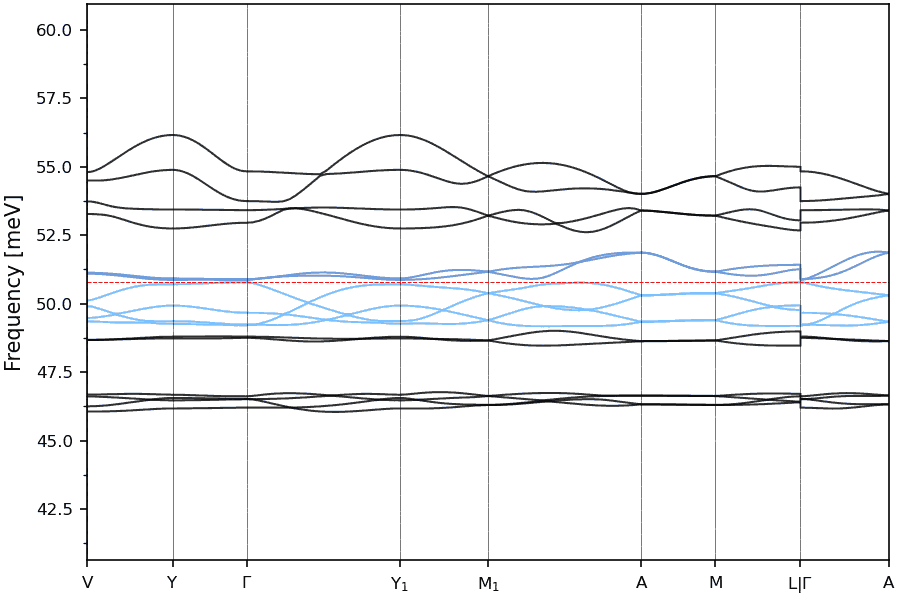} & \includegraphics[width=0.5\textwidth,angle=0]{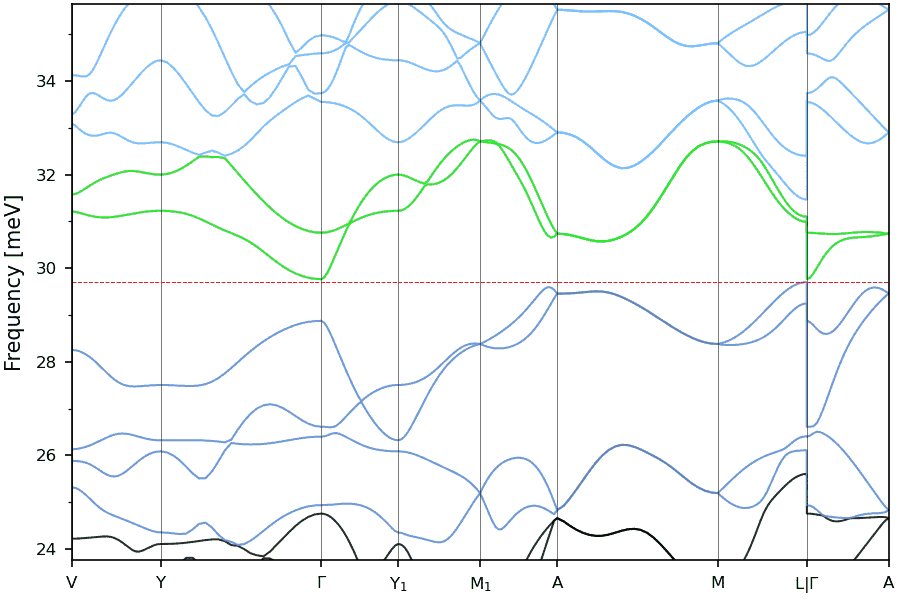} \\
\scriptsize{\mpidweb{18109}, SG 52 (\sgsymb{52}), \ce{RbDy(SO$_4$)$_2$}} & \scriptsize{\mpidweb{28874}, SG 60 (\sgsymb{60}), \ce{Hg$_2$P$_3$Br}} \\
\includegraphics[width=0.5\textwidth,angle=0]{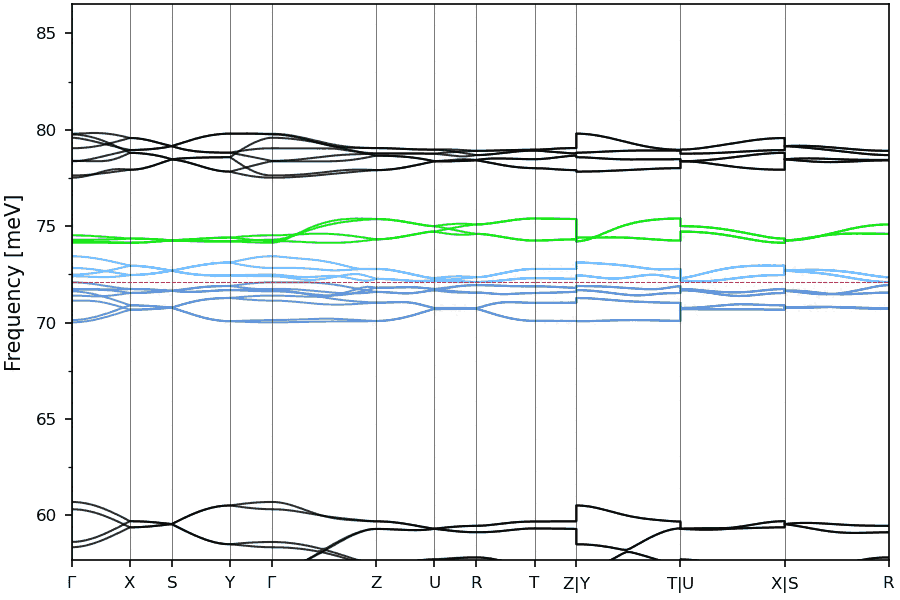} & \includegraphics[width=0.5\textwidth,angle=0]{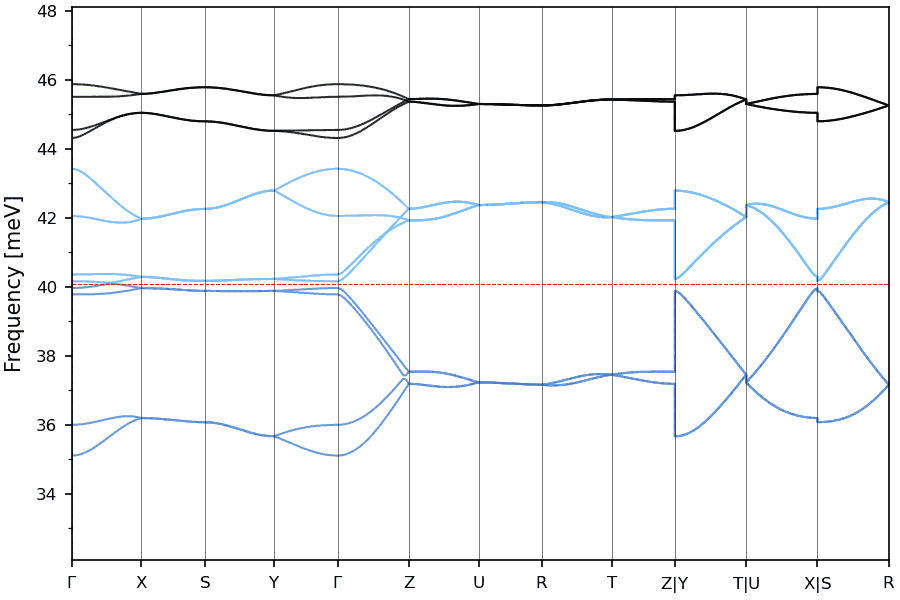} \\
\end{tabular}
\caption{Band structures of the ``ideal'' topological phonon bands with cumulative strong topology on the PhononDB@kyoto-u. (part 12/27)}
\label{fig:12_besttop_kyoto}
\end{figure}

\begin{figure}[h]
\centering
\begin{tabular}{cc}
\scriptsize{\mpidweb{554264}, SG 60 (\sgsymb{60}), \ce{Ba$_3$Tb$_2$(PS$_4$)$_4$}} & \scriptsize{\mpidweb{559171}, SG 60 (\sgsymb{60}), \ce{Ba$_3$Ho$_2$(PS$_4$)$_4$}} \\
\includegraphics[width=0.5\textwidth,angle=0]{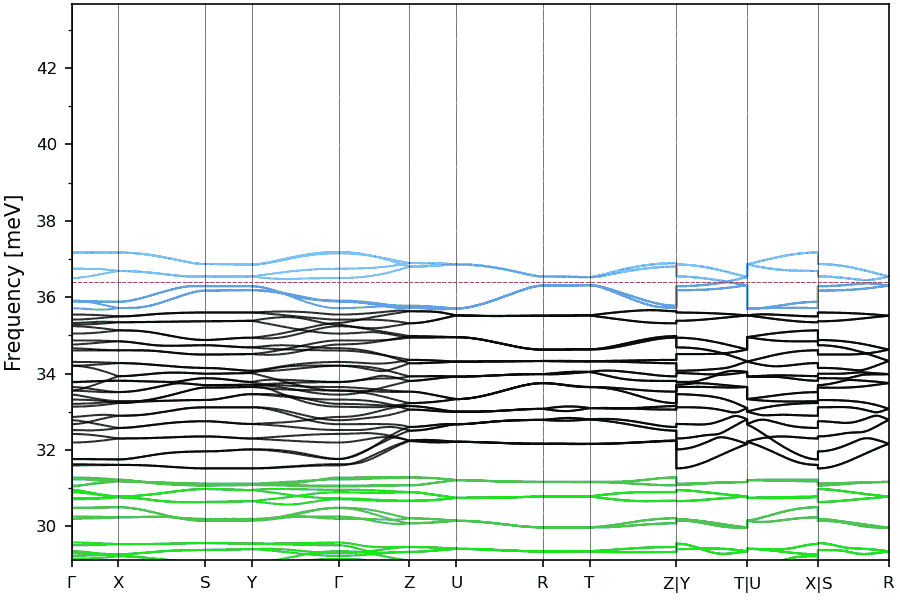} & \includegraphics[width=0.5\textwidth,angle=0]{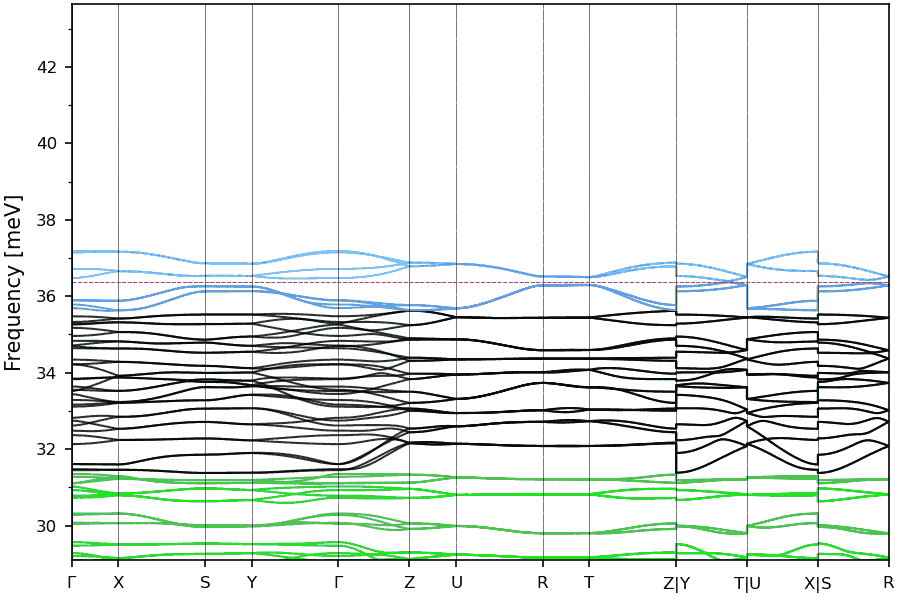} \\
\scriptsize{\mpidweb{560534}, SG 60 (\sgsymb{60}), \ce{Ba$_3$Er$_2$(PS$_4$)$_4$}} & \scriptsize{\mpidweb{560798}, SG 60 (\sgsymb{60}), \ce{Ba$_3$Dy$_2$(PS$_4$)$_4$}} \\
\includegraphics[width=0.5\textwidth,angle=0]{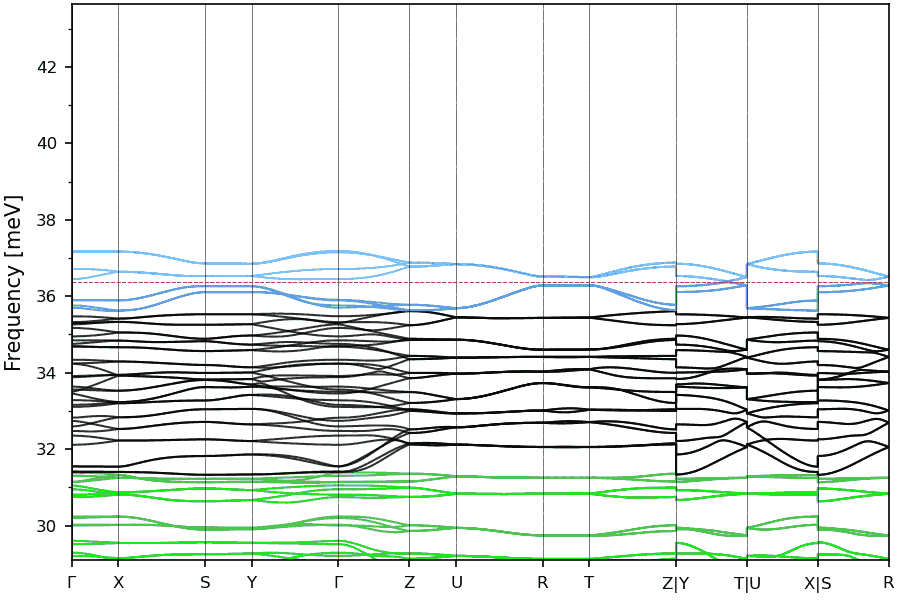} & \includegraphics[width=0.5\textwidth,angle=0]{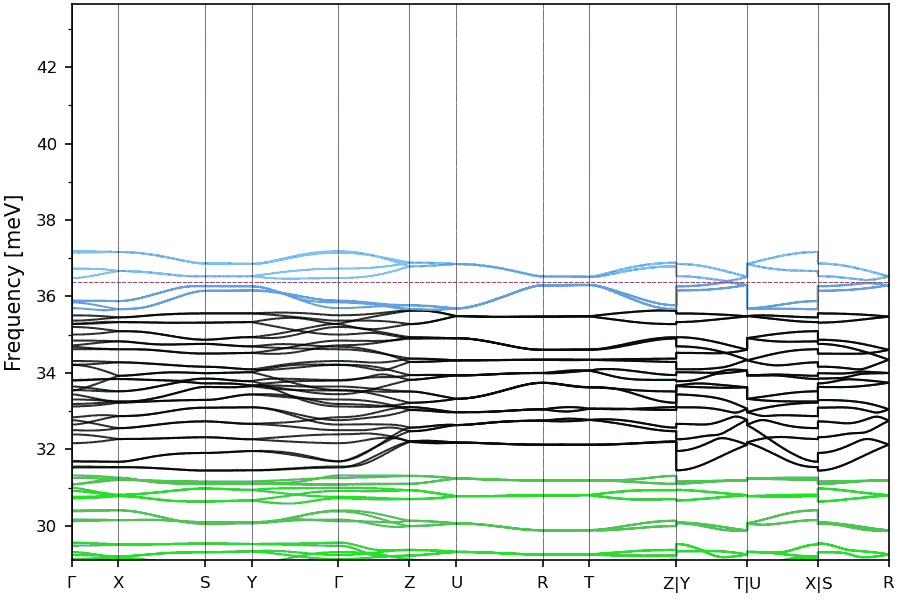} \\
\scriptsize{\mpidweb{14423}, SG 70 (\sgsymb{70}), \ce{Ca(GaS$_2$)$_2$}} & \scriptsize{\mpidweb{14424}, SG 70 (\sgsymb{70}), \ce{Sr(AlS$_2$)$_2$}} \\
\includegraphics[width=0.5\textwidth,angle=0]{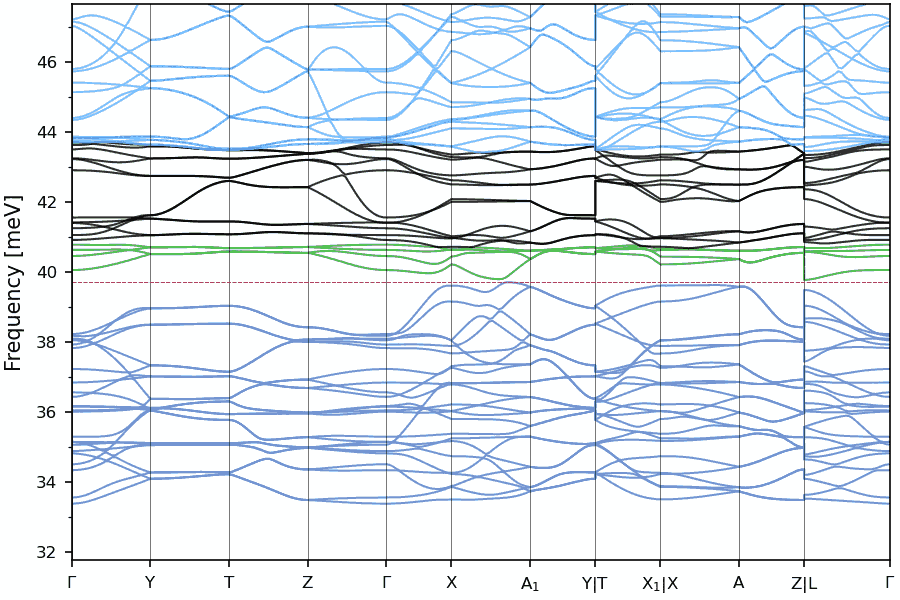} & \includegraphics[width=0.5\textwidth,angle=0]{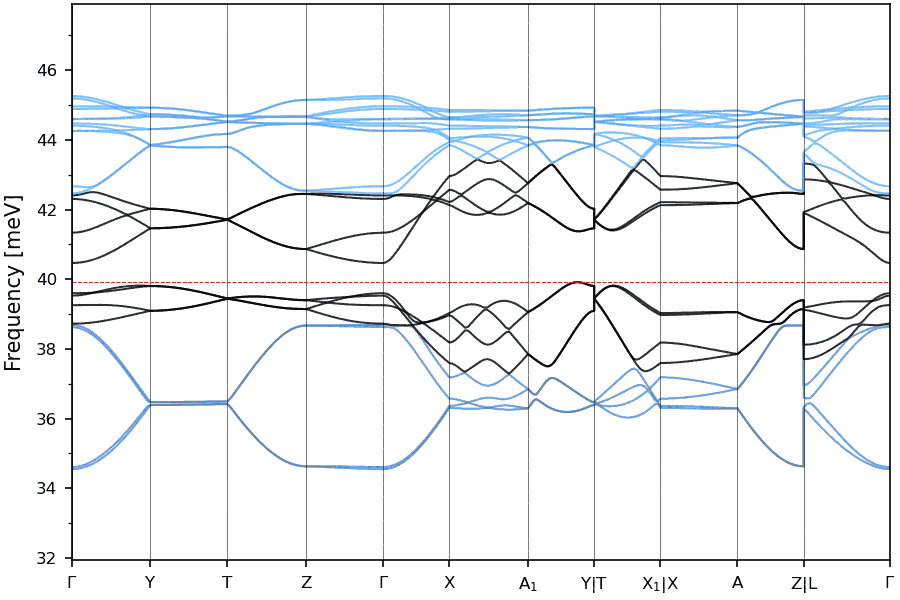} \\
\end{tabular}
\caption{Band structures of the ``ideal'' topological phonon bands with cumulative strong topology on the PhononDB@kyoto-u. (part 13/27)}
\label{fig:13_besttop_kyoto}
\end{figure}

\begin{figure}[h]
\centering
\begin{tabular}{cc}
\scriptsize{\mpidweb{561799}, SG 70 (\sgsymb{70}), \ce{Yb(GaS$_2$)$_2$}} & \scriptsize{\mpidweb{43030}, SG 81 (\sgsymb{81}), \ce{Na$_2$Li$_2$Al$_3$Si$_3$ClO$_12$}} \\
\includegraphics[width=0.5\textwidth,angle=0]{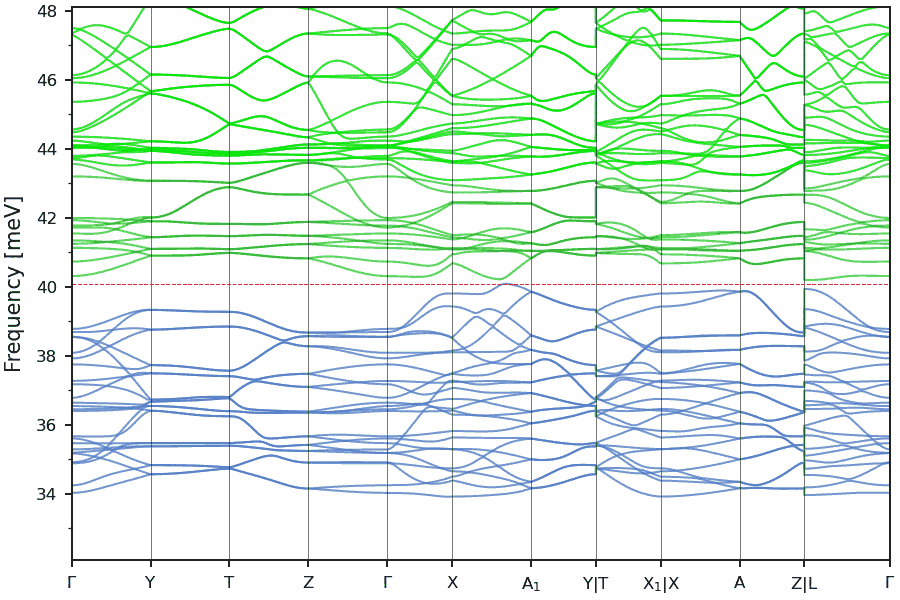} & \includegraphics[width=0.5\textwidth,angle=0]{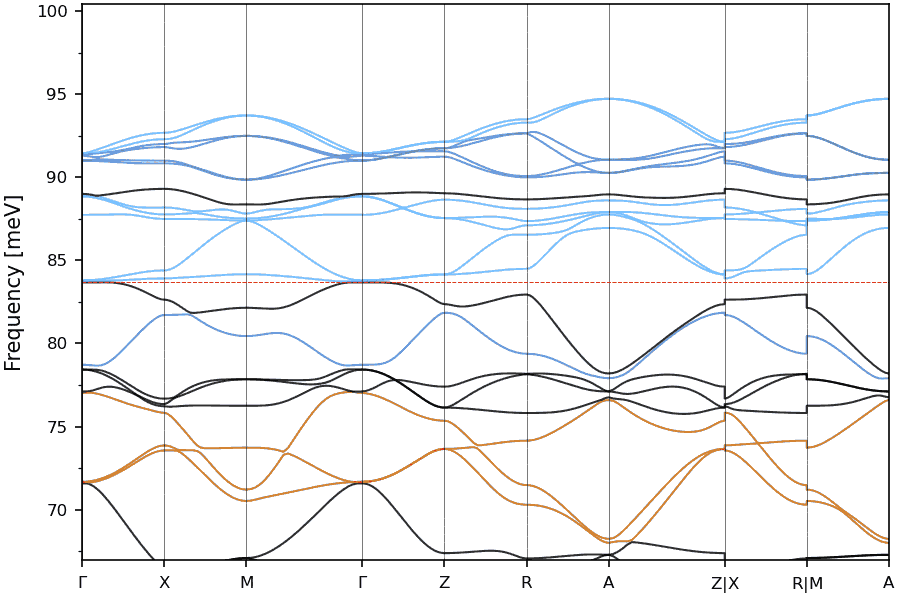} \\
\scriptsize{\mpidweb{3277}, SG 82 (\sgsymb{82}), \ce{BAsO$_4$}} & \scriptsize{\mpidweb{4809}, SG 82 (\sgsymb{82}), \ce{Ga$_2$HgS$_4$}} \\
\includegraphics[width=0.5\textwidth,angle=0]{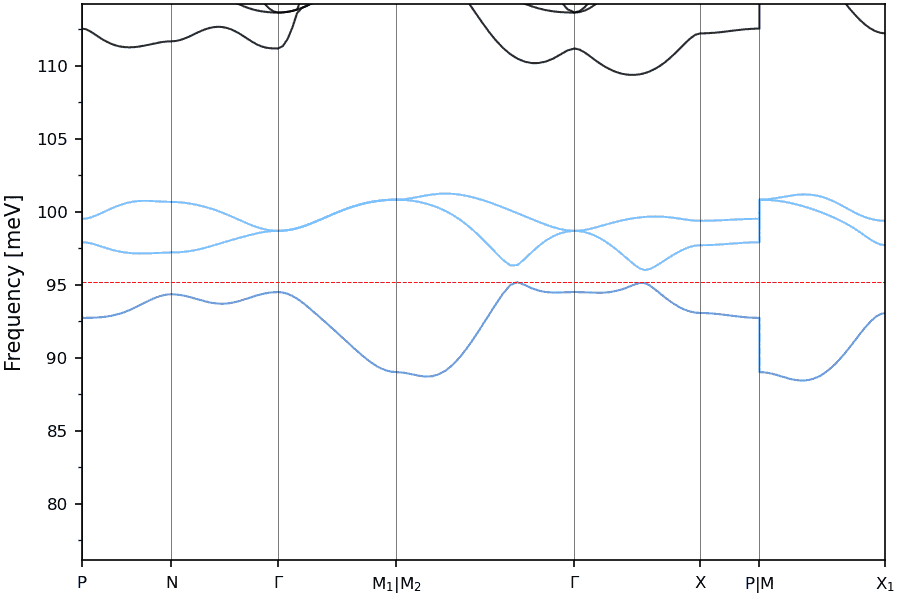} & \includegraphics[width=0.5\textwidth,angle=0]{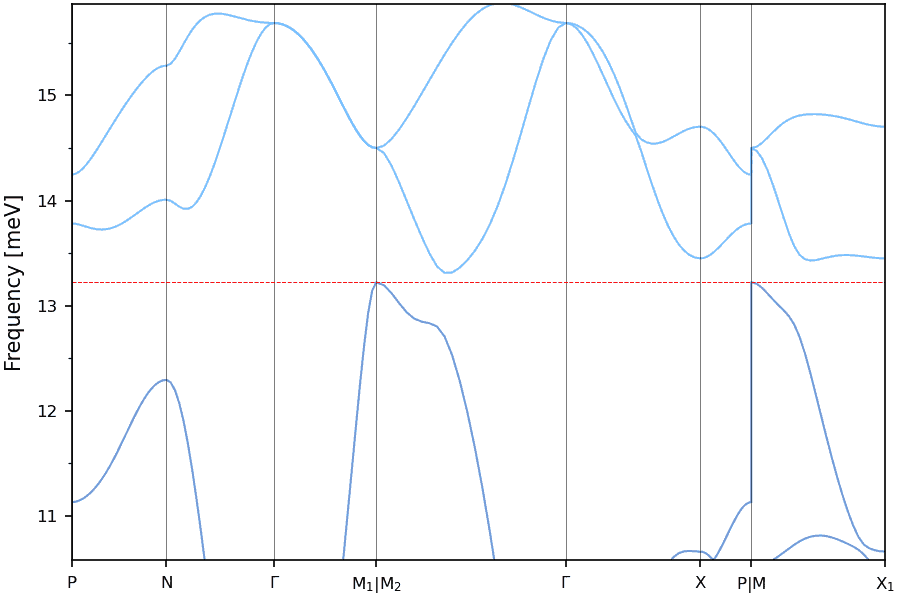} \\
\scriptsize{\mpidweb{5046}, SG 82 (\sgsymb{82}), \ce{BeSO$_4$}} & \scriptsize{\mpidweb{7907}, SG 82 (\sgsymb{82}), \ce{Al$_2$ZnSe$_4$}} \\
\includegraphics[width=0.5\textwidth,angle=0]{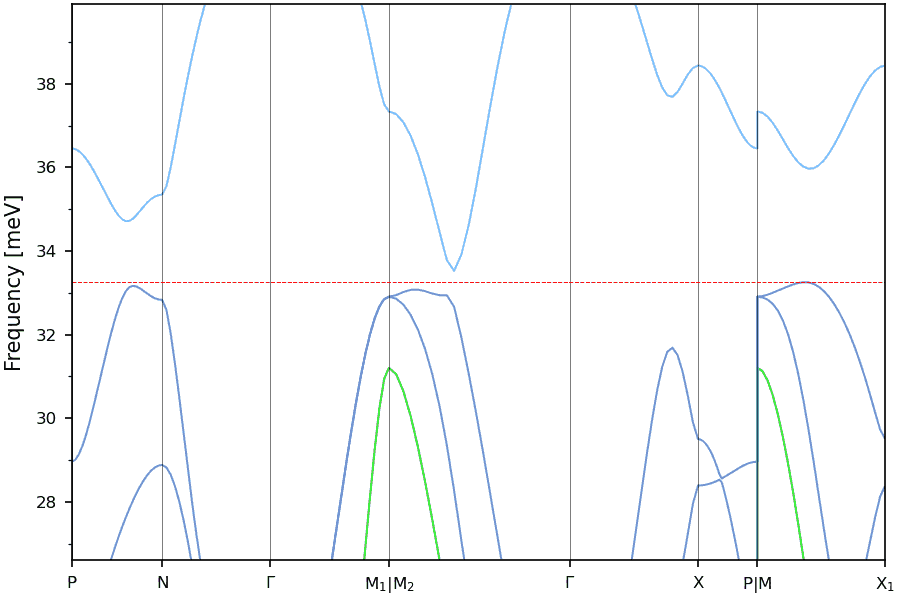} & \includegraphics[width=0.5\textwidth,angle=0]{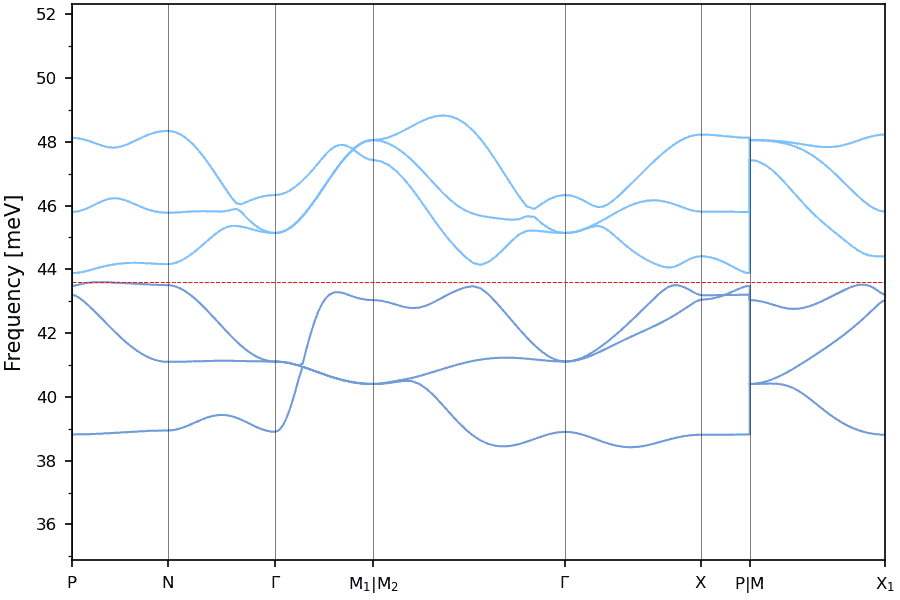} \\
\end{tabular}
\caption{Band structures of the ``ideal'' topological phonon bands with cumulative strong topology on the PhononDB@kyoto-u. (part 14/27)}
\label{fig:14_besttop_kyoto}
\end{figure}

\begin{figure}[h]
\centering
\begin{tabular}{cc}
\scriptsize{\mpidweb{7908}, SG 82 (\sgsymb{82}), \ce{Al$_2$ZnTe$_4$}} & \scriptsize{\mpidweb{8874}, SG 82 (\sgsymb{82}), \ce{LiSiBO$_4$}} \\
\includegraphics[width=0.5\textwidth,angle=0]{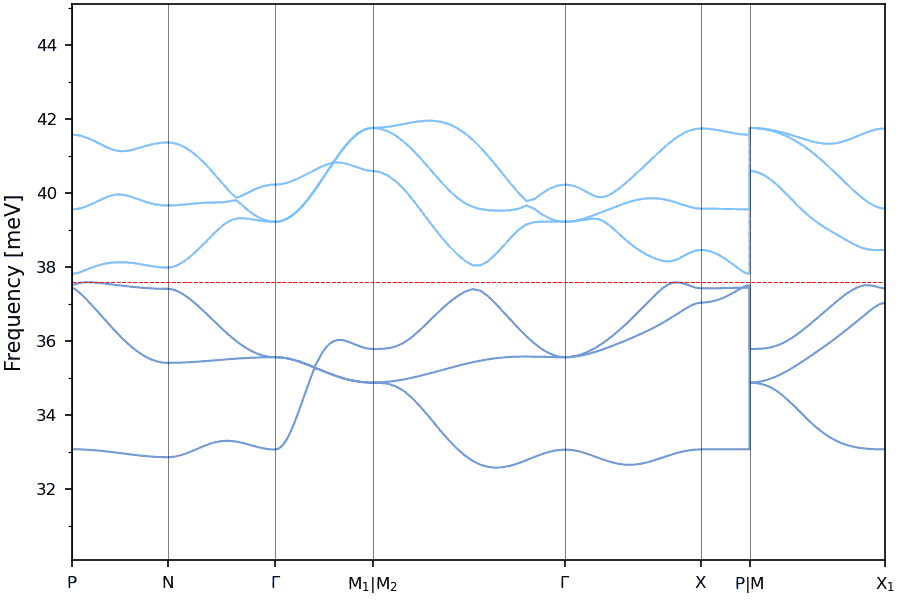} & \includegraphics[width=0.5\textwidth,angle=0]{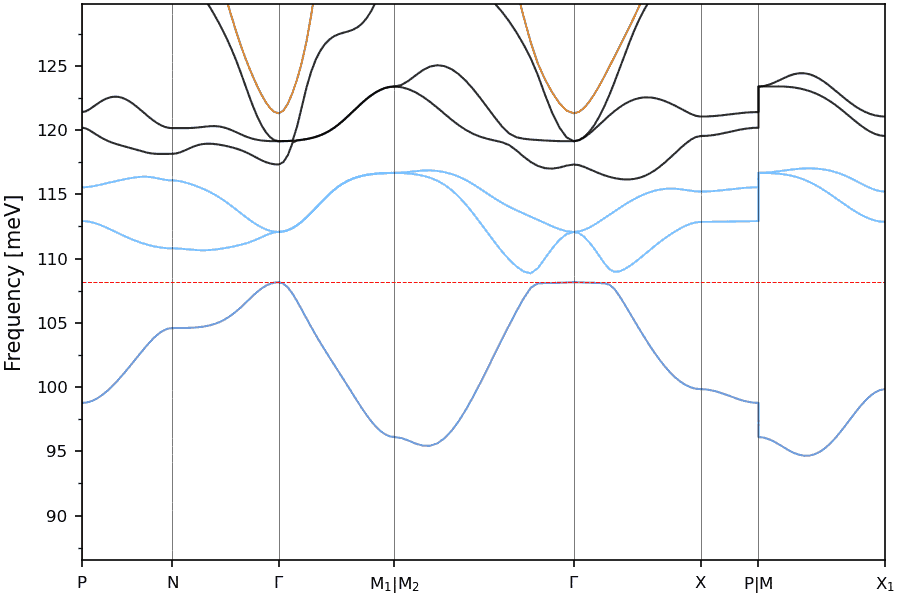} \\
\scriptsize{\mpidweb{20731}, SG 82 (\sgsymb{82}), \ce{In$_2$HgSe$_4$}} & \scriptsize{\mpidweb{28719}, SG 82 (\sgsymb{82}), \ce{Ba(AuF$_4$)$_2$}} \\
\includegraphics[width=0.5\textwidth,angle=0]{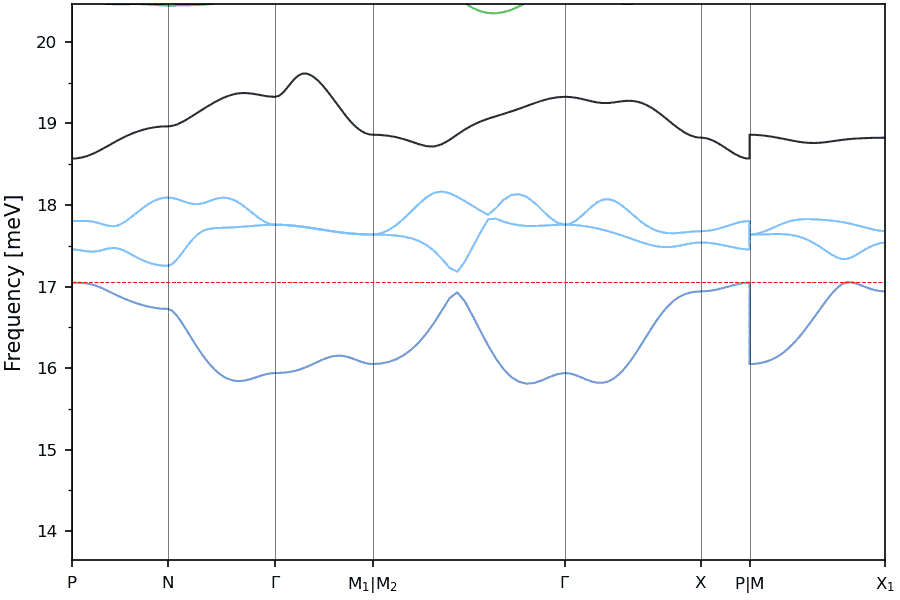} & \includegraphics[width=0.5\textwidth,angle=0]{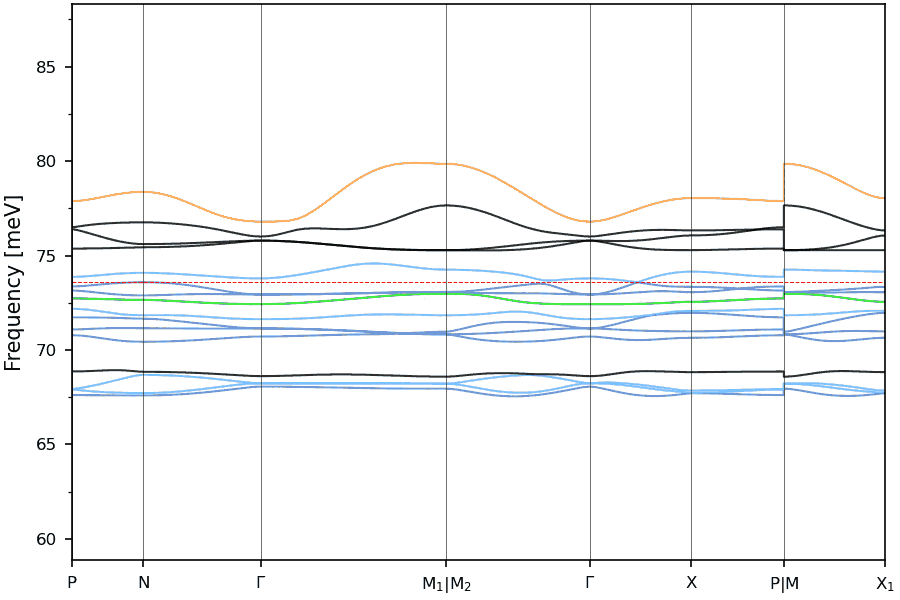} \\
\scriptsize{\mpidweb{37829}, SG 82 (\sgsymb{82}), \ce{La$_5$TlSe$_8$}} & \scriptsize{\mpidweb{530149}, SG 82 (\sgsymb{82}), \ce{Ca$_12$Al$_14$O$_33$}} \\
\includegraphics[width=0.5\textwidth,angle=0]{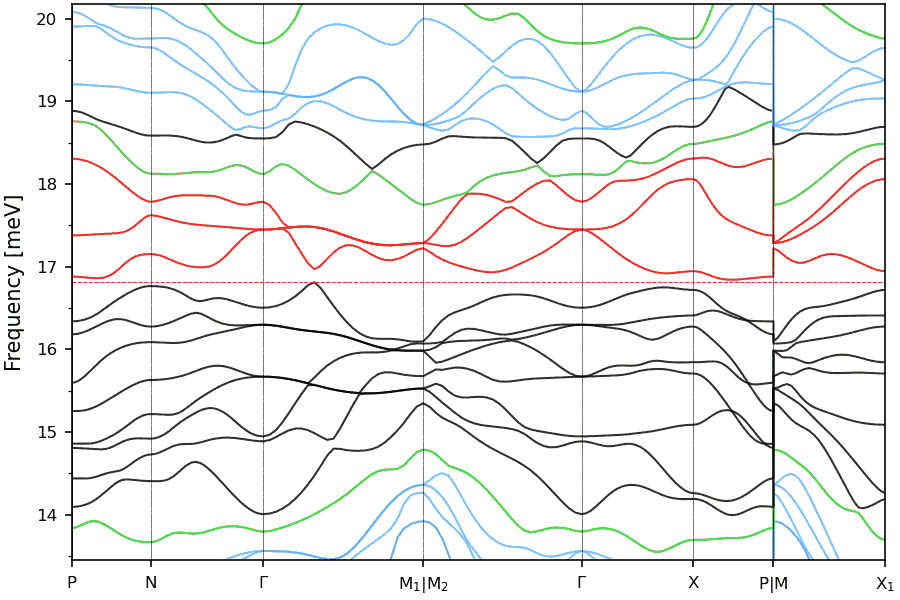} & \includegraphics[width=0.5\textwidth,angle=0]{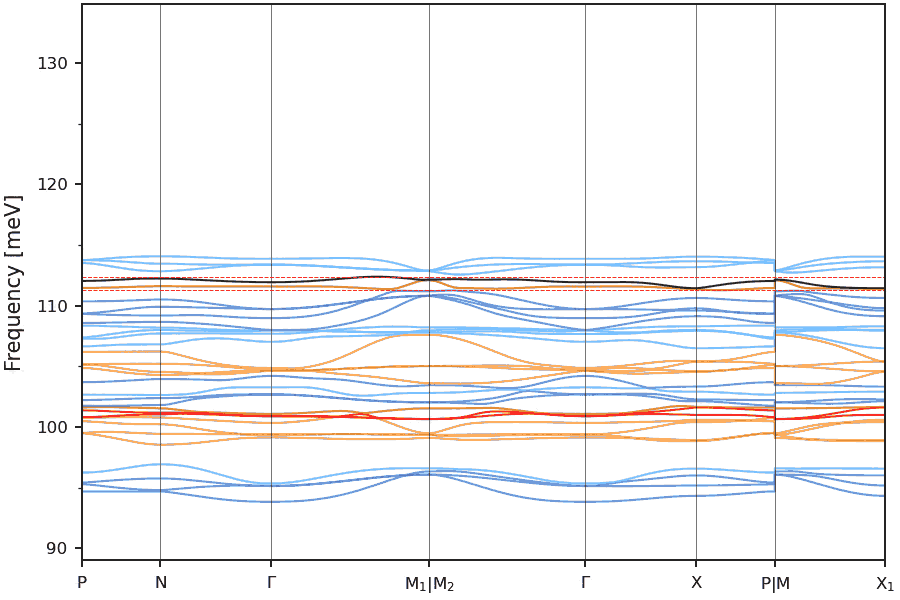} \\
\end{tabular}
\caption{Band structures of the ``ideal'' topological phonon bands with cumulative strong topology on the PhononDB@kyoto-u. (part 15/27)}
\label{fig:15_besttop_kyoto}
\end{figure}

\begin{figure}[h]
\centering
\begin{tabular}{cc}
\scriptsize{\mpidweb{5558}, SG 88 (\sgsymb{88}), \ce{NaReO$_4$}} & \scriptsize{\mpidweb{752451}, SG 88 (\sgsymb{88}), \ce{Zn(CuO$_2$)$_2$}} \\
\includegraphics[width=0.5\textwidth,angle=0]{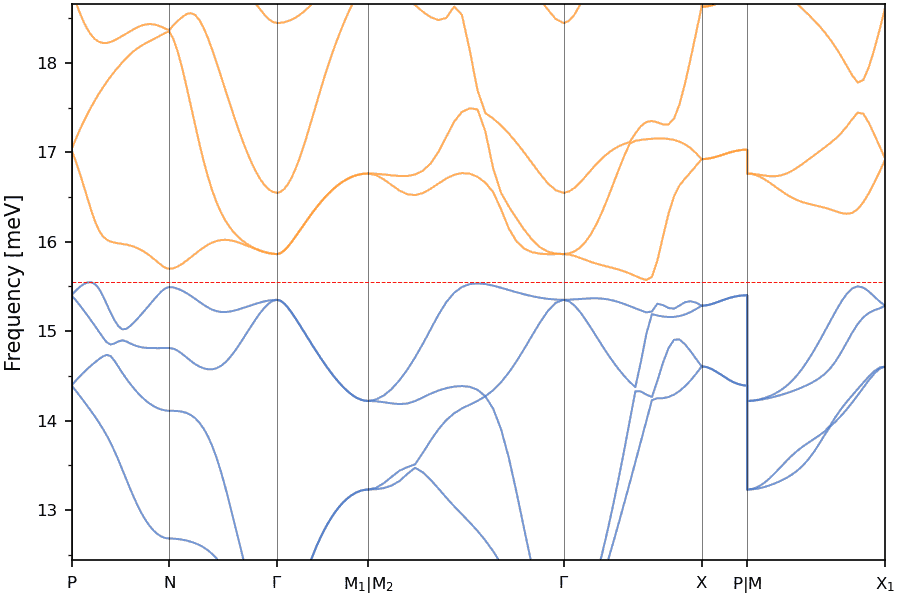} & \includegraphics[width=0.5\textwidth,angle=0]{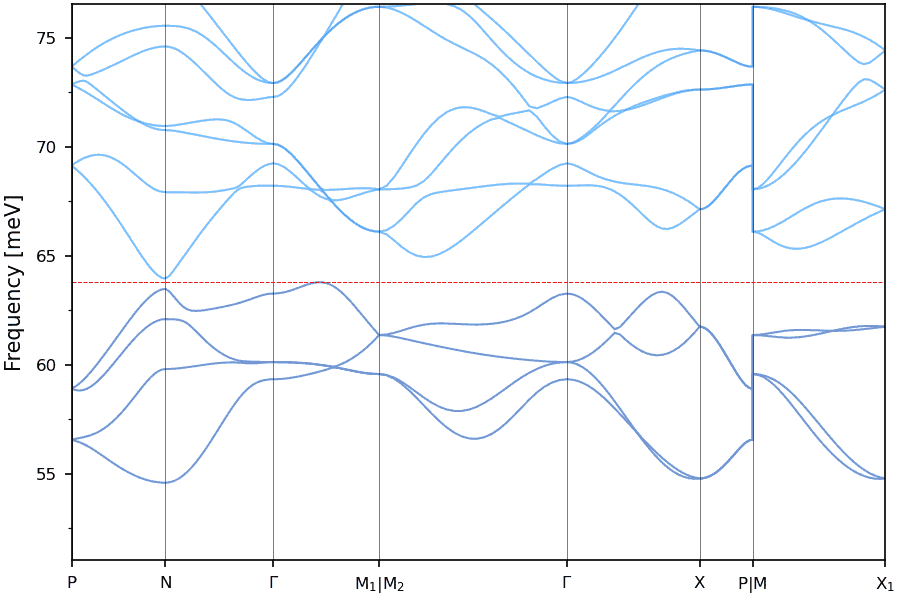} \\
\scriptsize{\mpidweb{559298}, SG 130 (\sgsymb{130}), \ce{La$_5$Sb$_4$(O$_4$F)$_3$}} & \scriptsize{\mpidweb{560785}, SG 130 (\sgsymb{130}), \ce{K$_2$Y$_4$Sn$_2$S$_11$}} \\
\includegraphics[width=0.5\textwidth,angle=0]{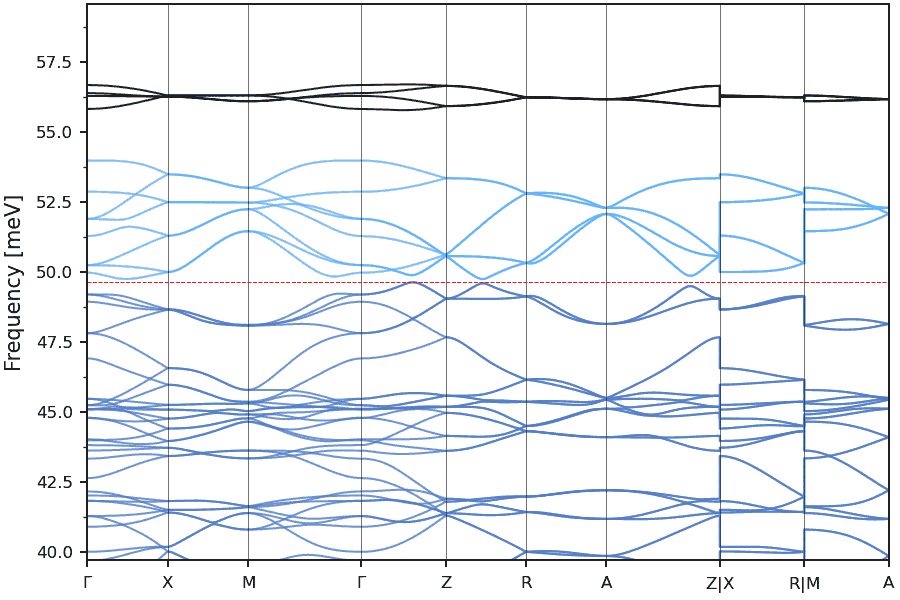} & \includegraphics[width=0.5\textwidth,angle=0]{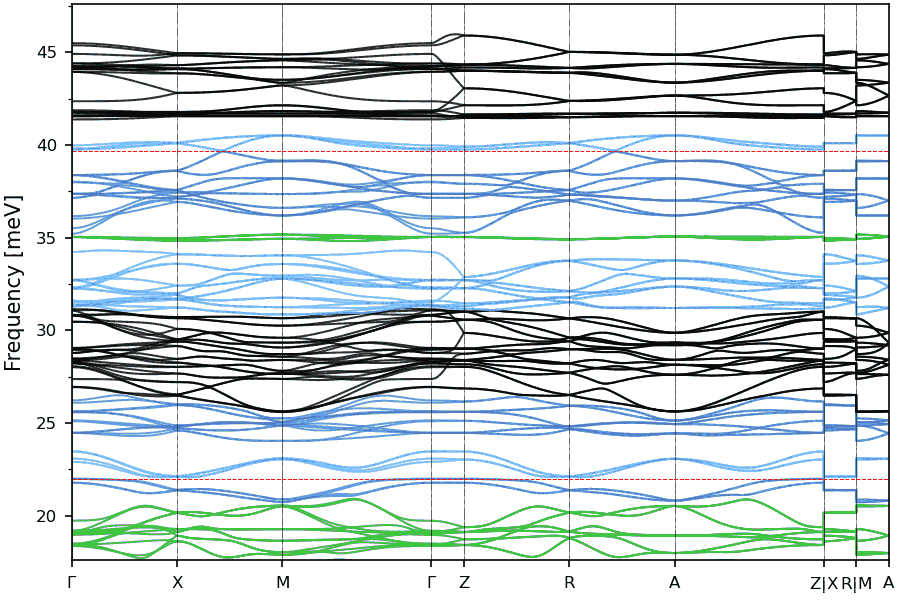} \\
\scriptsize{\mpidweb{27559}, SG 147 (\sgsymb{147}), \ce{Na$_14$Cd$_2$O$_9$}} & \scriptsize{\mpidweb{554822}, SG 147 (\sgsymb{147}), \ce{CsSb(PO$_4$)$_2$}} \\
\includegraphics[width=0.5\textwidth,angle=0]{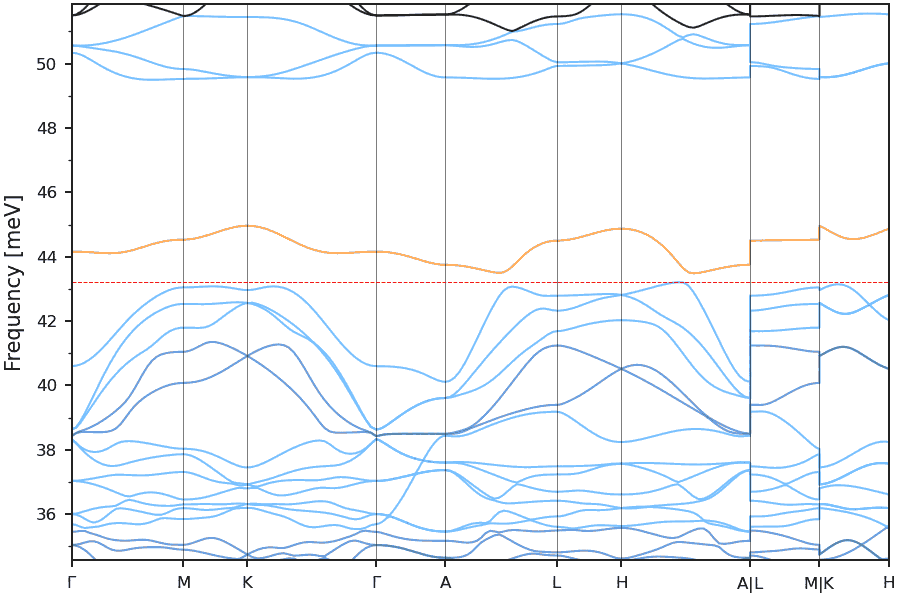} & \includegraphics[width=0.5\textwidth,angle=0]{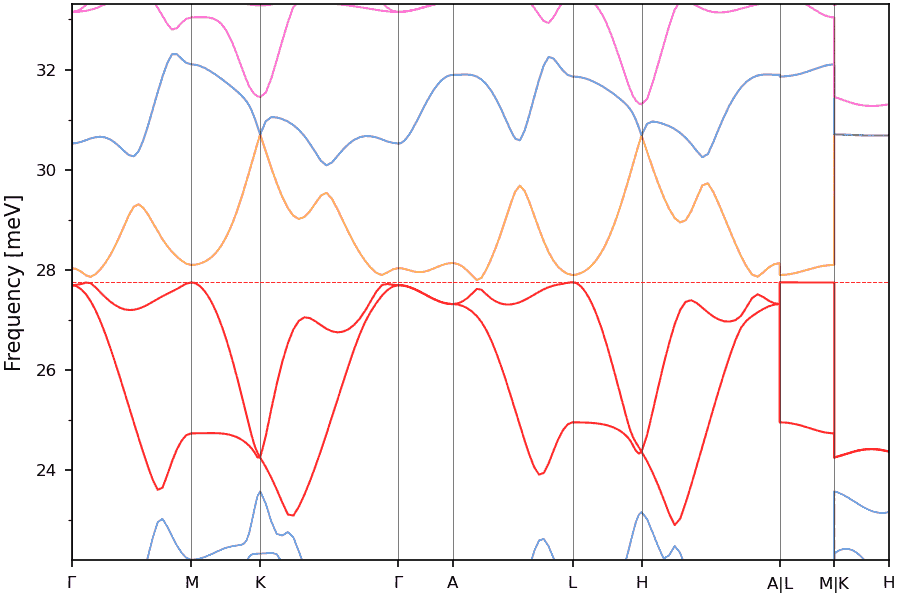} \\
\end{tabular}
\caption{Band structures of the ``ideal'' topological phonon bands with cumulative strong topology on the PhononDB@kyoto-u. (part 16/27)}
\label{fig:16_besttop_kyoto}
\end{figure}

\begin{figure}[h]
\centering
\begin{tabular}{cc}
\scriptsize{\mpidweb{560993}, SG 147 (\sgsymb{147}), \ce{CaAl$_12$Si$_4$O$_27$}} & \scriptsize{\mpidweb{849796}, SG 147 (\sgsymb{147}), \ce{NaZn$_4$H$_18$SClO$_16$}} \\
\includegraphics[width=0.5\textwidth,angle=0]{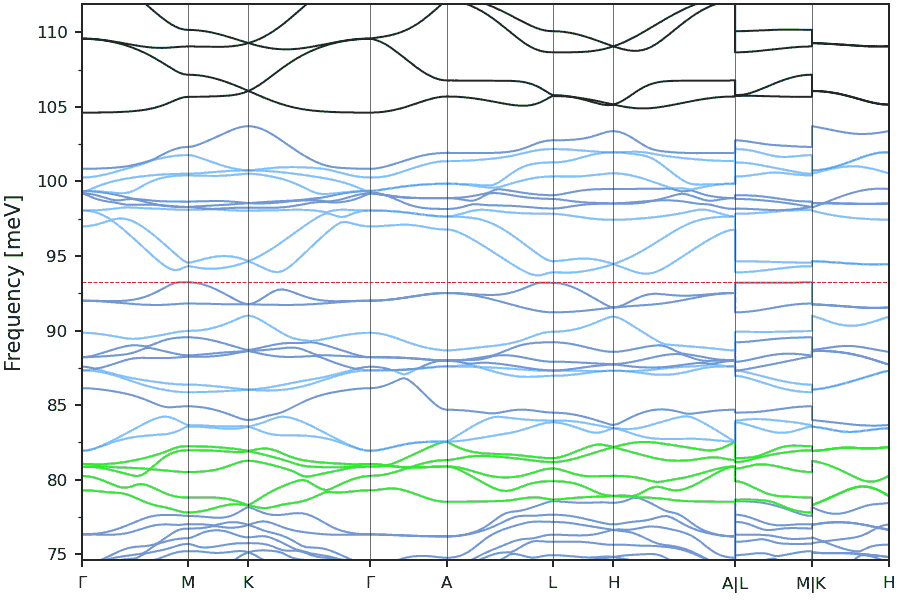} & \includegraphics[width=0.5\textwidth,angle=0]{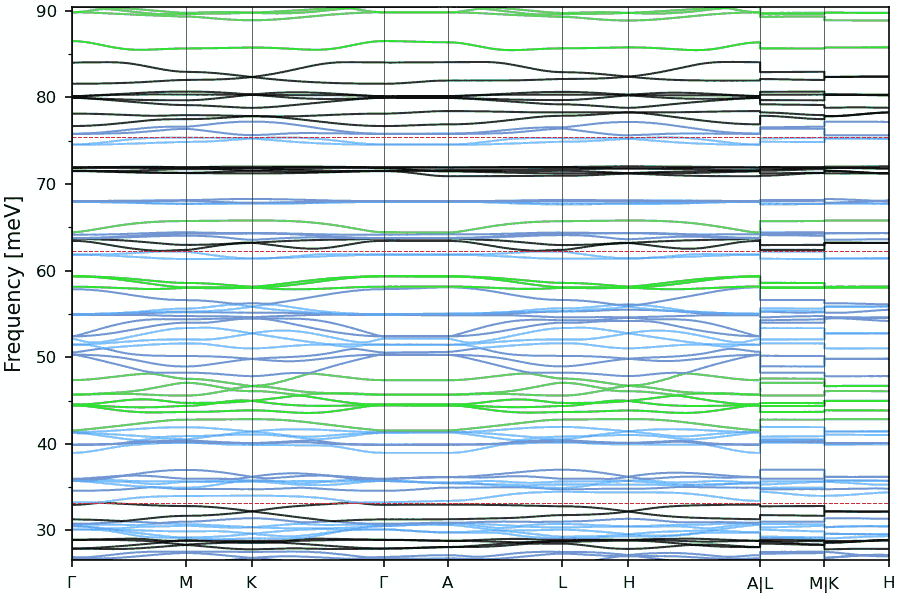} \\
\scriptsize{\mpidweb{2994}, SG 148 (\sgsymb{148}), \ce{Co$_2$(GeTe)$_3$}} & \scriptsize{\mpidweb{3273}, SG 148 (\sgsymb{148}), \ce{Si$_5$P$_6$O$_25$}} \\
\includegraphics[width=0.5\textwidth,angle=0]{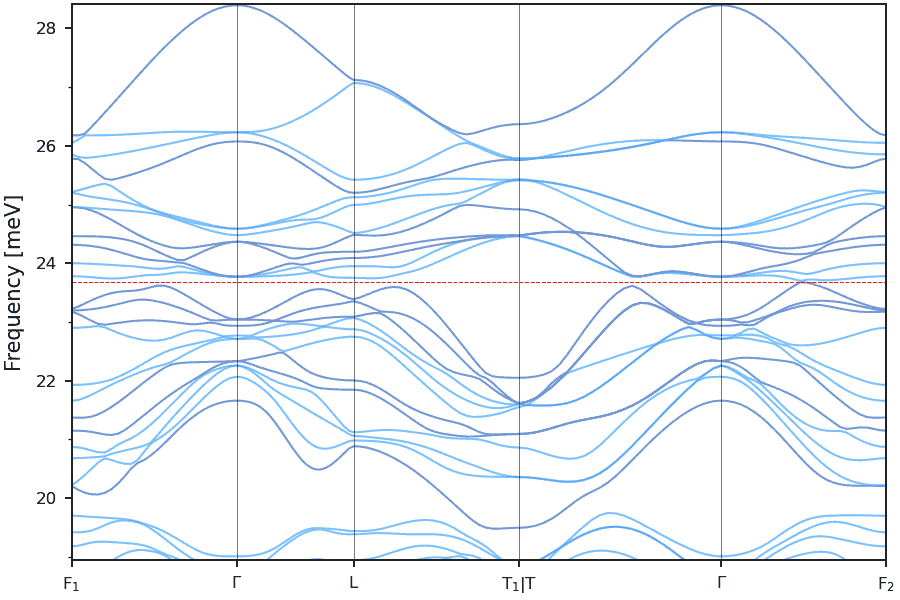} & \includegraphics[width=0.5\textwidth,angle=0]{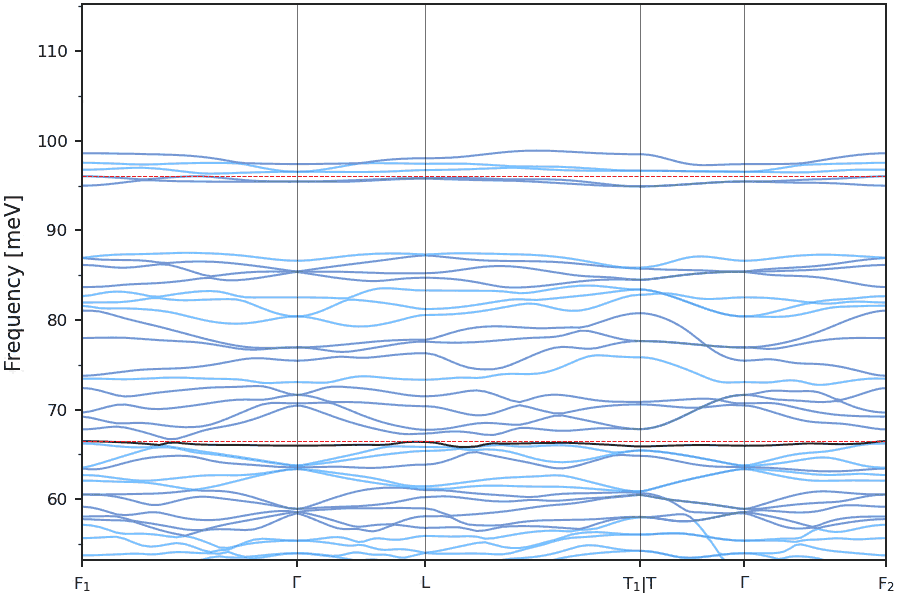} \\
\scriptsize{\mpidweb{3771}, SG 148 (\sgsymb{148}), \ce{MgTiO$_3$}} & \scriptsize{\mpidweb{4482}, SG 148 (\sgsymb{148}), \ce{NaSbO$_3$}} \\
\includegraphics[width=0.5\textwidth,angle=0]{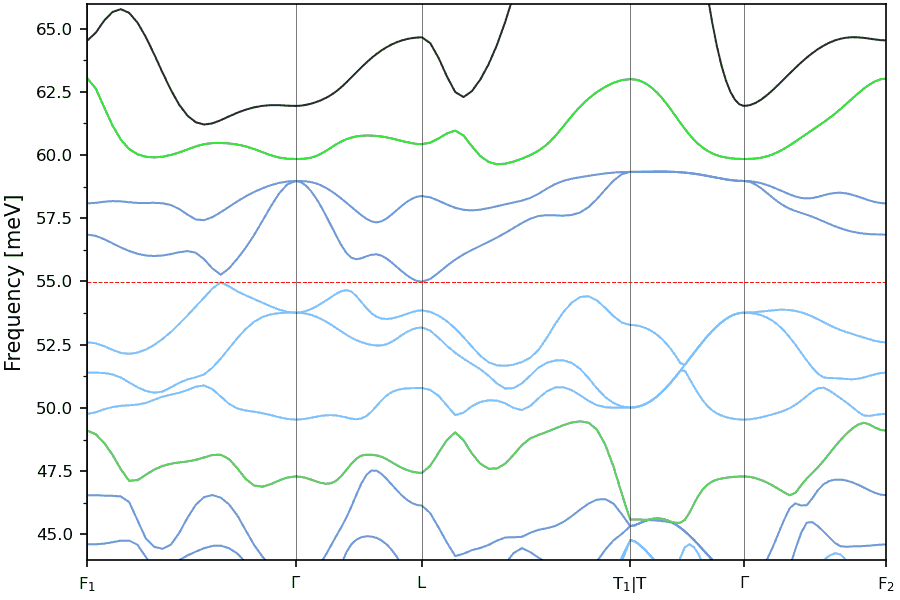} & \includegraphics[width=0.5\textwidth,angle=0]{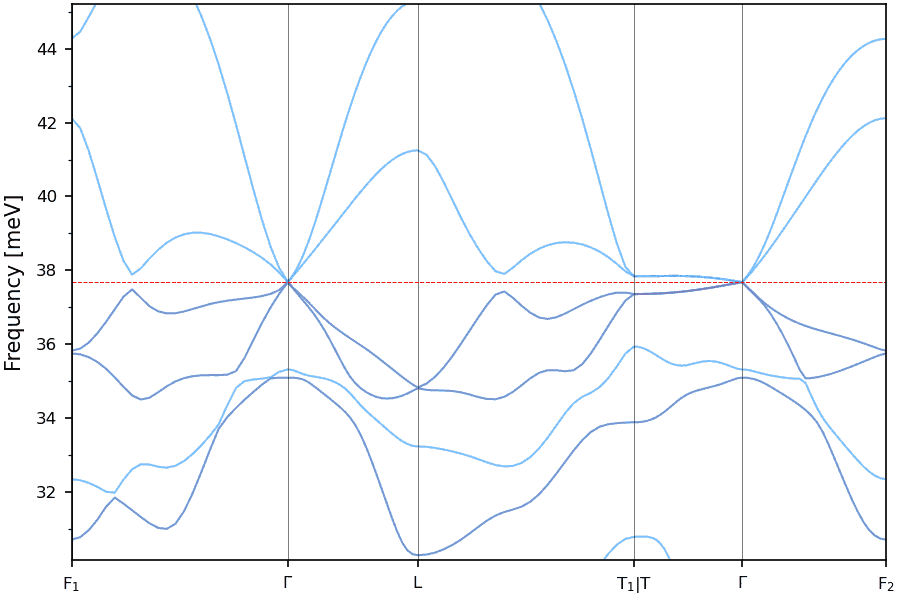} \\
\end{tabular}
\caption{Band structures of the ``ideal'' topological phonon bands with cumulative strong topology on the PhononDB@kyoto-u. (part 17/27)}
\label{fig:17_besttop_kyoto}
\end{figure}

\begin{figure}[h]
\centering
\begin{tabular}{cc}
\scriptsize{\mpidweb{5909}, SG 148 (\sgsymb{148}), \ce{Zn$_2$GeO$_4$}} & \scriptsize{\mpidweb{6031}, SG 148 (\sgsymb{148}), \ce{Li$_16$Nb$_2$N$_8$O}} \\
\includegraphics[width=0.5\textwidth,angle=0]{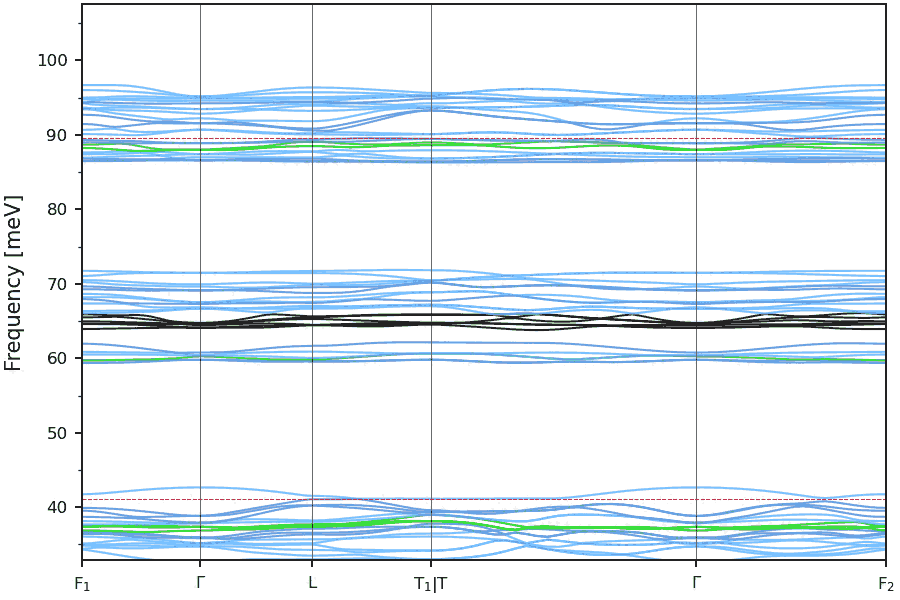} & \includegraphics[width=0.5\textwidth,angle=0]{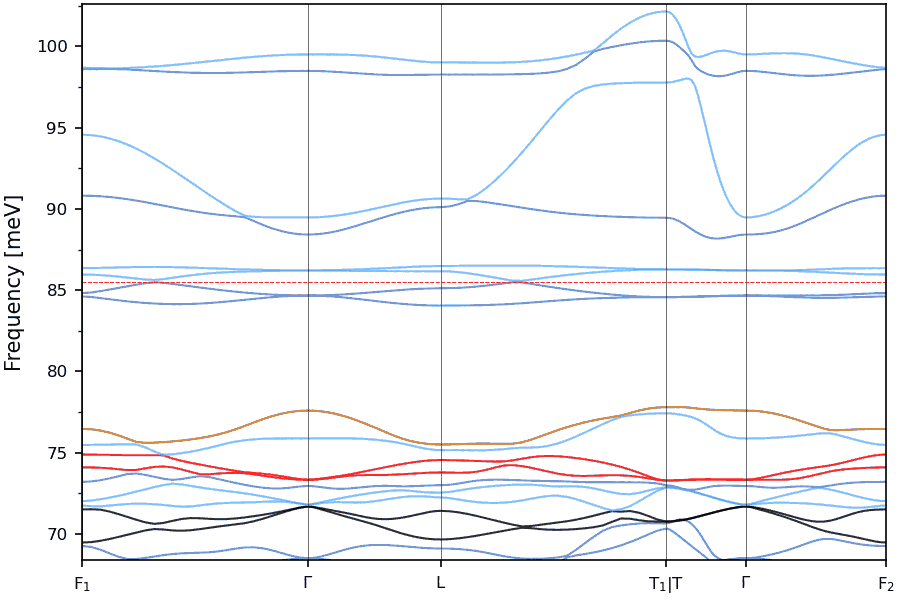} \\
\scriptsize{\mpidweb{6425}, SG 148 (\sgsymb{148}), \ce{Li$_3$In$_2$(PO$_4$)$_3$}} & \scriptsize{\mpidweb{8275}, SG 148 (\sgsymb{148}), \ce{CdGeO$_3$}} \\
\includegraphics[width=0.5\textwidth,angle=0]{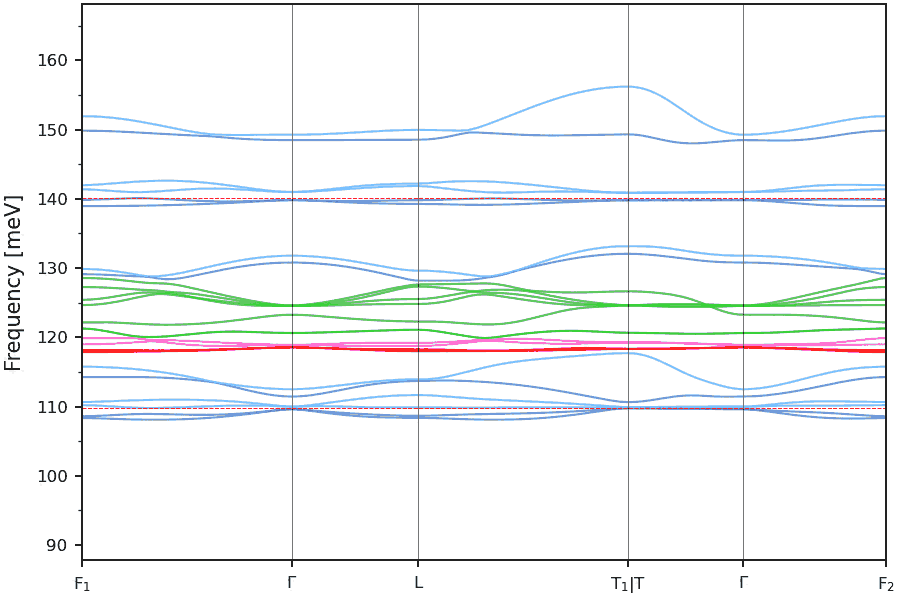} & \includegraphics[width=0.5\textwidth,angle=0]{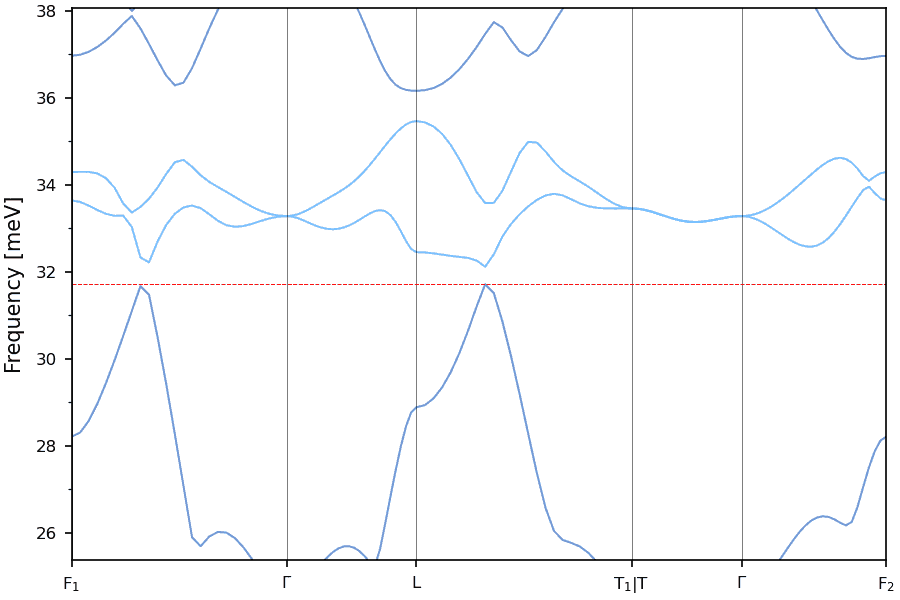} \\
\scriptsize{\mpidweb{8285}, SG 148 (\sgsymb{148}), \ce{ZnGeO$_3$}} & \scriptsize{\mpidweb{8702}, SG 148 (\sgsymb{148}), \ce{LiGaGeO$_4$}} \\
\includegraphics[width=0.5\textwidth,angle=0]{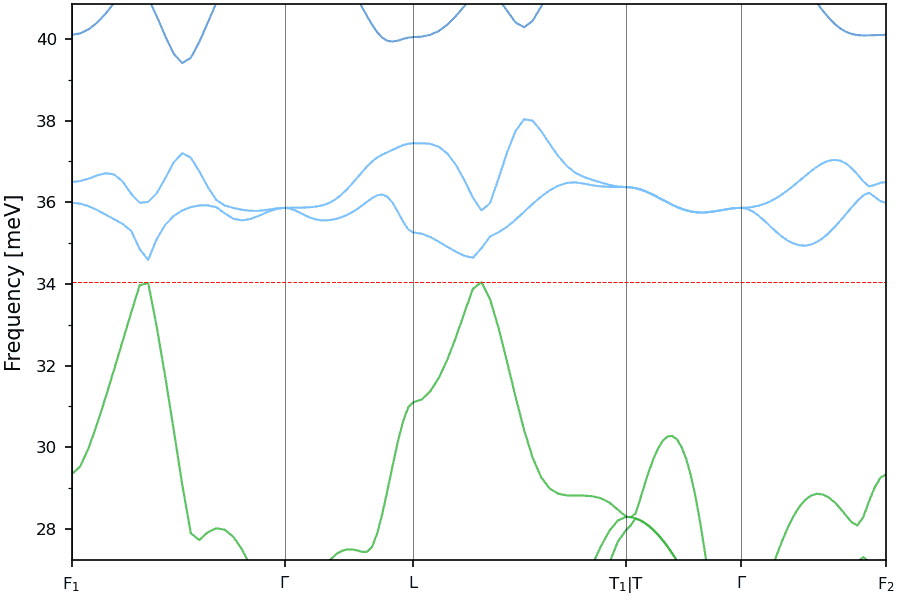} & \includegraphics[width=0.5\textwidth,angle=0]{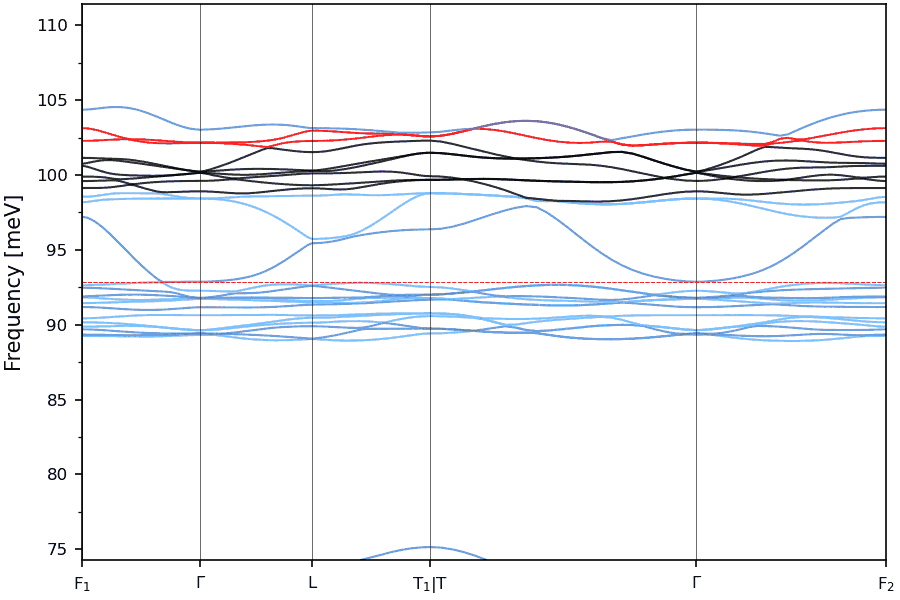} \\
\end{tabular}%
\caption{Band structures of the ``ideal'' topological phonon bands with cumulative strong topology on the PhononDB@kyoto-u. (part 27/27)}
\label{fig:27_besttop_kyoto}
\end{figure}

\clearpage
\subsubsection{Band structures of the best symmetry-enforced phonon band crossings}

\begin{figure}[h]
\centering
%
\caption{Band structures of the ``ideal'' symmetry-enforced phonon band nodes on the Material Project. (part 1/10)}
\label{fig:1_bestsm_mp}
\end{figure}

\begin{figure}[h]
\centering
%
\caption{Band structures of the ``ideal'' symmetry-enforced phonon band nodes on the Material Project. (part 2/10)}
\label{fig:2_bestsm_mp}
\end{figure}

\begin{figure}[h]
\centering
%
\caption{Band structures of the ``ideal'' symmetry-enforced phonon band nodes on the Material Project. (part 3/10)}
\label{fig:3_bestsm_mp}
\end{figure}

\begin{figure}[h]
\centering
%
\caption{Band structures of the ``ideal'' symmetry-enforced phonon band nodes on the Material Project. (part 10/10)}
\label{fig:10_bestsm_mp}
\end{figure}
\clearpage
\begin{figure}[h]
\centering
%
\caption{Band structures of the ``ideal'' symmetry-enforced band crossing in the PhononDB@kyoto-u. (part 1/132)}
\label{fig:1_bestsm_kyoto}
\end{figure}

\begin{figure}[h]
\centering
%
\caption{Band structures of the ``ideal'' symmetry-enforced band crossing in the PhononDB@kyoto-u. (part 2/132)}
\label{fig:2_bestsm_kyoto}
\end{figure}

\begin{figure}[h]
\centering
%
\caption{Band structures of the ``ideal'' symmetry-enforced band crossing in the PhononDB@kyoto-u. (part 3/132)}
\label{fig:3_bestsm_kyoto}
\end{figure}

\begin{figure}[h]
\centering
%
\caption{Band structures of the ``ideal'' symmetry-enforced band crossing in the PhononDB@kyoto-u. (part 132/132)}
\label{fig:132_bestsm_kyoto}
\end{figure}

\clearpage

\section{List of phonon materials}\label{app:phononmaterials}

This Appendix gives the complete lists of materials that we considered for the \webTQCphonon and thus this article, for each source of data namely  \webMP~in \siref{app:phononmaterialsmp} and \webphonondb(PhononDB@kyoto-u)
in \siref{app:phononmaterialskyoto}. Each entry in these lists contains the material structural chemical formula, SG, identifications (MPID and ICSD) and a summary of the non-atomic properties.

\subsection{Phonon materials from Materials Project}\label{app:phononmaterialsmp}

\LTcapwidth=1.0\textwidth
\renewcommand\arraystretch{1.0}


\subsection{Phonon materials from the Kyoto phonon database}\label{app:phononmaterialskyoto}

\LTcapwidth=1.0\textwidth
\renewcommand\arraystretch{1.0}
%

\clearpage
\bibliography{./main}

\begin{thebibliography}{62}%
\makeatletter
\providecommand \@ifxundefined [1]{%
 \@ifx{#1\undefined}
}%
\providecommand \@ifnum [1]{%
 \ifnum #1\expandafter \@firstoftwo
 \else \expandafter \@secondoftwo
 \fi
}%
\providecommand \@ifx [1]{%
 \ifx #1\expandafter \@firstoftwo
 \else \expandafter \@secondoftwo
 \fi
}%
\providecommand \natexlab [1]{#1}%
\providecommand \enquote  [1]{``#1''}%
\providecommand \bibnamefont  [1]{#1}%
\providecommand \bibfnamefont [1]{#1}%
\providecommand \citenamefont [1]{#1}%
\providecommand \href@noop [0]{\@secondoftwo}%
\providecommand \href [0]{\begingroup \@sanitize@url \@href}%
\providecommand \@href[1]{\@@startlink{#1}\@@href}%
\providecommand \@@href[1]{\endgroup#1\@@endlink}%
\providecommand \@sanitize@url [0]{\catcode `\\12\catcode `\$12\catcode
  `\&12\catcode `\#12\catcode `\^12\catcode `\_12\catcode `\%12\relax}%
\providecommand \@@startlink[1]{}%
\providecommand \@@endlink[0]{}%
\providecommand \url  [0]{\begingroup\@sanitize@url \@url }%
\providecommand \@url [1]{\endgroup\@href {#1}{\urlprefix }}%
\providecommand \urlprefix  [0]{URL }%
\providecommand \Eprint [0]{\href }%
\providecommand \doibase [0]{http://dx.doi.org/}%
\providecommand \selectlanguage [0]{\@gobble}%
\providecommand \bibinfo  [0]{\@secondoftwo}%
\providecommand \bibfield  [0]{\@secondoftwo}%
\providecommand \translation [1]{[#1]}%
\providecommand \BibitemOpen [0]{}%
\providecommand \bibitemStop [0]{}%
\providecommand \bibitemNoStop [0]{.\EOS\space}%
\providecommand \EOS [0]{\spacefactor3000\relax}%
\providecommand \BibitemShut  [1]{\csname bibitem#1\endcsname}%
\let\auto@bib@innerbib\@empty
\bibitem [{\citenamefont {Fu}\ \emph {et~al.}(2007)\citenamefont {Fu},
  \citenamefont {Kane},\ and\ \citenamefont {Mele}}]{fu2007}%
  \BibitemOpen
  \bibfield  {author} {\bibinfo {author} {\bibfnamefont {Liang}\ \bibnamefont
  {Fu}}, \bibinfo {author} {\bibfnamefont {C.~L.}\ \bibnamefont {Kane}}, \ and\
  \bibinfo {author} {\bibfnamefont {E.~J.}\ \bibnamefont {Mele}},\ }\bibfield
  {title} {\enquote {\bibinfo {title} {Topological insulators in three
  dimensions},}\ }\href {\doibase 10.1103/PhysRevLett.98.106803} {\bibfield
  {journal} {\bibinfo  {journal} {Phys. Rev. Lett.}\ }\textbf {\bibinfo
  {volume} {98}},\ \bibinfo {pages} {106803} (\bibinfo {year}
  {2007})}\BibitemShut {NoStop}%
\bibitem [{\citenamefont {Bradlyn}\ \emph {et~al.}(2017)\citenamefont
  {Bradlyn}, \citenamefont {Elcoro}, \citenamefont {Cano}, \citenamefont
  {Vergniory}, \citenamefont {Wang}, \citenamefont {Felser}, \citenamefont
  {Aroyo},\ and\ \citenamefont {Bernevig}}]{bradlyn_topological_2017}%
  \BibitemOpen
  \bibfield  {author} {\bibinfo {author} {\bibfnamefont {Barry}\ \bibnamefont
  {Bradlyn}}, \bibinfo {author} {\bibfnamefont {L.}~\bibnamefont {Elcoro}},
  \bibinfo {author} {\bibfnamefont {Jennifer}\ \bibnamefont {Cano}}, \bibinfo
  {author} {\bibfnamefont {M.~G.}\ \bibnamefont {Vergniory}}, \bibinfo {author}
  {\bibfnamefont {Zhijun}\ \bibnamefont {Wang}}, \bibinfo {author}
  {\bibfnamefont {C.}~\bibnamefont {Felser}}, \bibinfo {author} {\bibfnamefont
  {M.~I.}\ \bibnamefont {Aroyo}}, \ and\ \bibinfo {author} {\bibfnamefont
  {B.~Andrei}\ \bibnamefont {Bernevig}},\ }\bibfield  {title} {\enquote
  {\bibinfo {title} {Topological quantum chemistry},}\ }\href
  {http://www.nature.com/nature/journal/v547/n7663/full/nature23268.html}
  {\bibfield  {journal} {\bibinfo  {journal} {Nature}\ }\textbf {\bibinfo
  {volume} {547}},\ \bibinfo {pages} {298–305} (\bibinfo {year}
  {2017})}\BibitemShut {NoStop}%
\bibitem [{\citenamefont {Elcoro}\ \emph {et~al.}(2020)\citenamefont {Elcoro},
  \citenamefont {Wieder}, \citenamefont {Song}, \citenamefont {Xu},
  \citenamefont {Bradlyn},\ and\ \citenamefont {Bernevig}}]{MTQC}%
  \BibitemOpen
  \bibfield  {author} {\bibinfo {author} {\bibfnamefont {Luis}\ \bibnamefont
  {Elcoro}}, \bibinfo {author} {\bibfnamefont {Benjamin~J}\ \bibnamefont
  {Wieder}}, \bibinfo {author} {\bibfnamefont {Zhida}\ \bibnamefont {Song}},
  \bibinfo {author} {\bibfnamefont {Yuanfeng}\ \bibnamefont {Xu}}, \bibinfo
  {author} {\bibfnamefont {Barry}\ \bibnamefont {Bradlyn}}, \ and\ \bibinfo
  {author} {\bibfnamefont {B~Andrei}\ \bibnamefont {Bernevig}},\ }\bibfield
  {title} {\enquote {\bibinfo {title} {Magnetic topological quantum
  chemistry},}\ }\href@noop {} {\bibfield  {journal} {\bibinfo  {journal}
  {arXiv preprint arXiv:2010.00598}\ } (\bibinfo {year} {2020})}\BibitemShut
  {NoStop}%
\bibitem [{\citenamefont {Po}\ \emph {et~al.}(2017{\natexlab{a}})\citenamefont
  {Po}, \citenamefont {Vishwanath},\ and\ \citenamefont
  {Watanabe}}]{po_symmetry-based_2017}%
  \BibitemOpen
  \bibfield  {author} {\bibinfo {author} {\bibfnamefont {Hoi~Chun}\
  \bibnamefont {Po}}, \bibinfo {author} {\bibfnamefont {Ashvin}\ \bibnamefont
  {Vishwanath}}, \ and\ \bibinfo {author} {\bibfnamefont {Haruki}\ \bibnamefont
  {Watanabe}},\ }\bibfield  {title} {\enquote {\bibinfo {title} {Symmetry-based
  indicators of band topology in the 230 space groups},}\ }\href {\doibase
  10.1038/s41467-017-00133-2} {\bibfield  {journal} {\bibinfo  {journal}
  {Nature Communications}\ }\textbf {\bibinfo {volume} {8}},\ \bibinfo {pages}
  {50} (\bibinfo {year} {2017}{\natexlab{a}})}\BibitemShut {NoStop}%
\bibitem [{\citenamefont {Watanabe}\ \emph {et~al.}(2018)\citenamefont
  {Watanabe}, \citenamefont {Po},\ and\ \citenamefont
  {Vishwanath}}]{watanabe2018structure}%
  \BibitemOpen
  \bibfield  {author} {\bibinfo {author} {\bibfnamefont {Haruki}\ \bibnamefont
  {Watanabe}}, \bibinfo {author} {\bibfnamefont {Hoi~Chun}\ \bibnamefont {Po}},
  \ and\ \bibinfo {author} {\bibfnamefont {Ashvin}\ \bibnamefont
  {Vishwanath}},\ }\bibfield  {title} {\enquote {\bibinfo {title} {Structure
  and topology of band structures in the 1651 magnetic space groups},}\
  }\href@noop {} {\bibfield  {journal} {\bibinfo  {journal} {Science advances}\
  }\textbf {\bibinfo {volume} {4}},\ \bibinfo {pages} {eaat8685} (\bibinfo
  {year} {2018})}\BibitemShut {NoStop}%
\bibitem [{\citenamefont {Kruthoff}\ \emph {et~al.}(2017)\citenamefont
  {Kruthoff}, \citenamefont {de~Boer}, \citenamefont {van Wezel}, \citenamefont
  {Kane},\ and\ \citenamefont {Slager}}]{SlagerSymmetry}%
  \BibitemOpen
  \bibfield  {author} {\bibinfo {author} {\bibfnamefont {Jorrit}\ \bibnamefont
  {Kruthoff}}, \bibinfo {author} {\bibfnamefont {Jan}\ \bibnamefont {de~Boer}},
  \bibinfo {author} {\bibfnamefont {Jasper}\ \bibnamefont {van Wezel}},
  \bibinfo {author} {\bibfnamefont {Charles~L.}\ \bibnamefont {Kane}}, \ and\
  \bibinfo {author} {\bibfnamefont {Robert-Jan}\ \bibnamefont {Slager}},\
  }\bibfield  {title} {\enquote {\bibinfo {title} {Topological classification
  of crystalline insulators through band structure combinatorics},}\ }\href
  {\doibase 10.1103/PhysRevX.7.041069} {\bibfield  {journal} {\bibinfo
  {journal} {Phys. Rev. X}\ }\textbf {\bibinfo {volume} {7}},\ \bibinfo {pages}
  {041069} (\bibinfo {year} {2017})}\BibitemShut {NoStop}%
\bibitem [{\citenamefont {Song}\ \emph {et~al.}(2017)\citenamefont {Song},
  \citenamefont {Fang},\ and\ \citenamefont
  {Fang}}]{song_d-2-dimensional_2017}%
  \BibitemOpen
  \bibfield  {author} {\bibinfo {author} {\bibfnamefont {Zhida}\ \bibnamefont
  {Song}}, \bibinfo {author} {\bibfnamefont {Zhong}\ \bibnamefont {Fang}}, \
  and\ \bibinfo {author} {\bibfnamefont {Chen}\ \bibnamefont {Fang}},\
  }\bibfield  {title} {\enquote {\bibinfo {title} {(d-2)-{Dimensional} {Edge}
  {States} of {Rotation} {Symmetry} {Protected} {Topological} {States}},}\
  }\href {\doibase 10.1103/PhysRevLett.119.246402} {\bibfield  {journal}
  {\bibinfo  {journal} {Physical Review Letters}\ }\textbf {\bibinfo {volume}
  {119}},\ \bibinfo {pages} {246402} (\bibinfo {year} {2017})}\BibitemShut
  {NoStop}%
\bibitem [{\citenamefont {Vergniory}\ \emph {et~al.}(2019)\citenamefont
  {Vergniory}, \citenamefont {Elcoro}, \citenamefont {Felser}, \citenamefont
  {Regnault}, \citenamefont {Bernevig},\ and\ \citenamefont
  {Wang}}]{vergniory_complete_2019}%
  \BibitemOpen
  \bibfield  {author} {\bibinfo {author} {\bibfnamefont {M.~G.}\ \bibnamefont
  {Vergniory}}, \bibinfo {author} {\bibfnamefont {L.}~\bibnamefont {Elcoro}},
  \bibinfo {author} {\bibfnamefont {Claudia}\ \bibnamefont {Felser}}, \bibinfo
  {author} {\bibfnamefont {Nicolas}\ \bibnamefont {Regnault}}, \bibinfo
  {author} {\bibfnamefont {B.~Andrei}\ \bibnamefont {Bernevig}}, \ and\
  \bibinfo {author} {\bibfnamefont {Zhijun}\ \bibnamefont {Wang}},\ }\bibfield
  {title} {\enquote {\bibinfo {title} {A complete catalogue of high-quality
  topological materials},}\ }\href {\doibase 10.1038/s41586-019-0954-4}
  {\bibfield  {journal} {\bibinfo  {journal} {Nature}\ }\textbf {\bibinfo
  {volume} {566}},\ \bibinfo {pages} {480--485} (\bibinfo {year}
  {2019})}\BibitemShut {NoStop}%
\bibitem [{\citenamefont {Zhang}\ \emph
  {et~al.}(2019{\natexlab{a}})\citenamefont {Zhang}, \citenamefont {Jiang},
  \citenamefont {Song}, \citenamefont {Huang}, \citenamefont {He},
  \citenamefont {Fang}, \citenamefont {Weng},\ and\ \citenamefont
  {Fang}}]{zhang2019catalogue}%
  \BibitemOpen
  \bibfield  {author} {\bibinfo {author} {\bibfnamefont {Tiantian}\
  \bibnamefont {Zhang}}, \bibinfo {author} {\bibfnamefont {Yi}~\bibnamefont
  {Jiang}}, \bibinfo {author} {\bibfnamefont {Zhida}\ \bibnamefont {Song}},
  \bibinfo {author} {\bibfnamefont {He}~\bibnamefont {Huang}}, \bibinfo
  {author} {\bibfnamefont {Yuqing}\ \bibnamefont {He}}, \bibinfo {author}
  {\bibfnamefont {Zhong}\ \bibnamefont {Fang}}, \bibinfo {author}
  {\bibfnamefont {Hongming}\ \bibnamefont {Weng}}, \ and\ \bibinfo {author}
  {\bibfnamefont {Chen}\ \bibnamefont {Fang}},\ }\bibfield  {title} {\enquote
  {\bibinfo {title} {Catalogue of topological electronic materials},}\
  }\href@noop {} {\bibfield  {journal} {\bibinfo  {journal} {Nature}\ }\textbf
  {\bibinfo {volume} {566}},\ \bibinfo {pages} {475–479} (\bibinfo {year}
  {2019}{\natexlab{a}})}\BibitemShut {NoStop}%
\bibitem [{\citenamefont {Tang}\ \emph {et~al.}(2019)\citenamefont {Tang},
  \citenamefont {Po}, \citenamefont {Vishwanath},\ and\ \citenamefont
  {Wan}}]{tang2019comprehensive}%
  \BibitemOpen
  \bibfield  {author} {\bibinfo {author} {\bibfnamefont {Feng}\ \bibnamefont
  {Tang}}, \bibinfo {author} {\bibfnamefont {Hoi~Chun}\ \bibnamefont {Po}},
  \bibinfo {author} {\bibfnamefont {Ashvin}\ \bibnamefont {Vishwanath}}, \ and\
  \bibinfo {author} {\bibfnamefont {Xiangang}\ \bibnamefont {Wan}},\ }\bibfield
   {title} {\enquote {\bibinfo {title} {Comprehensive search for topological
  materials using symmetry indicators},}\ }\href@noop {} {\bibfield  {journal}
  {\bibinfo  {journal} {Nature}\ }\textbf {\bibinfo {volume} {566}},\ \bibinfo
  {pages} {486–489} (\bibinfo {year} {2019})}\BibitemShut {NoStop}%
\bibitem [{\citenamefont {Xu}\ \emph {et~al.}(2020)\citenamefont {Xu},
  \citenamefont {Elcoro}, \citenamefont {Song}, \citenamefont {Wieder},
  \citenamefont {Vergniory}, \citenamefont {Regnault}, \citenamefont {Chen},
  \citenamefont {Felser},\ and\ \citenamefont {Bernevig}}]{xu2020high}%
  \BibitemOpen
  \bibfield  {author} {\bibinfo {author} {\bibfnamefont {Yuanfeng}\
  \bibnamefont {Xu}}, \bibinfo {author} {\bibfnamefont {Luis}\ \bibnamefont
  {Elcoro}}, \bibinfo {author} {\bibfnamefont {Zhi-Da}\ \bibnamefont {Song}},
  \bibinfo {author} {\bibfnamefont {Benjamin~J}\ \bibnamefont {Wieder}},
  \bibinfo {author} {\bibfnamefont {MG}~\bibnamefont {Vergniory}}, \bibinfo
  {author} {\bibfnamefont {Nicolas}\ \bibnamefont {Regnault}}, \bibinfo
  {author} {\bibfnamefont {Yulin}\ \bibnamefont {Chen}}, \bibinfo {author}
  {\bibfnamefont {Claudia}\ \bibnamefont {Felser}}, \ and\ \bibinfo {author}
  {\bibfnamefont {B~Andrei}\ \bibnamefont {Bernevig}},\ }\bibfield  {title}
  {\enquote {\bibinfo {title} {High-throughput calculations of magnetic
  topological materials},}\ }\href@noop {} {\bibfield  {journal} {\bibinfo
  {journal} {Nature}\ }\textbf {\bibinfo {volume} {586}},\ \bibinfo {pages}
  {702--707} (\bibinfo {year} {2020})}\BibitemShut {NoStop}%
\bibitem [{\citenamefont {Vergniory}\ \emph {et~al.}(2022)\citenamefont
  {Vergniory}, \citenamefont {Wieder}, \citenamefont {Elcoro}, \citenamefont
  {Parkin}, \citenamefont {Felser}, \citenamefont {Bernevig},\ and\
  \citenamefont {Regnault}}]{Vergniory2021}%
  \BibitemOpen
  \bibfield  {author} {\bibinfo {author} {\bibfnamefont {Maia~G.}\ \bibnamefont
  {Vergniory}}, \bibinfo {author} {\bibfnamefont {Benjamin~J.}\ \bibnamefont
  {Wieder}}, \bibinfo {author} {\bibfnamefont {Luis}\ \bibnamefont {Elcoro}},
  \bibinfo {author} {\bibfnamefont {Stuart S.~P.}\ \bibnamefont {Parkin}},
  \bibinfo {author} {\bibfnamefont {Claudia}\ \bibnamefont {Felser}}, \bibinfo
  {author} {\bibfnamefont {B.~Andrei}\ \bibnamefont {Bernevig}}, \ and\
  \bibinfo {author} {\bibfnamefont {Nicolas}\ \bibnamefont {Regnault}},\
  }\bibfield  {title} {\enquote {\bibinfo {title} {All topological bands of all
  nonmagnetic stoichiometric materials},}\ }\href {\doibase
  10.1126/science.abg9094} {\bibfield  {journal} {\bibinfo  {journal}
  {Science}\ }\textbf {\bibinfo {volume} {376}},\ \bibinfo {pages} {eabg9094}
  (\bibinfo {year} {2022})}\BibitemShut {NoStop}%
\bibitem [{\citenamefont {Bergerhoff}\ \emph {et~al.}(1983)\citenamefont
  {Bergerhoff}, \citenamefont {Hundt}, \citenamefont {Sievers},\ and\
  \citenamefont {Brown}}]{ICSD}%
  \BibitemOpen
  \bibfield  {author} {\bibinfo {author} {\bibfnamefont {G.}~\bibnamefont
  {Bergerhoff}}, \bibinfo {author} {\bibfnamefont {R.}~\bibnamefont {Hundt}},
  \bibinfo {author} {\bibfnamefont {R.}~\bibnamefont {Sievers}}, \ and\
  \bibinfo {author} {\bibfnamefont {I.~D.}\ \bibnamefont {Brown}},\ }\bibfield
  {title} {\enquote {\bibinfo {title} {The inorganic crystal structure data
  base},}\ }\href {\doibase 10.1021/ci00038a003} {\bibfield  {journal}
  {\bibinfo  {journal} {Journal of Chemical Information and Computer Sciences}\
  }\textbf {\bibinfo {volume} {23}},\ \bibinfo {pages} {66--69} (\bibinfo
  {year} {1983})}\BibitemShut {NoStop}%
\bibitem [{\citenamefont {Zhang}\ \emph {et~al.}(2018)\citenamefont {Zhang},
  \citenamefont {Song}, \citenamefont {Alexandradinata}, \citenamefont {Weng},
  \citenamefont {Fang}, \citenamefont {Lu},\ and\ \citenamefont
  {Fang}}]{PhysRevLett.120.016401}%
  \BibitemOpen
  \bibfield  {author} {\bibinfo {author} {\bibfnamefont {Tiantian}\
  \bibnamefont {Zhang}}, \bibinfo {author} {\bibfnamefont {Zhida}\ \bibnamefont
  {Song}}, \bibinfo {author} {\bibfnamefont {A.}~\bibnamefont
  {Alexandradinata}}, \bibinfo {author} {\bibfnamefont {Hongming}\ \bibnamefont
  {Weng}}, \bibinfo {author} {\bibfnamefont {Chen}\ \bibnamefont {Fang}},
  \bibinfo {author} {\bibfnamefont {Ling}\ \bibnamefont {Lu}}, \ and\ \bibinfo
  {author} {\bibfnamefont {Zhong}\ \bibnamefont {Fang}},\ }\bibfield  {title}
  {\enquote {\bibinfo {title} {Double-weyl phonons in transition-metal
  monosilicides},}\ }\href {\doibase 10.1103/PhysRevLett.120.016401} {\bibfield
   {journal} {\bibinfo  {journal} {Phys. Rev. Lett.}\ }\textbf {\bibinfo
  {volume} {120}},\ \bibinfo {pages} {016401} (\bibinfo {year}
  {2018})}\BibitemShut {NoStop}%
\bibitem [{\citenamefont {Miao}\ \emph {et~al.}(2018)\citenamefont {Miao},
  \citenamefont {Zhang}, \citenamefont {Wang}, \citenamefont {Meyers},
  \citenamefont {Said}, \citenamefont {Wang}, \citenamefont {Shi},
  \citenamefont {Weng}, \citenamefont {Fang},\ and\ \citenamefont
  {Dean}}]{PhysRevLett.121.035302}%
  \BibitemOpen
  \bibfield  {author} {\bibinfo {author} {\bibfnamefont {H.}~\bibnamefont
  {Miao}}, \bibinfo {author} {\bibfnamefont {T.~T.}\ \bibnamefont {Zhang}},
  \bibinfo {author} {\bibfnamefont {L.}~\bibnamefont {Wang}}, \bibinfo {author}
  {\bibfnamefont {D.}~\bibnamefont {Meyers}}, \bibinfo {author} {\bibfnamefont
  {A.~H.}\ \bibnamefont {Said}}, \bibinfo {author} {\bibfnamefont {Y.~L.}\
  \bibnamefont {Wang}}, \bibinfo {author} {\bibfnamefont {Y.~G.}\ \bibnamefont
  {Shi}}, \bibinfo {author} {\bibfnamefont {H.~M.}\ \bibnamefont {Weng}},
  \bibinfo {author} {\bibfnamefont {Z.}~\bibnamefont {Fang}}, \ and\ \bibinfo
  {author} {\bibfnamefont {M.~P.~M.}\ \bibnamefont {Dean}},\ }\bibfield
  {title} {\enquote {\bibinfo {title} {Observation of double weyl phonons in
  parity-breaking fesi},}\ }\href {\doibase 10.1103/PhysRevLett.121.035302}
  {\bibfield  {journal} {\bibinfo  {journal} {Phys. Rev. Lett.}\ }\textbf
  {\bibinfo {volume} {121}},\ \bibinfo {pages} {035302} (\bibinfo {year}
  {2018})}\BibitemShut {NoStop}%
\bibitem [{\citenamefont {Zhang}\ \emph
  {et~al.}(2019{\natexlab{b}})\citenamefont {Zhang}, \citenamefont {Miao},
  \citenamefont {Wang}, \citenamefont {Lin}, \citenamefont {Cao}, \citenamefont
  {Fabbris}, \citenamefont {Said}, \citenamefont {Liu}, \citenamefont {Lei},
  \citenamefont {Fang}, \citenamefont {Weng},\ and\ \citenamefont
  {Dean}}]{PhysRevLett.123.245302}%
  \BibitemOpen
  \bibfield  {author} {\bibinfo {author} {\bibfnamefont {T.~T.}\ \bibnamefont
  {Zhang}}, \bibinfo {author} {\bibfnamefont {H.}~\bibnamefont {Miao}},
  \bibinfo {author} {\bibfnamefont {Q.}~\bibnamefont {Wang}}, \bibinfo {author}
  {\bibfnamefont {J.~Q.}\ \bibnamefont {Lin}}, \bibinfo {author} {\bibfnamefont
  {Y.}~\bibnamefont {Cao}}, \bibinfo {author} {\bibfnamefont {G.}~\bibnamefont
  {Fabbris}}, \bibinfo {author} {\bibfnamefont {A.~H.}\ \bibnamefont {Said}},
  \bibinfo {author} {\bibfnamefont {X.}~\bibnamefont {Liu}}, \bibinfo {author}
  {\bibfnamefont {H.~C.}\ \bibnamefont {Lei}}, \bibinfo {author} {\bibfnamefont
  {Z.}~\bibnamefont {Fang}}, \bibinfo {author} {\bibfnamefont {H.~M.}\
  \bibnamefont {Weng}}, \ and\ \bibinfo {author} {\bibfnamefont {M.~P.~M.}\
  \bibnamefont {Dean}},\ }\bibfield  {title} {\enquote {\bibinfo {title}
  {Phononic helical nodal lines with $\mathcal{PT}$ protection in
  ${\mathrm{mob}}_{2}$},}\ }\href {\doibase 10.1103/PhysRevLett.123.245302}
  {\bibfield  {journal} {\bibinfo  {journal} {Phys. Rev. Lett.}\ }\textbf
  {\bibinfo {volume} {123}},\ \bibinfo {pages} {245302} (\bibinfo {year}
  {2019}{\natexlab{b}})}\BibitemShut {NoStop}%
\bibitem [{\citenamefont {Peng}\ \emph
  {et~al.}(2022{\natexlab{a}})\citenamefont {Peng}, \citenamefont {Bouhon},
  \citenamefont {Monserrat},\ and\ \citenamefont {Slager}}]{peng2022phonons}%
  \BibitemOpen
  \bibfield  {author} {\bibinfo {author} {\bibfnamefont {Bo}~\bibnamefont
  {Peng}}, \bibinfo {author} {\bibfnamefont {Adrien}\ \bibnamefont {Bouhon}},
  \bibinfo {author} {\bibfnamefont {Bartomeu}\ \bibnamefont {Monserrat}}, \
  and\ \bibinfo {author} {\bibfnamefont {Robert-Jan}\ \bibnamefont {Slager}},\
  }\bibfield  {title} {\enquote {\bibinfo {title} {Phonons as a platform for
  non-abelian braiding and its manifestation in layered silicates},}\
  }\href@noop {} {\bibfield  {journal} {\bibinfo  {journal} {Nature
  communications}\ }\textbf {\bibinfo {volume} {13}},\ \bibinfo {pages} {1--15}
  (\bibinfo {year} {2022}{\natexlab{a}})}\BibitemShut {NoStop}%
\bibitem [{\citenamefont {Peng}\ \emph
  {et~al.}(2022{\natexlab{b}})\citenamefont {Peng}, \citenamefont {Bouhon},
  \citenamefont {Slager},\ and\ \citenamefont {Monserrat}}]{peng2022}%
  \BibitemOpen
  \bibfield  {author} {\bibinfo {author} {\bibfnamefont {Bo}~\bibnamefont
  {Peng}}, \bibinfo {author} {\bibfnamefont {Adrien}\ \bibnamefont {Bouhon}},
  \bibinfo {author} {\bibfnamefont {Robert-Jan}\ \bibnamefont {Slager}}, \ and\
  \bibinfo {author} {\bibfnamefont {Bartomeu}\ \bibnamefont {Monserrat}},\
  }\bibfield  {title} {\enquote {\bibinfo {title} {Multigap topology and
  non-abelian braiding of phonons from first principles},}\ }\href {\doibase
  10.1103/PhysRevB.105.085115} {\bibfield  {journal} {\bibinfo  {journal}
  {Phys. Rev. B}\ }\textbf {\bibinfo {volume} {105}},\ \bibinfo {pages}
  {085115} (\bibinfo {year} {2022}{\natexlab{b}})}\BibitemShut {NoStop}%
\bibitem [{\citenamefont {Liu}\ \emph {et~al.}(2021)\citenamefont {Liu},
  \citenamefont {Jin}, \citenamefont {Chen},\ and\ \citenamefont
  {Xu}}]{liu2021straight}%
  \BibitemOpen
  \bibfield  {author} {\bibinfo {author} {\bibfnamefont {Guang}\ \bibnamefont
  {Liu}}, \bibinfo {author} {\bibfnamefont {Yuanjun}\ \bibnamefont {Jin}},
  \bibinfo {author} {\bibfnamefont {Zhongjia}\ \bibnamefont {Chen}}, \ and\
  \bibinfo {author} {\bibfnamefont {Hu}~\bibnamefont {Xu}},\ }\bibfield
  {title} {\enquote {\bibinfo {title} {Symmetry-enforced straight nodal-line
  phonons},}\ }\href {\doibase 10.1103/PhysRevB.104.024304} {\bibfield
  {journal} {\bibinfo  {journal} {Phys. Rev. B}\ }\textbf {\bibinfo {volume}
  {104}},\ \bibinfo {pages} {024304} (\bibinfo {year} {2021})}\BibitemShut
  {NoStop}%
\bibitem [{\citenamefont {Zhu}\ \emph {et~al.}(2022)\citenamefont {Zhu},
  \citenamefont {Wu}, \citenamefont {Zhao}, \citenamefont {Chen}, \citenamefont
  {Zhang},\ and\ \citenamefont {Yang}}]{zhu2022symmetry}%
  \BibitemOpen
  \bibfield  {author} {\bibinfo {author} {\bibfnamefont {Jiaojiao}\
  \bibnamefont {Zhu}}, \bibinfo {author} {\bibfnamefont {Weikang}\ \bibnamefont
  {Wu}}, \bibinfo {author} {\bibfnamefont {Jianzhou}\ \bibnamefont {Zhao}},
  \bibinfo {author} {\bibfnamefont {Hao}\ \bibnamefont {Chen}}, \bibinfo
  {author} {\bibfnamefont {Lifa}\ \bibnamefont {Zhang}}, \ and\ \bibinfo
  {author} {\bibfnamefont {Shengyuan~A}\ \bibnamefont {Yang}},\ }\bibfield
  {title} {\enquote {\bibinfo {title} {Symmetry-enforced nodal chain
  phonons},}\ }\href@noop {} {\bibfield  {journal} {\bibinfo  {journal} {npj
  Quantum Materials}\ }\textbf {\bibinfo {volume} {7}},\ \bibinfo {pages}
  {1--6} (\bibinfo {year} {2022})}\BibitemShut {NoStop}%
\bibitem [{\citenamefont {Li}\ \emph {et~al.}(2021)\citenamefont {Li},
  \citenamefont {Liu}, \citenamefont {Baronett}, \citenamefont {Liu},
  \citenamefont {Wang}, \citenamefont {Li}, \citenamefont {Chen}, \citenamefont
  {Li}, \citenamefont {Zhu},\ and\ \citenamefont {Chen}}]{li2021computation}%
  \BibitemOpen
  \bibfield  {author} {\bibinfo {author} {\bibfnamefont {Jiangxu}\ \bibnamefont
  {Li}}, \bibinfo {author} {\bibfnamefont {Jiaxi}\ \bibnamefont {Liu}},
  \bibinfo {author} {\bibfnamefont {Stanley~A}\ \bibnamefont {Baronett}},
  \bibinfo {author} {\bibfnamefont {Mingfeng}\ \bibnamefont {Liu}}, \bibinfo
  {author} {\bibfnamefont {Lei}\ \bibnamefont {Wang}}, \bibinfo {author}
  {\bibfnamefont {Ronghan}\ \bibnamefont {Li}}, \bibinfo {author}
  {\bibfnamefont {Yun}\ \bibnamefont {Chen}}, \bibinfo {author} {\bibfnamefont
  {Dianzhong}\ \bibnamefont {Li}}, \bibinfo {author} {\bibfnamefont {Qiang}\
  \bibnamefont {Zhu}}, \ and\ \bibinfo {author} {\bibfnamefont {Xing-Qiu}\
  \bibnamefont {Chen}},\ }\bibfield  {title} {\enquote {\bibinfo {title}
  {Computation and data driven discovery of topological phononic materials},}\
  }\href@noop {} {\bibfield  {journal} {\bibinfo  {journal} {Nature
  communications}\ }\textbf {\bibinfo {volume} {12}},\ \bibinfo {pages} {1--12}
  (\bibinfo {year} {2021})}\BibitemShut {NoStop}%
\bibitem [{\citenamefont {Ma\~nes}(2020)}]{manes_fragile_2019}%
  \BibitemOpen
  \bibfield  {author} {\bibinfo {author} {\bibfnamefont {Juan~L.}\ \bibnamefont
  {Ma\~nes}},\ }\bibfield  {title} {\enquote {\bibinfo {title} {Fragile phonon
  topology on the honeycomb lattice with time-reversal symmetry},}\ }\href
  {\doibase 10.1103/PhysRevB.102.024307} {\bibfield  {journal} {\bibinfo
  {journal} {Phys. Rev. B}\ }\textbf {\bibinfo {volume} {102}},\ \bibinfo
  {pages} {024307} (\bibinfo {year} {2020})}\BibitemShut {NoStop}%
\bibitem [{\citenamefont {Petretto}\ \emph {et~al.}(2018)\citenamefont
  {Petretto}, \citenamefont {Dwaraknath}, \citenamefont {Miranda},
  \citenamefont {Winston}, \citenamefont {Giantomassi}, \citenamefont
  {Van~Setten}, \citenamefont {Gonze}, \citenamefont {Persson}, \citenamefont
  {Hautier},\ and\ \citenamefont {Rignanese}}]{petretto2018high}%
  \BibitemOpen
  \bibfield  {author} {\bibinfo {author} {\bibfnamefont {Guido}\ \bibnamefont
  {Petretto}}, \bibinfo {author} {\bibfnamefont {Shyam}\ \bibnamefont
  {Dwaraknath}}, \bibinfo {author} {\bibfnamefont {Henrique~PC}\ \bibnamefont
  {Miranda}}, \bibinfo {author} {\bibfnamefont {Donald}\ \bibnamefont
  {Winston}}, \bibinfo {author} {\bibfnamefont {Matteo}\ \bibnamefont
  {Giantomassi}}, \bibinfo {author} {\bibfnamefont {Michiel~J}\ \bibnamefont
  {Van~Setten}}, \bibinfo {author} {\bibfnamefont {Xavier}\ \bibnamefont
  {Gonze}}, \bibinfo {author} {\bibfnamefont {Kristin~A}\ \bibnamefont
  {Persson}}, \bibinfo {author} {\bibfnamefont {Geoffroy}\ \bibnamefont
  {Hautier}}, \ and\ \bibinfo {author} {\bibfnamefont {Gian-Marco}\
  \bibnamefont {Rignanese}},\ }\bibfield  {title} {\enquote {\bibinfo {title}
  {High-throughput density-functional perturbation theory phonons for inorganic
  materials},}\ }\href@noop {} {\bibfield  {journal} {\bibinfo  {journal}
  {Scientific data}\ }\textbf {\bibinfo {volume} {5}},\ \bibinfo {pages}
  {1--12} (\bibinfo {year} {2018})}\BibitemShut {NoStop}%
\bibitem [{\citenamefont {Song}\ \emph
  {et~al.}(2020{\natexlab{a}})\citenamefont {Song}, \citenamefont {Elcoro},\
  and\ \citenamefont {Bernevig}}]{song2020}%
  \BibitemOpen
  \bibfield  {author} {\bibinfo {author} {\bibfnamefont {Zhi-Da}\ \bibnamefont
  {Song}}, \bibinfo {author} {\bibfnamefont {Luis}\ \bibnamefont {Elcoro}}, \
  and\ \bibinfo {author} {\bibfnamefont {B.~Andrei}\ \bibnamefont {Bernevig}},\
  }\bibfield  {title} {\enquote {\bibinfo {title} {Twisted bulk-boundary
  correspondence of fragile topology},}\ }\href {\doibase
  10.1126/science.aaz7650} {\bibfield  {journal} {\bibinfo  {journal}
  {Science}\ }\textbf {\bibinfo {volume} {367}},\ \bibinfo {pages} {794--797}
  (\bibinfo {year} {2020}{\natexlab{a}})}\BibitemShut {NoStop}%
\bibitem [{\citenamefont {Xu}\ \emph {et~al.}(2021{\natexlab{a}})\citenamefont
  {Xu}, \citenamefont {Elcoro}, \citenamefont {Li}, \citenamefont {Song},
  \citenamefont {Regnault}, \citenamefont {Yang}, \citenamefont {Sun},
  \citenamefont {Parkin}, \citenamefont {Felser},\ and\ \citenamefont
  {Bernevig}}]{xu2021three}%
  \BibitemOpen
  \bibfield  {author} {\bibinfo {author} {\bibfnamefont {Yuanfeng}\
  \bibnamefont {Xu}}, \bibinfo {author} {\bibfnamefont {Luis}\ \bibnamefont
  {Elcoro}}, \bibinfo {author} {\bibfnamefont {Guowei}\ \bibnamefont {Li}},
  \bibinfo {author} {\bibfnamefont {Zhi-Da}\ \bibnamefont {Song}}, \bibinfo
  {author} {\bibfnamefont {Nicolas}\ \bibnamefont {Regnault}}, \bibinfo
  {author} {\bibfnamefont {Qun}\ \bibnamefont {Yang}}, \bibinfo {author}
  {\bibfnamefont {Yan}\ \bibnamefont {Sun}}, \bibinfo {author} {\bibfnamefont
  {Stuart}\ \bibnamefont {Parkin}}, \bibinfo {author} {\bibfnamefont {Claudia}\
  \bibnamefont {Felser}}, \ and\ \bibinfo {author} {\bibfnamefont {B~Andrei}\
  \bibnamefont {Bernevig}},\ }\bibfield  {title} {\enquote {\bibinfo {title}
  {Three-dimensional real space invariants, obstructed atomic insulators and a
  new principle for active catalytic sites},}\ }\href@noop {} {\bibfield
  {journal} {\bibinfo  {journal} {arXiv preprint arXiv:2111.02433}\ } (\bibinfo
  {year} {2021}{\natexlab{a}})}\BibitemShut {NoStop}%
\bibitem [{\citenamefont {Togo}\ and\ \citenamefont {Tanaka}(2015)}]{phonopy}%
  \BibitemOpen
  \bibfield  {author} {\bibinfo {author} {\bibfnamefont {A}~\bibnamefont
  {Togo}}\ and\ \bibinfo {author} {\bibfnamefont {I}~\bibnamefont {Tanaka}},\
  }\bibfield  {title} {\enquote {\bibinfo {title} {First principles phonon
  calculations in materials science},}\ }\href@noop {} {\bibfield  {journal}
  {\bibinfo  {journal} {Scr. Mater.}\ }\textbf {\bibinfo {volume} {108}},\
  \bibinfo {pages} {1--5} (\bibinfo {year} {2015})}\BibitemShut {NoStop}%
\bibitem [{\citenamefont {Gonze}\ \emph {et~al.}(2002)\citenamefont {Gonze},
  \citenamefont {Beuken}, \citenamefont {Caracas}, \citenamefont {Detraux},
  \citenamefont {Fuchs}, \citenamefont {Rignanese}, \citenamefont {Sindic},
  \citenamefont {Verstraete}, \citenamefont {Zerah}, \citenamefont {Jollet}
  \emph {et~al.}}]{gonze2002first}%
  \BibitemOpen
  \bibfield  {author} {\bibinfo {author} {\bibfnamefont {Xavier}\ \bibnamefont
  {Gonze}}, \bibinfo {author} {\bibfnamefont {J-M}\ \bibnamefont {Beuken}},
  \bibinfo {author} {\bibfnamefont {Razvan}\ \bibnamefont {Caracas}}, \bibinfo
  {author} {\bibfnamefont {F}~\bibnamefont {Detraux}}, \bibinfo {author}
  {\bibfnamefont {M}~\bibnamefont {Fuchs}}, \bibinfo {author} {\bibfnamefont
  {G-M}\ \bibnamefont {Rignanese}}, \bibinfo {author} {\bibfnamefont {Luc}\
  \bibnamefont {Sindic}}, \bibinfo {author} {\bibfnamefont {Matthieu}\
  \bibnamefont {Verstraete}}, \bibinfo {author} {\bibfnamefont {G}~\bibnamefont
  {Zerah}}, \bibinfo {author} {\bibfnamefont {F}~\bibnamefont {Jollet}},  \emph
  {et~al.},\ }\bibfield  {title} {\enquote {\bibinfo {title} {First-principles
  computation of material properties: the abinit software project},}\
  }\href@noop {} {\bibfield  {journal} {\bibinfo  {journal} {Computational
  Materials Science}\ }\textbf {\bibinfo {volume} {25}},\ \bibinfo {pages}
  {478--492} (\bibinfo {year} {2002})}\BibitemShut {NoStop}%
\bibitem [{\citenamefont {Gonze}\ \emph {et~al.}(2009)\citenamefont {Gonze},
  \citenamefont {Amadon}, \citenamefont {Anglade}, \citenamefont {Beuken},
  \citenamefont {Bottin}, \citenamefont {Boulanger}, \citenamefont {Bruneval},
  \citenamefont {Caliste}, \citenamefont {Caracas}, \citenamefont
  {C{\^o}t{\'e}} \emph {et~al.}}]{gonze2009abinit}%
  \BibitemOpen
  \bibfield  {author} {\bibinfo {author} {\bibfnamefont {Xavier}\ \bibnamefont
  {Gonze}}, \bibinfo {author} {\bibfnamefont {Bernard}\ \bibnamefont {Amadon}},
  \bibinfo {author} {\bibfnamefont {P-M}\ \bibnamefont {Anglade}}, \bibinfo
  {author} {\bibfnamefont {J-M}\ \bibnamefont {Beuken}}, \bibinfo {author}
  {\bibfnamefont {Fran{\c{c}}ois}\ \bibnamefont {Bottin}}, \bibinfo {author}
  {\bibfnamefont {Paul}\ \bibnamefont {Boulanger}}, \bibinfo {author}
  {\bibfnamefont {Fabien}\ \bibnamefont {Bruneval}}, \bibinfo {author}
  {\bibfnamefont {Damien}\ \bibnamefont {Caliste}}, \bibinfo {author}
  {\bibfnamefont {Razvan}\ \bibnamefont {Caracas}}, \bibinfo {author}
  {\bibfnamefont {Michel}\ \bibnamefont {C{\^o}t{\'e}}},  \emph {et~al.},\
  }\bibfield  {title} {\enquote {\bibinfo {title} {Abinit: First-principles
  approach to material and nanosystem properties},}\ }\href@noop {} {\bibfield
  {journal} {\bibinfo  {journal} {Computer Physics Communications}\ }\textbf
  {\bibinfo {volume} {180}},\ \bibinfo {pages} {2582--2615} (\bibinfo {year}
  {2009})}\BibitemShut {NoStop}%
\bibitem [{\citenamefont {Gonze}\ \emph {et~al.}(2016)\citenamefont {Gonze},
  \citenamefont {Jollet}, \citenamefont {Araujo}, \citenamefont {Adams},
  \citenamefont {Amadon}, \citenamefont {Applencourt}, \citenamefont {Audouze},
  \citenamefont {Beuken}, \citenamefont {Bieder}, \citenamefont {Bokhanchuk}
  \emph {et~al.}}]{gonze2016recent}%
  \BibitemOpen
  \bibfield  {author} {\bibinfo {author} {\bibfnamefont {Xavier}\ \bibnamefont
  {Gonze}}, \bibinfo {author} {\bibfnamefont {Fran{\c{c}}ois}\ \bibnamefont
  {Jollet}}, \bibinfo {author} {\bibfnamefont {F~Abreu}\ \bibnamefont
  {Araujo}}, \bibinfo {author} {\bibfnamefont {Donat}\ \bibnamefont {Adams}},
  \bibinfo {author} {\bibfnamefont {Bernard}\ \bibnamefont {Amadon}}, \bibinfo
  {author} {\bibfnamefont {Thomas}\ \bibnamefont {Applencourt}}, \bibinfo
  {author} {\bibfnamefont {Christophe}\ \bibnamefont {Audouze}}, \bibinfo
  {author} {\bibfnamefont {J-M}\ \bibnamefont {Beuken}}, \bibinfo {author}
  {\bibfnamefont {Jordan}\ \bibnamefont {Bieder}}, \bibinfo {author}
  {\bibfnamefont {A}~\bibnamefont {Bokhanchuk}},  \emph {et~al.},\ }\bibfield
  {title} {\enquote {\bibinfo {title} {Recent developments in the abinit
  software package},}\ }\href@noop {} {\bibfield  {journal} {\bibinfo
  {journal} {Computer Physics Communications}\ }\textbf {\bibinfo {volume}
  {205}},\ \bibinfo {pages} {106--131} (\bibinfo {year} {2016})}\BibitemShut
  {NoStop}%
\bibitem [{\citenamefont {Wu}\ \emph {et~al.}(2018)\citenamefont {Wu},
  \citenamefont {Zhang}, \citenamefont {Song}, \citenamefont {Troyer},\ and\
  \citenamefont {Soluyanov}}]{WU2017}%
  \BibitemOpen
  \bibfield  {author} {\bibinfo {author} {\bibfnamefont {QuanSheng}\
  \bibnamefont {Wu}}, \bibinfo {author} {\bibfnamefont {ShengNan}\ \bibnamefont
  {Zhang}}, \bibinfo {author} {\bibfnamefont {Hai-Feng}\ \bibnamefont {Song}},
  \bibinfo {author} {\bibfnamefont {Matthias}\ \bibnamefont {Troyer}}, \ and\
  \bibinfo {author} {\bibfnamefont {Alexey~A.}\ \bibnamefont {Soluyanov}},\
  }\bibfield  {title} {\enquote {\bibinfo {title} {Wanniertools : An
  open-source software package for novel topological materials},}\ }\href
  {\doibase https://doi.org/10.1016/j.cpc.2017.09.033} {\bibfield  {journal}
  {\bibinfo  {journal} {Computer Physics Communications}\ }\textbf {\bibinfo
  {volume} {224}},\ \bibinfo {pages} {405 -- 416} (\bibinfo {year}
  {2018})}\BibitemShut {NoStop}%
\bibitem [{\citenamefont {Xu}\ \emph {et~al.}(2021{\natexlab{b}})\citenamefont
  {Xu}, \citenamefont {Elcoro}, \citenamefont {Song}, \citenamefont
  {Vergniory}, \citenamefont {Felser}, \citenamefont {Parkin}, \citenamefont
  {Regnault}, \citenamefont {Ma{\~n}es},\ and\ \citenamefont
  {Bernevig}}]{xu2021filling}%
  \BibitemOpen
  \bibfield  {author} {\bibinfo {author} {\bibfnamefont {Yuanfeng}\
  \bibnamefont {Xu}}, \bibinfo {author} {\bibfnamefont {Luis}\ \bibnamefont
  {Elcoro}}, \bibinfo {author} {\bibfnamefont {Zhi-Da}\ \bibnamefont {Song}},
  \bibinfo {author} {\bibfnamefont {MG}~\bibnamefont {Vergniory}}, \bibinfo
  {author} {\bibfnamefont {Claudia}\ \bibnamefont {Felser}}, \bibinfo {author}
  {\bibfnamefont {Stuart~SP}\ \bibnamefont {Parkin}}, \bibinfo {author}
  {\bibfnamefont {Nicolas}\ \bibnamefont {Regnault}}, \bibinfo {author}
  {\bibfnamefont {Juan~L}\ \bibnamefont {Ma{\~n}es}}, \ and\ \bibinfo {author}
  {\bibfnamefont {B~Andrei}\ \bibnamefont {Bernevig}},\ }\bibfield  {title}
  {\enquote {\bibinfo {title} {Filling-enforced obstructed atomic
  insulators},}\ }\href@noop {} {\bibfield  {journal} {\bibinfo  {journal}
  {arXiv preprint arXiv:2106.10276}\ } (\bibinfo {year}
  {2021}{\natexlab{b}})}\BibitemShut {NoStop}%
\bibitem [{\citenamefont {Gao}\ \emph {et~al.}(2021)\citenamefont {Gao},
  \citenamefont {Qian}, \citenamefont {Jia}, \citenamefont {Guo}, \citenamefont
  {Fang}, \citenamefont {Liu}, \citenamefont {Weng},\ and\ \citenamefont
  {Wang}}]{gao2021unconventional}%
  \BibitemOpen
  \bibfield  {author} {\bibinfo {author} {\bibfnamefont {Jiacheng}\
  \bibnamefont {Gao}}, \bibinfo {author} {\bibfnamefont {Yuting}\ \bibnamefont
  {Qian}}, \bibinfo {author} {\bibfnamefont {Huaxian}\ \bibnamefont {Jia}},
  \bibinfo {author} {\bibfnamefont {Zhaopeng}\ \bibnamefont {Guo}}, \bibinfo
  {author} {\bibfnamefont {Zhong}\ \bibnamefont {Fang}}, \bibinfo {author}
  {\bibfnamefont {Miao}\ \bibnamefont {Liu}}, \bibinfo {author} {\bibfnamefont
  {Hongming}\ \bibnamefont {Weng}}, \ and\ \bibinfo {author} {\bibfnamefont
  {Zhijun}\ \bibnamefont {Wang}},\ }\bibfield  {title} {\enquote {\bibinfo
  {title} {Unconventional materials: the mismatch between electronic charge
  centers andatomic positions},}\ }\href@noop {} {\bibfield  {journal}
  {\bibinfo  {journal} {arXiv preprint arXiv:2106.08035}\ } (\bibinfo {year}
  {2021})}\BibitemShut {NoStop}%
\bibitem [{\citenamefont {Baroni}\ \emph {et~al.}(2001)\citenamefont {Baroni},
  \citenamefont {de~Gironcoli}, \citenamefont {Dal~Corso},\ and\ \citenamefont
  {Giannozzi}}]{RevModPhys.73.515}%
  \BibitemOpen
  \bibfield  {author} {\bibinfo {author} {\bibfnamefont {Stefano}\ \bibnamefont
  {Baroni}}, \bibinfo {author} {\bibfnamefont {Stefano}\ \bibnamefont
  {de~Gironcoli}}, \bibinfo {author} {\bibfnamefont {Andrea}\ \bibnamefont
  {Dal~Corso}}, \ and\ \bibinfo {author} {\bibfnamefont {Paolo}\ \bibnamefont
  {Giannozzi}},\ }\bibfield  {title} {\enquote {\bibinfo {title} {Phonons and
  related crystal properties from density-functional perturbation theory},}\
  }\href {\doibase 10.1103/RevModPhys.73.515} {\bibfield  {journal} {\bibinfo
  {journal} {Rev. Mod. Phys.}\ }\textbf {\bibinfo {volume} {73}},\ \bibinfo
  {pages} {515--562} (\bibinfo {year} {2001})}\BibitemShut {NoStop}%
\bibitem [{\citenamefont {Gonze}\ and\ \citenamefont
  {Lee}(1997)}]{gonze1997dynamical}%
  \BibitemOpen
  \bibfield  {author} {\bibinfo {author} {\bibfnamefont {Xavier}\ \bibnamefont
  {Gonze}}\ and\ \bibinfo {author} {\bibfnamefont {Changyol}\ \bibnamefont
  {Lee}},\ }\bibfield  {title} {\enquote {\bibinfo {title} {Dynamical matrices,
  born effective charges, dielectric permittivity tensors, and interatomic
  force constants from density-functional perturbation theory},}\ }\href@noop
  {} {\bibfield  {journal} {\bibinfo  {journal} {Physical Review B}\ }\textbf
  {\bibinfo {volume} {55}},\ \bibinfo {pages} {10355} (\bibinfo {year}
  {1997})}\BibitemShut {NoStop}%
\bibitem [{\citenamefont {Wang}\ \emph {et~al.}(2010)\citenamefont {Wang},
  \citenamefont {Wang}, \citenamefont {Wang}, \citenamefont {Mei},
  \citenamefont {Shang}, \citenamefont {Chen},\ and\ \citenamefont
  {Liu}}]{wang2010mixed}%
  \BibitemOpen
  \bibfield  {author} {\bibinfo {author} {\bibfnamefont {Yuedong}\ \bibnamefont
  {Wang}}, \bibinfo {author} {\bibfnamefont {JJ}~\bibnamefont {Wang}}, \bibinfo
  {author} {\bibfnamefont {WY}~\bibnamefont {Wang}}, \bibinfo {author}
  {\bibfnamefont {ZG}~\bibnamefont {Mei}}, \bibinfo {author} {\bibfnamefont
  {SL}~\bibnamefont {Shang}}, \bibinfo {author} {\bibfnamefont
  {LQ}~\bibnamefont {Chen}}, \ and\ \bibinfo {author} {\bibfnamefont
  {ZK}~\bibnamefont {Liu}},\ }\bibfield  {title} {\enquote {\bibinfo {title} {A
  mixed-space approach to first-principles calculations of phonon frequencies
  for polar materials},}\ }\href@noop {} {\bibfield  {journal} {\bibinfo
  {journal} {Journal of Physics: Condensed Matter}\ }\textbf {\bibinfo {volume}
  {22}},\ \bibinfo {pages} {202201} (\bibinfo {year} {2010})}\BibitemShut
  {NoStop}%
\bibitem [{\citenamefont {Song}\ \emph
  {et~al.}(2018{\natexlab{a}})\citenamefont {Song}, \citenamefont {Zhang},\
  and\ \citenamefont {Fang}}]{song_diagnosis_2018}%
  \BibitemOpen
  \bibfield  {author} {\bibinfo {author} {\bibfnamefont {Zhida}\ \bibnamefont
  {Song}}, \bibinfo {author} {\bibfnamefont {Tiantian}\ \bibnamefont {Zhang}},
  \ and\ \bibinfo {author} {\bibfnamefont {Chen}\ \bibnamefont {Fang}},\
  }\bibfield  {title} {\enquote {\bibinfo {title} {Diagnosis for {Nonmagnetic}
  {Topological} {Semimetals} in the {Absence} of {Spin}-{Orbital}
  {Coupling}},}\ }\href {\doibase 10.1103/PhysRevX.8.031069} {\bibfield
  {journal} {\bibinfo  {journal} {Physical Review X}\ }\textbf {\bibinfo
  {volume} {8}},\ \bibinfo {pages} {031069} (\bibinfo {year}
  {2018}{\natexlab{a}})}\BibitemShut {NoStop}%
\bibitem [{\citenamefont {Bradlyn}\ \emph {et~al.}(2016)\citenamefont
  {Bradlyn}, \citenamefont {Cano}, \citenamefont {Wang}, \citenamefont
  {Vergniory}, \citenamefont {Felser}, \citenamefont {Cava},\ and\
  \citenamefont {Bernevig}}]{bradlyn2016beyond}%
  \BibitemOpen
  \bibfield  {author} {\bibinfo {author} {\bibfnamefont {Barry}\ \bibnamefont
  {Bradlyn}}, \bibinfo {author} {\bibfnamefont {Jennifer}\ \bibnamefont
  {Cano}}, \bibinfo {author} {\bibfnamefont {Zhijun}\ \bibnamefont {Wang}},
  \bibinfo {author} {\bibfnamefont {MG}~\bibnamefont {Vergniory}}, \bibinfo
  {author} {\bibfnamefont {C}~\bibnamefont {Felser}}, \bibinfo {author}
  {\bibfnamefont {Robert~Joseph}\ \bibnamefont {Cava}}, \ and\ \bibinfo
  {author} {\bibfnamefont {B~Andrei}\ \bibnamefont {Bernevig}},\ }\bibfield
  {title} {\enquote {\bibinfo {title} {Beyond dirac and weyl fermions:
  Unconventional quasiparticles in conventional crystals},}\ }\href@noop {}
  {\bibfield  {journal} {\bibinfo  {journal} {Science}\ }\textbf {\bibinfo
  {volume} {353}},\ \bibinfo {pages} {aaf5037} (\bibinfo {year}
  {2016})}\BibitemShut {NoStop}%
\bibitem [{\citenamefont {Fang}\ \emph
  {et~al.}(2015{\natexlab{a}})\citenamefont {Fang}, \citenamefont {Chen},
  \citenamefont {Kee},\ and\ \citenamefont {Fu}}]{fang2015}%
  \BibitemOpen
  \bibfield  {author} {\bibinfo {author} {\bibfnamefont {Chen}\ \bibnamefont
  {Fang}}, \bibinfo {author} {\bibfnamefont {Yige}\ \bibnamefont {Chen}},
  \bibinfo {author} {\bibfnamefont {Hae-Young}\ \bibnamefont {Kee}}, \ and\
  \bibinfo {author} {\bibfnamefont {Liang}\ \bibnamefont {Fu}},\ }\bibfield
  {title} {\enquote {\bibinfo {title} {Topological nodal line semimetals with
  and without spin-orbital coupling},}\ }\href {\doibase
  10.1103/PhysRevB.92.081201} {\bibfield  {journal} {\bibinfo  {journal} {Phys.
  Rev. B}\ }\textbf {\bibinfo {volume} {92}},\ \bibinfo {pages} {081201}
  (\bibinfo {year} {2015}{\natexlab{a}})}\BibitemShut {NoStop}%
\bibitem [{\citenamefont {Ahn}\ \emph {et~al.}(2018)\citenamefont {Ahn},
  \citenamefont {Kim}, \citenamefont {Kim},\ and\ \citenamefont
  {Yang}}]{bjyang2018}%
  \BibitemOpen
  \bibfield  {author} {\bibinfo {author} {\bibfnamefont {Junyeong}\
  \bibnamefont {Ahn}}, \bibinfo {author} {\bibfnamefont {Dongwook}\
  \bibnamefont {Kim}}, \bibinfo {author} {\bibfnamefont {Youngkuk}\
  \bibnamefont {Kim}}, \ and\ \bibinfo {author} {\bibfnamefont {Bohm-Jung}\
  \bibnamefont {Yang}},\ }\bibfield  {title} {\enquote {\bibinfo {title} {Band
  topology and linking structure of nodal line semimetals with ${Z}_{2}$
  monopole charges},}\ }\href {\doibase 10.1103/PhysRevLett.121.106403}
  {\bibfield  {journal} {\bibinfo  {journal} {Phys. Rev. Lett.}\ }\textbf
  {\bibinfo {volume} {121}},\ \bibinfo {pages} {106403} (\bibinfo {year}
  {2018})}\BibitemShut {NoStop}%
\bibitem [{\citenamefont {Sancho}\ \emph {et~al.}(1985)\citenamefont {Sancho},
  \citenamefont {Sancho}, \citenamefont {Sancho},\ and\ \citenamefont
  {Rubio}}]{sancho1985highly}%
  \BibitemOpen
  \bibfield  {author} {\bibinfo {author} {\bibfnamefont {MP~Lopez}\
  \bibnamefont {Sancho}}, \bibinfo {author} {\bibfnamefont {JM~Lopez}\
  \bibnamefont {Sancho}}, \bibinfo {author} {\bibfnamefont {JM~Lopez}\
  \bibnamefont {Sancho}}, \ and\ \bibinfo {author} {\bibfnamefont
  {J}~\bibnamefont {Rubio}},\ }\bibfield  {title} {\enquote {\bibinfo {title}
  {Highly convergent schemes for the calculation of bulk and surface green
  functions},}\ }\href@noop {} {\bibfield  {journal} {\bibinfo  {journal}
  {Journal of Physics F: Metal Physics}\ }\textbf {\bibinfo {volume} {15}},\
  \bibinfo {pages} {851} (\bibinfo {year} {1985})}\BibitemShut {NoStop}%
\bibitem [{\citenamefont {Song}\ \emph
  {et~al.}(2018{\natexlab{b}})\citenamefont {Song}, \citenamefont {Zhang},
  \citenamefont {Fang},\ and\ \citenamefont {Fang}}]{song_quantitative_2018}%
  \BibitemOpen
  \bibfield  {author} {\bibinfo {author} {\bibfnamefont {Zhida}\ \bibnamefont
  {Song}}, \bibinfo {author} {\bibfnamefont {Tiantian}\ \bibnamefont {Zhang}},
  \bibinfo {author} {\bibfnamefont {Zhong}\ \bibnamefont {Fang}}, \ and\
  \bibinfo {author} {\bibfnamefont {Chen}\ \bibnamefont {Fang}},\ }\bibfield
  {title} {\enquote {\bibinfo {title} {Quantitative mappings between symmetry
  and topology in solids},}\ }\href {\doibase 10.1038/s41467-018-06010-w}
  {\bibfield  {journal} {\bibinfo  {journal} {Nature Communications}\ }\textbf
  {\bibinfo {volume} {9}},\ \bibinfo {pages} {3530} (\bibinfo {year}
  {2018}{\natexlab{b}})}\BibitemShut {NoStop}%
\bibitem [{\citenamefont {Cano}\ \emph {et~al.}(2018)\citenamefont {Cano},
  \citenamefont {Bradlyn}, \citenamefont {Wang}, \citenamefont {Elcoro},
  \citenamefont {Vergniory}, \citenamefont {Felser}, \citenamefont {Aroyo},\
  and\ \citenamefont {Bernevig}}]{cano_building_2018}%
  \BibitemOpen
  \bibfield  {author} {\bibinfo {author} {\bibfnamefont {Jennifer}\
  \bibnamefont {Cano}}, \bibinfo {author} {\bibfnamefont {Barry}\ \bibnamefont
  {Bradlyn}}, \bibinfo {author} {\bibfnamefont {Zhijun}\ \bibnamefont {Wang}},
  \bibinfo {author} {\bibfnamefont {L.}~\bibnamefont {Elcoro}}, \bibinfo
  {author} {\bibfnamefont {M.~G.}\ \bibnamefont {Vergniory}}, \bibinfo {author}
  {\bibfnamefont {C.}~\bibnamefont {Felser}}, \bibinfo {author} {\bibfnamefont
  {M.~I.}\ \bibnamefont {Aroyo}}, \ and\ \bibinfo {author} {\bibfnamefont
  {B.~Andrei}\ \bibnamefont {Bernevig}},\ }\bibfield  {title} {\enquote
  {\bibinfo {title} {Building blocks of topological quantum chemistry:
  {Elementary} band representations},}\ }\href {\doibase
  10.1103/PhysRevB.97.035139} {\bibfield  {journal} {\bibinfo  {journal}
  {Physical Review B}\ }\textbf {\bibinfo {volume} {97}},\ \bibinfo {pages}
  {035139} (\bibinfo {year} {2018})}\BibitemShut {NoStop}%
\bibitem [{\citenamefont {Elcoro}\ \emph {et~al.}(2017)\citenamefont {Elcoro},
  \citenamefont {Bradlyn}, \citenamefont {Wang}, \citenamefont {Vergniory},
  \citenamefont {Cano}, \citenamefont {Felser}, \citenamefont {Bernevig},
  \citenamefont {Orobengoa}, \citenamefont {de~la Flor},\ and\ \citenamefont
  {Aroyo}}]{elcoro_double_2017}%
  \BibitemOpen
  \bibfield  {author} {\bibinfo {author} {\bibfnamefont {Luis}\ \bibnamefont
  {Elcoro}}, \bibinfo {author} {\bibfnamefont {Barry}\ \bibnamefont {Bradlyn}},
  \bibinfo {author} {\bibfnamefont {Zhijun}\ \bibnamefont {Wang}}, \bibinfo
  {author} {\bibfnamefont {Maia~G.}\ \bibnamefont {Vergniory}}, \bibinfo
  {author} {\bibfnamefont {Jennifer}\ \bibnamefont {Cano}}, \bibinfo {author}
  {\bibfnamefont {Claudia}\ \bibnamefont {Felser}}, \bibinfo {author}
  {\bibfnamefont {B.~Andrei}\ \bibnamefont {Bernevig}}, \bibinfo {author}
  {\bibfnamefont {Danel}\ \bibnamefont {Orobengoa}}, \bibinfo {author}
  {\bibfnamefont {Gemma}\ \bibnamefont {de~la Flor}}, \ and\ \bibinfo {author}
  {\bibfnamefont {Mois~I.}\ \bibnamefont {Aroyo}},\ }\bibfield  {title}
  {\enquote {\bibinfo {title} {{Double crystallographic groups and their
  representations on the Bilbao Crystallographic Server}},}\ }\href {\doibase
  10.1107/S1600576717011712} {\bibfield  {journal} {\bibinfo  {journal}
  {Journal of Applied Crystallography}\ }\textbf {\bibinfo {volume} {50}},\
  \bibinfo {pages} {1457--1477} (\bibinfo {year} {2017})}\BibitemShut {NoStop}%
\bibitem [{\citenamefont {Vergniory}\ \emph {et~al.}(2017)\citenamefont
  {Vergniory}, \citenamefont {Elcoro}, \citenamefont {Wang}, \citenamefont
  {Cano}, \citenamefont {Felser}, \citenamefont {Aroyo}, \citenamefont
  {Bernevig},\ and\ \citenamefont {Bradlyn}}]{vergniory_graph_2017}%
  \BibitemOpen
  \bibfield  {author} {\bibinfo {author} {\bibfnamefont {M.~G.}\ \bibnamefont
  {Vergniory}}, \bibinfo {author} {\bibfnamefont {L.}~\bibnamefont {Elcoro}},
  \bibinfo {author} {\bibfnamefont {Zhijun}\ \bibnamefont {Wang}}, \bibinfo
  {author} {\bibfnamefont {Jennifer}\ \bibnamefont {Cano}}, \bibinfo {author}
  {\bibfnamefont {C.}~\bibnamefont {Felser}}, \bibinfo {author} {\bibfnamefont
  {M.~I.}\ \bibnamefont {Aroyo}}, \bibinfo {author} {\bibfnamefont {B.~Andrei}\
  \bibnamefont {Bernevig}}, \ and\ \bibinfo {author} {\bibfnamefont {Barry}\
  \bibnamefont {Bradlyn}},\ }\bibfield  {title} {\enquote {\bibinfo {title}
  {Graph theory data for topological quantum chemistry},}\ }\href {\doibase
  10.1103/PhysRevE.96.023310} {\bibfield  {journal} {\bibinfo  {journal} {Phys.
  Rev. E}\ }\textbf {\bibinfo {volume} {96}},\ \bibinfo {pages} {023310}
  (\bibinfo {year} {2017})}\BibitemShut {NoStop}%
\bibitem [{\citenamefont {Zak}(1980)}]{zak1980symmetry}%
  \BibitemOpen
  \bibfield  {author} {\bibinfo {author} {\bibfnamefont {J}~\bibnamefont
  {Zak}},\ }\bibfield  {title} {\enquote {\bibinfo {title} {Symmetry
  specification of bands in solids},}\ }\href@noop {} {\bibfield  {journal}
  {\bibinfo  {journal} {Physical Review Letters}\ }\textbf {\bibinfo {volume}
  {45}},\ \bibinfo {pages} {1025} (\bibinfo {year} {1980})}\BibitemShut
  {NoStop}%
\bibitem [{\citenamefont {Zak}(1981)}]{zak1981band}%
  \BibitemOpen
  \bibfield  {author} {\bibinfo {author} {\bibfnamefont {J}~\bibnamefont
  {Zak}},\ }\bibfield  {title} {\enquote {\bibinfo {title} {Band
  representations and symmetry types of bands in solids},}\ }\href@noop {}
  {\bibfield  {journal} {\bibinfo  {journal} {Physical Review B}\ }\textbf
  {\bibinfo {volume} {23}},\ \bibinfo {pages} {2824} (\bibinfo {year}
  {1981})}\BibitemShut {NoStop}%
\bibitem [{\citenamefont {Zak}(1982)}]{zak1982band}%
  \BibitemOpen
  \bibfield  {author} {\bibinfo {author} {\bibfnamefont {J}~\bibnamefont
  {Zak}},\ }\bibfield  {title} {\enquote {\bibinfo {title} {Band
  representations of space groups},}\ }\href@noop {} {\bibfield  {journal}
  {\bibinfo  {journal} {Physical Review B}\ }\textbf {\bibinfo {volume} {26}},\
  \bibinfo {pages} {3010} (\bibinfo {year} {1982})}\BibitemShut {NoStop}%
\bibitem [{\citenamefont {Hohenberg}\ and\ \citenamefont
  {Kohn}(1964)}]{Hohenberg-PR64}%
  \BibitemOpen
  \bibfield  {author} {\bibinfo {author} {\bibfnamefont {P.}~\bibnamefont
  {Hohenberg}}\ and\ \bibinfo {author} {\bibfnamefont {W.}~\bibnamefont
  {Kohn}},\ }\bibfield  {title} {\enquote {\bibinfo {title} {Inhomogeneous
  electron gas},}\ }\href {\doibase 10.1103/PhysRev.136.B864} {\bibfield
  {journal} {\bibinfo  {journal} {Phys. Rev.}\ }\textbf {\bibinfo {volume}
  {136}},\ \bibinfo {pages} {B864--B871} (\bibinfo {year} {1964})}\BibitemShut
  {NoStop}%
\bibitem [{\citenamefont {Kohn}\ and\ \citenamefont {Sham}(1965)}]{Kohn-PR65}%
  \BibitemOpen
  \bibfield  {author} {\bibinfo {author} {\bibfnamefont {W.}~\bibnamefont
  {Kohn}}\ and\ \bibinfo {author} {\bibfnamefont {L.~J.}\ \bibnamefont
  {Sham}},\ }\bibfield  {title} {\enquote {\bibinfo {title} {Self-consistent
  equations including exchange and correlation effects},}\ }\href {\doibase
  10.1103/PhysRev.140.A1133} {\bibfield  {journal} {\bibinfo  {journal} {Phys.
  Rev.}\ }\textbf {\bibinfo {volume} {140}},\ \bibinfo {pages} {A1133--A1138}
  (\bibinfo {year} {1965})}\BibitemShut {NoStop}%
\bibitem [{\citenamefont {Kresse}\ and\ \citenamefont
  {Furthmüller}(1996)}]{vasp1}%
  \BibitemOpen
  \bibfield  {author} {\bibinfo {author} {\bibfnamefont {G.}~\bibnamefont
  {Kresse}}\ and\ \bibinfo {author} {\bibfnamefont {J.}~\bibnamefont
  {Furthmüller}},\ }\bibfield  {title} {\enquote {\bibinfo {title} {Efficiency
  of ab-initio total energy calculations for metals and semiconductors using a
  plane-wave basis set},}\ }\href {\doibase
  https://doi.org/10.1016/0927-0256(96)00008-0} {\bibfield  {journal} {\bibinfo
   {journal} {Computational Materials Science}\ }\textbf {\bibinfo {volume}
  {6}},\ \bibinfo {pages} {15 -- 50} (\bibinfo {year} {1996})}\BibitemShut
  {NoStop}%
\bibitem [{\citenamefont {Kresse}\ and\ \citenamefont
  {Hafner}(1993)}]{PhysRevB.48.13115}%
  \BibitemOpen
  \bibfield  {author} {\bibinfo {author} {\bibfnamefont {G.}~\bibnamefont
  {Kresse}}\ and\ \bibinfo {author} {\bibfnamefont {J.}~\bibnamefont
  {Hafner}},\ }\bibfield  {title} {\enquote {\bibinfo {title} {\textit{Ab
  initio} molecular dynamics for open-shell transition metals},}\ }\href
  {\doibase 10.1103/PhysRevB.48.13115} {\bibfield  {journal} {\bibinfo
  {journal} {Phys. Rev. B}\ }\textbf {\bibinfo {volume} {48}},\ \bibinfo
  {pages} {13115--13118} (\bibinfo {year} {1993})}\BibitemShut {NoStop}%
\bibitem [{\citenamefont {Kresse}\ and\ \citenamefont {Joubert}(1999)}]{paw1}%
  \BibitemOpen
  \bibfield  {author} {\bibinfo {author} {\bibfnamefont {G.}~\bibnamefont
  {Kresse}}\ and\ \bibinfo {author} {\bibfnamefont {D.}~\bibnamefont
  {Joubert}},\ }\bibfield  {title} {\enquote {\bibinfo {title} {From ultrasoft
  pseudopotentials to the projector augmented-wave method},}\ }\href {\doibase
  10.1103/PhysRevB.59.1758} {\bibfield  {journal} {\bibinfo  {journal} {Phys.
  Rev. B}\ }\textbf {\bibinfo {volume} {59}},\ \bibinfo {pages} {1758--1775}
  (\bibinfo {year} {1999})}\BibitemShut {NoStop}%
\bibitem [{\citenamefont {Perdew}\ \emph {et~al.}(1996)\citenamefont {Perdew},
  \citenamefont {Burke},\ and\ \citenamefont
  {Ernzerhof}}]{PhysRevLett.77.3865}%
  \BibitemOpen
  \bibfield  {author} {\bibinfo {author} {\bibfnamefont {John~P.}\ \bibnamefont
  {Perdew}}, \bibinfo {author} {\bibfnamefont {Kieron}\ \bibnamefont {Burke}},
  \ and\ \bibinfo {author} {\bibfnamefont {Matthias}\ \bibnamefont
  {Ernzerhof}},\ }\bibfield  {title} {\enquote {\bibinfo {title} {Generalized
  gradient approximation made simple},}\ }\href {\doibase
  10.1103/PhysRevLett.77.3865} {\bibfield  {journal} {\bibinfo  {journal}
  {Phys. Rev. Lett.}\ }\textbf {\bibinfo {volume} {77}},\ \bibinfo {pages}
  {3865--3868} (\bibinfo {year} {1996})}\BibitemShut {NoStop}%
\bibitem [{\citenamefont {Po}\ \emph {et~al.}(2017{\natexlab{b}})\citenamefont
  {Po}, \citenamefont {Vishwanath},\ and\ \citenamefont
  {Watanabe}}]{po2017symmetry}%
  \BibitemOpen
  \bibfield  {author} {\bibinfo {author} {\bibfnamefont {Hoi~Chun}\
  \bibnamefont {Po}}, \bibinfo {author} {\bibfnamefont {Ashvin}\ \bibnamefont
  {Vishwanath}}, \ and\ \bibinfo {author} {\bibfnamefont {Haruki}\ \bibnamefont
  {Watanabe}},\ }\bibfield  {title} {\enquote {\bibinfo {title} {Symmetry-based
  indicators of band topology in the 230 space groups},}\ }\href@noop {}
  {\bibfield  {journal} {\bibinfo  {journal} {Nature communications}\ }\textbf
  {\bibinfo {volume} {8}},\ \bibinfo {pages} {50} (\bibinfo {year}
  {2017}{\natexlab{b}})}\BibitemShut {NoStop}%
\bibitem [{\citenamefont {Kresse}\ and\ \citenamefont
  {Furthm\"uller}(1996)}]{VASP1996}%
  \BibitemOpen
  \bibfield  {author} {\bibinfo {author} {\bibfnamefont {G.}~\bibnamefont
  {Kresse}}\ and\ \bibinfo {author} {\bibfnamefont {J.}~\bibnamefont
  {Furthm\"uller}},\ }\bibfield  {title} {\enquote {\bibinfo {title} {Efficient
  iterative schemes for ab initio total-energy calculations using a plane-wave
  basis set},}\ }\href {\doibase 10.1103/PhysRevB.54.11169} {\bibfield
  {journal} {\bibinfo  {journal} {Phys. Rev. B}\ }\textbf {\bibinfo {volume}
  {54}},\ \bibinfo {pages} {11169--11186} (\bibinfo {year} {1996})}\BibitemShut
  {NoStop}%
\bibitem [{\citenamefont {Kim}\ \emph {et~al.}(2015)\citenamefont {Kim},
  \citenamefont {Wieder}, \citenamefont {Kane},\ and\ \citenamefont
  {Rappe}}]{Kim2015}%
  \BibitemOpen
  \bibfield  {author} {\bibinfo {author} {\bibfnamefont {Youngkuk}\
  \bibnamefont {Kim}}, \bibinfo {author} {\bibfnamefont {Benjamin~J.}\
  \bibnamefont {Wieder}}, \bibinfo {author} {\bibfnamefont {C.~L.}\
  \bibnamefont {Kane}}, \ and\ \bibinfo {author} {\bibfnamefont {Andrew~M.}\
  \bibnamefont {Rappe}},\ }\bibfield  {title} {\enquote {\bibinfo {title}
  {Dirac line nodes in inversion-symmetric crystals},}\ }\href {\doibase
  10.1103/PhysRevLett.115.036806} {\bibfield  {journal} {\bibinfo  {journal}
  {Phys. Rev. Lett.}\ }\textbf {\bibinfo {volume} {115}},\ \bibinfo {pages}
  {036806} (\bibinfo {year} {2015})}\BibitemShut {NoStop}%
\bibitem [{\citenamefont {Fang}\ \emph
  {et~al.}(2015{\natexlab{b}})\citenamefont {Fang}, \citenamefont {Chen},
  \citenamefont {Kee},\ and\ \citenamefont {Fu}}]{Chen2015}%
  \BibitemOpen
  \bibfield  {author} {\bibinfo {author} {\bibfnamefont {Chen}\ \bibnamefont
  {Fang}}, \bibinfo {author} {\bibfnamefont {Yige}\ \bibnamefont {Chen}},
  \bibinfo {author} {\bibfnamefont {Hae-Young}\ \bibnamefont {Kee}}, \ and\
  \bibinfo {author} {\bibfnamefont {Liang}\ \bibnamefont {Fu}},\ }\bibfield
  {title} {\enquote {\bibinfo {title} {Topological nodal line semimetals with
  and without spin-orbital coupling},}\ }\href {\doibase
  10.1103/PhysRevB.92.081201} {\bibfield  {journal} {\bibinfo  {journal} {Phys.
  Rev. B}\ }\textbf {\bibinfo {volume} {92}},\ \bibinfo {pages} {081201}
  (\bibinfo {year} {2015}{\natexlab{b}})}\BibitemShut {NoStop}%
\bibitem [{\citenamefont {Song}\ \emph
  {et~al.}(2020{\natexlab{b}})\citenamefont {Song}, \citenamefont {Elcoro},
  \citenamefont {Xu}, \citenamefont {Regnault},\ and\ \citenamefont
  {Bernevig}}]{song_fragile_2019}%
  \BibitemOpen
  \bibfield  {author} {\bibinfo {author} {\bibfnamefont {Zhi-Da}\ \bibnamefont
  {Song}}, \bibinfo {author} {\bibfnamefont {Luis}\ \bibnamefont {Elcoro}},
  \bibinfo {author} {\bibfnamefont {Yuan-Feng}\ \bibnamefont {Xu}}, \bibinfo
  {author} {\bibfnamefont {Nicolas}\ \bibnamefont {Regnault}}, \ and\ \bibinfo
  {author} {\bibfnamefont {B.~Andrei}\ \bibnamefont {Bernevig}},\ }\bibfield
  {title} {\enquote {\bibinfo {title} {Fragile phases as affine monoids:
  Classification and material examples},}\ }\href {\doibase
  10.1103/PhysRevX.10.031001} {\bibfield  {journal} {\bibinfo  {journal} {Phys.
  Rev. X}\ }\textbf {\bibinfo {volume} {10}},\ \bibinfo {pages} {031001}
  (\bibinfo {year} {2020}{\natexlab{b}})}\BibitemShut {NoStop}%
\bibitem [{\citenamefont {Song}\ \emph {et~al.}(2019)\citenamefont {Song},
  \citenamefont {Wang}, \citenamefont {Shi}, \citenamefont {Li}, \citenamefont
  {Fang},\ and\ \citenamefont {Bernevig}}]{song2019all}%
  \BibitemOpen
  \bibfield  {author} {\bibinfo {author} {\bibfnamefont {Zhida}\ \bibnamefont
  {Song}}, \bibinfo {author} {\bibfnamefont {Zhijun}\ \bibnamefont {Wang}},
  \bibinfo {author} {\bibfnamefont {Wujun}\ \bibnamefont {Shi}}, \bibinfo
  {author} {\bibfnamefont {Gang}\ \bibnamefont {Li}}, \bibinfo {author}
  {\bibfnamefont {Chen}\ \bibnamefont {Fang}}, \ and\ \bibinfo {author}
  {\bibfnamefont {B~Andrei}\ \bibnamefont {Bernevig}},\ }\bibfield  {title}
  {\enquote {\bibinfo {title} {All magic angles in twisted bilayer graphene are
  topological},}\ }\href@noop {} {\bibfield  {journal} {\bibinfo  {journal}
  {Physical review letters}\ }\textbf {\bibinfo {volume} {123}},\ \bibinfo
  {pages} {036401} (\bibinfo {year} {2019})}\BibitemShut {NoStop}%
\bibitem [{\citenamefont {Ahn}\ \emph {et~al.}(2019)\citenamefont {Ahn},
  \citenamefont {Park},\ and\ \citenamefont {Yang}}]{ahn2019failure}%
  \BibitemOpen
  \bibfield  {author} {\bibinfo {author} {\bibfnamefont {Junyeong}\
  \bibnamefont {Ahn}}, \bibinfo {author} {\bibfnamefont {Sungjoon}\
  \bibnamefont {Park}}, \ and\ \bibinfo {author} {\bibfnamefont {Bohm-Jung}\
  \bibnamefont {Yang}},\ }\bibfield  {title} {\enquote {\bibinfo {title}
  {Failure of nielsen-ninomiya theorem and fragile topology in two-dimensional
  systems with space-time inversion symmetry: application to twisted bilayer
  graphene at magic angle},}\ }\href@noop {} {\bibfield  {journal} {\bibinfo
  {journal} {Physical Review X}\ }\textbf {\bibinfo {volume} {9}},\ \bibinfo
  {pages} {021013} (\bibinfo {year} {2019})}\BibitemShut {NoStop}%
\bibitem [{\citenamefont {Po}\ \emph {et~al.}(2019)\citenamefont {Po},
  \citenamefont {Zou}, \citenamefont {Senthil},\ and\ \citenamefont
  {Vishwanath}}]{PhysRevB.99.195455}%
  \BibitemOpen
  \bibfield  {author} {\bibinfo {author} {\bibfnamefont {Hoi~Chun}\
  \bibnamefont {Po}}, \bibinfo {author} {\bibfnamefont {Liujun}\ \bibnamefont
  {Zou}}, \bibinfo {author} {\bibfnamefont {T.}~\bibnamefont {Senthil}}, \ and\
  \bibinfo {author} {\bibfnamefont {Ashvin}\ \bibnamefont {Vishwanath}},\
  }\bibfield  {title} {\enquote {\bibinfo {title} {Faithful tight-binding
  models and fragile topology of magic-angle bilayer graphene},}\ }\href
  {\doibase 10.1103/PhysRevB.99.195455} {\bibfield  {journal} {\bibinfo
  {journal} {Phys. Rev. B}\ }\textbf {\bibinfo {volume} {99}},\ \bibinfo
  {pages} {195455} (\bibinfo {year} {2019})}\BibitemShut {NoStop}%
\bibitem [{\citenamefont {Regnault}\ \emph {et~al.}(2022)\citenamefont
  {Regnault}, \citenamefont {Xu}, \citenamefont {Li}, \citenamefont {Ma},
  \citenamefont {Jovanovic}, \citenamefont {Yazdani}, \citenamefont {Parkin},
  \citenamefont {Felser}, \citenamefont {Schoop}, \citenamefont {Ong},
  \citenamefont {Cava}, \citenamefont {Elcoro}, \citenamefont {Song},\ and\
  \citenamefont {Bernevig}}]{Regnault2022}%
  \BibitemOpen
  \bibfield  {author} {\bibinfo {author} {\bibfnamefont {Nicolas}\ \bibnamefont
  {Regnault}}, \bibinfo {author} {\bibfnamefont {Yuanfeng}\ \bibnamefont {Xu}},
  \bibinfo {author} {\bibfnamefont {Ming-Rui}\ \bibnamefont {Li}}, \bibinfo
  {author} {\bibfnamefont {Da-Shuai}\ \bibnamefont {Ma}}, \bibinfo {author}
  {\bibfnamefont {Milena}\ \bibnamefont {Jovanovic}}, \bibinfo {author}
  {\bibfnamefont {Ali}\ \bibnamefont {Yazdani}}, \bibinfo {author}
  {\bibfnamefont {Stuart S.~P.}\ \bibnamefont {Parkin}}, \bibinfo {author}
  {\bibfnamefont {Claudia}\ \bibnamefont {Felser}}, \bibinfo {author}
  {\bibfnamefont {Leslie~M.}\ \bibnamefont {Schoop}}, \bibinfo {author}
  {\bibfnamefont {N.~Phuan}\ \bibnamefont {Ong}}, \bibinfo {author}
  {\bibfnamefont {Robert~J.}\ \bibnamefont {Cava}}, \bibinfo {author}
  {\bibfnamefont {Luis}\ \bibnamefont {Elcoro}}, \bibinfo {author}
  {\bibfnamefont {Zhi-Da}\ \bibnamefont {Song}}, \ and\ \bibinfo {author}
  {\bibfnamefont {B.~Andrei}\ \bibnamefont {Bernevig}},\ }\bibfield  {title}
  {\enquote {\bibinfo {title} {Catalogue of flat-band stoichiometric
  materials},}\ }\href {\doibase 10.1038/s41586-022-04519-1} {\bibfield
  {journal} {\bibinfo  {journal} {Nature}\ }\textbf {\bibinfo {volume} {603}},\
  \bibinfo {pages} {824--828} (\bibinfo {year} {2022})}\BibitemShut {NoStop}%
\end{thebibliography}%
\end{document}